\documentclass{article}
\usepackage{amssymb}
\usepackage{amsmath,amsthm,amsfonts,amscd,mathrsfs,dsfont,diagbox,color}
\usepackage{multirow}
\usepackage{authblk}
\usepackage{lineno}
\usepackage{footmisc}
\usepackage{lipsum}
\usepackage{bm,longtable}

\evensidemargin=\oddsidemargin \textwidth=159mm \textheight=223mm
\newtheorem{lemma}{Lemma}[section]
\newtheorem{theorem}{Theorem}[section]
\newtheorem{corollary}[theorem]{Corollary}
\newtheorem{example}{Example}
\newtheorem{remark}{Remark}[section]

\newtheorem{tab}{Table}

\newcommand\blfootnote[1]{%
\begingroup
\renewcommand\thefootnote{}\footnote{#1}%
\addtocounter{footnote}{-1}%
\endgroup
}

\begin{document}

\begin{center}
{\LARGE \bf Quantum $k$-uniform states for heterogeneous systems from irredundant mixed orthogonal arrays}
\end{center}

\bigskip

\begin{center}
{\large Shanqi Pang\footnote{\label{note1}College of Mathematics and Information Science, Henan Normal University, Xinxiang, 453007, China}, Xiao Zhang\footref{note1}, Shao-Ming Fei\footnote{School of Mathematical Science, Capital Normal University, Beijing, 100048, China}, Zhu-Jun Zheng\footnote{School of Mathematics, South China University of Technology, Guangzhou, 510641, China} \footnote{Laboratory of Quantum Science and Engineering, South China University of Technology, Guangzhou, 510641, China} \blfootnote{Correspondence: Xiao Zhang, zhangxiao28176@163.com}}
\end{center}

\bigskip
Quantum multipartite entangled states play significant roles in quantum information processing. By using difference schemes and orthogonal partitions, we construct a series of infinite classes of irredundant mixed orthogonal arrays (IrMOAs) and thus provide positive answers to two open problems. The first is the extension of the method for constructing homogeneous systems from orthogonal arrays (OAs) to heterogeneous multipartite systems with different individual levels. The second is the existence of $k$-uniform states in heterogeneous quantum systems. We present explicit constructions of two and three-uniform states for arbitrary heterogeneous multipartite systems with coprime individual levels, and characterize the entangled states in heterogeneous systems  consisting of subsystems with nonprime power dimensions as well. Moreover, we obtain infinite classes of $k$-uniform states for heterogeneous multipartite systems for any $k\geq2$. The non-existence of a class of IrMOAs is also proved.


\section{Introduction}

Quantum entanglement has been used as a resource to experimentally demonstrate various modern quantum technologies. Genuinely multipartite quantum states are particularly useful in quantum information theory \cite{goy18,fei19,wangyl15,wangyl14,goy14,arn13,sco04,bou97,zhao04,rie04,roos04,lo12, jozsa03,hel12}. Recently considerable progress has been achieved in the construction and characterization of $k$-uniform states for homogeneous systems \cite{goy18,goy14,gao18,song17,huber17,wang18,zha2013,goy16,goy15,npj,feng17,limao19}. Among the constructions of $k$-uniform states for homogeneous systems, Goyeneche et al. \cite{goy14} provided a link between an irredundant orthogonal array (IrOA) and quantum $k$-uniform state and constructed two-uniform states for an arbitrary number of $N\geq 6$ qubits by using known Hadamard matrices. Based on Hamming distances of orthogonal arrays (OAs) with difference schemes and orthogonal partitions, Pang et al. \cite{npj} explicitly constructed infinite classes of $k$-uniform states for $k=2,3$. Furthermore, by using the product construction \cite{chen13}, Bush's construction, binary double-error-correcting BCH codes and expansive replacement method \cite{hss}, Pang et al. \cite{jpa} constructed infinitely classes of $k$-uniform states for $k\geq4$. In addition, $k$-uniform states can be constructed from mutually orthogonal Latin squares and Latin cubes \cite{goy15}, graph states \cite{helwig} and quantum combinatorial designs \cite{goy18}. Based on symmetric matrices and the concatenation of algebraic geometry codes, Feng et al. \cite{feng17} gave an explicit construction of $k$-uniform quantum state when $k$ tends to infinity. However, for heterogeneous systems despite some related nice results \cite{goy16, yucs07,yucs08,miya04,chen06,chenl06,wangs13}, little is known about the $k$-uniform states, especially for five-partite systems or $k\ge 3$.

A highly entangled quantum state of heterogeneous multipartite systems composed of $N>2$ parties is said to be $k$-uniform if every reduction to $k$ parties is maximally mixed \cite{goy16}. If an $r\times N$ array having $n_i$ columns with $d_i$ levels, where $i=1,2,\ldots,l$, $l$ is an integer, $N=\sum_{i=1}^ln_i$, and $d_i\neq d_j$ for $i\neq j$, satisfies all possible $k$-tuples appeared as a row equally often in any $r\times k$ submatrix, then it is a mixed OA, written as {\rm MOA}$(r,N,d_1^{n_1}d_2^{n_2}\cdots d_l^{n_l},k)$ or {\rm MOA}$(r,d_1^{n_1}d_2^{n_2}\cdots d_l^{n_l},k)$. If $l=1$, then it is called a symmetrical OA, written as {\rm OA}$(r,N,d_1,k)$ \cite{hss}. An {\rm MOA}$(r,N,d_1^{n_1}d_2^{n_2}\cdots d_l^{n_l},k)$ is said to be an IrMOA if all of its rows in any $r\times(N-k)$ subarray are different \cite{goy16}. Let $\mathbb{C}^{d}$ be a $d$-dimensional Hilbert space. An IrMOA$(r,N,d_1^{n_1}d_2^{n_2}\cdots d_l^{n_l},k)$ has been shown to lead to a $k$-uniform state which belongs to a Hilbert space $(\mathbb{C}^{d_1})^{\otimes n_1}\otimes (\mathbb{C}^{d_2})^{\otimes n_2}\otimes\cdots \otimes (\mathbb{C}^{d_l})^{\otimes n_l}$ \cite{goy16}. If
$$L=\left(\begin{array}{cccc}
a_1^1&a_2^1&\cdots&a_N^1\\
a_1^2&a_2^2&\cdots&a_N^2\\
\vdots&\vdots&\cdots&\vdots\\
a_1^r&a_2^r&\cdots&a_N^r
\end{array}\right)
$$
is an {\rm IrMOA}$(r,N,d_1^{n_1}d_2^{n_2}\cdots d_l^{n_l},k)$, then the superposition of $r$ product states,
$$|\phi_{d_1^{n_1}d_2^{n_2}\cdots d_l^{n_l}}\rangle=|a_1^1a_2^1\ldots a_N^1\rangle+|a_1^2a_2^2\ldots a_N^2\rangle+\cdots+|a_1^ra_2^r\ldots a_N^r\rangle,$$
is a $k$-uniform state of the heterogeneous system $d_1^{n_1}\times d_2^{n_2}\times\cdots \times d_l^{n_l}$.

Although the characterization of quantum $k$-uniform states in heterogeneous systems is notoriously hard, the quantum $k$-uniform states in heterogeneous play fundamental roles in quantum information processing such as quantum teleportation \cite{bou97,zhao04,rie04,roos04}, quantum key distribution \cite{lo12}, dense coding and error correcting codes \cite{arn13,sco04} and quantum computation \cite{jozsa03}. An absolutely maximally entangled (AME) state of heterogeneous system consisting $N$ subsystems requires that all the reductions to $\lfloor\frac{N}{2}\rfloor$ parties are maximally mixed \cite{goy16}. The $k$-uniform states include AME states as the special ones, which play a critical role in obtaining certain classes of multipartite protocols and have close connection to holography. As stated in \cite{npj}, the higher the uniformity of the multipartite entangled states, the more advantages they offer. Remarkably, the subsystems of more than two levels can improve the security of some quantum information protocols \cite{cerf02} and enhance the capacity of quantum channels \cite{fuji03} and the efficiency of quantum gates \cite{ralph07}. A genuinely tripartite entangled state consisting of one qubit and two qutrits had been produced experimentally \cite{malik16}. The heterogeneous systems enable one to implement quantum steering more efficiently. One may expect that multipartite entangled states of heterogeneous systems will be implemented experimentally too in quantum information processing in the near future.

These researches have motivated further studies on protecting entanglement under decoherence \cite{xiao13,xiao14} and finding $k$-uniform states with higher uniformity in heterogeneous systems. However, the theory of quantum entanglement in heterogeneous systems is far from satisfactory. In this article, we aim to solve two open problems. One is the extension of the method for constructing homogeneous systems from OAs to heterogeneous systems \cite{goy14}. The second open problem is the existence of $k$-uniform states for heterogeneous quantum systems \cite{goy16}.

The OAs have been used for designing experiments to systematically plan statistical data collection. As is often the case, OAs can be very useful for quantum information theory \cite{npj,wer01,it06,pang18is}. Recently, many new construction methods of MOAs with high strength have been provided \cite{h97,ss,zhang01,pang17,yin10,yin11,pang15,pang17amas,pang18l,pang17xu,cstm}. Such new developments in MOAs make it possible to obtain infinitely many new $k$-uniform states in heterogeneous systems from IrMOAs.

In this paper, we generalize difference scheme method and orthogonal partition method for constructing IrOAs \cite{npj} to IrMOAs. In addition, the expansive replacement method is introduced for constructing more IrMOAs. As a result, we obtain a series of infinite classes of IrMOAs and thus provide positive answers to the above two open problems. In particular, we find several infinite classes of examples of three-uniform states for heterogeneous systems. We not only present explicit constructions of two and three-uniform states for heterogeneous multipartite systems consisting of subsystems with coprime levels, but also characterize entanglement states in heterogeneous systems  consisting of subsystems with nonprime power dimensions. Moreover, we obtain infinite classes of $k$-uniform states for heterogeneous multipartite systems for every $k\geq2$. Finally, we prove that the non-existence of {\rm IrMOA}$(r,5,d_1\times d_2\times d_3\times d_4\times d_5,2)$ under certain conditions.

This paper is organized as follows. In Sect. 2 we introduce some concepts and related Lemmas. In Subsect. 3.1, by using difference schemes and orthogonal partitions, we find several infinite classes of two and three-uniform states for heterogeneous systems with coprime levels. In Subsect. 3.2, by using expansive replacement method, we construct several new infinite classes of $k$-uniform states for arbitrary heterogeneous multipartite systems consisting of subsystems with coprime and nonprime power dimensions or with different powers of a prime for any $k\geq2$. Section 3 provides positive answers to the two above-mentioned open problems and the proof of the non-existence of {\rm IrMOA}$(r,5,d_1\times d_2\times d_3\times d_4\times d_5,2)$ under certain conditions. Section 4 draws the concluding remarks. Proofs of some lemmas, theorems and corollaries are presented in Appendix A. In Appendix B, we give further examples of $k$-uniform states for heterogeneous systems. All tables are relegated to Appendix C. The IrMOAs constructed in Example \ref{56} are summarized in Supplementary information.

\section{Preliminaries}

We first introduce some notations, concepts and lemmas that will be used in this paper.
Let $A^T$ be the transpose of matrix $A$ and \textbf{(d)}$=(0,1,\ldots,d-1)^T$. Let \textbf{0}$_{r}$ and \textbf{1}$_{r}$ denote the $r\times 1$ vectors of $0s$ and $1s$, respectively. The Kronecker product $A\otimes B$ is defined in \cite{hss} and the Kronecker sum $A\oplus B$ is the Kronecker product with multiplication replaced by a binary operation on a group $G$. Let $H_n$ be a Hadamard matrix of order $n$ with elements from a finite field $F_2=\{0,1\}$. {\rm HD}$(A)$ represents all the values of the Hamming distances \cite{hss} between two distinct rows of a matrix $A$. The minimal distance of $A$, written as {\rm MD}$(A)$, is defined as the minimal value of {\rm HD}$(A)$. For simplicity, we introduce the following notation:\\
$(A_{[1, 2, \ldots, u]}, r)=\left(\begin{array}{c}
A_{1}\otimes \bm{1}_{r}\\
A_{2}\otimes \bm{1}_{r}\\
\cdots\\
A_{u}\otimes \bm{1}_{r}\\
\end{array}\right)$
and $(r, A_{[1,2, \ldots, u]})=\left(\begin{array}{c}
\bm{1}_{r}\otimes A_{1}\\
\bm{1}_{r}\otimes A_{2}\\
\cdots\\
\bm{1}_{r}\otimes A_{u}\\
\end{array}\right)$ for matrix $A_j(j=1,2,\ldots,u)$ and positive integers $u$ and $r$.

\begin{lemma}\label{hdmoa}
Suppose that $A$ is an ${\rm MOA}(r,N,d_1^{n_1}d_2^{n_2}\cdots d_l^{n_l},2)$ and that $B$ is a difference scheme $D(r,r,d)$. Then, $C=[A\oplus \bm{0}_d,B\oplus$\rm \textbf{(d)}$]$ is an ${\rm MOA}(dr,N+r,d_1^{n_1}d_2^{n_2}\cdots d_l^{n_l}d^r,2)$ and ${\rm MD}(C)={\rm min}\{r,$ ${\rm MD}(A)+r-\frac{r}{d}\}$.
\end{lemma}

\begin{lemma}\label{Mdel}
The existence of an {\rm MOA}$(r,N,d_1^{n_1}d_2^{n_2}\cdots d_l^{n_l},k)$ with minimal distance $w\geq k+1$ implies the existence of an {\rm IrMOA}$(r,N',d_1^{x_1}d_2^{x_2}\cdots d_l^{x_l},k)$ for $N-w+k+1 \leq N' \leq N$ with $0\leq x_i\leq n_i$ and $1\leq i\leq l$.
\end{lemma}

By using orthogonal partition, we present a method for constructing symmetrical OAs whose Hamming distances can be determined in Ref. \cite{npj}. The following lemma extends the result for constructing MOAs.

\begin{lemma}\label{hd}
Let $\{A_1, A_2, \ldots, A_{u}\}$ be an orthogonal partition of strength 1 of $A={\rm OA}(r', N',d',k')$, and let $\{B_1, B_2, \ldots, B_{v}\}$ be an orthogonal partition of strength 1 of $B={\rm OA}(r'', N'', d'', k'')$ for $r'=d'u,r''=d''v$, $u\leq v$, and $k',k''\geq3$. Suppose that ${\rm MD}(A)= w_1$ and ${\rm MD}(B)=w_2$. Let $h=l.c.m.\{u,v\}$. Then, the matrix $M=(\bm{1}_{\frac h u}\otimes (A_{[1, 2, \ldots, u]}, d''),\ \bm{1}_{\frac h v}\otimes (d',\ B_{[1, 2, \ldots, v]}))$ is an {\rm MOA}$(d'd''h, N'+N'', d'^{N'}d''^{N''}, 3)$, and\\
\centerline{$
{\rm MD}(M)\geq
\left\{
\begin{array}{l}
{\rm min}\{w_1+w_2,N',N''\}, \ if \ u=v, \\
{\rm min}\{N',w_2\},  \ if\ u|v, \ u<v,\\
{\rm min}\{w_1,w_2\}, \ otherwise.
\end{array}
\right.
$}
\end{lemma}

\begin{lemma} \label{tihuanff} (Expansive replacement method).
Suppose $A$ is an MOA of strength $k$ with column 1 having $d_1$ levels and that $B$ also is an MOA of strength $k$ with $d_1$ rows. After making a one-to-one mapping between the levels of column 1 in $A$ and the rows of $B$, if each level of column 1 in $A$ is replaced by the corresponding row from $B$, we can obtain an MOA of strength $k$.
\end{lemma}

\section{Quantum $k$-uniform states of heterogeneous systems}

\subsection{Uniform states from orthogonal partition and difference schemes}

By using known MOAs and MOAs constructed from orthogonal partition, difference schemes and Hamming distances, we construct abundant infinite classes of two and three-uniform states for heterogeneous multipartite systems of coprime levels and three-uniform states of the system $6^4\times2^n$ from an {\rm IrMOA}$(r,4+n,6^42^n,3)$ for $n\geq13$.

\begin{theorem}\label{232}
If a $D(r,r,2)$ and an ${\rm MOA}(r,a+b,3^a2^b,2)$ exist for $a\geq1$, $b\geq2$, then there exist an {\rm IrMOA}$(r',M+N,3^M2^N,2)$ and two-uniform states of the system $3^M\times2^N$ for $1\leq M\leq a$ and every $N\geq\frac{r}{2}+3$. In particular, we have an {\rm IrMOA}$(r',M+N,3^M2^N,2)$ and two-uniform states of the system $3^M\times 2^N$ for $1\leq M\leq 2^m3$ and $N\geq2^{m-1}3^2+3$ with $m\geq2$.
\end{theorem}

Let $A_0={\rm MOA}(12,5,3^12^4,2)$ in \cite{warren} and $B_0=D(12,12,2)$ in \cite{sloane}. Then we can obtain $A_1=[A_0\oplus \bm{0}_2,B_0\oplus\bm{(2)}]={\rm MOA}(24,17,3^12^{16},2)$ by Lemma \ref{hdmoa}. Consequently, an IrMOA$(24,1+N_1,3^12^{N_1},2)$ exists for $9\leq N_1\leq 16$ by deleting $0\leq j\leq 7$ columns from $A_1$ by Lemma \ref{Mdel}. In fact, by deleting any $j$ $(0\leq j\leq 3)$ 2-level columns in $A_1$, we obtain an IrMOA$(24,1+l,3^12^l,2)$ for $13\leq l \leq 16$. However, for $3< j\le 7$, we need to first delete all the 2-level columns in $A_0\oplus\bm{0}_2$, then delete any $j-4$ columns in $B_0\oplus\bm{(2)}$ and obtain an IrMOA$(24,1+l,3^12^l,2)$ for $9\leq l < 13$. In addition, we have an IrMOA$(48,1+N_2,3^12^{N_2},2)$ for $15\leq N_2\leq 40$ from $A_2=[A_1\oplus\bm{0}_2,B_1\oplus\bm{(2)}]={\rm MOA}(48,41,3^12^{40},2)$, where $B_1=D(24,24,2)=D(12,12,2)\oplus H_2$ and $H_2$ is a Hadamard matrix of order 2. Similarly, we can also obtain IrMOAs from $A_3,A_4$ and so on. Therefore, we can obtain an {\rm IrMOA}$(r',1+N,3^12^N,2)$ and two-uniform states of the system $3^1\times2^N$ for every $N\geq9$; especially, we can delete eight columns from $A_1$ to obtain a new {\rm IrMOA}$(24,9,3^12^8,2)$ and two-uniform state of the system $3^1\times2^8$.

Investigation of entanglement in heterogeneous systems was recently performed in several particular cases, e.g., for three-partite systems, $2\times2\times n=2^2\times n$ \cite{yucs07,yucs08} and $2\times n_1\times n_2$ \cite{miya04,chen06,chenl06}, and for four-partite systems, $2^3\times n$ \cite{wangs13}. In addition, Goyeneche et al. \cite{goy16} constructed one-uniform states and two-uniform states of the system $d^N\times p_1^1\times\cdots\times p_m^1$ for $m$ distinct primes $p_1, \ldots,p_m$ and a prime power $d$. However, in Theorem \ref{232}, by using $A_0={\rm MOA}(r,a+b,3^a2^b,2)$, $B_0=D(r,r,2)$ and $A_1=[A_0\oplus \bm{0}_2,B_0\oplus\bm{(2)}]$, we can obtain an IrMOA$(r',M+N,3^M2^N,2)$ for $M>1$ and $N>1$ which can produce two-uniform states of the system $3^M\times2^N$. These states cannot be obtained from existing methods.

It is obvious that for any given $M\geq1$, there is an $n_M$ such that an {\rm IrMOA}$(r,M+N,3^M2^N,2)$ and two-uniform states of the system $3^M\times 2^N$ exist for every $N\geq n_M$.

The smaller the number of subsystems is, the more difficult to find the state in heterogeneous systems is. The method of Theorem \ref{232} is recursive. By further analyzing and using the results in each step of Theorem \ref{232}, we can obtain more entanglement states with less subsystems.

\begin{corollary}\label{3142}
(1) There exist an {\rm IrMOA}$(r,1+N,3^12^N,2)$ and two-uniform states of the system $3^1\times2^{N}$ for every $N\geq 8$.

(2) There exist an {\rm IrMOA}$(r,2+N,3^22^N,2)$ and two-uniform states of the system $3^2\times2^{N}$ for every $N\geq 12$. There exist an {\rm IrMOA}$(r,3+N,3^32^N,2)$ and two-uniform states of the system $3^3\times2^{N}$ for every $N\geq 11$, and an {\rm IrMOA}$(r,4+N,3^42^N,2)$ and two-uniform states of the system $3^4\times2^{N}$ for every $N\geq 10$.

\end{corollary}

As an application of Corollary \ref{3142}, some two-uniform states are given in Example \ref{3132} of Appendix B.

We will extend the construction of two-uniform states of the system $3^M\times2^N$ in Theorem \ref{232} to that of the system $d^M\times2^N$ for any $d>3$.

\begin{theorem}
If a $D(r,r,2)$ and an ${\rm MOA}(r,a+b,d^a2^b,2)$ exist for $d>3$, $a\geq1$, and $b\geq2$, then there exist an {\rm IrMOA}$(r',M+N,d^M2^N,2)$ and two-uniform states of the system $d^M\times2^N$ for $1\leq M\leq a$ and every $N\geq\frac{r}{2}+3$.
\end{theorem}

Interestingly, the corresponding states obtained from the above results are not separable \cite{h4}. In fact, these states are genuinely entangled. In addition, by computing Hamming distance of known MOAs, we also construct more IrMOAs and the corresponding two-uniform states. For example, the ${\rm MOA}(28,13,7^12^{12},2)$ \cite{warren} with MD$=5$ yields an IrMOA$(28,1+n,7^12^{n},2)$ for $n=10,11,12$ by deleting any $j(j=0,1,2)$ 2-level columns of $A_1$, respectively. By deleting the last $j(j=0,1,\ldots,9)$ 2-level columns of ${\rm MOA}(60,24,5^12^{23},2)$ \cite{warren}, respectively, we can obtain IrMOA$(60,1+n,5^12^{n},2)$ for $n=14,15,\ldots,23$. Further, deleting the first 2-level column from the IrMOA$(60,15,5^12^{14},2)$ generates an IrMOA$(60,$ $14,5^12^{13},2)$. Deleting the last $j(j=0,1,\ldots,6)$ 2-level columns of ${\rm MOA}(60,20,5^13^12^{18},2)$ \cite{warren}, we can construct an IrMOA$(60,2+n,5^13^12^{n},2)$ for $n=12,13,\ldots,18$. Moreover, two-uniform states of the systems $7^1\times2^{10}$, $5^1\times2^{14}$, $5^1\times2^{13}$, $5^1\times3^1\times2^{15}$ and $5^1\times3^1\times2^{12}$ are given in Example \ref{72} of Appendix B.

Now we consider construction of three-uniform states of heterogeneous systems by Lemma \ref{hd}. From two OAs $A=D_3(18,5,3)\oplus (3)$ and $B=D_3(36\cdot2^{h_1},36\cdot2^{h_1},2)\oplus(2)$, we can obtain an IrMOA$(216\cdot2^{h_1},m+n_{h_1},3^m2^{n_{h_1}},3)$ for $4\leq m\leq 5$ and $18\cdot2^{h_1}+4\leq n_{h_1}\leq 36\cdot2^{h_1}$ with every $h_1\geq0$. From two OAs $A$ and $C=D_3(108\cdot2^{h_2},108\cdot2^{h_2},2)\oplus(2)$, we can construct an IrMOA$(648\cdot2^{h_2},m+n_{h_2}',3^m2^{n_{h_2}'},3)$ for $4\leq m\leq 5$ and $54\cdot2^{h_2}+4\leq n_{h_2}'\leq 108\cdot2^{h_2}$ with every $h_2\geq0$. Therefore, we can obtain the following result.

\begin{theorem}\label{216}
There exist an {\rm IrMOA}$(r,5+n,3^52^n,3)$, an {\rm IrMOA}$(r,4+n,3^42^n,3)$ and three-uniform states of the systems $3^5\times2^n$ and $3^4\times2^n$ for $n\geq16$.
\end{theorem}

Using Theorem \ref{216}, we construct three-uniform states $|\phi_{3^42^{22}}\rangle$, $|\phi_{3^52^{16}}\rangle$, and $|\phi_{3^42^{16}}\rangle$ in Example \ref{342} of Appendix B. The following result generalizes Theorem \ref{216}.

\begin{theorem}\label{d23}
Let $d>4$ be an odd prime power. If $D_3(4d^2,4d^2,2)$ and $D_3(12d^2,12d^2,2)$ exist, then there exist an {\rm IrMOA}$(r,m+n,d^m2^n,3)$ and three-uniform states of the system $d^m\times2^n$ for $4\leq m\leq d$, $2d^2+4 \leq n\leq 4d^2$, and $n\geq4d^2+4$.
\end{theorem}

By using Theorem \ref{d23} and Lemma \ref{hd}, three-uniform states of the system $5^m\times2^n$ can be obtained for $m=4$, $n\geq14$ and $m=5$, $n\geq13$. Let $d=5$ in Theorem \ref{d23}. We can obtain an {\rm IrMOA}$(r,m+n,5^m2^n,3)$ for $4\leq m\leq 5$, $54\leq n\leq100$ and $n\geq104$. The following IrMOAs can be obtained from Lemma \ref{hd}. From $D_3(25,5,5)$ and $D_3(200,200,2)=H_{100}\oplus H_2$ where $H_{100}$ in \cite{sloane}, we can obtain an {\rm IrMOA}$(2000,205,5^52^{200},3)$. By deleting the last 97, 98, and 99 columns, we have an {\rm IrMOA}$(2000,m+n,5^m2^n,3)$ for $4\leq m\leq 5$ and $n=101,102,103$. From $D_3(25,5,5)$ and $D_3(100,100,2)=H_{100}$ in \cite{sloane}, we can obtain an {\rm IrMOA}$(1000,105,5^52^{100},3)$. By deleting the last few 2-level columns, an {\rm IrMOA}$(1000,m+n,5^m2^n,3)$ for $m=4$, $n=14,15,\ldots,53$ and $m=5$, $n=13,14,\ldots,53$ can be constructed.

By arguments similar to those used in the proof of Theorem \ref{d23}, we can obtain an {\rm IrMOA}$(r,4+n,6^42^n,3)$ and three-uniform states of the system $6^4\times2^n$ for $n\geq13$. From $A=D_3(36,4,6)\oplus\bm{(6)}$, $B_0=D_3(36,36,2)\oplus\bm{(2)}$ and $C_0=D_3(108,108,2)\oplus\bm{(2)}$, we can obtain an {\rm IrMOA}$(r,4+n,6^42^n,3)$ for $22\leq n\leq36$ and $n\geq40$, including {\rm IrMOA}$(432,40,6^42^{36},3)$ and {\rm IrMOA}$(864,76,6^42^{72},3)$. The {\rm IrMOA}$(432,40,6^42^{36},3)$ can produce an {\rm IrMOA}$(432,4+n,6^42^n,3)$ for $13\leq n\leq 21$ and deleting the last $j(j=33,34,35)$ 2-level columns of the {\rm IrMOA}$(864,76,6^42^{72},3)$ can generate an {\rm IrMOA}$(864,4+n,6^42^n,3)$ for $n=37,38,39$.

It is difficult to construct $k$-uniform states for heterogeneous systems because of the lack of a suitable mathematical tool. By using orthogonal partition method, we can validly avoid depending on Galois fields. The results obtained give a positive answer to two open problems. On the one hand, we generalize the construction method from OAs for homogeneous systems to heterogeneous systems consisting of subsystems with coprime levels. On the other hand, we have found several infinite classes of $k$-uniform states of heterogeneous quantum systems. Now, we will further solve the two problems by using the expansive replacement method.

\subsection{Uniform states from difference schemes and the expansive replacement method}

By using expansive replacement method and difference schemes method, we obtain several new infinite classes of $k$-uniform states of heterogeneous multipartite systems for an arbitrary number of subsystems with coprime and nonprime power dimensions or with different powers of a prime.

\begin{theorem}\label{tihuan}
There is an ${\rm IrMOA}(r,N-s+\sum_{w=s+1}^N(m_w-1){u_w}+\sum_{w=1}^s\sum_{i=1}^{m_w}v_{iw},d_{s+1}^{1-u_{s+1}}d_{s+2}^{1-u_{s+2}}\cdots$ $ d_N^{1-u_N}d_{11}^{v_{11}}\cdots d_{m_11}^{v_{m_11}}\cdots d_{1s}^{v_{1s}} \cdots d_{m_ss}^{v_{m_ss}}d_{1(s+1)}^{u_{s+1}}\cdots d_{m_{s+1}(s+1)}^{u_{s+1}}\cdots d_{1N}^{u_N}\cdots d_{m_NN}^{u_N},k)$ and $k$-uniform states of the system $d_{s+1}^{1-u_{s+1}}\times d_{s+2}^{1-u_{s+2}}\times\cdots \times d_N^{1-u_N}\times d_{11}^{v_{11}}\times\cdots \times d_{m_11}^{v_{m_11}}\times\cdots \times d_{1s}^{v_{1s}} \times \cdots \times d_{m_ss}^{v_{m_ss}}\times d_{1(s+1)}^{u_{s+1}}\times \cdots \times d_{m_{s+1}(s+1)}^{u_{s+1}}\times \cdots \times d_{1N}^{u_N}\times\cdots \times d_{m_NN}^{u_N}$ for any integers $0\leq v_{1w}, \ldots, v_{m_ww}\leq 1(w=1,2,\ldots,s)$ and $0\leq u_w\leq1(w=s+1,s+2,\ldots,N)$, if there exists an ${\rm MOA}(r,N,d_1^1d_2^1\cdots d_N^1,k)$ such that the minimal Hamming distance of its $N-s$ columns subarray ${\rm MD}({\rm MOA}(r,N-s,d_{s+1}^1d_{s+2}^1\cdots d_N^1,k))\geq k+1$, and $N$ MOAs $B_{w}={\rm MOA}(d_w,m_{w},d_{1w}^1\cdots d_{m_{w}{w}}^1,k)$ for $w=1,2,\ldots,N$ such that {\rm MD}$(B_w)\ge 1$ for $w\geq s+1$ once $u_w=1$.
\end{theorem}

\begin{theorem}\label{ctihuan}
If there exists $A={\rm OA}(r,N,d,k)$ and $B_w={\rm MOA}(d,m_w,d_{1w}^1\cdots d_{m_ww}^1,k)$ for $w=1,2,\ldots,t$, then we have the followings:

(1) When {\rm MD}$(A)=k+1$ and {\rm MD}$(B_w)\ge 1$ for each $w$, there is an ${\rm IrMOA}(r,N+\sum_{j=1}^t(m_j-1){n_j},d^{N-(n_1+n_2\cdots+n_t)}d_{11}^{n_1}\cdots d_{m_11}^{n_1}\cdots d_{1t}^{n_t}\cdots d_{m_tt}^{n_t},k)$ and $k$-uniform states of the system $d^{N-(n_1+n_2\cdots+n_t)}\times d_{11}^{n_1}\times\cdots \times d_{m_11}^{n_1}\times\cdots \times d_{1t}^{n_t}\times\cdots \times d_{m_tt}^{n_t}$ for any non-negative integers $1\leq n_1+n_2\cdots+n_t\leq N$.

(2) When {\rm MD}$(${\rm OA}$(r,N-s,d,k))\ge k+1$, there is an {\rm IrMOA}$(r,N-s+\sum_{j=1}^t\sum_{i=1}^{m_j}v_{ij},d^{N-s}d_{11}^{v_{11}}$ $\cdots d_{m_11}^{v_{m_11}}\cdots d_{1t}^{v_{1t}}\cdots d_{m_tt}^{v_{m_tt}},k)$ and $k$-uniform states of the system $d^{N-s}\times d_{11}^{v_{11}}\times \cdots d_{m_11}^{v_{m_11}}\times \cdots d_{1t}^{v_{1t}}\times\cdots \times d_{m_tt}^{v_{m_tt}}$ for any non-negative integers $1\leq n_1+\cdots+n_t\leq s$ and $0\leq v_{1w}, \ldots, v_{m_ww}\leq n_w$ and $\sum_{j=1}^t\sum_{i=1}^{m_j}v_{ij}\geq1$.

(3) When {\rm MD}$(${\rm OA}$(r,N-s,d,k))\ge k+1$ and {\rm MD}$(B_w)\ge 1$ for each $w$, there is an {\rm IrMOA}$(r,N-s+\sum_{j=1}^t[(m_j-1){u_j}+\sum_{i=1}^{m_j}v_{ij}],d^{N-s-(u_1+u_2\cdots+u_t)}d_{11}^{u_1+v_{11}}\cdots d_{m_11}^{u_1+v_{m_11}}$ $\cdots d_{1t}^{u_t+v_{1t}}\cdots d_{m_tt}^{u_t+v_{m_tt}},k)$ and $k$-uniform states of the system for any non-negative integers $1\leq u_1+u_2\cdots+u_t\leq N-s$ and $1\leq n_1+n_2\cdots+n_t\leq s$, and $0\leq v_{1w}, \ldots, v_{m_ww}\leq n_w$.
\end{theorem}

From Theorem \ref{ctihuan} and IrOAs with strength two and three in Ref. \cite{npj}, we can construct two and three-uniform states of heterogeneous systems. For example, we have an {\rm IrOA}$(r_N,N,6,3)$ for $N=8$ and every $N\geq12$. By replacing $n_1$ 6-level columns by an ${\rm MOA}(6,2,3^12^1,2)$, we can obtain an {\rm IrMOA}$(r_N,N+n_1,6^{N-n_1}3^{n_1}2^{n_1},3)$ and three-uniform states of the system $6^{N-n_1}\times3^{n_1}\times2^{n_1}$ with $1\leq n_1< N$ for $N=8$ and every $N\geq12$ as follows. When $N=8$ and $n_1=1,2,\ldots,8$, we can obtain three-uniform states of the systems $6^7\times3^1\times2^1$, $6^6\times3^2\times2^2$, $\ldots$, $6^1\times3^7\times2^7$ and $3^8\times2^8$ consisting of $N'=9,10,\ldots,16$ subsystems, respectively. When $N=12$ and $n_1=1,2,\ldots,12$, we can obtain three-uniform states of the systems $6^{11}\times3^1\times2^1$, $6^{10}\times3^2\times2^2$, $\ldots$ and $3^{12}\times2^{12}$ consisting of $N'=13,14,\ldots,24$ subsystems, respectively. For every $N\geq13$ and $n_1=1,2,\ldots,N$, then $N'=N+1,N+2,\ldots,2N$. So we can obtain three-uniform states of heterogeneous systems consisting of $N'$ subsystems for every $N'\geq9$. Similarly, we can construct the two and three-uniform states of heterogeneous systems in Table \ref{23tai} (see Appendix C).

It is much more challenging to construct AME states in heterogeneous systems than in homogeneous systems because the heterogeneous systems are unruly and lack of efficient mathematical tools. Interestingly, from an {\rm IrMOA}, we can obtain an ${\rm AME}$ state sometimes. For example, in Table \ref{23tai}, the three-uniform states of seven subsystems and two-uniform states of five subsystems are AME states of heterogeneous systems. An {\rm IrMOA}$(6,3,6^13^12^1,1)$ can produce an ${\rm AME}$ state in $\mathbb{C}^6\otimes\mathbb{C}^3\otimes\mathbb{C}^2$.

To further explain Table \ref{23tai}, we give Examples \ref{43} and \ref{12} in Appendix B. The resulting two and three-uniform states consisting of $N'\leq22$ heterogeneous subsystems from Table \ref{23tai} are presented in Table \ref{23juti} in Appendix C.

The following result indicates that for every $k\geq1$, we can construct an IrMOA with non-prime power levels and corresponding $k$-uniform states of heterogeneous systems.

\begin{theorem}\label{2k}
For every $k\geq1$ and any non-negative integers $1\leq n_1+n_2\cdots+n_t\leq 2k$, there exist an {\rm IrMOA}$(d^k,d^{2k-(n_1+n_2\cdots+n_t)}d_{11}^{n_1}\cdots d_{m_11}^{n_1}\cdots d_{1t}^{n_t}\cdots d_{m_tt}^{n_t},k)$ and $k$-uniform states of the system $d^{2k-(n_1+n_2\cdots+n_t)}\times d_{11}^{n_1}\times \cdots \times d_{m_11}^{n_1}\times \cdots \times d_{1t}^{n_t}\cdots \times d_{m_tt}^{n_t}$, where $d=d_{1w}\cdots d_{m_ww}$ for $w=1,2,\ldots,t$ and $d_{11},\cdots,d_{m_11}$ are $m_1$ distinct prime powers and each $d_{u1}\geq 2k-1$.
\end{theorem}

For a given $k$, there are infinitely many IrMOAs and $k$-uniform states of heterogeneous systems, since there are infinitely many primes. We construct a large number of four-uniform states in Example \ref{56} in Appendix C to illustrate an application of Theorem \ref{2k}. AME states can be applied in designing holographic quantum codes \cite{pasta}. Very interestingly, the above four-uniform states of nine subsystems are AME states of heterogeneous systems.

Let $s=0$ and $s=1$ for $u_2=\ldots=u_N=0$ in Theorem \ref{tihuan}. Starting from a difference scheme $D(N,M,d)$, we can obtain the following theorem which allows us to obtain two-uniform states of a heterogeneous system having subsystems with a non-prime power number of levels and generalizes the result in \cite{goy16}.

\begin{theorem}\label{czgouzao}
Suppose that $D(N,M,d)$ is a difference scheme and that $B$ is an ${\rm MOA}(N,m,p_1^1p_2^1\cdots p_m^1,$ $2)$. Let $A=[A_1,A_2]=[{\bf(N)} \oplus { \bf{0}}_d,D(N,M,d)\oplus \bf{(d)}]$. Then, we have that:

(1) If {\rm MD}$(A)=3$ and {\rm MD}$(B)\ge 1$, then there is an {\rm IrMOA}$(dN,M+m,d^Mp_1^1p_2^1\cdots p_m^1,2)$ and two-uniform states of the system $d^M\times p_1^1\times p_2^1\times \cdots \times p_m^1$.

(2) If {\rm MD}$(A_2)\geq3$, then there is an {\rm IrMOA}$(dN,M+m,d^Mp_1^1p_2^1\cdots p_m^1,2)$ and two-uniform states of the system $d^M\times p_1^1\times p_2^1\times \cdots \times p_m^1$.
\end{theorem}

Especially, if $M=N$, then ${\rm HD}(D(N,M,d))=N-\frac{N}{d}$. So the {\rm OA}$(Nd,N,d,2)=D(N,N,d)\oplus{\bf (d)}$ has two Hamming distances $N$ and $N-\frac{N}{d}$ \cite{zyl09}. In Theorem \ref{czgouzao}. If $d=N$ and $d$ is a prime, we can only obtain IrOAs. Otherwise, consider the case $M=N$. Only if $d=2$ and $N=4$, we can obtain {\rm MD}$(A)=3$ and {\rm IrMOA}$(8,5,4^12^4,2)$; for the other cases, we have {\rm MD}$(A_2)\geq3$. Moreover, a difference scheme $D(d^n,d^n,d)$ exists for $n\geq 1$ \cite{hss}. Then we have the following corollary.

\begin{corollary}\label{dn}
If $d$ is a prime power and $B={\rm MOA}(d^n,m,p_1^1p_2^1\cdots p_m^1,2)$ with ${\rm MD}(B)\ge 1$, then an {\rm IrMOA}$(d^{n+1},d^n+m,d^{d^n}p_1^1p_2^1\cdots p_m^1,2)$ exists.
\end{corollary}

Tables \ref{24runs}, \ref{36runs} and \ref{72runs} (see Appendix C) provide some IrMOAs with 24, 36 and 72 rows, respectively, and corresponding two-uniform states constructed from Theorems \ref{tihuan} and \ref{czgouzao} and Corollary \ref{dn}. Especially, Table \ref{72runs} only provides the IrMOAs obtained from the MOA$(72,7,12^16^6,2)=[\bm{(12)}\oplus \bm{0}_6, D(12,6,6)\oplus\bm{(6)}]$.

\begin{remark}
Besides the IrMOAs in Tables \ref{36runs} and \ref{72runs}, other numerous types of IrMOAs of runsize $\geq36$ can be also obtained from the known MOAs in \cite{warren} by using the expansive replacement method. For instance, we can construct more IrMOAs with coprime levels and corresponding two-uniform states.
\end{remark}

The non-existence of a class of IrMOAs is discussed.

\begin{theorem}\label{ame5}
Let $d_1,d_2,d_3,d_4$ and $d_5$ be not all identical integers. If $d_i\neq d_j$, their greatest common divisor is 1. Then there is no {\rm IrMOA}$(r,5,d_1\times d_2\times d_3\times d_4\times d_5,2)$ except {\rm IrMOA}$(ab^2,5,a^1b^4,2)$ for $a<b$.
\end{theorem}

However, an ${\rm IrMOA}(18,5,2^13^4,2)$ and an {\rm IrMOA}$(r,6,3^12^5,2)$ do not exist.

\section{Conclusions and discussions}

We have presented positive answers to two open problems raised in \cite{goy14,goy16}. First, we have generalized the method for constructing homogeneous systems from IrOAs to heterogeneous systems with different individual levels from IrMOAs. Then, we have addressed the existence of a series of infinite classes of $k$-uniform states for heterogeneous quantum systems. In particular, we found several infinite classes of examples of three-uniform states for such systems. Our results can be summarized as follows.

(1) By using known MOAs and MOAs constructed from difference schemes, Hamming distances, we obtained abundant infinite classes of two-uniform states of heterogeneous multipartite systems. For example, for any given $a\geq1$, there is an $n_a$ such that two-uniform states $|\phi_{3^a2^b}\rangle$ exist for every $b\geq n_a$. For any given $M\geq1$, there is an $n_M$ such that an {\rm IrMOA}$(r,M+N,d^M2^N,2)$ and two-uniform states of the system $d^M\times2^N$ exist for any $d\geq4$ and every $N\geq n_M$ if the Hadamard conjecture holds.

(2) By using orthogonal partition, difference schemes, and Hamming distances, we obtained additional infinite classes of three-uniform states of heterogeneous systems.

(3) By using Hamming distances and the expansive replacement method, we obtained infinite classes of two and three-uniform states of heterogeneous multipartite systems for an arbitrary number of subsystems with coprime and nonprime power levels or with different powers of a prime levels.

(4) By using the constructed IrMOAs, Hamming distances, and the expansive replacement method, we obtained infinite classes of $k$-uniform states of heterogeneous multipartite systems for every $k\geq2$.

It is worth mentioning that finding OAs and MOAs with some factors having a non-prime power number of levels is more difficult than finding ones in which all of the factors have a prime power number of levels. In particular, it is more challenging to construct such irredundant OAs and MOAs. Characterizing entanglement in heterogeneous systems consisting of at least one subsystem with nonprime power dimension is more complex than doing so in homogeneous systems, as the Galois fields do not exist in nonprime power dimensions. However, our method can efficiently avoid dependence on the Galois fields to obtain a large number of infinite classes of such entanglement states.

Furthermore, our methods are effective and robust. They do not require the calculation of the rank of a tensor with three indices over any finite field that is NP-complete with respect to the dimension of the tensor \cite{goy16,hast90}. These states obtained in this paper may be useful for experimental implementations and facilitate the quantification of entanglement in some multipartite heterogeneous systems \cite{hm01}. Moreover, remarkable progress is expected to be made in the field of QECCs over mixed alphabets \cite{goy16,wangz13} by applying the results presented herein.

As stated in \cite{goy14,goy16}, many open problems remain unresolved with regard to the construction and characterization of entanglement in multipartite quantum systems, such as the problem of whether $N$-qubit states exist in which all $k$-body reduced densities are maximally mixed for $k<\lfloor\frac{N}{2}\rfloor$ in \cite{huber17}. These problems are central to quantum error correction. The results presented herein will facilitate further investigations on such related open problems. For example, we can construct IrOA$(r,N,2^m,k)$ for $k\geq4$ \cite{hss}, and obtain such states by using expansive replacement method.

Although in Table \ref{23tai} (see Appendix C), AME states of seven-partite and five-partite heterogeneous systems are obtained, the knowledge about the existence and non-existence of AME states of subsystems with coprime levels is still limited. The present results we obtained will lay a foundation for obtaining AME states for heterogeneous subsystems from IrMOAs. One can investigate the nonexistence of AME states by using the nonexistence of symmetric matrix in \cite{feng17} and IrMOAs. Our results would highlight further investigations on the properties of quantum multipartite entanglement.

\bigskip
\noindent{\bf Acknowledgments}: This work is supported by the NSF of China under
Grant Nos. 11971004 and 12075159, the Key Project of Beijing Municipal Commission of Education (Grant No. KZ201810028042), Beijing Natural Science Foundation (Z190005), Academy for Multidisciplinary Studies, Capital Normal University, Shenzhen Institute for Quantum Science and Engineering, Southern University of Science and Technology (Grant No.SIQSE202005), Key Research and Development Project of Guangdong Province (Grant No.2020B0303300001), and the Guangdong Basic and Applied Basic Research Foundation (Grant No.2020B1515310016).

\section*{Appendix A: Proofs of some lemmas, theorems and corollaries}

\ \ \ \ \ {\bf Proof of Lemma \ref{hdmoa}.}

It follows from \cite{zyl09} and \cite{pang15} that MD$(B\oplus$\textbf{(d)}$)=r-\frac{r}{d}$ and that $C$ is an MOA, respectively. Suppose that $c_1$ and $c_2$ are any two rows of $C$. Let $a_j$ and $b_j$ denote rows of $A$ and $B$, respectively, for $j=1,2$. Now, consider the following two cases.

(1) If $c_1=(a_1,b_1)$ and $c_2=(a_1,i+b_1)$ for $i=1,2,\ldots,d-1$, then HD$(c_1,c_2)=r$.

(2) If $c_1=(a_1,b_1)$ and $c_2=(a_2,i+b_2)$ for $i=0,1,\ldots,d-1$, then HD$(c_1,c_2)\geq {\rm MD}(A)+r-\frac{r}{d}$. When HD$(a_1,a_2)={\rm MD}(A)$, HD$(c_1,c_2)={\rm MD}(A)+r-\frac{r}{d}$.

Therefore, ${\rm MD}(C)={\rm min}\{r,{\rm MD}(A)+r-\frac{r}{d}\}$. \hfill $\blacksquare$

{\bf Proof of Lemma \ref{hd}.}

It follows from \cite{pang18l} that the matrix $M$ is the MOA desired. For the Hamming distance between any two rows of $M$, we proceed with three cases.

(1) If $u=v$, then the matrix
$M=((A_{[1, 2, \ldots, u]}, d''),\ (d',\ B_{[1, 2, \ldots, u]}))$.

If any two rows $x$ and $y$ are in $(A_{i}\otimes \bm{1}_{d''}, \bm{1}_{d'}\otimes B_{i})$ for some $i\in\{1,2,\ldots,u\}$, then there are three values $N'$, $N''$, and $N'+N''$ for ${\rm HD}(x,y)$. If $x$ is from $(A_{i}\otimes \bm{1}_{d''}, \bm{1}_{d'}\otimes B_{i})$ and $y$ is from $(A_{j}\otimes \bm{1}_{d''}, \bm{1}_{d'}\otimes B_{j})$ for $i,j\in\{1,2,\ldots,u\}$ and $i\neq j$, then ${\rm HD}(x,y)\geq w_1+w_2$. Therefore, ${\rm MD}(M)\geq {\rm min}\{w_1+w_2,N',N''\}$.

(2) If $u|v$ and $u<v$, then the matrix
$M=(\bm{1}_{\frac v u}\otimes (A_{[1, 2, \ldots, u]}, d''),\ (d',\ B_{[1, 2, \ldots, v]}))$.

If any two rows $x$ and $y$ are in $(A_{i}\otimes \bm{1}_{d''}, \bm{1}_{d'}\otimes B_{j})$ for some $(i,j)\in\{1,2,\ldots,u\}\times \{1,2,\ldots,v\}$, then ${\rm HD}(x,y)=N'$, $N''$, and $N'+N''$. If  $x$ is from $(A_{i_1}\otimes \bm{1}_{d''}, \bm{1}_{d'}\otimes B_{j_1})$ and $y$ is from $(A_{i_2}\otimes \bm{1}_{d''}, \bm{1}_{d'}\otimes B_{j_2})$ for  $j_1\neq j_2$, then ${\rm HD}(x,y)\geq w_2$ if $i_1=i_2$; otherwise, ${\rm HD}(x,y)\geq w_1+w_2$. Therefore, ${\rm MD}(M)\geq {\rm min}\{N',w_2\}$.

(3) If $u<v$ and $u\nmid v$, then the matrix $M=(\bm{1}_{\frac h u}\otimes (A_{[1, 2, \ldots, u]}, d''),\ \bm{1}_{\frac h v}\otimes (d',\ B_{[1, 2, \ldots, v]})).$

If any two rows $x$ and $y$ are in $(A_{i}\otimes \bm{1}_{d''}, \bm{1}_{d'}\otimes B_{j})$ for some $(i,j)\in \{1,2,\ldots,u\}\times \{1,2,\ldots,v\}$, then ${\rm HD}(x,y)=N'$, $N''$, and $N'+N''$. If $x$ is from $(A_{i_1}\otimes \bm{1}_{d''}, \bm{1}_{d'}\otimes B_{j_1})$ and $y$ is from $(A_{i_2}\otimes \bm{1}_{d''}, \bm{1}_{d'}\otimes B_{j_2})$ for $(i_1,j_1)\neq(i_2,j_2)$ and $i_1,i_2\in \{1,2,\ldots,u\},j_1,j_2\in \{1,2,\ldots,v\}$, then\\
\centerline{${\rm HD}(x,y)\geq
\left\{
\begin{array}{l}
w_2, \ if \ i_1=i_2, \ j_1\neq j_2, \\
w_1,  \ if \ i_1\neq i_2, \ j_1=j_2, \\
w_1+w_2, if \ i_1\neq i_2, \ j_1\neq j_2.
\end{array}
\right.
$} Therefore, ${\rm MD}(M)\geq {\rm min}\{w_1,w_2\}$. \hfill $\blacksquare$

{\bf Proof of Lemma \ref{tihuanff}.}

Suppose that $A={\rm MOA}(r,N,d_1^1d_2^1\cdots d_N^1,k)$ and $B={\rm MOA}(d_1,N_B,f_1^1f_2^1\cdots f_{N_B}^1,k)$. Then, we will prove that the resulting array $C$ is an MOA$(r,N-1+N_B,f_1^1f_2^1\cdots f_{N_B}^1d_2^1\cdots d_N^1,k)$. Consider the following two cases.

(1) If any $k$ columns are chosen from either the first $N_B$ columns or the last $N-1$ columns in $C$, then it is obvious that they constitute an MOA of strength $k$.

(2) For any $k$ columns of $C$, if their $k_1$ columns (i.e., $i_1$-th, $\ldots$, $i_{k_1}$-th) are chosen from the first $N_B$ columns and $k_2$ columns (i.e., $j_1$-th, $\ldots$, $j_{k_2}$-th) are chosen from the last $N-1$ columns, where $k_1+k_2=k$, then each $k$-tuple as a row occurs $\frac{r}{d_1d_{j_1}\cdots d_{j_{k_2}}}\cdot \frac{d_1}{f_{i_1}\cdots f_{i_{k_1}}}=\frac{r}{d_{j_1}\cdots d_{j_{k_2}}f_{i_1}\cdots f_{i_{k_1}}}$ times. The $k$ columns constitute an MOA of strength $k$.

Therefore, we complete the proof. \hfill $\blacksquare$

{\bf Proof of Theorem \ref{232}.}

Let $A_0={\rm MOA}(r,a+b,3^a2^b,2)$ and $B_0=D(r,r,2)$. Then by Lemma \ref{hdmoa}, we can obtain $A_1=[A_0\oplus \bm{0}_2,B_0\oplus\bm{(2)}]={\rm MOA}(2r,a+b+r,3^a2^{b+r},2)$ with ${\rm MD}(A_1)={\rm min}\{r,{\rm MD}(A_0)+\frac{r}{2}\}\geq\frac{r}{2}$. Consequently, we can construct an IrMOA$(2r,a+N_1,3^a2^{N_1},2)$ for $\frac{r}{2}+3\leq N_1\leq b+r$ by deleting $0\leq j\leq b+\frac{r}{2}-3$ columns from $A_1$. In fact, by deleting any $j$ $(0\leq j\leq \frac{r}{2}-3)$ 2-level columns in $A_1$, we obtain an IrMOA$(2r,a+l,3^a2^l,2)$ for $b+\frac{r}{2}+3\leq l \leq b+r$. However, for $\frac{r}{2}-3< j\le b+\frac{r}{2}-3$, we need to first delete all 2-level columns in $A_0\oplus\bm{0}_2$, then delete any $j-b$ columns in $B_0\oplus\bm{(2)}$ and obtain IrMOA$(2r,a+l,3^a2^l,2)$ for $\frac{r}{2}+3\leq l < b+\frac{r}{2}+3$.

Similarly, we have an IrMOA$(4r,a+N_2,3^a2^{N_2},2)$ for $r+3\leq N_2\leq b+3r$ from $A_2=[A_1\oplus\bm{0}_2,B_1\oplus\bm{(2)}]={\rm MOA}(4r,a+b+3r,3^a2^{b+3r},2)$, where $B_1=D(r,r,2)\oplus H_2$ and $H_2$ is a Hadamard matrix of order 2.

Let $H_{2^{n}}$ be a Hadamard matrix of order $2^{n}$ for $n\geq0$. Take $B_n=D(r,r,2)\oplus H_{2^{n}}=D(r\cdot2^{n},r\cdot2^{n},2)$. By mathematical induction, we have $A_{n+1}=[A_n\oplus\bm{0}_2,B_n\oplus\bm{(2)}]={\rm MOA}(r\cdot2^{n+1},a+b+r\cdot2^{n+1}-r,3^a2^{b+r\cdot2^{n+1}-r},2)$ and MD$(A_{n+1})\geq r\cdot2^{n-1}$. Then, by the above-mentioned similar methods, we can obtain an IrMOA$(r\cdot2^n,a+N_{n+1},3^a2^{N_{n+1}},2)$ for $r\cdot2^{n-1}+3\leq N_{n+1}\leq b+r\cdot2^{n+1}-r$.

It can easily be shown that $r\cdot2^{n}+3\leq N_{n+2} \leq b+r\cdot2^{n+2}-r$ and $b+r\cdot2^{n+1}-r-(r\cdot2^n+3)\geq-1$ for $n\geq0$ and $b\geq2$.

Therefore, we can obtain an {\rm IrMOA}$(r',M+N,3^M2^N,2)$ and two-uniform states of the system $3^M\times2^N$ for $1\leq M\leq a$ and every $N\geq\frac{r}{2}+3$.

In particular, let $m\geq2$. The $D(3,3,3)=\left(\begin{array}{c}
000\\
012\\
021\\
\end{array}\right)$ implies the existence of
a $D(2^m3,2^m3,3)$ \cite{hjg}. From Lemma \ref{hdmoa},  $D(2^m3,2^m3,2)$ and OA$(12,11,2,2)$, we have $M_m={\rm OA}(2^m3,2^m3-1,2,2)$ with MD$=2^{m-1}3$. Then, by Lemma \ref{hdmoa}, we have $A_0={\rm MOA}(2^m3^2,2^{m+1}3-1,3^{2^m3}2^{2^m3-1},2)=[M_m\oplus \bm{0}_3,D(2^m3,2^m3,3)\oplus\bm{(3)}]$ and $B_0=D(2^m3^2,2^m3^2,2)=D(36,36,2)\oplus H_{2^{m-2}}$.

Therefore, there exist an {\rm IrMOA}$(r,M+N,3^M2^N,2)$ and two-uniform states of the system $3^M\times 2^N$ for $1\leq M\leq 2^m3$ and  $N\geq2^{m-1}3^2+3$ with any $m\geq2$. \hfill $\blacksquare$

{\bf Proof of Corollary \ref{3142}.}

(1) Starting from the ${\rm MOA}(12,5,3^12^4,2)$ in \cite{warren} and the $D(12,12,2)$ in \cite{sloane}, we have an {\rm IrMOA}$(r,1+N,3^12^N,2)$ and two-uniform states of the system $3^1\times2^{N}$ for every $N\geq 9$ by Theorem \ref{232}. Moreover, in Table \ref{24runs} (see Appendix C), there exist an {\rm IrMOA}$(24,9,3^12^8,2)$ and two-uniform state of the system $3^1\times2^8$.

(2) Beginning with $A_0={\rm MOA}(36,20,3^42^{16},2)$ in \cite{warren} and $B_0=D(36,36,2)$ in \cite{sloane}, Theorem \ref{232} can produce an {\rm IrMOA}$(r,M+N_0,3^M2^{N_0},2)$ and two-uniform states of the system $3^M\times2^{N_0}$ for $2\leq M\leq4$ and every $N_0\geq 21$.

From MD$(A_0)=8$ and Lemma \ref{Mdel}, we have an IrMOA$(36,2+N_1,3^22^{N_1},2)$ for $13\leq N_1\leq16$. In particular, we can obtain an IrMOA$(36,14,3^22^{12},2)$. From the ${\rm MOA}(36,22,3^22^{20},2)$ in \cite{warren} with MD$=8$, we can obtain an IrMOA$(36,2+N_2,3^22^{N_2},2)$ for $15\leq N_2\leq20$. Therefore, there exist an {\rm IrMOA}$(r,2+N,3^22^N,2)$ and two-uniform states of the system $3^2\times2^{N}$ for every $N\geq 12$.

Similarly, we have an IrMOA$(36,3+m_1,3^32^{m_1},2)$ for $12\leq m_1\leq16$ and an IrMOA$(36,4+n_1,3^42^{n_1},2)$ for $11\leq n_1\leq16$ from $A_0$.

From Lemma \ref{hdmoa} and $A_0$ and $B_0$, we have $A_1=[A_0\oplus \bm{0}_2,B_0\oplus\bm{(2)}]={\rm MOA}(72,56,3^42^{52},2)$. By deleting the last 3-level column and the first 32, 33, 34, and 35 2-level columns from $A_1$, respectively, we can obtain an IrMOA$(72,3+m_2,3^32^{m_2},2)$ for $17\leq m_2 \leq 20$. If deleting only the first 32, 33, 34, and 35 2-level columns from $A_1$, respectively, we can obtain an IrMOA$(72,4+n_2,3^42^{n_2},2)$ for $17\leq n_2 \leq 20$.

By Lemma \ref{hdmoa}, we can obtain an ${\rm MOA}(36,23,3^{12}2^{11},2)=[{\rm OA}(12,11,2,2)\oplus\bm{0}_3,D(12,12,3)\oplus\bm{(3)}]$ with MD$=12$. Then, we have an ${\rm IrMOA}(36,14,3^l2^{14-l},2)$ for $l=3,4,\ldots,12$. In particular, we can obtain an IrMOA$(36,14,$ $3^32^{m_0},2)$ for $m_0=11$ and an IrMOA$(36,14,3^42^{n_0},2)$ for $n_0=10$.

Then, we finish the proof. \hfill $\blacksquare$

{\bf Proof of Theorem \ref{216}.}

Let $D_3(18,5,3)=(a_1^T,a_2^T,\ldots,a_{18}^T)^T=\left(\begin{array}{c}
0	1	1	1	0	0	0	0	2	2	2	2	0	1	1	2	1	0\\
0	1	1	0	2	1	1	0	2	2	0	0	2	2	0	1	2	2\\
0	1	2	1	2	1	0	2	1	0	1	2	0	0	2	1	2	1\\
0	1	2	0	0	0	1	2	1	0	0	1	1	1	1	0	0	2\\
1	0	2	2	1	2	2	0	1	2	0	2	0	1	1	1	2	2\\
\end{array}\right)^T.$
Take $D_3(36\cdot2^{h_1},36\cdot2^{h_1},2)=H_{36}\otimes H_{2^{h_1}}$ for $h_1\geq0$.

When $h_1=0$, $D_3(36,36,2)=(b_1^T,b_2^T,\ldots,b_{36}^T)^T=[\bm{0}_{36},{\rm OA}(36,35,2,2)]$, where the ${\rm OA}(36,35,2,2)$ can be found in \cite{warren}. Let $A={\rm OA}(54,5,3,3)=D_3(18,5,3)\oplus\bm{(3)}$ and $B={\rm OA}(72,36,2,3)=D_3(36,36,2)\oplus\bm{(2)}$. Then, $\{A_{i}=a_i\oplus\bm{(3)}|i=1,2,\ldots,18\}$ and $\{B_{j}=b_j\oplus\bm{(2)}|j=1,2,\ldots,36\}$ are orthogonal partitions of strength 1 of $A$ and $B$, respectively. So we can obtain $M=(\bm{1}_2\otimes (A_{[1, 2, \ldots, 18]}, 2),\ (3,\ B_{[1, 2, \ldots, 36]}))={\rm IrMOA}(216,41,3^52^{36},3)$ from Lemma \ref{hd}. As MD$(B)=18$ from \cite{npj}, deleting 1-14 2-level columns from $B$, we can obtain $B_1',\ldots,B_{14}'$, each of which has MD$\geq4$ and an orthogonal partition of strength 1. Then, from $A$ and each $B_l'$ and Lemma \ref{hd}, we can obtain an {\rm IrMOA}$(216,5+n_0,3^52^{n_0},3)$ for $n_0=22,23,\ldots,35$.

On the other hand, deleting a 3-level column from $A$, we can obtain $A'$ and its orthogonal partition of strength 1. Then, from $A'$ and each of $B_{14}',\ldots B_1',B$ and Lemma \ref{hd}, we have an {\rm IrMOA}$(216,4+n_0,3^42^{n_0},3)$ for $n_0=22,23,\ldots,36$, respectively.

In particular, from the {\rm IrMOA}$(216,27,3^52^{22},3)$ and the {\rm IrMOA}$(216,26,3^42^{22},3)$, we can further delete some columns to obtain an {\rm IrMOA}$(216,5+n_{0}^0,3^52^{n_{0}^0},3)$ and an {\rm IrMOA}$(216,4+{n_{0}^0},3^42^{n_{0}^0},3)$, respectively, for $n_{0}^0=16,17,\ldots,21$.

When $h_1\geq1$, we can obtain $M={\rm MOA}(216\cdot2^{h_1},5+36\cdot2^{h_1},3^52^{36\cdot2^{h_1}},3)$ from $A$ and $B_{h_1}=D_3(36\cdot2^{h_1},36\cdot2^{h_1},2)\oplus\bm{(2)}$ with MD$(B_{h_1})=18\cdot2^{h_1}$ by Ref. \cite{npj} and Lemma \ref{hd}. By arguments similar to the case of $h_1=0$, we can obtain an MOA$(216\cdot2^{h_1},m+n_{h_1},3^m2^{n_{h_1}},3)$ with MD$\geq{\rm min}\{m,18\cdot2^{h_1}-(36\cdot2^{h_1}-n_{h_1})\}=\{m,n_{h_1}-18\cdot2^{h_1}\}\geq4$ for $4\leq m\leq 5$ and $18\cdot2^{h_1}+4\leq n_{h_1}\leq 36\cdot2^{h_1}$. So it is an IrMOA. When $h_1=1$, then $40\leq n_1\leq 72$. In particular, by Lemma \ref{hd}, we can obtain MOA$(432,77,3^52^{72},3)$ from $A$ and $D_3(72,72,2)\oplus\bm{(2)}$, where the $D_3(72,72,2)$ is in \cite{sloane}. Deleting the last 33, 34, and 35 2-level columns, respectively, we can obtain an IrMOA$(432,m+n_{1}^0,3^m2^{n_{1}^0},3)$ for $4\leq m\leq 5$ and $n_{1}^0=37,38,39$.

Take $D_3(108\cdot2^{h_2},108\cdot2^{h_2},2)=H_{108}\oplus H_{2^{h_2}}$ for ${h_2}\geq0$. Then, we can obtain $M={\rm MOA}(648\cdot2^{h_2},5+108\cdot2^{h_2},3^52^{108\cdot2^{h_2}},3)$ from $A$ and $C_{h_2}=D_3(108\cdot2^{h_2},108\cdot2^{h_2},2)\oplus\bm{(2)}$ with MD$(C_{h_2})=54\cdot2^{h_2}$ by Ref. \cite{npj} and Lemma \ref{hd}. By arguments similar to the case of $h_1=0$, we can obtain an IrMOA$(648\cdot2^{h_2},m+n_{h_2}',3^m2^{n_{h_2}'},3)$ for $4\leq m\leq 5$ and $54\cdot2^{h_2}+4\leq n_{h_2}'\leq 108\cdot2^{h_2}$.

From $n_{h_1}$ and $n_{h_2}'$, we can obtain an IrMOA$(r,m+n,3^m2^n,3)$ for $4\leq m\leq 5$ and $n\geq16$. \hfill $\blacksquare$

{\bf Proof of Theorem \ref{d23}.}

If $D_3(4d^2,4d^2,2)$ and $D_3(12d^2,12d^2,2)$ exist, then we have $D_3(4d^2,4d^2,2)\oplus H_{2^h}$ and $D_3(12d^2,12d^2,2)\oplus H_{2^h}$ for $h\geq0$, i. e. $D_3(v,v,2)$ for $v=4\cdot2^hd^2$ and $v=12\cdot2^hd^2$. By Ref. \cite{chen17}, $D_3(d^2,d,d)$ exists for an odd prime power $d>4$. By Lemma \ref{hd}, $A=D_3(d^2,d,d)\oplus$\textbf{(d)} and $B=D_3(v,v,2)\oplus\bm{(2)}$ with MD$(B)=\frac{v}{2}$, we can obtain an ${\rm MOA}(2dv,d+v,d^d2^v,3)$ for $v=4\cdot2^hd^2$ and $v=12\cdot2^hd^2$. By selectively deleting some columns from the ${\rm MOA}(2dv,d+v,d^d2^v,3)$ such that an IrMOA$(2dv,m+n,d^m2^n,3)$ exists for $4\leq m\leq d$ and $\frac{v}{2}+4\leq n\leq v$.

For $4\leq m\leq d$, we have an IrMOA$(8\cdot2^hd^3,m+n_h,d^m2^{n_h},3)$ for $2^{h+1}d^2+4\leq n_h\leq 4\cdot2^hd^2$ and an IrMOA$(24\cdot2^hd^3,m+n_h,d^m2^{n_h},3)$ for $6\cdot2^hd^2+4\leq n_h\leq 12\cdot2^hd^2$, respectively.

Then, the desired result follows. \hfill $\blacksquare$

{\bf Proof of Theorem \ref{tihuan}.}

The expansive replacement method enables us to replace any $d_w^1$ in the ${\rm MOA}(r,N,d_1^1d_2^1\cdots d_N^1,k)$ by a subarray of $B_{w}$ for $w=1,2,\ldots,s$ or $B_{w}$ for $w=s+1,s+2,\ldots,N$, since ${\rm MD}({\rm MOA}(r,N-s,d_{s+1}^1d_{s+2}^1\cdots d_N^1,k)\geq k+1$, and MD$(B_w)\ge 1$ for $w=s+1,s+2,\ldots,N$ once $u_w=1$. Then, the resulting MOAs are irredundant and the corresponding states exist. \hfill $\blacksquare$

{\bf Proof of Theorem \ref{ctihuan}.}

This proof is analogous to the proof of Theorem \ref{tihuan}. \hfill $\blacksquare$

First, we use examples to illustrate a trivial MOA since it will be used in the proof of Theorem \ref{2k}. The matrix $[{\bf (7)}\otimes {\bf 1}_8,{\bf 1}_{7}\otimes {\bf (4)}\otimes {\bf 1}_{2}, {\bf 1}_{28}\otimes {\bf (2)}]$ is a trivial MOA$(56,3,7^14^12^1,3)$. Moreover, the array $[{\bf (28)}\otimes {\bf 1}_2,{\bf 1}_{28}\otimes (2)]$ is a trivial MOA$(56,2,28^12^1,3)$. The array $[{\bf (7)}\otimes {\bf 1}_8,{\bf 1}_{7}\otimes {\bf (2)}\otimes {\bf 1}_{4}, {\bf 1}_{14}\otimes {\bf (2)}\otimes {\bf 1}_{2},{\bf 1}_{28}\otimes {\bf (2)}]$ is also a trivial MOA$(56,4,7^12^3,3)$.

{\bf Proof of Theorem \ref{2k}.}

As $d_{11},\ldots,d_{m_11}$ are $m_1$ distinct prime powers and $d_{u1}\geq 2k-1$ for $u=1,2,\ldots,m_1$, we obtain an IrOA$(d^k,2k,d,k)$ from Refs. \cite{npj,hss,bushb}. For $w=1,2,\ldots,t$, $d=d_{1w}\cdots d_{m_ww}$, take an MOA$(d,m_w,d_{1w}^1$ $\cdots d_{m_ww}^1,k)$ to be a trivial ${\rm MOA}(d,m_w,d_{1w}^1\cdots d_{m_ww}^1,m_w)$. Then, the desired result holds by Theorem \ref{ctihuan}. \hfill $\blacksquare$

{\bf Proof of Theorem \ref{ame5}.}

We consider the following two cases.

(1) Assume that an IrMOA$(r,5,a^1b^4,2)$ exists for $a>b$. Then we have $ab^2|r$. Since any two rows in its any $r\times3$ subarray are different, $r\leq b^3$. So $r<ab^2$. A contradiction.

(2) If an MOA$(r,5,a^2b^3,2)$ exists, then $a^2b^2|r$. Therefore, it is not irredundant.

Similarly, there are no IrMOA$(r,5,a^2b^2c^1,2)$, IrMOA$(r,5,a^3b^1c^1,2)$, IrMOA$(r,5,a^2b^1c^1d^1,2)$ and\\ IrMOA$(r,5,a^1b^1c^1d^1e^1,2)$. \hfill $\blacksquare$

\section*{Appendix B: Some examples}

\begin{example}\label{3132}
Two-uniform states of the systems $3^1\times2^{10}$, $3^1\times2^9$, $3^2\times2^{13}$, $3^2\times2^{12}$, $3^3\times2^{11}$ and $3^4\times2^{10}$.

$|\phi_{3^12^{10}}\rangle=
|0	0	1	1	1	1	1	1	1	1	1\rangle+
|0	1	0	0	0	0	0	0	0	0	0\rangle+
|1	0	0	1	0	1	1	1	0	0	0\rangle+
|1	1	1	0	1	0	0	0	1	1	1\rangle+
|2	0	0	0	1	0	1	1	1	0	0\rangle+\\
|2	1	1	1	0	1	0	0	0	1	1\rangle+
|2	0	1	0	0	1	0	1	1	1	0\rangle+
|2	1	0	1	1	0	1	0	0	0	1\rangle+
|0	0	0	1	0	0	1	0	1	1	1\rangle+
|0	1	1	0	1	1	0	1	0	0	0\rangle+
|1	0	0	0	1	0	0	1	0	1	1\rangle+
|1	1	1	1	0	1	1	0	1	0	0\rangle+
|1	0	0	0	0	1	0	0	1	0	1\rangle+
|1	1	1	1	1	0	1	1	0	1	0\rangle+
|2	0	1	0	0	0	1	0	0	1	0\rangle+
|2	1	0	1	1	1	0	1	1	0	1\rangle+
|0	0	1	1	0	0	0	1	0	0	1\rangle+
|0	1	0	0	1	1	1	0	1	1	0\rangle+
|1	0	1	1	1	0	0	0	1	0	0\rangle+
|1	1	0	0	0	1	1	1	0	1	1\rangle+
|2	0	0	1	1	1	0	0	0	1	0\rangle+
|2	1	1	0	0	0	1	1	1	0	1\rangle+
|0	0	1	0	1	1	1	0	0	0	1\rangle+
|0	1	0	1	0	0	0	1	1	1	0\rangle$.

$|\phi_{3^12^9}\rangle=
|0	0	1	1	1	1	1	1	1	1\rangle+
|0	1	0	0	0	0	0	0	0	0\rangle+
|1	0	0	1	0	1	1	1	0	0\rangle+
|1	1	1	0	1	0	0	0	1	1\rangle+
|2	0	0	0	1	0	1	1	1	0\rangle+
|2	1	1	1	0	1	0	0	0	1\rangle+
|2	0	1	0	0	1	0	1	1	1\rangle+
|2	1	0	1	1	0	1	0	0	0\rangle+
|0	0	0	1	0	0	1	0	1	1\rangle+
|0	1	1	0	1	1	0	1	0	0\rangle+
|1	0	0	0	1	0	0	1	0	1\rangle+
|1	1	1	1	0	1	1	0	1	0\rangle+
|1	0	0	0	0	1	0	0	1	0\rangle+
|1	1	1	1	1	0	1	1	0	1\rangle+
|2	0	1	0	0	0	1	0	0	1\rangle+
|2	1	0	1	1	1	0	1	1	0\rangle+
|0	0	1	1	0	0	0	1	0	0\rangle+
|0	1	0	0	1	1	1	0	1	1\rangle+
|1	0	1	1	1	0	0	0	1	0\rangle+
|1	1	0	0	0	1	1	1	0	1\rangle+
|2	0	0	1	1	1	0	0	0	1\rangle+
|2	1	1	0	0	0	1	1	1	0\rangle+
|0	0	1	0	1	1	1	0	0	0\rangle+
|0	1	0	1	0	0	0	1	1	1\rangle$.

$|\phi_{3^22^{13}}\rangle=
|0	10	0	0	0	0	0	1	0	0	1	0	0	1	\rangle+
|0	00	0	0	0	0	1	0	0	0	0	0	1	0	\rangle+
|0	00	0	0	1	0	0	0	1	1	1	1	0	0	\rangle+
|0	10	0	0	1	1	1	1	0	1	1	1	1	1	\rangle+\\
|1	10	0	1	0	0	1	1	1	1	0	1	1	1	\rangle+
|2	10	0	1	0	1	0	1	0	1	0	1	0	0	\rangle+
|1	00	0	1	0	1	1	0	0	1	1	1	0	0	\rangle+
|1	10	0	1	1	1	0	0	1	0	1	0	1	0	\rangle+
|0	20	0	1	1	1	1	1	1	0	0	0	1	1	\rangle+
|1	00	1	0	0	1	0	0	1	0	0	0	0	1	\rangle+
|2	20	1	0	0	1	0	1	0	1	0	0	1	0	\rangle+
|2	10	1	0	1	0	1	0	1	0	1	1	1	1	\rangle+
|2	20	1	0	1	0	1	1	1	0	0	1	0	0	\rangle+
|1	20	1	0	1	1	0	0	0	0	0	1	1	1	\rangle+
|0	20	1	1	0	0	0	1	1	0	1	1	0	0	\rangle+
|1	20	1	1	0	1	1	0	1	1	1	0	0	1	\rangle+
|2	00	1	1	1	0	0	0	0	1	1	0	1	0	\rangle+
|2	00	1	1	1	0	1	1	0	1	0	0	0	1	\rangle+
|2	01	0	0	0	1	0	1	1	0	1	1	0	1	\rangle+
|2	01	0	0	0	1	1	0	1	0	0	1	1	0	\rangle+
|1	01	0	0	1	0	0	1	1	1	0	0	1	1	\rangle+
|1	21	0	0	1	0	1	1	1	1	1	0	0	0	\rangle+
|0	21	0	0	1	1	1	0	0	1	0	0	0	0	\rangle+
|2	21	0	1	0	0	0	0	1	1	1	0	1	1	\rangle+
|1	21	0	1	0	0	0	1	0	0	0	1	1	0	\rangle+
|2	11	0	1	1	0	0	0	0	0	0	0	0	1	\rangle+
|2	21	0	1	1	1	1	0	0	0	1	1	0	1	\rangle+
|0	21	1	0	0	0	0	0	0	1	1	1	1	1	\rangle+
|1	11	1	0	0	0	1	0	0	1	0	1	0	1	\rangle+
|2	11	1	0	0	1	1	1	1	1	1	0	1	0	\rangle+
|1	11	1	0	1	1	0	1	0	0	1	0	0	0	\rangle+
|0	11	1	1	0	0	1	0	1	0	0	0	0	0	\rangle+
|0	01	1	1	0	1	1	1	0	0	1	0	1	1	\rangle+
|1	01	1	1	1	0	1	1	0	0	1	1	1	0	\rangle+
|0	11	1	1	1	1	0	0	1	1	0	1	1	0	\rangle+
|0	01	1	1	1	1	0	1	1	1	0	1	0	1	\rangle$.

$|\phi_{3^22^{12}}\rangle=
|0	10	0	0	0	0	0	1	0	0	1	0	0	\rangle+
|0	00	0	0	0	0	1	0	0	0	0	0	1	\rangle+
|0	00	0	0	1	0	0	0	1	1	1	1	0	\rangle+
|0	10	0	0	1	1	1	1	0	1	1	1	1	\rangle+\\
|1	10	0	1	0	0	1	1	1	1	0	1	1	\rangle+
|2	10	0	1	0	1	0	1	0	1	0	1	0	\rangle+
|1	00	0	1	0	1	1	0	0	1	1	1	0	\rangle+
|1	10	0	1	1	1	0	0	1	0	1	0	1	\rangle+
|0	20	0	1	1	1	1	1	1	0	0	0	1	\rangle+
|1	00	1	0	0	1	0	0	1	0	0	0	0	\rangle+
|2	20	1	0	0	1	0	1	0	1	0	0	1	\rangle+
|2	10	1	0	1	0	1	0	1	0	1	1	1	\rangle+
|2	20	1	0	1	0	1	1	1	0	0	1	0	\rangle+
|1	20	1	0	1	1	0	0	0	0	0	1	1	\rangle+
|0	20	1	1	0	0	0	1	1	0	1	1	0	\rangle+
|1	20	1	1	0	1	1	0	1	1	1	0	0	\rangle+
|2	00	1	1	1	0	0	0	0	1	1	0	1	\rangle+
|2	00	1	1	1	0	1	1	0	1	0	0	0	\rangle+
|2	01	0	0	0	1	0	1	1	0	1	1	0	\rangle+
|2	01	0	0	0	1	1	0	1	0	0	1	1	\rangle+
|1	01	0	0	1	0	0	1	1	1	0	0	1	\rangle+
|1	21	0	0	1	0	1	1	1	1	1	0	0	\rangle+
|0	21	0	0	1	1	1	0	0	1	0	0	0	\rangle+
|2	21	0	1	0	0	0	0	1	1	1	0	1	\rangle+
|1	21	0	1	0	0	0	1	0	0	0	1	1	\rangle+
|2	11	0	1	1	0	0	0	0	0	0	0	0	\rangle+
|2	21	0	1	1	1	1	0	0	0	1	1	0	\rangle+
|0	21	1	0	0	0	0	0	0	1	1	1	1	\rangle+
|1	11	1	0	0	0	1	0	0	1	0	1	0	\rangle+
|2	11	1	0	0	1	1	1	1	1	1	0	1	\rangle+
|1	11	1	0	1	1	0	1	0	0	1	0	0	\rangle+
|0	11	1	1	0	0	1	0	1	0	0	0	0	\rangle+
|0	01	1	1	0	1	1	1	0	0	1	0	1	\rangle+
|1	01	1	1	1	0	1	1	0	0	1	1	1	\rangle+
|0	11	1	1	1	1	0	0	1	1	0	1	1	\rangle+
|0	01	1	1	1	1	0	1	1	1	0	1	0	\rangle$.

$|\phi_{3^32^{11}}\rangle=
|0	1	2	0	0	0	0	0	0	1	0	0	1	0\rangle+
|0	0	0	0	0	0	0	0	1	0	0	0	0	0\rangle+
|0	0	0	0	0	0	1	0	0	0	1	1	1	1\rangle+
|0	1	2	0	0	0	1	1	1	1	0	1	1	1\rangle+\\
|1	1	1	0	0	1	0	0	1	1	1	1	0	1\rangle+
|2	1	0	0	0	1	0	1	0	1	0	1	0	1\rangle+
|1	0	2	0	0	1	0	1	1	0	0	1	1	1\rangle+
|1	1	1	0	0	1	1	1	0	0	1	0	1	0\rangle+
|0	2	1	0	0	1	1	1	1	1	1	0	0	0\rangle+
|1	0	2	0	1	0	0	1	0	0	1	0	0	0\rangle+
|2	2	2	0	1	0	0	1	0	1	0	1	0	0\rangle+
|2	1	0	0	1	0	1	0	1	0	1	0	1	1\rangle+
|2	2	2	0	1	0	1	0	1	1	1	0	0	1\rangle+
|1	2	0	0	1	0	1	1	0	0	0	0	0	1\rangle+
|0	2	1	0	1	1	0	0	0	1	1	0	1	1\rangle+
|1	2	0	0	1	1	0	1	1	0	1	1	1	0\rangle+
|2	0	1	0	1	1	1	0	0	0	0	1	1	0\rangle+
|2	0	1	0	1	1	1	0	1	1	0	1	0	0\rangle+
|2	0	1	1	0	0	0	1	0	1	1	0	1	1\rangle+
|2	0	1	1	0	0	0	1	1	0	1	0	0	1\rangle+
|1	0	2	1	0	0	1	0	0	1	1	1	0	0\rangle+
|1	2	0	1	0	0	1	0	1	1	1	1	1	0\rangle+
|0	2	1	1	0	0	1	1	1	0	0	1	0	0\rangle+
|2	2	2	1	0	1	0	0	0	0	1	1	1	0\rangle+
|1	2	0	1	0	1	0	0	0	1	0	0	0	1\rangle+
|2	1	0	1	0	1	1	0	0	0	0	0	0	0\rangle+
|2	2	2	1	0	1	1	1	1	0	0	0	1	1\rangle+
|0	2	1	1	1	0	0	0	0	0	0	1	1	1\rangle+
|1	1	1	1	1	0	0	0	1	0	0	1	0	1\rangle+
|2	1	0	1	1	0	0	1	1	1	1	1	1	0\rangle+
|1	1	1	1	1	0	1	1	0	1	0	0	1	0\rangle+
|0	1	2	1	1	1	0	0	1	0	1	0	0	0\rangle+
|0	0	0	1	1	1	0	1	1	1	0	0	1	0\rangle+
|1	0	2	1	1	1	1	0	1	1	0	0	1	1\rangle+
|0	1	2	1	1	1	1	1	0	0	1	1	0	1\rangle+
|0	0	0	1	1	1	1	1	0	1	1	1	0	1\rangle$.

$|\phi_{3^42^{10}}\rangle=
|0	1	2	1	0	0	0	0	0	0	1	0	0	1\rangle+
|0	0	0	0	0	0	0	0	0	1	0	0	0	0\rangle+
|0	0	0	0	0	0	0	1	0	0	0	1	1	1\rangle+
|0	1	2	1	0	0	0	1	1	1	1	0	1	1\rangle+\\
|1	1	1	0	0	0	1	0	0	1	1	1	1	0\rangle+
|2	1	0	2	0	0	1	0	1	0	1	0	1	0\rangle+
|1	0	2	2	0	0	1	0	1	1	0	0	1	1\rangle+
|1	1	1	0	0	0	1	1	1	0	0	1	0	1\rangle+
|0	2	1	2	0	0	1	1	1	1	1	1	0	0\rangle+
|1	0	2	2	0	1	0	0	1	0	0	1	0	0\rangle+
|2	2	2	0	0	1	0	0	1	0	1	0	1	0\rangle+
|2	1	0	2	0	1	0	1	0	1	0	1	0	1\rangle+
|2	2	2	0	0	1	0	1	0	1	1	1	0	0\rangle+
|1	2	0	1	0	1	0	1	1	0	0	0	0	0\rangle+
|0	2	1	2	0	1	1	0	0	0	1	1	0	1\rangle+
|1	2	0	1	0	1	1	0	1	1	0	1	1	1\rangle+
|2	0	1	1	0	1	1	1	0	0	0	0	1	1\rangle+
|2	0	1	1	0	1	1	1	0	1	1	0	1	0\rangle+
|2	0	1	1	1	0	0	0	1	0	1	1	0	1\rangle+
|2	0	1	1	1	0	0	0	1	1	0	1	0	0\rangle+
|1	0	2	2	1	0	0	1	0	0	1	1	1	0\rangle+
|1	2	0	1	1	0	0	1	0	1	1	1	1	1\rangle+
|0	2	1	2	1	0	0	1	1	1	0	0	1	0\rangle+
|2	2	2	0	1	0	1	0	0	0	0	1	1	1\rangle+
|1	2	0	1	1	0	1	0	0	0	1	0	0	0\rangle+
|2	1	0	2	1	0	1	1	0	0	0	0	0	0\rangle+
|2	2	2	0	1	0	1	1	1	1	0	0	0	1\rangle+
|0	2	1	2	1	1	0	0	0	0	0	0	1	1\rangle+
|1	1	1	0	1	1	0	0	0	1	0	0	1	0\rangle+
|2	1	0	2	1	1	0	0	1	1	1	1	1	1\rangle+
|1	1	1	0	1	1	0	1	1	0	1	0	0	1\rangle+
|0	1	2	1	1	1	1	0	0	1	0	1	0	0\rangle+
|0	0	0	0	1	1	1	0	1	1	1	0	0	1\rangle+
|1	0	2	2	1	1	1	1	0	1	1	0	0	1\rangle+
|0	1	2	1	1	1	1	1	1	0	0	1	1	0\rangle+
|0	0	0	0	1	1	1	1	1	0	1	1	1	0\rangle$.

\end{example}

\begin{example}\label{72}
Two-uniform states of the systems $7^1\times2^{10}$, $5^1\times2^{14}$, $5^1\times2^{13}$, $5^1\times3^1\times2^{15}$, $5^1\times3^1\times2^{12}$.

$|\phi_{7^12^{10}}\rangle=
|0	0	0	0	0	0	0	0	0	0	0\rangle+
|0	0	0	0	0	0	0	1	1	1	1\rangle+
|3	0	0	0	1	1	0	1	0	0	1\rangle+
|4	0	0	0	1	1	1	0	1	1	0\rangle+
|5	0	0	1	0	0	1	1	0	1	0\rangle+\\
|6	0	0	1	0	1	1	0	1	1	1\rangle+
|2	0	0	1	1	1	1	1	0	0	1\rangle+
|1	0	1	0	0	1	1	0	0	1	1\rangle+
|5	0	1	0	1	0	1	0	1	0	0\rangle+
|6	0	1	0	1	0	1	1	1	0	1\rangle+
|2	0	1	1	0	1	0	0	1	0	0\rangle+
|4	0	1	1	0	1	0	1	0	1	0\rangle+
|3	0	1	1	1	0	0	0	0	1	1\rangle+
|1	0	1	1	1	0	0	1	1	0	0\rangle+
|3	1	0	0	0	1	1	0	1	0	0\rangle+
|2	1	0	0	1	0	1	1	0	1	0\rangle+
|1	1	0	0	1	1	0	1	1	1	0\rangle+
|1	1	0	1	0	0	1	0	0	0	1\rangle+
|5	1	0	1	0	1	0	1	1	0	1\rangle+
|6	1	0	1	1	0	0	0	0	1	0\rangle+
|4	1	0	1	1	0	0	0	1	0	1\rangle+
|2	1	1	0	0	0	0	0	1	1	1\rangle+
|4	1	1	0	0	0	1	1	0	0	1\rangle+
|6	1	1	0	0	1	0	1	0	0	0\rangle+
|5	1	1	0	1	1	0	0	0	1	1\rangle+
|3	1	1	1	0	0	1	1	1	1	0\rangle+
|0	1	1	1	1	1	1	0	0	0	0\rangle+
|0	1	1	1	1	1	1	1	1	1	1\rangle$.

$|\phi_{5^12^{14}}\rangle=
|2	0	0	0	0	0	0	0	0	0	0	0	0	0	0\rangle+
|4	0	0	0	0	0	0	0	0	0	0	1	1	1	0\rangle+
|3	0	0	0	0	0	0	0	0	0	1	1	1	0	1\rangle+
|1	0	0	0	1	0	0	1	0	1	1	1	0	0	0\rangle+\\
|0	0	0	0	1	0	0	1	0	1	1	1	1	1	1\rangle+
|4	0	0	0	1	1	1	0	1	1	0	1	0	0	0\rangle+
|2	0	0	0	1	1	1	0	1	1	0	1	1	0	1\rangle+
|3	0	0	0	1	1	1	0	1	1	1	0	1	1	0\rangle+
|0	0	0	1	0	0	1	0	1	1	1	0	0	0	0\rangle+
|1	0	0	1	0	0	1	0	1	1	1	0	1	1	1\rangle+
|1	0	0	1	0	1	1	1	0	0	0	1	0	0	0\rangle+
|0	0	0	1	0	1	1	1	0	0	0	1	1	1	1\rangle+
|3	0	0	1	1	1	0	1	1	0	0	0	0	1	1\rangle+
|4	0	0	1	1	1	0	1	1	0	1	0	1	0	0\rangle+
|2	0	0	1	1	1	0	1	1	0	1	1	0	1	1\rangle+
|4	0	1	0	0	0	1	1	1	0	0	0	1	1	1\rangle+
|2	0	1	0	0	0	1	1	1	0	1	0	0	0	1\rangle+
|3	0	1	0	0	0	1	1	1	0	1	1	0	1	0\rangle+
|0	0	1	0	0	1	0	1	1	1	0	0	0	0	1\rangle+
|1	0	1	0	0	1	0	1	1	1	0	0	1	1	0\rangle+
|0	0	1	0	1	1	1	0	0	0	1	0	0	0	1\rangle+
|1	0	1	0	1	1	1	0	0	0	1	0	1	1	0\rangle+
|2	0	1	1	0	1	0	0	0	1	0	0	1	1	1\rangle+
|3	0	1	1	0	1	0	0	0	1	1	0	0	0	1\rangle+
|4	0	1	1	0	1	0	0	0	1	1	1	0	1	0\rangle+
|1	0	1	1	1	0	0	0	1	0	0	1	0	0	1\rangle+
|0	0	1	1	1	0	0	0	1	0	0	1	1	1	0\rangle+
|4	0	1	1	1	0	1	1	0	1	0	0	0	0	0\rangle+
|3	0	1	1	1	0	1	1	0	1	0	1	1	1	0\rangle+
|2	0	1	1	1	0	1	1	0	1	1	1	1	0	1\rangle+
|1	1	0	0	0	1	0	0	1	0	1	1	0	1	0\rangle+
|0	1	0	0	0	1	0	0	1	0	1	1	1	0	1\rangle+
|4	1	0	0	0	1	1	1	0	1	0	0	0	1	1\rangle+
|2	1	0	0	0	1	1	1	0	1	1	0	1	0	0\rangle+
|3	1	0	0	0	1	1	1	0	1	1	1	0	1	1\rangle+
|0	1	0	0	1	0	1	1	1	0	0	0	0	1	0\rangle+
|1	1	0	0	1	0	1	1	1	0	0	0	1	0	1\rangle+
|2	1	0	1	0	0	0	1	1	1	0	1	0	0	0\rangle+
|3	1	0	1	0	0	0	1	1	1	0	1	1	0	1\rangle+
|4	1	0	1	0	0	0	1	1	1	1	0	1	1	0\rangle+
|3	1	0	1	1	0	1	0	0	0	0	0	1	1	1\rangle+
|4	1	0	1	1	0	1	0	0	0	1	0	0	0	1\rangle+
|2	1	0	1	1	0	1	0	0	0	1	1	0	1	0\rangle+
|0	1	0	1	1	1	0	0	0	1	0	0	0	1	0\rangle+
|1	1	0	1	1	1	0	0	0	1	0	0	1	0	1\rangle+
|1	1	1	0	0	0	1	0	0	1	0	1	0	1	1\rangle+
|0	1	1	0	0	0	1	0	0	1	0	1	1	0	0\rangle+
|2	1	1	0	1	0	0	0	1	1	0	0	0	1	1\rangle+
|3	1	1	0	1	0	0	0	1	1	1	0	1	0	0\rangle+
|4	1	1	0	1	0	0	0	1	1	1	1	0	1	1\rangle+
|3	1	1	0	1	1	0	1	0	0	0	1	0	0	0\rangle+
|4	1	1	0	1	1	0	1	0	0	0	1	1	0	1\rangle+
|2	1	1	0	1	1	0	1	0	0	1	0	1	1	0\rangle+
|0	1	1	1	0	0	0	1	0	0	1	0	0	1	1\rangle+
|1	1	1	1	0	0	0	1	0	0	1	0	1	0	0\rangle+
|3	1	1	1	0	1	1	0	1	0	0	0	0	0	0\rangle+
|2	1	1	1	0	1	1	0	1	0	0	1	1	1	0\rangle+
|4	1	1	1	0	1	1	0	1	0	1	1	1	0	1\rangle+
|1	1	1	1	1	1	1	1	1	1	1	1	0	1	1\rangle+
|0	1	1	1	1	1	1	1	1	1	1	1	1	0	0\rangle$.

$|\phi_{5^12^{13}}\rangle=
|2	0	0	0	0	0	0	0	0	0	0	0	0	0\rangle+
|4	0	0	0	0	0	0	0	0	0	1	1	1	0\rangle+
|3	0	0	0	0	0	0	0	0	1	1	1	0	1\rangle+
|1	0	0	1	0	0	1	0	1	1	1	0	0	0\rangle+\\
|0	0	0	1	0	0	1	0	1	1	1	1	1	1\rangle+
|4	0	0	1	1	1	0	1	1	0	1	0	0	0\rangle+
|2	0	0	1	1	1	0	1	1	0	1	1	0	1\rangle+
|3	0	0	1	1	1	0	1	1	1	0	1	1	0\rangle+
|0	0	1	0	0	1	0	1	1	1	0	0	0	0\rangle+
|1	0	1	0	0	1	0	1	1	1	0	1	1	1\rangle+
|1	0	1	0	1	1	1	0	0	0	1	0	0	0\rangle+
|0	0	1	0	1	1	1	0	0	0	1	1	1	1\rangle+
|3	0	1	1	1	0	1	1	0	0	0	0	1	1\rangle+
|4	0	1	1	1	0	1	1	0	1	0	1	0	0\rangle+
|2	0	1	1	1	0	1	1	0	1	1	0	1	1\rangle+
|4	1	0	0	0	1	1	1	0	0	0	1	1	1\rangle+
|2	1	0	0	0	1	1	1	0	1	0	0	0	1\rangle+
|3	1	0	0	0	1	1	1	0	1	1	0	1	0\rangle+
|0	1	0	0	1	0	1	1	1	0	0	0	0	1\rangle+
|1	1	0	0	1	0	1	1	1	0	0	1	1	0\rangle+
|0	1	0	1	1	1	0	0	0	1	0	0	0	1\rangle+
|1	1	0	1	1	1	0	0	0	1	0	1	1	0\rangle+
|2	1	1	0	1	0	0	0	1	0	0	1	1	1\rangle+
|3	1	1	0	1	0	0	0	1	1	0	0	0	1\rangle+
|4	1	1	0	1	0	0	0	1	1	1	0	1	0\rangle+
|1	1	1	1	0	0	0	1	0	0	1	0	0	1\rangle+
|0	1	1	1	0	0	0	1	0	0	1	1	1	0\rangle+
|4	1	1	1	0	1	1	0	1	0	0	0	0	0\rangle+
|3	1	1	1	0	1	1	0	1	0	1	1	1	0\rangle+
|2	1	1	1	0	1	1	0	1	1	1	1	0	1\rangle+
|1	0	0	0	1	0	0	1	0	1	1	0	1	0\rangle+
|0	0	0	0	1	0	0	1	0	1	1	1	0	1\rangle+
|4	0	0	0	1	1	1	0	1	0	0	0	1	1\rangle+
|2	0	0	0	1	1	1	0	1	1	0	1	0	0\rangle+
|3	0	0	0	1	1	1	0	1	1	1	0	1	1\rangle+
|0	0	0	1	0	1	1	1	0	0	0	0	1	0\rangle+
|1	0	0	1	0	1	1	1	0	0	0	1	0	1\rangle+
|2	0	1	0	0	0	1	1	1	0	1	0	0	0\rangle+
|3	0	1	0	0	0	1	1	1	0	1	1	0	1\rangle+
|4	0	1	0	0	0	1	1	1	1	0	1	1	0\rangle+
|3	0	1	1	0	1	0	0	0	0	0	1	1	1\rangle+
|4	0	1	1	0	1	0	0	0	1	0	0	0	1\rangle+
|2	0	1	1	0	1	0	0	0	1	1	0	1	0\rangle+
|0	0	1	1	1	0	0	0	1	0	0	0	1	0\rangle+
|1	0	1	1	1	0	0	0	1	0	0	1	0	1\rangle+
|1	1	0	0	0	1	0	0	1	0	1	0	1	1\rangle+
|0	1	0	0	0	1	0	0	1	0	1	1	0	0\rangle+
|2	1	0	1	0	0	0	1	1	0	0	0	1	1\rangle+
|3	1	0	1	0	0	0	1	1	1	0	1	0	0\rangle+
|4	1	0	1	0	0	0	1	1	1	1	0	1	1\rangle+
|3	1	0	1	1	0	1	0	0	0	1	0	0	0\rangle+
|4	1	0	1	1	0	1	0	0	0	1	1	0	1\rangle+
|2	1	0	1	1	0	1	0	0	1	0	1	1	0\rangle+
|0	1	1	0	0	0	1	0	0	1	0	0	1	1\rangle+
|1	1	1	0	0	0	1	0	0	1	0	1	0	0\rangle+
|3	1	1	0	1	1	0	1	0	0	0	0	0	0\rangle+
|2	1	1	0	1	1	0	1	0	0	1	1	1	0\rangle+
|4	1	1	0	1	1	0	1	0	1	1	1	0	1\rangle+
|1	1	1	1	1	1	1	1	1	1	1	0	1	1\rangle+
|0	1	1	1	1	1	1	1	1	1	1	1	0	0\rangle$.

$|\phi_{5^13^12^{15}}\rangle=
|4	0	0	0	0	0	0	1	0	1	0	1	0	0	0	0	0\rangle+
|3	1	0	0	0	0	0	1	1	1	1	0	0	1	0	1	1\rangle+
|4	2	0	0	0	0	1	1	0	1	1	1	0	1	1	0	1\rangle+\\
|2	1	0	0	0	1	0	0	0	1	0	0	0	1	0	1	1\rangle+
|4	1	0	0	0	1	0	0	1	1	1	0	0	0	1	0	1\rangle+
|3	1	0	0	0	1	1	0	0	0	1	1	1	0	0	0	0\rangle+
|3	2	0	0	0	1	1	0	1	0	0	1	0	1	0	0	1\rangle+\\
|1	2	0	0	0	1	1	1	1	0	1	1	1	0	1	1	0\rangle+
|0	0	0	0	1	0	0	0	1	0	1	1	1	0	0	1	1\rangle+
|1	0	0	0	1	0	0	1	0	0	0	1	0	0	0	0	0\rangle+
|2	0	0	0	1	0	1	1	0	1	0	0	1	1	1	1	1\rangle+\\
|3	0	0	0	1	1	0	1	0	0	1	0	0	0	1	0	0\rangle+
|3	0	0	0	1	1	1	0	0	1	1	1	1	1	1	0	1\rangle+
|4	0	0	0	1	1	1	0	1	1	1	0	1	0	0	1	0\rangle+
|1	1	0	0	1	1	1	1	1	1	0	1	1	1	1	1	0\rangle+\\
|0	1	0	1	0	0	0	0	0	1	0	1	1	0	1	1	1\rangle+
|2	2	0	1	0	0	0	1	1	1	1	1	1	0	1	1	0\rangle+
|0	2	0	1	0	0	1	0	0	0	1	0	0	0	1	0	1\rangle+
|4	1	0	1	0	0	1	1	0	0	0	0	1	0	0	1	0\rangle+\\
|3	2	0	1	0	0	1	1	1	0	1	1	0	1	1	1	0\rangle+
|4	2	0	1	0	1	0	0	0	0	0	0	1	1	0	1	0\rangle+
|2	2	0	1	0	1	1	0	1	1	0	0	1	1	0	0	1\rangle+
|0	2	0	1	1	0	0	0	0	0	1	1	1	1	0	1	0\rangle+\\
|0	0	0	1	1	0	0	1	1	1	0	0	0	1	1	0	0\rangle+
|1	2	0	1	1	0	1	0	1	1	0	0	0	1	0	0	0\rangle+
|2	1	0	1	1	0	1	1	1	0	0	1	1	0	1	0	1\rangle+
|1	1	0	1	1	1	0	0	0	0	0	1	0	0	0	1	1\rangle+\\
|1	0	0	1	1	1	0	0	1	0	1	0	0	1	1	1	1\rangle+
|2	0	0	1	1	1	0	1	1	0	1	0	1	1	0	0	1\rangle+
|0	1	0	1	1	1	1	1	0	1	0	0	0	0	1	0	0\rangle+
|1	2	1	0	0	0	0	0	0	0	0	0	1	1	1	0	1\rangle+\\
|1	1	1	0	0	0	0	0	0	1	1	0	1	1	1	0	0\rangle+
|1	0	1	0	0	0	1	1	0	1	0	1	1	1	0	1	1\rangle+
|2	0	1	0	0	1	0	0	1	0	0	1	0	0	1	1	0\rangle+
|0	1	1	0	0	1	0	1	1	0	1	0	1	1	0	1	0\rangle+\\
|0	2	1	0	0	1	0	1	1	1	0	0	0	0	1	1	1\rangle+
|0	0	1	0	0	1	1	1	0	0	0	0	1	0	0	0	1\rangle+
|4	1	1	0	1	0	0	0	1	0	0	1	1	1	1	0	0\rangle+
|2	2	1	0	1	0	0	1	0	0	1	0	0	1	0	0	0\rangle+\\
|3	2	1	0	1	0	0	1	0	1	1	0	1	0	1	1	1\rangle+
|4	2	1	0	1	0	1	0	1	0	0	0	0	0	1	1	0\rangle+
|0	1	1	0	1	0	1	0	1	0	1	1	0	1	0	0	1\rangle+
|2	1	1	0	1	0	1	0	1	1	1	0	0	0	0	1	0\rangle+\\
|2	2	1	0	1	1	1	0	0	0	0	1	0	0	1	1	1\rangle+
|0	2	1	0	1	1	1	1	1	1	0	1	1	1	0	0	0\rangle+
|3	0	1	1	0	0	0	0	1	1	0	1	0	1	0	1	0\rangle+
|2	0	1	1	0	0	1	0	0	1	1	1	0	0	0	0	0\rangle+\\
|3	0	1	1	0	0	1	1	1	0	0	0	1	0	0	1	1\rangle+
|1	1	1	1	0	0	1	1	1	0	1	0	0	0	0	0	1\rangle+
|2	1	1	1	0	1	0	1	0	0	1	1	1	1	1	0	0\rangle+
|4	0	1	1	0	1	0	1	1	0	0	1	0	1	1	0	1\rangle+\\
|0	0	1	1	0	1	1	0	0	1	1	1	0	1	1	1	0\rangle+
|1	0	1	1	0	1	1	0	1	1	1	0	1	0	1	0	0\rangle+
|3	1	1	1	1	0	0	0	1	1	0	1	1	0	1	0	1\rangle+
|4	0	1	1	1	0	1	0	0	0	1	0	1	1	1	1	1\rangle+\\
|3	2	1	1	1	1	0	0	0	1	0	0	1	0	0	0	0\rangle+
|1	2	1	1	1	1	0	1	0	1	1	1	0	0	0	1	1\rangle+
|4	2	1	1	1	1	0	1	1	1	1	1	1	0	0	0	1\rangle+
|3	1	1	1	1	1	1	1	0	0	0	0	0	1	1	1	0\rangle+\\
|4	1	1	1	1	1	1	1	0	1	1	1	0	1	0	1	1\rangle$.

$|\phi_{5^13^12^{12}}\rangle=
|4	0	0	0	0	0	0	1	0	1	0	1	0	0\rangle+
|3	1	0	0	0	0	0	1	1	1	1	0	0	1\rangle+
|4	2	0	0	0	0	1	1	0	1	1	1	0	1\rangle+
|2	1	0	0	0	1	0	0	0	1	0	0	0	1\rangle+\\
|4	1	0	0	0	1	0	0	1	1	1	0	0	0\rangle+
|3	1	0	0	0	1	1	0	0	0	1	1	1	0\rangle+
|3	2	0	0	0	1	1	0	1	0	0	1	0	1\rangle+
|1	2	0	0	0	1	1	1	1	0	1	1	1	0\rangle+
|0	0	0	0	1	0	0	0	1	0	1	1	1	0\rangle+
|1	0	0	0	1	0	0	1	0	0	0	1	0	0\rangle+
|2	0	0	0	1	0	1	1	0	1	0	0	1	1\rangle+
|3	0	0	0	1	1	0	1	0	0	1	0	0	0\rangle+
|3	0	0	0	1	1	1	0	0	1	1	1	1	1\rangle+
|4	0	0	0	1	1	1	0	1	1	1	0	1	0\rangle+
|1	1	0	0	1	1	1	1	1	1	0	1	1	1\rangle+
|0	1	0	1	0	0	0	0	0	1	0	1	1	0\rangle+
|2	2	0	1	0	0	0	1	1	1	1	1	1	0\rangle+
|0	2	0	1	0	0	1	0	0	0	1	0	0	0\rangle+
|4	1	0	1	0	0	1	1	0	0	0	0	1	0\rangle+
|3	2	0	1	0	0	1	1	1	0	1	1	0	1\rangle+
|4	2	0	1	0	1	0	0	0	0	0	0	1	1\rangle+
|2	2	0	1	0	1	1	0	1	1	0	0	1	1\rangle+
|0	2	0	1	1	0	0	0	0	0	1	1	1	1\rangle+
|0	0	0	1	1	0	0	1	1	1	0	0	0	1\rangle+
|1	2	0	1	1	0	1	0	1	1	0	0	0	1\rangle+
|2	1	0	1	1	0	1	1	1	0	0	1	1	0\rangle+
|1	1	0	1	1	1	0	0	0	0	0	1	0	0\rangle+
|1	0	0	1	1	1	0	0	1	0	1	0	0	1\rangle+
|2	0	0	1	1	1	0	1	1	0	1	0	1	1\rangle+
|0	1	0	1	1	1	1	1	0	1	0	0	0	0\rangle+
|1	2	1	0	0	0	0	0	0	0	0	0	1	1\rangle+
|1	1	1	0	0	0	0	0	0	1	1	0	1	1\rangle+
|1	0	1	0	0	0	1	1	0	1	0	1	1	1\rangle+
|2	0	1	0	0	1	0	0	1	0	0	1	0	0\rangle+
|0	1	1	0	0	1	0	1	1	0	1	0	1	1\rangle+
|0	2	1	0	0	1	0	1	1	1	0	0	0	0\rangle+
|0	0	1	0	0	1	1	1	0	0	0	0	1	0\rangle+
|4	1	1	0	1	0	0	0	1	0	0	1	1	1\rangle+
|2	2	1	0	1	0	0	1	0	0	1	0	0	1\rangle+
|3	2	1	0	1	0	0	1	0	1	1	0	1	0\rangle+
|4	2	1	0	1	0	1	0	1	0	0	0	0	0\rangle+
|0	1	1	0	1	0	1	0	1	0	1	1	0	1\rangle+
|2	1	1	0	1	0	1	0	1	1	1	0	0	0\rangle+
|2	2	1	0	1	1	1	0	0	0	0	1	0	0\rangle+
|0	2	1	0	1	1	1	1	1	1	0	1	1	1\rangle+
|3	0	1	1	0	0	0	0	1	1	0	1	0	1\rangle+
|2	0	1	1	0	0	1	0	0	1	1	1	0	0\rangle+
|3	0	1	1	0	0	1	1	1	0	0	0	1	0\rangle+
|1	1	1	1	0	0	1	1	1	0	1	0	0	0\rangle+
|2	1	1	1	0	1	0	1	0	0	1	1	1	1\rangle+
|4	0	1	1	0	1	0	1	1	0	0	1	0	1\rangle+
|0	0	1	1	0	1	1	0	0	1	1	1	0	1\rangle+
|1	0	1	1	0	1	1	0	1	1	1	0	1	0\rangle+
|3	1	1	1	1	0	0	0	1	1	0	1	1	0\rangle+
|4	0	1	1	1	0	1	0	0	0	1	0	1	1\rangle+
|3	2	1	1	1	1	0	0	0	1	0	0	1	0\rangle+
|1	2	1	1	1	1	0	1	0	1	1	1	0	0\rangle+
|4	2	1	1	1	1	0	1	1	1	1	1	1	0\rangle+
|3	1	1	1	1	1	1	1	0	0	0	0	0	1\rangle+
|4	1	1	1	1	1	1	1	0	1	1	1	0	1\rangle$.
\end{example}

\begin{example}\label{342}
Three three-uniform states obtained from Theorem \ref{216}.

$|\phi_{3^42^{22}}\rangle=
|0	0	0	1	0	0	0	0	0	0	0	0	0	0	0	0	0	0	0	0	0	0	0	0	 0	 0\rangle+
|1	1	1	2	0	0	0	0	0	0	0	0	0	0	0	0	0	0	0	0	0	0	0	0	 0	 0\rangle+\\
|2	2	2	0	0	0	0	0	0	0	0	0	0	0	0	0	0	0	0	0	0	0	0	0	 0	 0\rangle+
|0	0	0	1	1	1	1	1	1	1	1	1	1	1	1	1	1	1	1	1	1	1	1	1	 1	 1\rangle+
|1	1	1	2	1	1	1	1	1	1	1	1	1	1	1	1	1	1	1	1	1	1	1	1	 1	 1\rangle+\\
|2	2	2	0	1	1	1	1	1	1	1	1	1	1	1	1	1	1	1	1	1	1	1	1	 1	 1\rangle+
|1	1	1	0	0	0	0	0	0	1	0	0	1	1	0	0	0	0	1	0	1	1	0	1	 1	 1\rangle+
|2	2	2	1	0	0	0	0	0	1	0	0	1	1	0	0	0	0	1	0	1	1	0	1	 1	 1\rangle+\\
|0	0	0	2	0	0	0	0	0	1	0	0	1	1	0	0	0	0	1	0	1	1	0	1	 1	 1\rangle+
|1	1	1	0	1	1	1	1	1	0	1	1	0	0	1	1	1	1	0	1	0	0	1	0	 0	 0\rangle+
|2	2	2	1	1	1	1	1	1	0	1	1	0	0	1	1	1	1	0	1	0	0	1	0	 0	 0\rangle+\\
|0	0	0	2	1	1	1	1	1	0	1	1	0	0	1	1	1	1	0	1	0	0	1	0	 0	 0\rangle+
|1	2	2	2	0	0	0	0	0	1	1	1	0	0	0	0	0	1	0	1	0	1	1	1	 1	 1\rangle+
|2	0	0	0	0	0	0	0	0	1	1	1	0	0	0	0	0	1	0	1	0	1	1	1	 1	 1\rangle+\\
|0	1	1	1	0	0	0	0	0	1	1	1	0	0	0	0	0	1	0	1	0	1	1	1	 1	 1\rangle+
|1	2	2	2	1	1	1	1	1	0	0	0	1	1	1	1	1	0	1	0	1	0	0	0	 0	 0\rangle+
|2	0	0	0	1	1	1	1	1	0	0	0	1	1	1	1	1	0	1	0	1	0	0	0	 0	 0\rangle+\\
|0	1	1	1	1	1	1	1	1	0	0	0	1	1	1	1	1	0	1	0	1	0	0	0	 0	 0\rangle+
|0	1	0	2	0	0	0	0	1	1	1	0	0	0	0	0	1	0	1	1	1	1	0	0	 0	 1\rangle+
|1	2	1	0	0	0	0	0	1	1	1	0	0	0	0	0	1	0	1	1	1	1	0	0	 0	 1\rangle+\\
|2	0	2	1	0	0	0	0	1	1	1	0	0	0	0	0	1	0	1	1	1	1	0	0	 0	 1\rangle+
|0	1	0	2	1	1	1	1	0	0	0	1	1	1	1	1	0	1	0	0	0	0	1	1	 1	 0\rangle+
|1	2	1	0	1	1	1	1	0	0	0	1	1	1	1	1	0	1	0	0	0	0	1	1	 1	 0\rangle+\\
|2	0	2	1	1	1	1	1	0	0	0	1	1	1	1	1	0	1	0	0	0	0	1	1	 1	 0\rangle+
|2	2	0	1	0	0	0	1	0	1	0	1	0	1	0	1	1	1	1	0	1	0	1	1	 1	 0\rangle+
|0	0	1	2	0	0	0	1	0	1	0	1	0	1	0	1	1	1	1	0	1	0	1	1	 1	 0\rangle+\\
|1	1	2	0	0	0	0	1	0	1	0	1	0	1	0	1	1	1	1	0	1	0	1	1	 1	 0\rangle+
|2	2	0	1	1	1	1	0	1	0	1	0	1	0	1	0	0	0	0	1	0	1	0	0	 0	 1\rangle+
|0	0	1	2	1	1	1	0	1	0	1	0	1	0	1	0	0	0	0	1	0	1	0	0	 0	 1\rangle+\\
|1	1	2	0	1	1	1	0	1	0	1	0	1	0	1	0	0	0	0	1	0	1	0	0	 0	 1\rangle+
|1	1	0	2	0	0	0	1	1	0	1	1	1	0	1	1	0	0	1	0	1	0	1	0	 0	 1\rangle+
|2	2	1	0	0	0	0	1	1	0	1	1	1	0	1	1	0	0	1	0	1	0	1	0	 0	 1\rangle+\\
|0	0	2	1	0	0	0	1	1	0	1	1	1	0	1	1	0	0	1	0	1	0	1	0	 0	 1\rangle+
|1	1	0	2	1	1	1	0	0	1	0	0	0	1	0	0	1	1	0	1	0	1	0	1	 1	 0\rangle+
|2	2	1	0	1	1	1	0	0	1	0	0	0	1	0	0	1	1	0	1	0	1	0	1	 1	 0\rangle+\\
|0	0	2	1	1	1	1	0	0	1	0	0	0	1	0	0	1	1	0	1	0	1	0	1	 1	 0\rangle+
|1	0	1	2	0	0	0	1	1	0	1	1	1	1	0	0	0	0	0	1	0	0	1	1	 1	 0\rangle+
|2	1	2	0	0	0	0	1	1	0	1	1	1	1	0	0	0	0	0	1	0	0	1	1	 1	 0\rangle+\\
|0	2	0	1	0	0	0	1	1	0	1	1	1	1	0	0	0	0	0	1	0	0	1	1	 1	 0\rangle+
|1	0	1	2	1	1	1	0	0	1	0	0	0	0	1	1	1	1	1	0	1	1	0	0	 0	 1\rangle+
|2	1	2	0	1	1	1	0	0	1	0	0	0	0	1	1	1	1	1	0	1	1	0	0	 0	 1\rangle+\\
|0	2	0	1	1	1	1	0	0	1	0	0	0	0	1	1	1	1	1	0	1	1	0	0	 0	 1\rangle+
|0	2	2	0	0	0	0	1	1	0	1	1	1	1	0	1	1	1	0	0	1	1	0	0	 0	 1\rangle+
|1	0	0	1	0	0	0	1	1	0	1	1	1	1	0	1	1	1	0	0	1	1	0	0	 0	 1\rangle+\\
|2	1	1	2	0	0	0	1	1	0	1	1	1	1	0	1	1	1	0	0	1	1	0	0	 0	 1\rangle+
|0	2	2	0	1	1	1	0	0	1	0	0	0	0	1	0	0	0	1	1	0	0	1	1	 1	 0\rangle+
|1	0	0	1	1	1	1	0	0	1	0	0	0	0	1	0	0	0	1	1	0	0	1	1	 1	 0\rangle+\\
|2	1	1	2	1	1	1	0	0	1	0	0	0	0	1	0	0	0	1	1	0	0	1	1	 1	 0\rangle+
|2	1	1	1	0	0	0	1	1	1	0	0	0	0	1	1	1	1	0	1	0	0	0	0	 1	 1\rangle+
|0	2	2	2	0	0	0	1	1	1	0	0	0	0	1	1	1	1	0	1	0	0	0	0	 1	 1\rangle+\\
|1	0	0	0	0	0	0	1	1	1	0	0	0	0	1	1	1	1	0	1	0	0	0	0	 1	 1\rangle+
|2	1	1	1	1	1	1	0	0	0	1	1	1	1	0	0	0	0	1	0	1	1	1	1	 0	 0\rangle+
|0	2	2	2	1	1	1	0	0	0	1	1	1	1	0	0	0	0	1	0	1	1	1	1	 0	 0\rangle+\\
|1	0	0	0	1	1	1	0	0	0	1	1	1	1	0	0	0	0	1	0	1	1	1	1	 0	 0\rangle+
|2	0	0	2	0	0	1	0	0	0	0	0	1	1	1	1	1	0	0	0	1	1	1	1	 0	 1\rangle+
|0	1	1	0	0	0	1	0	0	0	0	0	1	1	1	1	1	0	0	0	1	1	1	1	 0	 1\rangle+\\
|1	2	2	1	0	0	1	0	0	0	0	0	1	1	1	1	1	0	0	0	1	1	1	1	 0	 1\rangle+
|2	0	0	2	1	1	0	1	1	1	1	1	0	0	0	0	0	1	1	1	0	0	0	0	 1	 0\rangle+
|0	1	1	0	1	1	0	1	1	1	1	1	0	0	0	0	0	1	1	1	0	0	0	0	 1	 0\rangle+\\
|1	2	2	1	1	1	0	1	1	1	1	1	0	0	0	0	0	1	1	1	0	0	0	0	 1	 0\rangle+
|0	1	0	0	0	0	1	0	0	0	1	1	0	0	1	1	1	0	1	0	0	1	0	0	 1	 0\rangle+
|1	2	1	1	0	0	1	0	0	0	1	1	0	0	1	1	1	0	1	0	0	1	0	0	 1	 0\rangle+\\
|2	0	2	2	0	0	1	0	0	0	1	1	0	0	1	1	1	0	1	0	0	1	0	0	 1	 0\rangle+
|0	1	0	0	1	1	0	1	1	1	0	0	1	1	0	0	0	1	0	1	1	0	1	1	 0	 1\rangle+
|1	2	1	1	1	1	0	1	1	1	0	0	1	1	0	0	0	1	0	1	1	0	1	1	 0	 1\rangle+\\
|2	0	2	2	1	1	0	1	1	1	0	0	1	1	0	0	0	1	0	1	1	0	1	1	 0	 1\rangle+
|0	2	1	2	0	0	1	0	0	1	1	1	1	0	1	1	1	1	0	1	0	0	1	1	 0	 1\rangle+
|1	0	2	0	0	0	1	0	0	1	1	1	1	0	1	1	1	1	0	1	0	0	1	1	 0	 1\rangle+\\
|2	1	0	1	0	0	1	0	0	1	1	1	1	0	1	1	1	1	0	1	0	0	1	1	 0	 1\rangle+
|0	2	1	2	1	1	0	1	1	0	0	0	0	1	0	0	0	0	1	0	1	1	0	0	 1	 0\rangle+
|1	0	2	0	1	1	0	1	1	0	0	0	0	1	0	0	0	0	1	0	1	1	0	0	 1	 0\rangle+\\
|2	1	0	1	1	1	0	1	1	0	0	0	0	1	0	0	0	0	1	0	1	1	0	0	 1	 0\rangle+
|2	0	1	0	0	0	1	0	1	0	1	0	0	1	0	1	0	1	1	1	1	0	1	0	 1	 0\rangle+
|0	1	2	1	0	0	1	0	1	0	1	0	0	1	0	1	0	1	1	1	1	0	1	0	 1	 0\rangle+\\
|1	2	0	2	0	0	1	0	1	0	1	0	0	1	0	1	0	1	1	1	1	0	1	0	 1	 0\rangle+
|2	0	1	0	1	1	0	1	0	1	0	1	1	0	1	0	1	0	0	0	0	1	0	1	 0	 1\rangle+
|0	1	2	1	1	1	0	1	0	1	0	1	1	0	1	0	1	0	0	0	0	1	0	1	 0	 1\rangle+\\
|1	2	0	2	1	1	0	1	0	1	0	1	1	0	1	0	1	0	0	0	0	1	0	1	 0	 1\rangle+
|2	0	1	1	0	0	1	0	1	1	1	0	1	1	1	0	0	1	1	0	0	1	1	1	 0	 0\rangle+
|0	1	2	2	0	0	1	0	1	1	1	0	1	1	1	0	0	1	1	0	0	1	1	1	 0	 0\rangle+\\
|1	2	0	0	0	0	1	0	1	1	1	0	1	1	1	0	0	1	1	0	0	1	1	1	 0	 0\rangle+
|2	0	1	1	1	1	0	1	0	0	0	1	0	0	0	1	1	0	0	1	1	0	0	0	 1	 1\rangle+
|0	1	2	2	1	1	0	1	0	0	0	1	0	0	0	1	1	0	0	1	1	0	0	0	 1	 1\rangle+\\
|1	2	0	0	1	1	0	1	0	0	0	1	0	0	0	1	1	0	0	1	1	0	0	0	 1	 1\rangle+
|0	2	1	1	0	0	1	1	0	0	0	1	0	0	1	0	0	1	0	1	1	1	0	1	 0	 0\rangle+
|1	0	2	2	0	0	1	1	0	0	0	1	0	0	1	0	0	1	0	1	1	1	0	1	 0	 0\rangle+\\
|2	1	0	0	0	0	1	1	0	0	0	1	0	0	1	0	0	1	0	1	1	1	0	1	 0	 0\rangle+
|0	2	1	1	1	1	0	0	1	1	1	0	1	1	0	1	1	0	1	0	0	0	1	0	 1	 1\rangle+
|1	0	2	2	1	1	0	0	1	1	1	0	1	1	0	1	1	0	1	0	0	0	1	0	 1	 1\rangle+\\
|2	1	0	0	1	1	0	0	1	1	1	0	1	1	0	1	1	0	1	0	0	0	1	0	 1	 1\rangle+
|1	1	0	1	0	0	1	1	0	1	0	1	1	1	0	1	0	0	1	1	0	0	0	0	 0	 0\rangle+
|2	2	1	2	0	0	1	1	0	1	0	1	1	1	0	1	0	0	1	1	0	0	0	0	 0	 0\rangle+\\
|0	0	2	0	0	0	1	1	0	1	0	1	1	1	0	1	0	0	1	1	0	0	0	0	 0	 0\rangle+
|1	1	0	1	1	1	0	0	1	0	1	0	0	0	1	0	1	1	0	0	1	1	1	1	 1	 1\rangle+
|2	2	1	2	1	1	0	0	1	0	1	0	0	0	1	0	1	1	0	0	1	1	1	1	 1	 1\rangle+\\
|0	0	2	0	1	1	0	0	1	0	1	0	0	0	1	0	1	1	0	0	1	1	1	1	 1	 1\rangle+
|2	2	0	2	0	0	1	1	1	0	0	0	0	0	1	0	1	0	1	0	0	0	1	1	 1	 1\rangle+
|0	0	1	0	0	0	1	1	1	0	0	0	0	0	1	0	1	0	1	0	0	0	1	1	 1	 1\rangle+\\
|1	1	2	1	0	0	1	1	1	0	0	0	0	0	1	0	1	0	1	0	0	0	1	1	 1	 1\rangle+
|2	2	0	2	1	1	0	0	0	1	1	1	1	1	0	1	0	1	0	1	1	1	0	0	 0	 0\rangle+
|0	0	1	0	1	1	0	0	0	1	1	1	1	1	0	1	0	1	0	1	1	1	0	0	 0	 0\rangle+\\
|1	1	2	1	1	1	0	0	0	1	1	1	1	1	0	1	0	1	0	1	1	1	0	0	 0	 0\rangle+
|2	1	2	2	0	0	1	1	1	1	0	0	1	1	1	0	1	1	0	1	1	1	0	0	 1	 0\rangle+
|0	2	0	0	0	0	1	1	1	1	0	0	1	1	1	0	1	1	0	1	1	1	0	0	 1	 0\rangle+\\
|1	0	1	1	0	0	1	1	1	1	0	0	1	1	1	0	1	1	0	1	1	1	0	0	 1	 0\rangle+
|2	1	2	2	1	1	0	0	0	0	1	1	0	0	0	1	0	0	1	0	0	0	1	1	 0	 1\rangle+
|0	2	0	0	1	1	0	0	0	0	1	1	0	0	0	1	0	0	1	0	0	0	1	1	 0	 1\rangle+\\
|1	0	1	1	1	1	0	0	0	0	1	1	0	0	0	1	0	0	1	0	0	0	1	1	 0	 1\rangle+
|0	0	0	1	0	1	0	0	0	0	0	1	0	1	1	1	0	0	0	1	1	1	1	0	 1	 1\rangle+
|1	1	1	2	0	1	0	0	0	0	0	1	0	1	1	1	0	0	0	1	1	1	1	0	 1	 1\rangle+\\
|2	2	2	0	0	1	0	0	0	0	0	1	0	1	1	1	0	0	0	1	1	1	1	0	 1	 1\rangle+
|0	0	0	1	1	0	1	1	1	1	1	0	1	0	0	0	1	1	1	0	0	0	0	1	 0	 0\rangle+
|1	1	1	2	1	0	1	1	1	1	1	0	1	0	0	0	1	1	1	0	0	0	0	1	 0	 0\rangle+\\
|2	2	2	0	1	0	1	1	1	1	1	0	1	0	0	0	1	1	1	0	0	0	0	1	 0	 0\rangle+
|1	1	1	0	0	1	0	0	0	1	0	1	1	0	1	0	1	1	1	0	1	0	1	0	 1	 0\rangle+
|2	2	2	1	0	1	0	0	0	1	0	1	1	0	1	0	1	1	1	0	1	0	1	0	 1	 0\rangle+\\
|0	0	0	2	0	1	0	0	0	1	0	1	1	0	1	0	1	1	1	0	1	0	1	0	 1	 0\rangle+
|1	1	1	0	1	0	1	1	1	0	1	0	0	1	0	1	0	0	0	1	0	1	0	1	 0	 1\rangle+
|2	2	2	1	1	0	1	1	1	0	1	0	0	1	0	1	0	0	0	1	0	1	0	1	 0	 1\rangle+\\
|0	0	0	2	1	0	1	1	1	0	1	0	0	1	0	1	0	0	0	1	0	1	0	1	 0	 1\rangle+
|1	2	2	2	0	1	0	0	1	0	0	0	0	1	1	1	0	1	0	0	0	0	0	1	 0	 0\rangle+
|2	0	0	0	0	1	0	0	1	0	0	0	0	1	1	1	0	1	0	0	0	0	0	1	 0	 0\rangle+\\
|0	1	1	1	0	1	0	0	1	0	0	0	0	1	1	1	0	1	0	0	0	0	0	1	 0	 0\rangle+
|1	2	2	2	1	0	1	1	0	1	1	1	1	0	0	0	1	0	1	1	1	1	1	0	 1	 1\rangle+
|2	0	0	0	1	0	1	1	0	1	1	1	1	0	0	0	1	0	1	1	1	1	1	0	 1	 1\rangle+\\
|0	1	1	1	1	0	1	1	0	1	1	1	1	0	0	0	1	0	1	1	1	1	1	0	 1	 1\rangle+
|0	1	0	2	0	1	0	0	1	1	0	0	1	0	0	1	1	0	0	1	1	0	1	1	 0	 0\rangle+
|1	2	1	0	0	1	0	0	1	1	0	0	1	0	0	1	1	0	0	1	1	0	1	1	 0	 0\rangle+\\
|2	0	2	1	0	1	0	0	1	1	0	0	1	0	0	1	1	0	0	1	1	0	1	1	 0	 0\rangle+
|0	1	0	2	1	0	1	1	0	0	1	1	0	1	1	0	0	1	1	0	0	1	0	0	 1	 1\rangle+
|1	2	1	0	1	0	1	1	0	0	1	1	0	1	1	0	0	1	1	0	0	1	0	0	 1	 1\rangle+\\
|2	0	2	1	1	0	1	1	0	0	1	1	0	1	1	0	0	1	1	0	0	1	0	0	 1	 1\rangle+
|2	2	0	1	0	1	0	0	1	1	1	1	0	1	1	1	1	0	1	1	0	1	0	1	 0	 0\rangle+
|0	0	1	2	0	1	0	0	1	1	1	1	0	1	1	1	1	0	1	1	0	1	0	1	 0	 0\rangle+\\
|1	1	2	0	0	1	0	0	1	1	1	1	0	1	1	1	1	0	1	1	0	1	0	1	 0	 0\rangle+
|2	2	0	1	1	0	1	1	0	0	0	0	1	0	0	0	0	1	0	0	1	0	1	0	 1	 1\rangle+
|0	0	1	2	1	0	1	1	0	0	0	0	1	0	0	0	0	1	0	0	1	0	1	0	 1	 1\rangle+\\
|1	1	2	0	1	0	1	1	0	0	0	0	1	0	0	0	0	1	0	0	1	0	1	0	 1	 1\rangle+
|1	1	0	2	0	1	0	1	0	0	1	0	1	0	0	1	1	1	1	0	0	1	0	1	 1	 0\rangle+
|2	2	1	0	0	1	0	1	0	0	1	0	1	0	0	1	1	1	1	0	0	1	0	1	 1	 0\rangle+\\
|0	0	2	1	0	1	0	1	0	0	1	0	1	0	0	1	1	1	1	0	0	1	0	1	 1	 0\rangle+
|1	1	0	2	1	0	1	0	1	1	0	1	0	1	1	0	0	0	0	1	1	0	1	0	 0	 1\rangle+
|2	2	1	0	1	0	1	0	1	1	0	1	0	1	1	0	0	0	0	1	1	0	1	0	 0	 1\rangle+\\
|0	0	2	1	1	0	1	0	1	1	0	1	0	1	1	0	0	0	0	1	1	0	1	0	 0	 1\rangle+
|1	0	1	2	0	1	0	1	0	0	1	0	1	0	1	0	0	1	1	1	1	0	0	1	 0	 1\rangle+
|2	1	2	0	0	1	0	1	0	0	1	0	1	0	1	0	0	1	1	1	1	0	0	1	 0	 1\rangle+\\
|0	2	0	1	0	1	0	1	0	0	1	0	1	0	1	0	0	1	1	1	1	0	0	1	 0	 1\rangle+
|1	0	1	2	1	0	1	0	1	1	0	1	0	1	0	1	1	0	0	0	0	1	1	0	 1	 0\rangle+
|2	1	2	0	1	0	1	0	1	1	0	1	0	1	0	1	1	0	0	0	0	1	1	0	 1	 0\rangle+\\
|0	2	0	1	1	0	1	0	1	1	0	1	0	1	0	1	1	0	0	0	0	1	1	0	 1	 0\rangle+
|0	2	2	0	0	1	0	1	0	0	1	0	1	1	1	0	1	0	0	1	0	1	1	0	 1	 0\rangle+
|1	0	0	1	0	1	0	1	0	0	1	0	1	1	1	0	1	0	0	1	0	1	1	0	 1	 0\rangle+\\
|2	1	1	2	0	1	0	1	0	0	1	0	1	1	1	0	1	0	0	1	0	1	1	0	 1	 0\rangle+
|0	2	2	0	1	0	1	0	1	1	0	1	0	0	0	1	0	1	1	0	1	0	0	1	 0	 1\rangle+
|1	0	0	1	1	0	1	0	1	1	0	1	0	0	0	1	0	1	1	0	1	0	0	1	 0	 1\rangle+\\
|2	1	1	2	1	0	1	0	1	1	0	1	0	0	0	1	0	1	1	0	1	0	0	1	 0	 1\rangle+
|2	1	1	1	0	1	0	1	1	1	0	1	0	1	1	0	0	1	1	0	0	1	1	0	 0	 1\rangle+
|0	2	2	2	0	1	0	1	1	1	0	1	0	1	1	0	0	1	1	0	0	1	1	0	 0	 1\rangle+\\
|1	0	0	0	0	1	0	1	1	1	0	1	0	1	1	0	0	1	1	0	0	1	1	0	 0	 1\rangle+
|2	1	1	1	1	0	1	0	0	0	1	0	1	0	0	1	1	0	0	1	1	0	0	1	 1	 0\rangle+
|0	2	2	2	1	0	1	0	0	0	1	0	1	0	0	1	1	0	0	1	1	0	0	1	 1	 0\rangle+\\
|1	0	0	0	1	0	1	0	0	0	1	0	1	0	0	1	1	0	0	1	1	0	0	1	 1	 0\rangle+
|2	0	0	2	0	1	1	0	0	0	0	1	1	1	0	0	1	1	1	1	0	0	0	0	 0	 1\rangle+
|0	1	1	0	0	1	1	0	0	0	0	1	1	1	0	0	1	1	1	1	0	0	0	0	 0	 1\rangle+\\
|1	2	2	1	0	1	1	0	0	0	0	1	1	1	0	0	1	1	1	1	0	0	0	0	 0	 1\rangle+
|2	0	0	2	1	0	0	1	1	1	1	0	0	0	1	1	0	0	0	0	1	1	1	1	 1	 0\rangle+
|0	1	1	0	1	0	0	1	1	1	1	0	0	0	1	1	0	0	0	0	1	1	1	1	 1	 0\rangle+\\
|1	2	2	1	1	0	0	1	1	1	1	0	0	0	1	1	0	0	0	0	1	1	1	1	 1	 0\rangle+
|0	1	0	0	0	1	1	0	1	0	0	0	1	0	0	1	0	1	1	1	0	1	1	0	 1	 1\rangle+
|1	2	1	1	0	1	1	0	1	0	0	0	1	0	0	1	0	1	1	1	0	1	1	0	 1	 1\rangle+\\
|2	0	2	2	0	1	1	0	1	0	0	0	1	0	0	1	0	1	1	1	0	1	1	0	 1	 1\rangle+
|0	1	0	0	1	0	0	1	0	1	1	1	0	1	1	0	1	0	0	0	1	0	0	1	 0	 0\rangle+
|1	2	1	1	1	0	0	1	0	1	1	1	0	1	1	0	1	0	0	0	1	0	0	1	 0	 0\rangle+\\
|2	0	2	2	1	0	0	1	0	1	1	1	0	1	1	0	1	0	0	0	1	0	0	1	 0	 0\rangle+
|0	2	1	2	0	1	1	0	1	0	1	1	0	1	0	0	1	1	0	0	1	0	0	1	 1	 1\rangle+
|1	0	2	0	0	1	1	0	1	0	1	1	0	1	0	0	1	1	0	0	1	0	0	1	 1	 1\rangle+\\
|2	1	0	1	0	1	1	0	1	0	1	1	0	1	0	0	1	1	0	0	1	0	0	1	 1	 1\rangle+
|0	2	1	2	1	0	0	1	0	1	0	0	1	0	1	1	0	0	1	1	0	1	1	0	 0	 0\rangle+
|1	0	2	0	1	0	0	1	0	1	0	0	1	0	1	1	0	0	1	1	0	1	1	0	 0	 0\rangle+\\
|2	1	0	1	1	0	0	1	0	1	0	0	1	0	1	1	0	0	1	1	0	1	1	0	 0	 0\rangle+
|2	0	1	0	0	1	1	0	1	1	1	1	1	0	1	0	0	0	0	0	1	0	0	0	 1	 0\rangle+
|0	1	2	1	0	1	1	0	1	1	1	1	1	0	1	0	0	0	0	0	1	0	0	0	 1	 0\rangle+\\
|1	2	0	2	0	1	1	0	1	1	1	1	1	0	1	0	0	0	0	0	1	0	0	0	 1	 0\rangle+
|2	0	1	0	1	0	0	1	0	0	0	0	0	1	0	1	1	1	1	1	0	1	1	1	 0	 1\rangle+
|0	1	2	1	1	0	0	1	0	0	0	0	0	1	0	1	1	1	1	1	0	1	1	1	 0	 1\rangle+\\
|1	2	0	2	1	0	0	1	0	0	0	0	0	1	0	1	1	1	1	1	0	1	1	1	 0	 1\rangle+
|2	0	1	1	0	1	1	1	0	1	1	0	0	0	0	1	0	1	0	0	1	1	1	0	 0	 0\rangle+
|0	1	2	2	0	1	1	1	0	1	1	0	0	0	0	1	0	1	0	0	1	1	1	0	 0	 0\rangle+\\
|1	2	0	0	0	1	1	1	0	1	1	0	0	0	0	1	0	1	0	0	1	1	1	0	 0	 0\rangle+
|2	0	1	1	1	0	0	0	1	0	0	1	1	1	1	0	1	0	1	1	0	0	0	1	 1	 1\rangle+
|0	1	2	2	1	0	0	0	1	0	0	1	1	1	1	0	1	0	1	1	0	0	0	1	 1	 1\rangle+\\
|1	2	0	0	1	0	0	0	1	0	0	1	1	1	1	0	1	0	1	1	0	0	0	1	 1	 1\rangle+
|0	2	1	1	0	1	1	1	0	1	1	0	0	1	0	0	1	0	0	0	0	0	1	0	 0	 1\rangle+
|1	0	2	2	0	1	1	1	0	1	1	0	0	1	0	0	1	0	0	0	0	0	1	0	 0	 1\rangle+\\
|2	1	0	0	0	1	1	1	0	1	1	0	0	1	0	0	1	0	0	0	0	0	1	0	 0	 1\rangle+
|0	2	1	1	1	0	0	0	1	0	0	1	1	0	1	1	0	1	1	1	1	1	0	1	 1	 0\rangle+
|1	0	2	2	1	0	0	0	1	0	0	1	1	0	1	1	0	1	1	1	1	1	0	1	 1	 0\rangle+\\
|2	1	0	0	1	0	0	0	1	0	0	1	1	0	1	1	0	1	1	1	1	1	0	1	 1	 0\rangle+
|1	1	0	1	0	1	1	1	0	1	1	0	0	1	1	1	0	0	1	1	1	0	0	1	 1	 1\rangle+
|2	2	1	2	0	1	1	1	0	1	1	0	0	1	1	1	0	0	1	1	1	0	0	1	 1	 1\rangle+\\
|0	0	2	0	0	1	1	1	0	1	1	0	0	1	1	1	0	0	1	1	1	0	0	1	 1	 1\rangle+
|1	1	0	1	1	0	0	0	1	0	0	1	1	0	0	0	1	1	0	0	0	1	1	0	 0	 0\rangle+
|2	2	1	2	1	0	0	0	1	0	0	1	1	0	0	0	1	1	0	0	0	1	1	0	 0	 0\rangle+\\
|0	0	2	0	1	0	0	0	1	0	0	1	1	0	0	0	1	1	0	0	0	1	1	0	 0	 0\rangle+
|2	2	0	2	0	1	1	1	1	0	0	1	0	0	0	0	1	0	1	1	1	1	1	1	 0	 0\rangle+
|0	0	1	0	0	1	1	1	1	0	0	1	0	0	0	0	1	0	1	1	1	1	1	1	 0	 0\rangle+\\
|1	1	2	1	0	1	1	1	1	0	0	1	0	0	0	0	1	0	1	1	1	1	1	1	 0	 0\rangle+
|2	2	0	2	1	0	0	0	0	1	1	0	1	1	1	1	0	1	0	0	0	0	0	0	 1	 1\rangle+
|0	0	1	0	1	0	0	0	0	1	1	0	1	1	1	1	0	1	0	0	0	0	0	0	 1	 1\rangle+\\
|1	1	2	1	1	0	0	0	0	1	1	0	1	1	1	1	0	1	0	0	0	0	0	0	 1	 1\rangle+
|2	1	2	2	0	1	1	1	1	1	0	1	1	0	0	1	0	0	0	0	0	1	0	1	 1	 1\rangle+
|0	2	0	0	0	1	1	1	1	1	0	1	1	0	0	1	0	0	0	0	0	1	0	1	 1	 1\rangle+\\
|1	0	1	1	0	1	1	1	1	1	0	1	1	0	0	1	0	0	0	0	0	1	0	1	 1	 1\rangle+
|2	1	2	2	1	0	0	0	0	0	1	0	0	1	1	0	1	1	1	1	1	0	1	0	 0	 0\rangle+
|0	2	0	0	1	0	0	0	0	0	1	0	0	1	1	0	1	1	1	1	1	0	1	0	 0	 0\rangle+\\
|1	0	1	1	1	0	0	0	0	0	1	0	0	1	1	0	1	1	1	1	1	0	1	0	 0	 0\rangle.$

$|\phi_{3^52^{16}}\rangle=
|0	0	0	0	1	0	0	0	0	0	0	0	0	0	0	0	0	0	0	0	0\rangle+
|1	1	1	1	2	0	0	0	0	0	0	0	0	0	0	0	0	0	0	0	0\rangle+
|2	2	2	2	0	0	0	0	0	0	0	0	0	0	0	0	0	0	0	0	0\rangle+\\
|0	0	0	0	1	1	1	1	1	1	1	1	1	1	1	1	1	1	1	1	1\rangle+
|1	1	1	1	2	1	1	1	1	1	1	1	1	1	1	1	1	1	1	1	1\rangle+
|2	2	2	2	0	1	1	1	1	1	1	1	1	1	1	1	1	1	1	1	1\rangle+\\
|1	1	1	1	0	0	0	0	0	0	1	0	0	1	1	0	0	0	0	1	0\rangle+
|2	2	2	2	1	0	0	0	0	0	1	0	0	1	1	0	0	0	0	1	0\rangle+
|0	0	0	0	2	0	0	0	0	0	1	0	0	1	1	0	0	0	0	1	0\rangle+\\
|1	1	1	1	0	1	1	1	1	1	0	1	1	0	0	1	1	1	1	0	1\rangle+
|2	2	2	2	1	1	1	1	1	1	0	1	1	0	0	1	1	1	1	0	1\rangle+
|0	0	0	0	2	1	1	1	1	1	0	1	1	0	0	1	1	1	1	0	1\rangle+\\
|1	1	2	2	2	0	0	0	0	0	1	1	1	0	0	0	0	0	1	0	1\rangle+
|2	2	0	0	0	0	0	0	0	0	1	1	1	0	0	0	0	0	1	0	1\rangle+
|0	0	1	1	1	0	0	0	0	0	1	1	1	0	0	0	0	0	1	0	1\rangle+\\
|1	1	2	2	2	1	1	1	1	1	0	0	0	1	1	1	1	1	0	1	0\rangle+
|2	2	0	0	0	1	1	1	1	1	0	0	0	1	1	1	1	1	0	1	0\rangle+
|0	0	1	1	1	1	1	1	1	1	0	0	0	1	1	1	1	1	0	1	0\rangle+\\
|1	0	1	0	2	0	0	0	0	1	1	1	0	0	0	0	0	1	0	1	1\rangle+
|2	1	2	1	0	0	0	0	0	1	1	1	0	0	0	0	0	1	0	1	1\rangle+
|0	2	0	2	1	0	0	0	0	1	1	1	0	0	0	0	0	1	0	1	1\rangle+\\
|1	0	1	0	2	1	1	1	1	0	0	0	1	1	1	1	1	0	1	0	0\rangle+
|2	1	2	1	0	1	1	1	1	0	0	0	1	1	1	1	1	0	1	0	0\rangle+
|0	2	0	2	1	1	1	1	1	0	0	0	1	1	1	1	1	0	1	0	0\rangle+\\
|0	2	2	0	1	0	0	0	1	0	1	0	1	0	1	0	1	1	1	1	0\rangle+
|1	0	0	1	2	0	0	0	1	0	1	0	1	0	1	0	1	1	1	1	0\rangle+
|2	1	1	2	0	0	0	0	1	0	1	0	1	0	1	0	1	1	1	1	0\rangle+\\
|0	2	2	0	1	1	1	1	0	1	0	1	0	1	0	1	0	0	0	0	1\rangle+
|1	0	0	1	2	1	1	1	0	1	0	1	0	1	0	1	0	0	0	0	1\rangle+
|2	1	1	2	0	1	1	1	0	1	0	1	0	1	0	1	0	0	0	0	1\rangle+\\
|0	1	1	0	2	0	0	0	1	1	0	1	1	1	0	1	1	0	0	1	0\rangle+
|1	2	2	1	0	0	0	0	1	1	0	1	1	1	0	1	1	0	0	1	0\rangle+
|2	0	0	2	1	0	0	0	1	1	0	1	1	1	0	1	1	0	0	1	0\rangle+\\
|0	1	1	0	2	1	1	1	0	0	1	0	0	0	1	0	0	1	1	0	1\rangle+
|1	2	2	1	0	1	1	1	0	0	1	0	0	0	1	0	0	1	1	0	1\rangle+
|2	0	0	2	1	1	1	1	0	0	1	0	0	0	1	0	0	1	1	0	1\rangle+\\
|0	1	0	1	2	0	0	0	1	1	0	1	1	1	1	0	0	0	0	0	1\rangle+
|1	2	1	2	0	0	0	0	1	1	0	1	1	1	1	0	0	0	0	0	1\rangle+
|2	0	2	0	1	0	0	0	1	1	0	1	1	1	1	0	0	0	0	0	1\rangle+\\
|0	1	0	1	2	1	1	1	0	0	1	0	0	0	0	1	1	1	1	1	0\rangle+
|1	2	1	2	0	1	1	1	0	0	1	0	0	0	0	1	1	1	1	1	0\rangle+
|2	0	2	0	1	1	1	1	0	0	1	0	0	0	0	1	1	1	1	1	0\rangle+\\
|0	0	2	2	0	0	0	0	1	1	0	1	1	1	1	0	1	1	1	0	0\rangle+
|1	1	0	0	1	0	0	0	1	1	0	1	1	1	1	0	1	1	1	0	0\rangle+
|2	2	1	1	2	0	0	0	1	1	0	1	1	1	1	0	1	1	1	0	0\rangle+\\
|0	0	2	2	0	1	1	1	0	0	1	0	0	0	0	1	0	0	0	1	1\rangle+
|1	1	0	0	1	1	1	1	0	0	1	0	0	0	0	1	0	0	0	1	1\rangle+
|2	2	1	1	2	1	1	1	0	0	1	0	0	0	0	1	0	0	0	1	1\rangle+\\
|2	2	1	1	1	0	0	0	1	1	1	0	0	0	0	1	1	1	1	0	1\rangle+
|0	0	2	2	2	0	0	0	1	1	1	0	0	0	0	1	1	1	1	0	1\rangle+
|1	1	0	0	0	0	0	0	1	1	1	0	0	0	0	1	1	1	1	0	1\rangle+\\
|2	2	1	1	1	1	1	1	0	0	0	1	1	1	1	0	0	0	0	1	0\rangle+
|0	0	2	2	2	1	1	1	0	0	0	1	1	1	1	0	0	0	0	1	0\rangle+
|1	1	0	0	0	1	1	1	0	0	0	1	1	1	1	0	0	0	0	1	0\rangle+\\
|2	2	0	0	2	0	0	1	0	0	0	0	0	1	1	1	1	1	0	0	0\rangle+
|0	0	1	1	0	0	0	1	0	0	0	0	0	1	1	1	1	1	0	0	0\rangle+
|1	1	2	2	1	0	0	1	0	0	0	0	0	1	1	1	1	1	0	0	0\rangle+\\
|2	2	0	0	2	1	1	0	1	1	1	1	1	0	0	0	0	0	1	1	1\rangle+
|0	0	1	1	0	1	1	0	1	1	1	1	1	0	0	0	0	0	1	1	1\rangle+
|1	1	2	2	1	1	1	0	1	1	1	1	1	0	0	0	0	0	1	1	1\rangle+\\
|2	0	1	0	0	0	0	1	0	0	0	1	1	0	0	1	1	1	0	1	0\rangle+
|0	1	2	1	1	0	0	1	0	0	0	1	1	0	0	1	1	1	0	1	0\rangle+
|1	2	0	2	2	0	0	1	0	0	0	1	1	0	0	1	1	1	0	1	0\rangle+\\
|2	0	1	0	0	1	1	0	1	1	1	0	0	1	1	0	0	0	1	0	1\rangle+
|0	1	2	1	1	1	1	0	1	1	1	0	0	1	1	0	0	0	1	0	1\rangle+
|1	2	0	2	2	1	1	0	1	1	1	0	0	1	1	0	0	0	1	0	1\rangle+\\
|2	0	2	1	2	0	0	1	0	0	1	1	1	1	0	1	1	1	1	0	1\rangle+
|0	1	0	2	0	0	0	1	0	0	1	1	1	1	0	1	1	1	1	0	1\rangle+
|1	2	1	0	1	0	0	1	0	0	1	1	1	1	0	1	1	1	1	0	1\rangle+\\
|2	0	2	1	2	1	1	0	1	1	0	0	0	0	1	0	0	0	0	1	0\rangle+
|0	1	0	2	0	1	1	0	1	1	0	0	0	0	1	0	0	0	0	1	0\rangle+
|1	2	1	0	1	1	1	0	1	1	0	0	0	0	1	0	0	0	0	1	0\rangle+\\
|0	2	0	1	0	0	0	1	0	1	0	1	0	0	1	0	1	0	1	1	1\rangle+
|1	0	1	2	1	0	0	1	0	1	0	1	0	0	1	0	1	0	1	1	1\rangle+
|2	1	2	0	2	0	0	1	0	1	0	1	0	0	1	0	1	0	1	1	1\rangle+\\
|0	2	0	1	0	1	1	0	1	0	1	0	1	1	0	1	0	1	0	0	0\rangle+
|1	0	1	2	1	1	1	0	1	0	1	0	1	1	0	1	0	1	0	0	0\rangle+
|2	1	2	0	2	1	1	0	1	0	1	0	1	1	0	1	0	1	0	0	0\rangle+\\
|1	2	0	1	1	0	0	1	0	1	1	1	0	1	1	1	0	0	1	1	0\rangle+
|2	0	1	2	2	0	0	1	0	1	1	1	0	1	1	1	0	0	1	1	0\rangle+
|0	1	2	0	0	0	0	1	0	1	1	1	0	1	1	1	0	0	1	1	0\rangle+\\
|1	2	0	1	1	1	1	0	1	0	0	0	1	0	0	0	1	1	0	0	1\rangle+
|2	0	1	2	2	1	1	0	1	0	0	0	1	0	0	0	1	1	0	0	1\rangle+
|0	1	2	0	0	1	1	0	1	0	0	0	1	0	0	0	1	1	0	0	1\rangle+\\
|1	0	2	1	1	0	0	1	1	0	0	0	1	0	0	1	0	0	1	0	1\rangle+
|2	1	0	2	2	0	0	1	1	0	0	0	1	0	0	1	0	0	1	0	1\rangle+
|0	2	1	0	0	0	0	1	1	0	0	0	1	0	0	1	0	0	1	0	1\rangle+\\
|1	0	2	1	1	1	1	0	0	1	1	1	0	1	1	0	1	1	0	1	0\rangle+
|2	1	0	2	2	1	1	0	0	1	1	1	0	1	1	0	1	1	0	1	0\rangle+
|0	2	1	0	0	1	1	0	0	1	1	1	0	1	1	0	1	1	0	1	0\rangle+\\
|2	1	1	0	1	0	0	1	1	0	1	0	1	1	1	0	1	0	0	1	1\rangle+
|0	2	2	1	2	0	0	1	1	0	1	0	1	1	1	0	1	0	0	1	1\rangle+
|1	0	0	2	0	0	0	1	1	0	1	0	1	1	1	0	1	0	0	1	1\rangle+\\
|2	1	1	0	1	1	1	0	0	1	0	1	0	0	0	1	0	1	1	0	0\rangle+
|0	2	2	1	2	1	1	0	0	1	0	1	0	0	0	1	0	1	1	0	0\rangle+
|1	0	0	2	0	1	1	0	0	1	0	1	0	0	0	1	0	1	1	0	0\rangle+\\
|1	2	2	0	2	0	0	1	1	1	0	0	0	0	0	1	0	1	0	1	0\rangle+
|2	0	0	1	0	0	0	1	1	1	0	0	0	0	0	1	0	1	0	1	0\rangle+
|0	1	1	2	1	0	0	1	1	1	0	0	0	0	0	1	0	1	0	1	0\rangle+\\
|1	2	2	0	2	1	1	0	0	0	1	1	1	1	1	0	1	0	1	0	1\rangle+
|2	0	0	1	0	1	1	0	0	0	1	1	1	1	1	0	1	0	1	0	1\rangle+
|0	1	1	2	1	1	1	0	0	0	1	1	1	1	1	0	1	0	1	0	1\rangle+\\
|0	2	1	2	2	0	0	1	1	1	1	0	0	1	1	1	0	1	1	0	1\rangle+
|1	0	2	0	0	0	0	1	1	1	1	0	0	1	1	1	0	1	1	0	1\rangle+
|2	1	0	1	1	0	0	1	1	1	1	0	0	1	1	1	0	1	1	0	1\rangle+\\
|0	2	1	2	2	1	1	0	0	0	0	1	1	0	0	0	1	0	0	1	0\rangle+
|1	0	2	0	0	1	1	0	0	0	0	1	1	0	0	0	1	0	0	1	0\rangle+
|2	1	0	1	1	1	1	0	0	0	0	1	1	0	0	0	1	0	0	1	0\rangle+\\
|0	0	0	0	1	0	1	0	0	0	0	0	1	0	1	1	1	0	0	0	1\rangle+
|1	1	1	1	2	0	1	0	0	0	0	0	1	0	1	1	1	0	0	0	1\rangle+
|2	2	2	2	0	0	1	0	0	0	0	0	1	0	1	1	1	0	0	0	1\rangle+\\
|0	0	0	0	1	1	0	1	1	1	1	1	0	1	0	0	0	1	1	1	0\rangle+
|1	1	1	1	2	1	0	1	1	1	1	1	0	1	0	0	0	1	1	1	0\rangle+
|2	2	2	2	0	1	0	1	1	1	1	1	0	1	0	0	0	1	1	1	0\rangle+\\
|1	1	1	1	0	0	1	0	0	0	1	0	1	1	0	1	0	1	1	1	0\rangle+
|2	2	2	2	1	0	1	0	0	0	1	0	1	1	0	1	0	1	1	1	0\rangle+
|0	0	0	0	2	0	1	0	0	0	1	0	1	1	0	1	0	1	1	1	0\rangle+\\
|1	1	1	1	0	1	0	1	1	1	0	1	0	0	1	0	1	0	0	0	1\rangle+
|2	2	2	2	1	1	0	1	1	1	0	1	0	0	1	0	1	0	0	0	1\rangle+
|0	0	0	0	2	1	0	1	1	1	0	1	0	0	1	0	1	0	0	0	1\rangle+\\
|1	1	2	2	2	0	1	0	0	1	0	0	0	0	1	1	1	0	1	0	0\rangle+
|2	2	0	0	0	0	1	0	0	1	0	0	0	0	1	1	1	0	1	0	0\rangle+
|0	0	1	1	1	0	1	0	0	1	0	0	0	0	1	1	1	0	1	0	0\rangle+\\
|1	1	2	2	2	1	0	1	1	0	1	1	1	1	0	0	0	1	0	1	1\rangle+
|2	2	0	0	0	1	0	1	1	0	1	1	1	1	0	0	0	1	0	1	1\rangle+
|0	0	1	1	1	1	0	1	1	0	1	1	1	1	0	0	0	1	0	1	1\rangle+\\
|1	0	1	0	2	0	1	0	0	1	1	0	0	1	0	0	1	1	0	0	1\rangle+
|2	1	2	1	0	0	1	0	0	1	1	0	0	1	0	0	1	1	0	0	1\rangle+
|0	2	0	2	1	0	1	0	0	1	1	0	0	1	0	0	1	1	0	0	1\rangle+\\
|1	0	1	0	2	1	0	1	1	0	0	1	1	0	1	1	0	0	1	1	0\rangle+
|2	1	2	1	0	1	0	1	1	0	0	1	1	0	1	1	0	0	1	1	0\rangle+
|0	2	0	2	1	1	0	1	1	0	0	1	1	0	1	1	0	0	1	1	0\rangle+\\
|0	2	2	0	1	0	1	0	0	1	1	1	1	0	1	1	1	1	0	1	1\rangle+
|1	0	0	1	2	0	1	0	0	1	1	1	1	0	1	1	1	1	0	1	1\rangle+
|2	1	1	2	0	0	1	0	0	1	1	1	1	0	1	1	1	1	0	1	1\rangle+\\
|0	2	2	0	1	1	0	1	1	0	0	0	0	1	0	0	0	0	1	0	0\rangle+
|1	0	0	1	2	1	0	1	1	0	0	0	0	1	0	0	0	0	1	0	0\rangle+
|2	1	1	2	0	1	0	1	1	0	0	0	0	1	0	0	0	0	1	0	0\rangle+\\
|0	1	1	0	2	0	1	0	1	0	0	1	0	1	0	0	1	1	1	1	0\rangle+
|1	2	2	1	0	0	1	0	1	0	0	1	0	1	0	0	1	1	1	1	0\rangle+
|2	0	0	2	1	0	1	0	1	0	0	1	0	1	0	0	1	1	1	1	0\rangle+\\
|0	1	1	0	2	1	0	1	0	1	1	0	1	0	1	1	0	0	0	0	1\rangle+
|1	2	2	1	0	1	0	1	0	1	1	0	1	0	1	1	0	0	0	0	1\rangle+
|2	0	0	2	1	1	0	1	0	1	1	0	1	0	1	1	0	0	0	0	1\rangle+\\
|0	1	0	1	2	0	1	0	1	0	0	1	0	1	0	1	0	0	1	1	1\rangle+
|1	2	1	2	0	0	1	0	1	0	0	1	0	1	0	1	0	0	1	1	1\rangle+
|2	0	2	0	1	0	1	0	1	0	0	1	0	1	0	1	0	0	1	1	1\rangle+\\
|0	1	0	1	2	1	0	1	0	1	1	0	1	0	1	0	1	1	0	0	0\rangle+
|1	2	1	2	0	1	0	1	0	1	1	0	1	0	1	0	1	1	0	0	0\rangle+
|2	0	2	0	1	1	0	1	0	1	1	0	1	0	1	0	1	1	0	0	0\rangle+\\
|0	0	2	2	0	0	1	0	1	0	0	1	0	1	1	1	0	1	0	0	1\rangle+
|1	1	0	0	1	0	1	0	1	0	0	1	0	1	1	1	0	1	0	0	1\rangle+
|2	2	1	1	2	0	1	0	1	0	0	1	0	1	1	1	0	1	0	0	1\rangle+\\
|0	0	2	2	0	1	0	1	0	1	1	0	1	0	0	0	1	0	1	1	0\rangle+
|1	1	0	0	1	1	0	1	0	1	1	0	1	0	0	0	1	0	1	1	0\rangle+
|2	2	1	1	2	1	0	1	0	1	1	0	1	0	0	0	1	0	1	1	0\rangle+\\
|2	2	1	1	1	0	1	0	1	1	1	0	1	0	1	1	0	0	1	1	0\rangle+
|0	0	2	2	2	0	1	0	1	1	1	0	1	0	1	1	0	0	1	1	0\rangle+
|1	1	0	0	0	0	1	0	1	1	1	0	1	0	1	1	0	0	1	1	0\rangle+\\
|2	2	1	1	1	1	0	1	0	0	0	1	0	1	0	0	1	1	0	0	1\rangle+
|0	0	2	2	2	1	0	1	0	0	0	1	0	1	0	0	1	1	0	0	1\rangle+
|1	1	0	0	0	1	0	1	0	0	0	1	0	1	0	0	1	1	0	0	1\rangle+\\
|2	2	0	0	2	0	1	1	0	0	0	0	1	1	1	0	0	1	1	1	1\rangle+
|0	0	1	1	0	0	1	1	0	0	0	0	1	1	1	0	0	1	1	1	1\rangle+
|1	1	2	2	1	0	1	1	0	0	0	0	1	1	1	0	0	1	1	1	1\rangle+\\
|2	2	0	0	2	1	0	0	1	1	1	1	0	0	0	1	1	0	0	0	0\rangle+
|0	0	1	1	0	1	0	0	1	1	1	1	0	0	0	1	1	0	0	0	0\rangle+
|1	1	2	2	1	1	0	0	1	1	1	1	0	0	0	1	1	0	0	0	0\rangle+\\
|2	0	1	0	0	0	1	1	0	1	0	0	0	1	0	0	1	0	1	1	1\rangle+
|0	1	2	1	1	0	1	1	0	1	0	0	0	1	0	0	1	0	1	1	1\rangle+
|1	2	0	2	2	0	1	1	0	1	0	0	0	1	0	0	1	0	1	1	1\rangle+\\
|2	0	1	0	0	1	0	0	1	0	1	1	1	0	1	1	0	1	0	0	0\rangle+
|0	1	2	1	1	1	0	0	1	0	1	1	1	0	1	1	0	1	0	0	0\rangle+
|1	2	0	2	2	1	0	0	1	0	1	1	1	0	1	1	0	1	0	0	0\rangle+\\
|2	0	2	1	2	0	1	1	0	1	0	1	1	0	1	0	0	1	1	0	0\rangle+
|0	1	0	2	0	0	1	1	0	1	0	1	1	0	1	0	0	1	1	0	0\rangle+
|1	2	1	0	1	0	1	1	0	1	0	1	1	0	1	0	0	1	1	0	0\rangle+\\
|2	0	2	1	2	1	0	0	1	0	1	0	0	1	0	1	1	0	0	1	1\rangle+
|0	1	0	2	0	1	0	0	1	0	1	0	0	1	0	1	1	0	0	1	1\rangle+
|1	2	1	0	1	1	0	0	1	0	1	0	0	1	0	1	1	0	0	1	1\rangle+\\
|0	2	0	1	0	0	1	1	0	1	1	1	1	1	0	1	0	0	0	0	0\rangle+
|1	0	1	2	1	0	1	1	0	1	1	1	1	1	0	1	0	0	0	0	0\rangle+
|2	1	2	0	2	0	1	1	0	1	1	1	1	1	0	1	0	0	0	0	0\rangle+\\
|0	2	0	1	0	1	0	0	1	0	0	0	0	0	1	0	1	1	1	1	1\rangle+
|1	0	1	2	1	1	0	0	1	0	0	0	0	0	1	0	1	1	1	1	1\rangle+
|2	1	2	0	2	1	0	0	1	0	0	0	0	0	1	0	1	1	1	1	1\rangle+\\
|1	2	0	1	1	0	1	1	1	0	1	1	0	0	0	0	1	0	1	0	0\rangle+
|2	0	1	2	2	0	1	1	1	0	1	1	0	0	0	0	1	0	1	0	0\rangle+
|0	1	2	0	0	0	1	1	1	0	1	1	0	0	0	0	1	0	1	0	0\rangle+\\
|1	2	0	1	1	1	0	0	0	1	0	0	1	1	1	1	0	1	0	1	1\rangle+
|2	0	1	2	2	1	0	0	0	1	0	0	1	1	1	1	0	1	0	1	1\rangle+
|0	1	2	0	0	1	0	0	0	1	0	0	1	1	1	1	0	1	0	1	1\rangle+\\
|1	0	2	1	1	0	1	1	1	0	1	1	0	0	1	0	0	1	0	0	0\rangle+
|2	1	0	2	2	0	1	1	1	0	1	1	0	0	1	0	0	1	0	0	0\rangle+
|0	2	1	0	0	0	1	1	1	0	1	1	0	0	1	0	0	1	0	0	0\rangle+\\
|1	0	2	1	1	1	0	0	0	1	0	0	1	1	0	1	1	0	1	1	1\rangle+
|2	1	0	2	2	1	0	0	0	1	0	0	1	1	0	1	1	0	1	1	1\rangle+
|0	2	1	0	0	1	0	0	0	1	0	0	1	1	0	1	1	0	1	1	1\rangle+\\
|2	1	1	0	1	0	1	1	1	0	1	1	0	0	1	1	1	0	0	1	1\rangle+
|0	2	2	1	2	0	1	1	1	0	1	1	0	0	1	1	1	0	0	1	1\rangle+
|1	0	0	2	0	0	1	1	1	0	1	1	0	0	1	1	1	0	0	1	1\rangle+\\
|2	1	1	0	1	1	0	0	0	1	0	0	1	1	0	0	0	1	1	0	0\rangle+
|0	2	2	1	2	1	0	0	0	1	0	0	1	1	0	0	0	1	1	0	0\rangle+
|1	0	0	2	0	1	0	0	0	1	0	0	1	1	0	0	0	1	1	0	0\rangle+\\
|1	2	2	0	2	0	1	1	1	1	0	0	1	0	0	0	0	1	0	1	1\rangle+
|2	0	0	1	0	0	1	1	1	1	0	0	1	0	0	0	0	1	0	1	1\rangle+
|0	1	1	2	1	0	1	1	1	1	0	0	1	0	0	0	0	1	0	1	1\rangle+\\
|1	2	2	0	2	1	0	0	0	0	1	1	0	1	1	1	1	0	1	0	0\rangle+
|2	0	0	1	0	1	0	0	0	0	1	1	0	1	1	1	1	0	1	0	0\rangle+
|0	1	1	2	1	1	0	0	0	0	1	1	0	1	1	1	1	0	1	0	0\rangle+\\
|0	2	1	2	2	0	1	1	1	1	1	0	1	1	0	0	1	0	0	0	0\rangle+
|1	0	2	0	0	0	1	1	1	1	1	0	1	1	0	0	1	0	0	0	0\rangle+
|2	1	0	1	1	0	1	1	1	1	1	0	1	1	0	0	1	0	0	0	0\rangle+\\
|0	2	1	2	2	1	0	0	0	0	0	1	0	0	1	1	0	1	1	1	1\rangle+
|1	0	2	0	0	1	0	0	0	0	0	1	0	0	1	1	0	1	1	1	1\rangle+
|2	1	0	1	1	1	0	0	0	0	0	1	0	0	1	1	0	1	1	1	1\rangle.$

$|\phi_{3^42^{16}}\rangle=
|0	0	0	1	0	0	0	0	0	0	0	0	0	0	0	0	0	0	0	0	\rangle+
|1	1	1	2	0	0	0	0	0	0	0	0	0	0	0	0	0	0	0	0	\rangle+
|2	2	2	0	0	0	0	0	0	0	0	0	0	0	0	0	0	0	0	0	\rangle+\\
|0	0	0	1	1	1	1	1	1	1	1	1	1	1	1	1	1	1	1	1	\rangle+
|1	1	1	2	1	1	1	1	1	1	1	1	1	1	1	1	1	1	1	1	\rangle+
|2	2	2	0	1	1	1	1	1	1	1	1	1	1	1	1	1	1	1	1	\rangle+\\
|1	1	1	0	0	0	0	0	0	1	0	0	1	1	0	0	0	0	1	0	\rangle+
|2	2	2	1	0	0	0	0	0	1	0	0	1	1	0	0	0	0	1	0	\rangle+
|0	0	0	2	0	0	0	0	0	1	0	0	1	1	0	0	0	0	1	0	\rangle+\\
|1	1	1	0	1	1	1	1	1	0	1	1	0	0	1	1	1	1	0	1	\rangle+
|2	2	2	1	1	1	1	1	1	0	1	1	0	0	1	1	1	1	0	1	\rangle+
|0	0	0	2	1	1	1	1	1	0	1	1	0	0	1	1	1	1	0	1	\rangle+\\
|1	2	2	2	0	0	0	0	0	1	1	1	0	0	0	0	0	1	0	1	\rangle+
|2	0	0	0	0	0	0	0	0	1	1	1	0	0	0	0	0	1	0	1	\rangle+
|0	1	1	1	0	0	0	0	0	1	1	1	0	0	0	0	0	1	0	1	\rangle+\\
|1	2	2	2	1	1	1	1	1	0	0	0	1	1	1	1	1	0	1	0	\rangle+
|2	0	0	0	1	1	1	1	1	0	0	0	1	1	1	1	1	0	1	0	\rangle+
|0	1	1	1	1	1	1	1	1	0	0	0	1	1	1	1	1	0	1	0	\rangle+\\
|0	1	0	2	0	0	0	0	1	1	1	0	0	0	0	0	1	0	1	1	\rangle+
|1	2	1	0	0	0	0	0	1	1	1	0	0	0	0	0	1	0	1	1	\rangle+
|2	0	2	1	0	0	0	0	1	1	1	0	0	0	0	0	1	0	1	1	\rangle+\\
|0	1	0	2	1	1	1	1	0	0	0	1	1	1	1	1	0	1	0	0	\rangle+
|1	2	1	0	1	1	1	1	0	0	0	1	1	1	1	1	0	1	0	0	\rangle+
|2	0	2	1	1	1	1	1	0	0	0	1	1	1	1	1	0	1	0	0	\rangle+\\
|2	2	0	1	0	0	0	1	0	1	0	1	0	1	0	1	1	1	1	0	\rangle+
|0	0	1	2	0	0	0	1	0	1	0	1	0	1	0	1	1	1	1	0	\rangle+
|1	1	2	0	0	0	0	1	0	1	0	1	0	1	0	1	1	1	1	0	\rangle+\\
|2	2	0	1	1	1	1	0	1	0	1	0	1	0	1	0	0	0	0	1	\rangle+
|0	0	1	2	1	1	1	0	1	0	1	0	1	0	1	0	0	0	0	1	\rangle+
|1	1	2	0	1	1	1	0	1	0	1	0	1	0	1	0	0	0	0	1	\rangle+\\
|1	1	0	2	0	0	0	1	1	0	1	1	1	0	1	1	0	0	1	0	\rangle+
|2	2	1	0	0	0	0	1	1	0	1	1	1	0	1	1	0	0	1	0	\rangle+
|0	0	2	1	0	0	0	1	1	0	1	1	1	0	1	1	0	0	1	0	\rangle+\\
|1	1	0	2	1	1	1	0	0	1	0	0	0	1	0	0	1	1	0	1	\rangle+
|2	2	1	0	1	1	1	0	0	1	0	0	0	1	0	0	1	1	0	1	\rangle+
|0	0	2	1	1	1	1	0	0	1	0	0	0	1	0	0	1	1	0	1	\rangle+\\
|1	0	1	2	0	0	0	1	1	0	1	1	1	1	0	0	0	0	0	1	\rangle+
|2	1	2	0	0	0	0	1	1	0	1	1	1	1	0	0	0	0	0	1	\rangle+
|0	2	0	1	0	0	0	1	1	0	1	1	1	1	0	0	0	0	0	1	\rangle+\\
|1	0	1	2	1	1	1	0	0	1	0	0	0	0	1	1	1	1	1	0	\rangle+
|2	1	2	0	1	1	1	0	0	1	0	0	0	0	1	1	1	1	1	0	\rangle+
|0	2	0	1	1	1	1	0	0	1	0	0	0	0	1	1	1	1	1	0	\rangle+\\
|0	2	2	0	0	0	0	1	1	0	1	1	1	1	0	1	1	1	0	0	\rangle+
|1	0	0	1	0	0	0	1	1	0	1	1	1	1	0	1	1	1	0	0	\rangle+
|2	1	1	2	0	0	0	1	1	0	1	1	1	1	0	1	1	1	0	0	\rangle+\\
|0	2	2	0	1	1	1	0	0	1	0	0	0	0	1	0	0	0	1	1	\rangle+
|1	0	0	1	1	1	1	0	0	1	0	0	0	0	1	0	0	0	1	1	\rangle+
|2	1	1	2	1	1	1	0	0	1	0	0	0	0	1	0	0	0	1	1	\rangle+\\
|2	1	1	1	0	0	0	1	1	1	0	0	0	0	1	1	1	1	0	1	\rangle+
|0	2	2	2	0	0	0	1	1	1	0	0	0	0	1	1	1	1	0	1	\rangle+
|1	0	0	0	0	0	0	1	1	1	0	0	0	0	1	1	1	1	0	1	\rangle+\\
|2	1	1	1	1	1	1	0	0	0	1	1	1	1	0	0	0	0	1	0	\rangle+
|0	2	2	2	1	1	1	0	0	0	1	1	1	1	0	0	0	0	1	0	\rangle+
|1	0	0	0	1	1	1	0	0	0	1	1	1	1	0	0	0	0	1	0	\rangle+\\
|2	0	0	2	0	0	1	0	0	0	0	0	1	1	1	1	1	0	0	0	\rangle+
|0	1	1	0	0	0	1	0	0	0	0	0	1	1	1	1	1	0	0	0	\rangle+
|1	2	2	1	0	0	1	0	0	0	0	0	1	1	1	1	1	0	0	0	\rangle+\\
|2	0	0	2	1	1	0	1	1	1	1	1	0	0	0	0	0	1	1	1	\rangle+
|0	1	1	0	1	1	0	1	1	1	1	1	0	0	0	0	0	1	1	1	\rangle+
|1	2	2	1	1	1	0	1	1	1	1	1	0	0	0	0	0	1	1	1	\rangle+\\
|0	1	0	0	0	0	1	0	0	0	1	1	0	0	1	1	1	0	1	0	\rangle+
|1	2	1	1	0	0	1	0	0	0	1	1	0	0	1	1	1	0	1	0	\rangle+
|2	0	2	2	0	0	1	0	0	0	1	1	0	0	1	1	1	0	1	0	\rangle+\\
|0	1	0	0	1	1	0	1	1	1	0	0	1	1	0	0	0	1	0	1	\rangle+
|1	2	1	1	1	1	0	1	1	1	0	0	1	1	0	0	0	1	0	1	\rangle+
|2	0	2	2	1	1	0	1	1	1	0	0	1	1	0	0	0	1	0	1	\rangle+\\
|0	2	1	2	0	0	1	0	0	1	1	1	1	0	1	1	1	1	0	1	\rangle+
|1	0	2	0	0	0	1	0	0	1	1	1	1	0	1	1	1	1	0	1	\rangle+
|2	1	0	1	0	0	1	0	0	1	1	1	1	0	1	1	1	1	0	1	\rangle+\\
|0	2	1	2	1	1	0	1	1	0	0	0	0	1	0	0	0	0	1	0	\rangle+
|1	0	2	0	1	1	0	1	1	0	0	0	0	1	0	0	0	0	1	0	\rangle+
|2	1	0	1	1	1	0	1	1	0	0	0	0	1	0	0	0	0	1	0	\rangle+\\
|2	0	1	0	0	0	1	0	1	0	1	0	0	1	0	1	0	1	1	1	\rangle+
|0	1	2	1	0	0	1	0	1	0	1	0	0	1	0	1	0	1	1	1	\rangle+
|1	2	0	2	0	0	1	0	1	0	1	0	0	1	0	1	0	1	1	1	\rangle+\\
|2	0	1	0	1	1	0	1	0	1	0	1	1	0	1	0	1	0	0	0	\rangle+
|0	1	2	1	1	1	0	1	0	1	0	1	1	0	1	0	1	0	0	0	\rangle+
|1	2	0	2	1	1	0	1	0	1	0	1	1	0	1	0	1	0	0	0	\rangle+\\
|2	0	1	1	0	0	1	0	1	1	1	0	1	1	1	0	0	1	1	0	\rangle+
|0	1	2	2	0	0	1	0	1	1	1	0	1	1	1	0	0	1	1	0	\rangle+
|1	2	0	0	0	0	1	0	1	1	1	0	1	1	1	0	0	1	1	0	\rangle+\\
|2	0	1	1	1	1	0	1	0	0	0	1	0	0	0	1	1	0	0	1	\rangle+
|0	1	2	2	1	1	0	1	0	0	0	1	0	0	0	1	1	0	0	1	\rangle+
|1	2	0	0	1	1	0	1	0	0	0	1	0	0	0	1	1	0	0	1	\rangle+\\
|0	2	1	1	0	0	1	1	0	0	0	1	0	0	1	0	0	1	0	1	\rangle+
|1	0	2	2	0	0	1	1	0	0	0	1	0	0	1	0	0	1	0	1	\rangle+
|2	1	0	0	0	0	1	1	0	0	0	1	0	0	1	0	0	1	0	1	\rangle+\\
|0	2	1	1	1	1	0	0	1	1	1	0	1	1	0	1	1	0	1	0	\rangle+
|1	0	2	2	1	1	0	0	1	1	1	0	1	1	0	1	1	0	1	0	\rangle+
|2	1	0	0	1	1	0	0	1	1	1	0	1	1	0	1	1	0	1	0	\rangle+\\
|1	1	0	1	0	0	1	1	0	1	0	1	1	1	0	1	0	0	1	1	\rangle+
|2	2	1	2	0	0	1	1	0	1	0	1	1	1	0	1	0	0	1	1	\rangle+
|0	0	2	0	0	0	1	1	0	1	0	1	1	1	0	1	0	0	1	1	\rangle+\\
|1	1	0	1	1	1	0	0	1	0	1	0	0	0	1	0	1	1	0	0	\rangle+
|2	2	1	2	1	1	0	0	1	0	1	0	0	0	1	0	1	1	0	0	\rangle+
|0	0	2	0	1	1	0	0	1	0	1	0	0	0	1	0	1	1	0	0	\rangle+\\
|2	2	0	2	0	0	1	1	1	0	0	0	0	0	1	0	1	0	1	0	\rangle+
|0	0	1	0	0	0	1	1	1	0	0	0	0	0	1	0	1	0	1	0	\rangle+
|1	1	2	1	0	0	1	1	1	0	0	0	0	0	1	0	1	0	1	0	\rangle+\\
|2	2	0	2	1	1	0	0	0	1	1	1	1	1	0	1	0	1	0	1	\rangle+
|0	0	1	0	1	1	0	0	0	1	1	1	1	1	0	1	0	1	0	1	\rangle+
|1	1	2	1	1	1	0	0	0	1	1	1	1	1	0	1	0	1	0	1	\rangle+\\
|2	1	2	2	0	0	1	1	1	1	0	0	1	1	1	0	1	1	0	1	\rangle+
|0	2	0	0	0	0	1	1	1	1	0	0	1	1	1	0	1	1	0	1	\rangle+
|1	0	1	1	0	0	1	1	1	1	0	0	1	1	1	0	1	1	0	1	\rangle+\\
|2	1	2	2	1	1	0	0	0	0	1	1	0	0	0	1	0	0	1	0	\rangle+
|0	2	0	0	1	1	0	0	0	0	1	1	0	0	0	1	0	0	1	0	\rangle+
|1	0	1	1	1	1	0	0	0	0	1	1	0	0	0	1	0	0	1	0	\rangle+\\
|0	0	0	1	0	1	0	0	0	0	0	1	0	1	1	1	0	0	0	1	\rangle+
|1	1	1	2	0	1	0	0	0	0	0	1	0	1	1	1	0	0	0	1	\rangle+
|2	2	2	0	0	1	0	0	0	0	0	1	0	1	1	1	0	0	0	1	\rangle+\\
|0	0	0	1	1	0	1	1	1	1	1	0	1	0	0	0	1	1	1	0	\rangle+
|1	1	1	2	1	0	1	1	1	1	1	0	1	0	0	0	1	1	1	0	\rangle+
|2	2	2	0	1	0	1	1	1	1	1	0	1	0	0	0	1	1	1	0	\rangle+\\
|1	1	1	0	0	1	0	0	0	1	0	1	1	0	1	0	1	1	1	0	\rangle+
|2	2	2	1	0	1	0	0	0	1	0	1	1	0	1	0	1	1	1	0	\rangle+
|0	0	0	2	0	1	0	0	0	1	0	1	1	0	1	0	1	1	1	0	\rangle+\\
|1	1	1	0	1	0	1	1	1	0	1	0	0	1	0	1	0	0	0	1	\rangle+
|2	2	2	1	1	0	1	1	1	0	1	0	0	1	0	1	0	0	0	1	\rangle+
|0	0	0	2	1	0	1	1	1	0	1	0	0	1	0	1	0	0	0	1	\rangle+\\
|1	2	2	2	0	1	0	0	1	0	0	0	0	1	1	1	0	1	0	0	\rangle+
|2	0	0	0	0	1	0	0	1	0	0	0	0	1	1	1	0	1	0	0	\rangle+
|0	1	1	1	0	1	0	0	1	0	0	0	0	1	1	1	0	1	0	0	\rangle+\\
|1	2	2	2	1	0	1	1	0	1	1	1	1	0	0	0	1	0	1	1	\rangle+
|2	0	0	0	1	0	1	1	0	1	1	1	1	0	0	0	1	0	1	1	\rangle+
|0	1	1	1	1	0	1	1	0	1	1	1	1	0	0	0	1	0	1	1	\rangle+\\
|0	1	0	2	0	1	0	0	1	1	0	0	1	0	0	1	1	0	0	1	\rangle+
|1	2	1	0	0	1	0	0	1	1	0	0	1	0	0	1	1	0	0	1	\rangle+
|2	0	2	1	0	1	0	0	1	1	0	0	1	0	0	1	1	0	0	1	\rangle+\\
|0	1	0	2	1	0	1	1	0	0	1	1	0	1	1	0	0	1	1	0	\rangle+
|1	2	1	0	1	0	1	1	0	0	1	1	0	1	1	0	0	1	1	0	\rangle+
|2	0	2	1	1	0	1	1	0	0	1	1	0	1	1	0	0	1	1	0	\rangle+\\
|2	2	0	1	0	1	0	0	1	1	1	1	0	1	1	1	1	0	1	1	\rangle+
|0	0	1	2	0	1	0	0	1	1	1	1	0	1	1	1	1	0	1	1	\rangle+
|1	1	2	0	0	1	0	0	1	1	1	1	0	1	1	1	1	0	1	1	\rangle+\\
|2	2	0	1	1	0	1	1	0	0	0	0	1	0	0	0	0	1	0	0	\rangle+
|0	0	1	2	1	0	1	1	0	0	0	0	1	0	0	0	0	1	0	0	\rangle+
|1	1	2	0	1	0	1	1	0	0	0	0	1	0	0	0	0	1	0	0	\rangle+\\
|1	1	0	2	0	1	0	1	0	0	1	0	1	0	0	1	1	1	1	0	\rangle+
|2	2	1	0	0	1	0	1	0	0	1	0	1	0	0	1	1	1	1	0	\rangle+
|0	0	2	1	0	1	0	1	0	0	1	0	1	0	0	1	1	1	1	0	\rangle+\\
|1	1	0	2	1	0	1	0	1	1	0	1	0	1	1	0	0	0	0	1	\rangle+
|2	2	1	0	1	0	1	0	1	1	0	1	0	1	1	0	0	0	0	1	\rangle+
|0	0	2	1	1	0	1	0	1	1	0	1	0	1	1	0	0	0	0	1	\rangle+\\
|1	0	1	2	0	1	0	1	0	0	1	0	1	0	1	0	0	1	1	1	\rangle+
|2	1	2	0	0	1	0	1	0	0	1	0	1	0	1	0	0	1	1	1	\rangle+
|0	2	0	1	0	1	0	1	0	0	1	0	1	0	1	0	0	1	1	1	\rangle+\\
|1	0	1	2	1	0	1	0	1	1	0	1	0	1	0	1	1	0	0	0	\rangle+
|2	1	2	0	1	0	1	0	1	1	0	1	0	1	0	1	1	0	0	0	\rangle+
|0	2	0	1	1	0	1	0	1	1	0	1	0	1	0	1	1	0	0	0	\rangle+\\
|0	2	2	0	0	1	0	1	0	0	1	0	1	1	1	0	1	0	0	1	\rangle+
|1	0	0	1	0	1	0	1	0	0	1	0	1	1	1	0	1	0	0	1	\rangle+
|2	1	1	2	0	1	0	1	0	0	1	0	1	1	1	0	1	0	0	1	\rangle+\\
|0	2	2	0	1	0	1	0	1	1	0	1	0	0	0	1	0	1	1	0	\rangle+
|1	0	0	1	1	0	1	0	1	1	0	1	0	0	0	1	0	1	1	0	\rangle+
|2	1	1	2	1	0	1	0	1	1	0	1	0	0	0	1	0	1	1	0	\rangle+\\
|2	1	1	1	0	1	0	1	1	1	0	1	0	1	1	0	0	1	1	0	\rangle+
|0	2	2	2	0	1	0	1	1	1	0	1	0	1	1	0	0	1	1	0	\rangle+
|1	0	0	0	0	1	0	1	1	1	0	1	0	1	1	0	0	1	1	0	\rangle+\\
|2	1	1	1	1	0	1	0	0	0	1	0	1	0	0	1	1	0	0	1	\rangle+
|0	2	2	2	1	0	1	0	0	0	1	0	1	0	0	1	1	0	0	1	\rangle+
|1	0	0	0	1	0	1	0	0	0	1	0	1	0	0	1	1	0	0	1	\rangle+\\
|2	0	0	2	0	1	1	0	0	0	0	1	1	1	0	0	1	1	1	1	\rangle+
|0	1	1	0	0	1	1	0	0	0	0	1	1	1	0	0	1	1	1	1	\rangle+
|1	2	2	1	0	1	1	0	0	0	0	1	1	1	0	0	1	1	1	1	\rangle+\\
|2	0	0	2	1	0	0	1	1	1	1	0	0	0	1	1	0	0	0	0	\rangle+
|0	1	1	0	1	0	0	1	1	1	1	0	0	0	1	1	0	0	0	0	\rangle+
|1	2	2	1	1	0	0	1	1	1	1	0	0	0	1	1	0	0	0	0	\rangle+\\
|0	1	0	0	0	1	1	0	1	0	0	0	1	0	0	1	0	1	1	1	\rangle+
|1	2	1	1	0	1	1	0	1	0	0	0	1	0	0	1	0	1	1	1	\rangle+
|2	0	2	2	0	1	1	0	1	0	0	0	1	0	0	1	0	1	1	1	\rangle+\\
|0	1	0	0	1	0	0	1	0	1	1	1	0	1	1	0	1	0	0	0	\rangle+
|1	2	1	1	1	0	0	1	0	1	1	1	0	1	1	0	1	0	0	0	\rangle+
|2	0	2	2	1	0	0	1	0	1	1	1	0	1	1	0	1	0	0	0	\rangle+\\
|0	2	1	2	0	1	1	0	1	0	1	1	0	1	0	0	1	1	0	0	\rangle+
|1	0	2	0	0	1	1	0	1	0	1	1	0	1	0	0	1	1	0	0	\rangle+
|2	1	0	1	0	1	1	0	1	0	1	1	0	1	0	0	1	1	0	0	\rangle+\\
|0	2	1	2	1	0	0	1	0	1	0	0	1	0	1	1	0	0	1	1	\rangle+
|1	0	2	0	1	0	0	1	0	1	0	0	1	0	1	1	0	0	1	1	\rangle+
|2	1	0	1	1	0	0	1	0	1	0	0	1	0	1	1	0	0	1	1	\rangle+\\
|2	0	1	0	0	1	1	0	1	1	1	1	1	0	1	0	0	0	0	0	\rangle+
|0	1	2	1	0	1	1	0	1	1	1	1	1	0	1	0	0	0	0	0	\rangle+
|1	2	0	2	0	1	1	0	1	1	1	1	1	0	1	0	0	0	0	0	\rangle+\\
|2	0	1	0	1	0	0	1	0	0	0	0	0	1	0	1	1	1	1	1	\rangle+
|0	1	2	1	1	0	0	1	0	0	0	0	0	1	0	1	1	1	1	1	\rangle+
|1	2	0	2	1	0	0	1	0	0	0	0	0	1	0	1	1	1	1	1	\rangle+\\
|2	0	1	1	0	1	1	1	0	1	1	0	0	0	0	1	0	1	0	0	\rangle+
|0	1	2	2	0	1	1	1	0	1	1	0	0	0	0	1	0	1	0	0	\rangle+
|1	2	0	0	0	1	1	1	0	1	1	0	0	0	0	1	0	1	0	0	\rangle+\\
|2	0	1	1	1	0	0	0	1	0	0	1	1	1	1	0	1	0	1	1	\rangle+
|0	1	2	2	1	0	0	0	1	0	0	1	1	1	1	0	1	0	1	1	\rangle+
|1	2	0	0	1	0	0	0	1	0	0	1	1	1	1	0	1	0	1	1	\rangle+\\
|0	2	1	1	0	1	1	1	0	1	1	0	0	1	0	0	1	0	0	0	\rangle+
|1	0	2	2	0	1	1	1	0	1	1	0	0	1	0	0	1	0	0	0	\rangle+
|2	1	0	0	0	1	1	1	0	1	1	0	0	1	0	0	1	0	0	0	\rangle+\\
|0	2	1	1	1	0	0	0	1	0	0	1	1	0	1	1	0	1	1	1	\rangle+
|1	0	2	2	1	0	0	0	1	0	0	1	1	0	1	1	0	1	1	1	\rangle+
|2	1	0	0	1	0	0	0	1	0	0	1	1	0	1	1	0	1	1	1	\rangle+\\
|1	1	0	1	0	1	1	1	0	1	1	0	0	1	1	1	0	0	1	1	\rangle+
|2	2	1	2	0	1	1	1	0	1	1	0	0	1	1	1	0	0	1	1	\rangle+
|0	0	2	0	0	1	1	1	0	1	1	0	0	1	1	1	0	0	1	1	\rangle+\\
|1	1	0	1	1	0	0	0	1	0	0	1	1	0	0	0	1	1	0	0	\rangle+
|2	2	1	2	1	0	0	0	1	0	0	1	1	0	0	0	1	1	0	0	\rangle+
|0	0	2	0	1	0	0	0	1	0	0	1	1	0	0	0	1	1	0	0	\rangle+\\
|2	2	0	2	0	1	1	1	1	0	0	1	0	0	0	0	1	0	1	1	\rangle+
|0	0	1	0	0	1	1	1	1	0	0	1	0	0	0	0	1	0	1	1	\rangle+
|1	1	2	1	0	1	1	1	1	0	0	1	0	0	0	0	1	0	1	1	\rangle+\\
|2	2	0	2	1	0	0	0	0	1	1	0	1	1	1	1	0	1	0	0	\rangle+
|0	0	1	0	1	0	0	0	0	1	1	0	1	1	1	1	0	1	0	0	\rangle+
|1	1	2	1	1	0	0	0	0	1	1	0	1	1	1	1	0	1	0	0	\rangle+\\
|2	1	2	2	0	1	1	1	1	1	0	1	1	0	0	1	0	0	0	0	\rangle+
|0	2	0	0	0	1	1	1	1	1	0	1	1	0	0	1	0	0	0	0	\rangle+
|1	0	1	1	0	1	1	1	1	1	0	1	1	0	0	1	0	0	0	0	\rangle+\\
|2	1	2	2	1	0	0	0	0	0	1	0	0	1	1	0	1	1	1	1	\rangle+
|0	2	0	0	1	0	0	0	0	0	1	0	0	1	1	0	1	1	1	1	\rangle+
|1	0	1	1	1	0	0	0	0	0	1	0	0	1	1	0	1	1	1	1	\rangle.$
\end{example}

\begin{example}\label{43}
Three-uniform states of the systems $4^5\times2^2$, $4^4\times2^4$, $4^3\times2^6$, $4^2\times2^8$ and $4^1\times2^{10}$.

$|\phi_{4^52^2}\rangle=
|0	0	0	0	0	0 0\rangle+
|0	1	2	2	0	0 1\rangle+
|0	2	3	3	0	1 0\rangle+
|0	3	1	1	0	1 1\rangle+
|0	2	1	2	1	0 0\rangle+
|0	3	3	0	1	0 1\rangle+
|0	0	2	1	1	1 0\rangle+
|0	1	0	3	1	1 1\rangle+
|0	3	2	3	2	0 0\rangle+
|0	2	0	1	2	0 1\rangle+
|0	1	1	0	2	1 0\rangle+
|0	0	3	2	2	1 1\rangle+
|0	1	3	1	3	0 0\rangle+
|0	0	1	3	3	0 1\rangle+
|0	3	0	2	3	1 0\rangle+
|0	2	2	0	3	1 1\rangle+
|1	1	1	1	1	0 1\rangle+
|1	0	3	3	1	0 0\rangle+
|1	3	2	2	1	1 1\rangle+
|1	2	0	0	1	1 0\rangle+
|1	3	0	3	0	0 1\rangle+
|1	2	2	1	0	0 0\rangle+
|1	1	3	0	0	1 1\rangle+
|1	0	1	2	0	1 0\rangle+
|1	2	3	2	3	0 1\rangle+
|1	3	1	0	3	0 0\rangle+
|1	0	0	1	3	1 1\rangle+
|1	1	2	3	3	1 0\rangle+
|1	0	2	0	2	0 1\rangle+
|1	1	0	2	2	0 0\rangle+
|1	2	1	3	2	1 1\rangle+
|1	3	3	1	2	1 0\rangle+
|2	2	2	2	2	1 0\rangle+
|2	3	0	0	2	1 1\rangle+
|2	0	1	1	2	0 0\rangle+
|2	1	3	3	2	0 1\rangle+
|2	0	3	0	3	1 0\rangle+
|2	1	1	2	3	1 1\rangle+
|2	2	0	3	3	0 0\rangle+
|2	3	2	1	3	0 1\rangle+
|2	1	0	1	0	1 0\rangle+
|2	0	2	3	0	1 1\rangle+
|2	3	3	2	0	0 0\rangle+
|2	2	1	0	0	0 1\rangle+
|2	3	1	3	1	1 0\rangle+
|2	2	3	1	1	1 1\rangle+
|2	1	2	0	1	0 0\rangle+
|2	0	0	2	1	0 1\rangle+
|3	3	3	3	3	1 1\rangle+
|3	2	1	1	3	1 0\rangle+
|3	1	0	0	3	0 1\rangle+
|3	0	2	2	3	0 0\rangle+
|3	1	2	1	2	1 1\rangle+
|3	0	0	3	2	1 0\rangle+
|3	3	1	2	2	0 1\rangle+
|3	2	3	0	2	0 0\rangle+
|3	0	1	0	1	1 1\rangle+
|3	1	3	2	1	1 0\rangle+
|3	2	2	3	1	0 1\rangle+
|3	3	0	1	1	0 0\rangle+
|3	2	0	2	0	1 1\rangle+
|3	3	2	0	0	1 0\rangle+
|3	0	3	1	0	0 1\rangle+
|3	1	1	3	0	0 0\rangle$.

$|\phi_{4^42^4}\rangle=
|0	0	0	0	0   0	0   0\rangle+
|0	1	2	2	0   0	0   1\rangle+
|0	2	3	3	0   0	1   0\rangle+
|0	3	1	1	0   0	1   1\rangle+
|0	2	1	2	0   1	0   0\rangle+
|0	3	3	0	0   1	0   1\rangle+
|0	0	2	1	0   1	1   0\rangle+
|0	1	0	3	0   1	1   1\rangle+
|0	3	2	3	1   0	0   0\rangle+
|0	2	0	1	1   0	0   1\rangle+
|0	1	1	0	1   0	1   0\rangle+
|0	0	3	2	1   0	1   1\rangle+
|0	1	3	1	1   1	0   0\rangle+
|0	0	1	3	1   1	0   1\rangle+
|0	3	0	2	1   1	1   0\rangle+
|0	2	2	0	1   1	1   1\rangle+
|1	1	1	1	0   1	0   1\rangle+
|1	0	3	3	0   1	0   0\rangle+
|1	3	2	2	0   1	1   1\rangle+
|1	2	0	0	0   1	1   0\rangle+
|1	3	0	3	0   0	0   1\rangle+
|1	2	2	1	0   0	0   0\rangle+
|1	1	3	0	0   0	1   1\rangle+
|1	0	1	2	0   0	1   0\rangle+
|1	2	3	2	1   1	0   1\rangle+
|1	3	1	0	1   1	0   0\rangle+
|1	0	0	1	1   1	1   1\rangle+
|1	1	2	3	1   1	1   0\rangle+
|1	0	2	0	1   0	0   1\rangle+
|1	1	0	2	1   0	0   0\rangle+
|1	2	1	3	1   0	1   1\rangle+
|1	3	3	1	1   0	1   0\rangle+
|2	2	2	2	1   0	1   0\rangle+
|2	3	0	0	1   0	1   1\rangle+
|2	0	1	1	1   0	0   0\rangle+
|2	1	3	3	1   0	0   1\rangle+
|2	0	3	0	1   1	1   0\rangle+
|2	1	1	2	1   1	1   1\rangle+
|2	2	0	3	1   1	0   0\rangle+
|2	3	2	1	1   1	0   1\rangle+
|2	1	0	1	0   0	1   0\rangle+
|2	0	2	3	0   0	1   1\rangle+
|2	3	3	2	0   0	0   0\rangle+
|2	2	1	0	0   0	0   1\rangle+
|2	3	1	3	0   1	1   0\rangle+
|2	2	3	1	0   1	1   1\rangle+
|2	1	2	0	0   1	0   0\rangle+
|2	0	0	2	0   1	0   1\rangle+
|3	3	3	3	1   1	1   1\rangle+
|3	2	1	1	1   1	1   0\rangle+
|3	1	0	0	1   1	0   1\rangle+
|3	0	2	2	1   1	0   0\rangle+
|3	1	2	1	1   0	1   1\rangle+
|3	0	0	3	1   0	1   0\rangle+
|3	3	1	2	1   0	0   1\rangle+
|3	2	3	0	1   0	0   0\rangle+
|3	0	1	0	0   1	1   1\rangle+
|3	1	3	2	0   1	1   0\rangle+
|3	2	2	3	0   1	0   1\rangle+
|3	3	0	1	0   1	0   0\rangle+
|3	2	0	2	0   0	1   1\rangle+
|3	3	2	0	0   0	1   0\rangle+
|3	0	3	1	0   0	0   1\rangle+
|3	1	1	3	0   0 	0   0\rangle$.

$|\phi_{4^32^6}\rangle=
|0	0	0	0   0	0   0	0   0\rangle+
|0	1	2	1   0	0   0	0   1\rangle+
|0	2	3	1   1	0   0	1   0\rangle+
|0	3	1	0   1	0   0	1   1\rangle+
|0	2	1	1   0	0   1	0   0\rangle+
|0	3	3	0   0	0   1	0   1\rangle+
|0	0	2	0   1	0   1	1   0\rangle+
|0	1	0	1   1	0   1	1   1\rangle+
|0	3	2	1   1	1   0	0   0\rangle+
|0	2	0	0   1	1   0	0   1\rangle+
|0	1	1	0   0	1   0	1   0\rangle+
|0	0	3	1   0	1   0	1   1\rangle+
|0	1	3	0   1	1   1	0   0\rangle+
|0	0	1	1   1	1   1	0   1\rangle+
|0	3	0	1   0	1   1	1   0\rangle+
|0	2	2	0   0	1   1	1   1\rangle+
|1	1	1	0   1	0   1	0   1\rangle+
|1	0	3	1   1	0   1	0   0\rangle+
|1	3	2	1   0	0   1	1   1\rangle+
|1	2	0	0   0	0   1	1   0\rangle+
|1	3	0	1   1	0   0	0   1\rangle+
|1	2	2	0   1	0   0	0   0\rangle+
|1	1	3	0   0	0   0	1   1\rangle+
|1	0	1	1   0	0   0	1   0\rangle+
|1	2	3	1   0	1   1	0   1\rangle+
|1	3	1	0   0	1   1	0   0\rangle+
|1	0	0	0   1	1   1	1   1\rangle+
|1	1	2	1   1	1   1	1   0\rangle+
|1	0	2	0   0	1   0	0   1\rangle+
|1	1	0	1   0	1   0	0   0\rangle+
|1	2	1	1   1	1   0	1   1\rangle+
|1	3	3	0   1	1   0	1   0\rangle+
|2	2	2	1   0	1   0	1   0\rangle+
|2	3	0	0   0	1   0	1   1\rangle+
|2	0	1	0   1	1   0	0   0\rangle+
|2	1	3	1   1	1   0	0   1\rangle+
|2	0	3	0   0	1   1	1   0\rangle+
|2	1	1	1   0	1   1	1   1\rangle+
|2	2	0	1   1	1   1	0   0\rangle+
|2	3	2	0   1	1   1	0   1\rangle+
|2	1	0	0   1	0   0	1   0\rangle+
|2	0	2	1   1	0   0	1   1\rangle+
|2	3	3	1   0	0   0	0   0\rangle+
|2	2	1	0   0	0   0	0   1\rangle+
|2	3	1	1   1	0   1	1   0\rangle+
|2	2	3	0   1	0   1	1   1\rangle+
|2	1	2	0   0	0   1	0   0\rangle+
|2	0	0	1   0	0   1	0   1\rangle+
|3	3	3	1   1	1   1	1   1\rangle+
|3	2	1	0   1	1   1	1   0\rangle+
|3	1	0	0   0	1   1	0   1\rangle+
|3	0	2	1   0	1   1	0   0\rangle+
|3	1	2	0   1	1   0	1   1\rangle+
|3	0	0	1   1	1   0	1   0\rangle+
|3	3	1	1   0	1   0	0   1\rangle+
|3	2	3	0   0	1   0	0   0\rangle+
|3	0	1	0   0	0   1	1   1\rangle+
|3	1	3	1   0	0   1	1   0\rangle+
|3	2	2	1   1	0   1	0   1\rangle+
|3	3	0	0   1	0   1	0   0\rangle+
|3	2	0	1   0	0   0	1   1\rangle+
|3	3	2	0   0	0   0	1   0\rangle+
|3	0	3	0   1	0   0	0   1\rangle+
|3	1	1	1   1 	0   0 	0   0\rangle$.

$|\phi_{4^22^8}\rangle=
|0	0	0    0	0   0	0   0	0   0\rangle+
|0	1	1    0	1   0	0   0	0   1\rangle+
|0	2	1    1	1   1	0   0	1   0\rangle+
|0	3	0    1	0   1	0   0	1   1\rangle+
|0	2	0    1	1   0	0   1	0   0\rangle+
|0	3	1    1	0   0	0   1	0   1\rangle+
|0	0	1    0	0   1	0   1	1   0\rangle+
|0	1	0    0	1   1	0   1	1   1\rangle+
|0	3	1    0	1   1	1   0	0   0\rangle+
|0	2	0    0	0   1	1   0	0   1\rangle+
|0	1	0    1	0   0	1   0	1   0\rangle+
|0	0	1    1	1   0	1   0	1   1\rangle+
|0	1	1    1	0   1	1   1	0   0\rangle+
|0	0	0    1	1   1	1   1	0   1\rangle+
|0	3	0    0	1   0	1   1	1   0\rangle+
|0	2	1    0	0   0	1   1	1   1\rangle+
|1	1	0    1	0   1	0   1	0   1\rangle+
|1	0	1    1	1   1	0   1	0   0\rangle+
|1	3	1    0	1   0	0   1	1   1\rangle+
|1	2	0    0	0   0	0   1	1   0\rangle+
|1	3	0    0	1   1	0   0	0   1\rangle+
|1	2	1    0	0   1	0   0	0   0\rangle+
|1	1	1    1	0   0	0   0	1   1\rangle+
|1	0	0    1	1   0	0   0	1   0\rangle+
|1	2	1    1	1   0	1   1	0   1\rangle+
|1	3	0    1	0   0	1   1	0   0\rangle+
|1	0	0    0	0   1	1   1	1   1\rangle+
|1	1	1    0	1   1	1   1	1   0\rangle+
|1	0	1    0	0   0	1   0	0   1\rangle+
|1	1	0    0	1   0	1   0	0   0\rangle+
|1	2	0    1	1   1	1   0	1   1\rangle+
|1	3	1    1	0   1	1   0	1   0\rangle+
|2	2	1    0	1   0	1   0	1   0\rangle+
|2	3	0    0	0   0	1   0	1   1\rangle+
|2	0	0    1	0   1	1   0	0   0\rangle+
|2	1	1    1	1   1	1   0	0   1\rangle+
|2	0	1    1	0   0	1   1	1   0\rangle+
|2	1	0    1	1   0	1   1	1   1\rangle+
|2	2	0    0	1   1	1   1	0   0\rangle+
|2	3	1    0	0   1	1   1	0   1\rangle+
|2	1	0    0	0   1	0   0	1   0\rangle+
|2	0	1    0	1   1	0   0	1   1\rangle+
|2	3	1    1	1   0	0   0	0   0\rangle+
|2	2	0    1	0   0	0   0	0   1\rangle+
|2	3	0    1	1   1	0   1	1   0\rangle+
|2	2	1    1	0   1	0   1	1   1\rangle+
|2	1	1    0	0   0	0   1	0   0\rangle+
|2	0	0    0	1   0	0   1	0   1\rangle+
|3	3	1    1	1   1	1   1	1   1\rangle+
|3	2	0    1	0   1	1   1	1   0\rangle+
|3	1	0    0	0   0	1   1	0   1\rangle+
|3	0	1    0	1   0	1   1	0   0\rangle+
|3	1	1    0	0   1	1   0	1   1\rangle+
|3	0	0    0	1   1	1   0	1   0\rangle+
|3	3	0    1	1   0	1   0	0   1\rangle+
|3	2	1    1	0   0	1   0	0   0\rangle+
|3	0	0    1	0   0	0   1	1   1\rangle+
|3	1	1    1	1   0	0   1	1   0\rangle+
|3	2	1    0	1   1	0   1	0   1\rangle+
|3	3	0    0	0   1	0   1	0   0\rangle+
|3	2	0    0	1   0	0   0	1   1\rangle+
|3	3	1    0	0   0	0   0	1   0\rangle+
|3	0	1    1	0   1	0   0	0   1\rangle+
|3	1	0    1 	1   1 	0   0 	0   0\rangle$.

$|\phi_{4^12^{10}}\rangle=
|0	0      0	0    0	0   0	0   0	0   0\rangle+
|0	0      1	1    0	1   0	0   0	0   1\rangle+
|0	1      0	1    1	1   1	0   0	1   0\rangle+
|0	1      1	0    1	0   1	0   0	1   1\rangle+
|0	1      0	0    1	1   0	0   1	0   0\rangle+\\
|0	1      1	1    1	0   0	0   1	0   1\rangle+
|0	0      0	1    0	0   1	0   1	1   0\rangle+
|0	0      1	0    0	1   1	0   1	1   1\rangle+
|0	1      1	1    0	1   1	1   0	0   0\rangle+
|0	1      0	0    0	0   1	1   0	0   1\rangle+
|0	0      1	0    1	0   0	1   0	1   0\rangle+
|0	0      0	1    1	1   0	1   0	1   1\rangle+
|0	0      1	1    1	0   1	1   1	0   0\rangle+
|0	0      0	0    1	1   1	1   1	0   1\rangle+
|0	1      1	0    0	1   0	1   1	1   0\rangle+
|0	1      0	1    0	0   0	1   1	1   1\rangle+
|1	0      1	0    1	0   1	0   1	0   1\rangle+
|1	0      0	1    1	1   1	0   1	0   0\rangle+
|1	1      1	1    0	1   0	0   1	1   1\rangle+
|1	1      0	0    0	0   0	0   1	1   0\rangle+
|1	1      1	0    0	1   1	0   0	0   1\rangle+
|1	1      0	1    0	0   1	0   0	0   0\rangle+
|1	0      1	1    1	0   0	0   0	1   1\rangle+
|1	0      0	0    1	1   0	0   0	1   0\rangle+
|1	1      0	1    1	1   0	1   1	0   1\rangle+
|1	1      1	0    1	0   0	1   1	0   0\rangle+
|1	0      0	0    0	0   1	1   1	1   1\rangle+
|1	0      1	1    0	1   1	1   1	1   0\rangle+
|1	0      0	1    0	0   0	1   0	0   1\rangle+
|1	0      1	0    0	1   0	1   0	0   0\rangle+
|1	1      0	0    1	1   1	1   0	1   1\rangle+
|1	1      1	1    1	0   1	1   0	1   0\rangle+
|2	1      0	1    0	1   0	1   0	1   0\rangle+
|2	1      1	0    0	0   0	1   0	1   1\rangle+
|2	0      0	0    1	0   1	1   0	0   0\rangle+
|2	0      1	1    1	1   1	1   0	0   1\rangle+
|2	0      0	1    1	0   0	1   1	1   0\rangle+
|2	0      1	0    1	1   0	1   1	1   1\rangle+
|2	1      0	0    0	1   1	1   1	0   0\rangle+
|2	1      1	1    0	0   1	1   1	0   1\rangle+
|2	0      1	0    0	0   1	0   0	1   0\rangle+
|2	0      0	1    0	1   1	0   0	1   1\rangle+
|2	1      1	1    1	1   0	0   0	0   0\rangle+
|2	1      0	0    1	0   0	0   0	0   1\rangle+
|2	1      1	0    1	1   1	0   1	1   0\rangle+
|2	1      0	1    1	0   1	0   1	1   1\rangle+
|2	0      1	1    0	0   0	0   1	0   0\rangle+
|2	0      0	0    0	1   0	0   1	0   1\rangle+
|3	1      1	1    1	1   1	1   1	1   1\rangle+
|3	1      0	0    1	0   1	1   1	1   0\rangle+
|3	0      1	0    0	0   0	1   1	0   1\rangle+
|3	0      0	1    0	1   0	1   1	0   0\rangle+
|3	0      1	1    0	0   1	1   0	1   1\rangle+
|3	0      0	0    0	1   1	1   0	1   0\rangle+
|3	1      1	0    1	1   0	1   0	0   1\rangle+
|3	1      0	1    1	0   0	1   0	0   0\rangle+
|3	0      0	0    1	0   0	0   1	1   1\rangle+
|3	0      1	1    1	1   0	0   1	1   0\rangle+
|3	1      0	1    0	1   1	0   1	0   1\rangle+
|3	1      1	0    0	0   1	0   1	0   0\rangle+
|3	1      0	0    0	1   0	0   0	1   1\rangle+
|3	1      1	1    0	0   0	0   0	1   0\rangle+
|3	0      0	1    1	0   1	0   0	0   1\rangle+
|3	0      1 	0    1 	1   1 	0   0 	0   0\rangle$.
\end{example}

\begin{example}\label{12}
By Table 1 in Ref. \cite{npj}, we have an {\rm IrOA}$(r_N,N,12,3)$ for $N=8$ and every $N\geq12$. Using ${\rm MOA}(12,2,6^12^1,3)$ $=(${\rm \textbf{(6)}} $\otimes {\bf 1}_2, {\bf 1}_6\otimes$ {\rm \textbf{(2)})}, ${\rm MOA}(12,2,4^13^1,3)=(${\rm \textbf{(4)}} $\otimes {\bf 1}_3, {\bf 1}_4\otimes$ {\rm \textbf{(3)})}, ${\rm MOA}(12,3,3^12^2,3)$, and Theorem \ref{ctihuan}, there exist an {\rm IrMOA}$(r_N,N+n_1+n_2+2n_3,12^{N-(n_1+n_2+n_3)}6^{n_1}2^{n_1}4^{n_2}3^{n_2}3^{n_3}2^{2n_3},3)$ and three-uniform states of the system $12^{N-(n_1+n_2+n_3)}\times6^{n_1}\times4^{n_2}\times3^{n_2+n_3}\times2^{n_1+2n_3}$ with $1\leq n_1+n_2+n_3\leq N$ for $N=8$ and every $N\geq12$.

For the case of $N=8$ and $r_N=12^4$, we can obtain 164 IrMOAs from 164 non-negative integer solutions to the inequation $1\leq n_1+n_2+n_3\leq 8$.

For example, when $n_1+n_2+n_3=1$, we can obtain 3 IrMOAs as follows.

$n_1=0,n_2=0,n_3=1$, {\rm IrMOA}$(r_8,10,12^73^12^2,3)$,

$n_1=0,n_2=1,n_3=0$, {\rm IrMOA}$(r_8,9,12^74^13^1,3)$,

$n_1=1,n_2=0,n_3=0$, {\rm IrMOA}$(r_8,9,12^76^12^1,3)$.

When $n_1+n_2+n_3=2$, we can obtain 6 IrMOAs as follows.

$n_1=0,n_2=0,n_3=2$, {\rm IrMOA}$(r_8,12,12^63^22^4,3)$,

$n_1=0,n_2=2,n_3=0$, {\rm IrMOA}$(r_8,10,12^64^23^2,3)$,

$n_1=2,n_2=0,n_3=0$, {\rm IrMOA}$(r_8,10,12^66^22^2,3)$,

$n_1=0,n_2=1,n_3=1$, {\rm IrMOA}$(r_8,11,12^64^13^22^2,3)$,

$n_1=1,n_2=0,n_3=1$, {\rm IrMOA}$(r_8,11,12^66^13^12^3,3)$,

$n_1=1,n_2=1,n_3=0$, {\rm IrMOA}$(r_8,10,12^66^14^13^12^1,3)$.

When $n_1+n_2+n_3=3$, we can obtain 10 IrMOAs as follows.

$n_1=0,n_2=0,n_3=3$, {\rm IrMOA}$(r_8,14,12^53^32^6,3)$,

$n_1=0,n_2=3,n_3=0$, {\rm IrMOA}$(r_8,11,12^54^33^3,3)$,

$n_1=3,n_2=0,n_3=0$, {\rm IrMOA}$(r_8,11,12^56^32^3,3)$,

$n_1=0,n_2=1,n_3=2$, {\rm IrMOA}$(r_8,13,12^54^13^32^4,3)$,

$n_1=1,n_2=0,n_3=2$, {\rm IrMOA}$(r_8,13,12^56^13^22^5,3)$,

$n_1=1,n_2=2,n_3=0$, {\rm IrMOA}$(r_8,11,12^56^14^23^22^1,3)$,

$n_1=0,n_2=2,n_3=1$, {\rm IrMOA}$(r_8,12,12^54^23^32^2,3)$,

$n_1=2,n_2=0,n_3=1$, {\rm IrMOA}$(r_8,12,12^56^23^12^4,3)$,

$n_1=2,n_2=1,n_3=0$, {\rm IrMOA}$(r_8,11,12^56^24^13^12^2,3)$,

$n_1=1,n_2=1,n_3=1$, {\rm IrMOA}$(r_8,12,12^56^14^13^22^3,3)$.

When $n_1+n_2+n_3=4$, we can obtain 15 IrMOAs as follows.

$n_1=0,n_2=0,n_3=4$, {\rm IrMOA}$(r_8,16,12^43^42^8,3)$,

$n_1=0,n_2=4,n_3=0$, {\rm IrMOA}$(r_8,12,12^44^43^4,3)$,

$n_1=4,n_2=0,n_3=0$, {\rm IrMOA}$(r_8,12,12^46^42^4,3)$,

$n_1=0,n_2=1,n_3=3$, {\rm IrMOA}$(r_8,15,12^44^13^42^6,3)$,

$n_1=1,n_2=0,n_3=3$, {\rm IrMOA}$(r_8,15,12^46^13^32^7,3)$,

$n_1=1,n_2=3,n_3=0$, {\rm IrMOA}$(r_8,12,12^46^14^33^32^1,3)$,

$n_1=0,n_2=3,n_3=1$, {\rm IrMOA}$(r_8,13,12^44^33^42^2,3)$,

$n_1=3,n_2=0,n_3=1$, {\rm IrMOA}$(r_8,13,12^46^33^12^5,3)$,

$n_1=3,n_2=1,n_3=0$, {\rm IrMOA}$(r_8,12,12^46^34^13^12^3,3)$,

$n_1=0,n_2=2,n_3=2$, {\rm IrMOA}$(r_8,14,12^44^23^42^4,3)$,

$n_1=2,n_2=0,n_3=2$, {\rm IrMOA}$(r_8,14,12^46^23^22^6,3)$,

$n_1=2,n_2=2,n_3=0$, {\rm IrMOA}$(r_8,12,12^46^24^23^22^2,3)$,

$n_1=1,n_2=1,n_3=2$, {\rm IrMOA}$(r_8,14,12^46^14^13^32^5,3)$,

$n_1=1,n_2=2,n_3=1$, {\rm IrMOA}$(r_8,13,12^46^14^23^32^3,3)$,

$n_1=2,n_2=1,n_3=1$, {\rm IrMOA}$(r_8,13,12^46^24^13^22^4,3)$.

Similarly, when $n_1+n_2+n_3=5,6,7,8$, we can obtain 21, 28, 36, and 45 IrMOAs, respectively.
\end{example}

\begin{example}\label{56}
Let $k=4$, $d_{11}=7$, $d_{21}=8$ in Theorem \ref{2k}. We can obtain an {\rm IrOA}$(56^4,8,56,4)$ and a four-uniform state of eight qudits ($d=56$) from an {\rm OA}$(7^4,8,7,4)$ and an {\rm OA}$(8^4,9,8,4)$. Take $d_{12}=7$, $d_{22}=4$, $d_{32}=2$ and $d_{13}=7$, $d_{23}=d_{33}=d_{43}=2$ and $d_{14}=14$, $d_{24}=4$ and $d_{15}=14$, $d_{25}=d_{35}=2$ and $d_{16}=28$, $d_{26}=2$, and $t=6$. Then, we can obtain the following results.

When $n_1=1$ and $n_2=\ldots=n_6=0$, we obtain an {\rm IrMOA}$(56^4,9,56^78^17^1,4)$ and a four-uniform state of the system $56^7\times 7^1\times 8^1$.

When $n_2=1$ and $n_1=n_3=\ldots=n_6=0$, we obtain an {\rm IrMOA}$(56^4,10,56^77^14^12^1,4)$ and a four-uniform state of the system $56^7\times 7^1\times 4^1\times 2^1$.

When $n_1=1$, $n_2=1$, and $n_3=\ldots=n_6=0$, we obtain an {\rm IrMOA}$(56^4,11,56^68^17^24^12^1,4)$ and a four-uniform state of the system $56^6\times8^1\times7^2\times4^1\times2^1$.

The inequation $1\leq n_1+n_2+\cdots+n_6\leq 8$ has many solutions of $n_i$ for $i=1,...,6$. From these solutions, we can obtain all IrMOAs, which are provided in the Supplementary information. Then, we have the corresponding four-uniform states.
\end{example}

\section*{Appendix C: Tables}

\begin{tab}\label{23tai}
\begin{center}
Two and three-uniform states of heterogeneous systems obtained\\
from {\rm IrOA}$(r,N,d,k)$ in Table 1 of Ref. \cite{npj}
{\begin{tabular}{|l|l|l|l|l|l|l|}\hline
  $k$&$d$&$t$ & $d_{uv}$             & $N$  & $k$-uniform states of      &  $N'=N+$\\
     &   &    & $(v=1,2,\ldots,t$    &      & the system                                    & $\sum_{j=1}^t(m_j-1){n_j}$ \\
     &   &    & $u=1,2,\ldots,m_t)$  &      &$d^{N-(n_1+\cdots+n_t)}\times d_{11}^{n_1}$ &  \\
     &   &    &                      &      &$\times \cdots \times d_{m_11}^{n_1}\times \cdots \times$ &  \\
     &   &    &                      &      &$d_{1t}^{n_t}\cdots \times d_{m_tt}^{n_t}$ &\\\hline
\multirow{15}{*}{$3$}
&4&1& $d_{11}=d_{21}=2$. &$N=6$,  &$4^{N-n_1}\times2^{2n_1}$   & $N'\geq7$\\
& & &   & $N\geq8$  &   & \\\cline{2-7}
&6&1& $d_{11}=3$, $d_{21}=2$. &$N=8$, & $6^{N-n_1}\times3^{n_1}\times2^{n_1}$ & $N'\geq9$\\
& & & & $N\geq12$ & & \\\cline{2-7}
&$8$&3& $d_{11}=4$, $d_{21}=2$; &$N\geq6$&$8^{N-(n_1+n_2+n_3)}\times$ &$N'\geq7$\\
&   & &$d_{12}=d_{22}=d_{32}=2$; & & $ 4^{n_1}\times 2^{n_1+3n_2+4n_3}$ & \\
&   & &$d_{13}=\ldots=d_{43}=2$. & & $ $ &   \\\cline{2-7}
&$9$&1& $d_{11}=d_{21}=3$. &$N\geq6$   & $9^{N-n_1}\times 3^{2n_1}$ & $N'\geq7$\\\cline{2-7}
&$10$&1& $d_{11}=5$, $d_{21}=2$. &$N=8$,   & $10^{N-n_1}\times 5^{n_1}\times2^{n_1}$ & $N'\geq9$\\
& & & & $N\geq12$ & & \\\cline{2-7}
&$12$&3& $d_{11}=6$, $d_{21}=2$; &$N=8$, & $12^{N-(n_1+n_2+n_3)}$ & $N'\geq9$\\
&&&$d_{12}=4$, $d_{22}=3$; &$N\geq12$& $\times6^{n_1}\times4^{n_2}$&\\
&&&$d_{13}=3$, $d_{23}=d_{33}=2$. & & $\times3^{n_2+n_3}\times2^{n_1+2n_3}$ &\\\cline{2-7}
&$p^n$&1& $d_{11}=p^{n_1'}$, $d_{21}=p^{n_2'}$, &$N\geq6$, & $d^{N-1}d_{11}^1d_{21}^1$ & $N'\geq7$\\
&&& $n_1'+n_2'=n$ & & & \\\cline{2-7}
&$d_1d_2$ &1& $d_{11}=d_1$, $d_{21}=d_2$, &$N=8$, & $d^{N-1}d_1^1d_{2}^1$; & $N'\geq9$\\
&&&& $N\geq12$&$d^6d_{1}^2d_{2}^2$; $d^5d_{1}^3d_{2}^3$;& \\
&&&& &$d^4d_{1}^4d_{2}^4$; $d^3d_{1}^5d_{2}^5$&\\\hline
\multirow{15}{*}{$2$}
&$4$&2& $d_{11}=d_{21}=2$; &$N\geq4$ & $4^{N-(n_1+n_2)}\times $ &$N'\geq5$\\
&   & & $d_{12}=d_{22}=d_{32}=2$. &  & $2^{2n_1+3n_2}$ & \\\cline{2-7}
&$6$&1 &$d_{11}=3$, $d_{21}=2$. &$N\geq5$ & $6^{N-n_1}\times 3^{n_1} \times2^{n_1}$ &$N'\geq6$\\\cline{2-7}
&$8$&9& $d_{11}=4$, $d_{21}=2$; &$N\geq4$     & $8^{N-(n_1+n_2+\cdots+n_9)}\times $        &$N'\geq5$\\
&&&$d_{12}=4$, $d_{22}=d_{32}=2$; &     & $4^{n_1+\cdots+n_4}\times 2^x$,   & \\
&&&$d_{13}=4$, $d_{23}=d_{33}=d_{43}=2$; & & $x=n_1+2n_2+$       & \\
&&&$d_{14}=4$, $d_{24}=\ldots=d_{54}=2$; & & $3n_3+4n_4+3n_5+$      & \\
&&&$d_{15}=d_{25}=d_{35}=2$; &     & $4n_6+5n_7+6n_8+$   &  \\
&&&$d_{16}=\ldots=d_{46}=2$; &     & $7n_9$   &  \\
&&&$d_{17}=\ldots=d_{57}=2$; &     &    &     \\
&&&$d_{18}=\ldots=d_{68}=2$; &     &    &     \\
&&&$d_{19}=\cdots=d_{79}=2$.  &  &    & \\\cline{2-7}
&$9$&3& $d_{11}=d_{21}=3$; &$N\geq4$  & $9^{N-(n_1+n_2+n_3)}\times $  &$N'\geq5$\\
&&&$d_{12}=d_{22}=d_{32}=3$; &     & $3^{2n_1+3n_2+4n_3}$    &     \\
&&&$d_{13}=\ldots=d_{43}=3$. &     &    &     \\\cline{2-7}
&$10$&1& $d_{11}=5$, $d_{21}=2$. &$N=4$, & $10^{N-n_1}\times 5^{n_1}\times2^{n_1}$ &$N'\geq5$\\
& & & & $N\geq6$ & & \\\cline{2-7}
&$12$&12& $d_{11}=6$, $d_{21}=2$; &$N\geq4$ & $12^{N-(n_1+n_2+\cdots+n_{12})}$ & $N'\geq5$\\
&&&$d_{12}=6$, $d_{22}=d_{32}=2$;  & &$\times6^{n_1+n_2}\times4^{n_3}\times$  & \\
&&&$d_{13}=4$, $d_{23}=3$;  & &$3^{n_3+n_4+n_5+n_6}\times2^x$  & \\
&&&$d_{14}=3$, $d_{24}=d_{34}=2$;&        & $x=n_1+2n_2+$ & \\
&&&$d_{15}=3$, $d_{25}=\ldots=d_{45}=2$;&        & $2n_4+3n_5+4n_6+$ &  \\
&&&$d_{16}=3$, $d_{26}=\cdots=d_{56}=2$;& &$6n_7+7n_8+8n_9+$    &\\
&&&$d_{17}=\cdots=d_{67}=2$;& &$9n_{10}+10n_{11}+ 11n_{12}$    & \\
&&&$d_{18}=\cdots=d_{78}=2$;& & & \\
&&&$d_{19}=\cdots=d_{89}=2$;& &    & \\
&&&$d_{1,10}=\cdots =d_{9,10}=2$;&&&\\
&&&$d_{1,11}=\cdots =d_{10,11}=2$;&&&\\
&&&$d_{1,12}=\cdots =d_{11,12}=2$;&&&\\\cline{2-7}
&$f_1f_2$ &1& $d_{11}=f_1$, $d_{21}=f_2$, &$N\geq4$, & $d^{N-1}f_1^1f_2^1$; & $N'\geq5$\\\hline
\end{tabular}}\\
Note: In the column headed ``$d$'', $p\geq4$ is a prime, $n\neq1$, $d_1d_2$ is not a prime power and $f_1f_2\neq6$.
\end{center}
\end{tab}

\newpage

\begin{tab}\label{23juti}
\begin{center}
Resulting two and three-uniform states consisting of $N'\leq22$ \\
heterogeneous subsystems from Table \ref{23tai}
{\begin{tabular}{|l|l|l|}\hline
$k$&$N'$& $k$-uniform states\\\hline
\multirow{8}{*}{$3$}
&7& $|\phi_{4^5\times 2^2}\rangle$, $|\phi_{8^5\times 4^1\times2^1}\rangle$, $|\phi_{9^5\times 3^2}\rangle$\\\cline{2-3}
&8& $|\phi_{4^4\times 2^4}\rangle$, $|\phi_{8^4\times 4^2\times2^2}\rangle$, $|\phi_{9^4\times 3^4}\rangle$\\\cline{2-3}
&9& $|\phi_{6^7\times3^1\times2^1}\rangle$, $|\phi_{4^3\times 2^6}\rangle$, $|\phi_{4^7\times 2^2}\rangle$, $|\phi_{8^3\times 4^3\times2^3}\rangle$, $|\phi_{9^3\times 3^6}\rangle$, $|\phi_{10^7\times 5^1\times2^1}\rangle$, $|\phi_{12^7\times 4^1\times3^1}\rangle$ \\\cline{2-3}
&10& $|\phi_{6^6\times 3^2\times2^2}\rangle$, $|\phi_{4^2\times 2^8}\rangle$, $|\phi_{4^6\times 2^4}\rangle$, $|\phi_{8^2\times 4^4\times2^4}\rangle$, $|\phi_{9^2\times 3^8}\rangle$, $|\phi_{10^6\times 5^2\times2^2}\rangle$, $|\phi_{12^6\times 4^2\times3^2}\rangle$\\\cline{2-3}
&11& $|\phi_{6^5\times 3^3\times2^3}\rangle$, $|\phi_{4^1\times 2^{10}}\rangle$, $|\phi_{4^5\times 2^6}\rangle$, $|\phi_{8^1\times 4^5\times2^5}\rangle$, $|\phi_{9^1\times 3^{10}}\rangle$, $|\phi_{10^5\times 5^3\times2^3}\rangle$, $|\phi_{12^5\times 4^3\times3^3}\rangle$ \\\cline{2-3}
&12& $|\phi_{6^4\times 3^4\times2^4}\rangle$, $|\phi_{4^4\times 2^8}\rangle$, $|\phi_{10^4\times 5^4\times2^4}\rangle$, $|\phi_{12^4\times 6^4\times2^4}\rangle$, $|\phi_{12^4\times 4^4\times3^4}\rangle$, $|\phi_{12^6\times 3^2\times2^4}\rangle$ \\\cline{2-3}
&13& $|\phi_{6^{11}\times3^1\times2^1}\rangle$, $|\phi_{4^3\times 2^{10}}\rangle$, $|\phi_{6^3\times 3^5\times2^5}\rangle$, $|\phi_{10^3\times 5^5\times2^5}\rangle$, $|\phi_{12^3\times 6^5\times2^5}\rangle$, $|\phi_{12^3\times 4^5\times3^5}\rangle$ \\\cline{2-3}
&14& $|\phi_{6^{12}\times3^1\times2^1}\rangle$, $|\phi_{4^2\times 2^{12}}\rangle$, $|\phi_{6^2\times 3^6\times2^6}\rangle$, $|\phi_{10^2\times 5^6\times2^6}\rangle$, $|\phi_{12^2\times 6^6\times2^6}\rangle$, $|\phi_{12^2\times 4^6\times3^6}\rangle$, $|\phi_{12^5\times 3^3\times2^6}\rangle$ \\\cline{2-3}
&15& $|\phi_{6^{13}\times3^1\times2^1}\rangle$, $|\phi_{4^1\times 2^{14}}\rangle$, $|\phi_{6^1\times 3^7\times2^7}\rangle$, $|\phi_{8^3\times 2^{12}}\rangle$, $|\phi_{10^1\times 5^7\times2^7}\rangle$, $|\phi_{12^1\times 6^7\times2^7}\rangle$, $|\phi_{12^1\times 4^7\times3^7}\rangle$ \\\cline{2-3}
&16& $|\phi_{6^{14}\times3^1\times2^1}\rangle$, $|\phi_{3^8\times2^8}\rangle$, $|\phi_{5^8\times2^8}\rangle$, $|\phi_{6^8\times2^8}\rangle$, $|\phi_{4^8\times3^8}\rangle$, $|\phi_{12^4\times 3^4\times2^8}\rangle$ \\\cline{2-3}
&17& $|\phi_{6^{15}\times3^1\times2^1}\rangle$, $|\phi_{10^7\times 5^5\times2^5}\rangle$, $|\phi_{12^7\times 6^5\times2^5}\rangle$ \\\cline{2-3}
&18& $|\phi_{6^{16}\times3^1\times2^1}\rangle$, $|\phi_{10^6\times 5^6\times2^6}\rangle$, $|\phi_{12^6\times 6^6\times2^6}\rangle$, $|\phi_{12^3\times 3^5\times2^{10}}\rangle$ \\\cline{2-3}
&19& $|\phi_{6^{17}\times3^1\times2^1}\rangle$, $|\phi_{10^5\times 5^7\times2^7}\rangle$, $|\phi_{12^5\times 6^7\times2^7}\rangle$ \\\cline{2-3}
&20& $|\phi_{6^{18}\times3^1\times2^1}\rangle$, $|\phi_{10^4\times 5^8\times2^8}\rangle$, $|\phi_{12^4\times 6^8\times2^8}\rangle$, $|\phi_{12^2\times 3^6\times2^{12}}\rangle$ \\\cline{2-3}
&21& $|\phi_{6^{19}\times3^1\times2^1}\rangle$, $|\phi_{10^3\times 5^9\times2^9}\rangle$, $|\phi_{12^3\times 6^9\times2^9}\rangle$ \\\cline{2-3}
&22& $|\phi_{6^{20}\times3^1\times2^1}\rangle$, $|\phi_{10^2\times 5^{10}\times2^{10}}\rangle$, $|\phi_{12^2\times 6^{10}\times2^{10}}\rangle$, $|\phi_{12^1\times 3^7\times2^{14}}\rangle$ \\\hline
\multirow{8}{*}{$2$}
&5& $|\phi_{12^3\times 4^1\times 3^1}\rangle$, $|\phi_{10^3\times 5^1\times2^1}\rangle$ \\\cline{2-3}
&6& $|\phi_{12^4\times 4^1\times 3^1}\rangle$, $|\phi_{4^3\times 2^3}\rangle$, $|\phi_{6^4\times 3^1\times2^1}\rangle$, $|\phi_{10^2\times 5^2\times2^2}\rangle$, $|\phi_{12^3\times 6^1\times2^2}\rangle$, $|\phi_{12^2\times 4^2\times3^2}\rangle$  \\\cline{2-3}
&7& $|\phi_{12^5\times 4^1\times 3^1}\rangle$, $|\phi_{4^4\times 2^3}\rangle$, $|\phi_{6^3\times 3^2\times2^2}\rangle$, $|\phi_{9^3\times 3^4}\rangle$, $|\phi_{10^1\times 5^3\times2^3}\rangle$, $|\phi_{12^1\times 4^3\times3^3}\rangle$ \\\cline{2-3}
&8& $|\phi_{12^6\times 4^1\times 3^1}\rangle$, $|\phi_{4^2\times 2^6}\rangle$, $|\phi_{6^2\times 3^3\times2^3}\rangle$, $|\phi_{8^3\times 4^1\times2^4}\rangle$, $|\phi_{12^2\times 6^2\times2^4}\rangle$, $|\phi_{12^3\times 3^1\times2^4}\rangle$ \\\cline{2-3}
&9& $|\phi_{12^7\times 4^1\times 3^1}\rangle$, $|\phi_{4^3\times 2^6}\rangle$, $|\phi_{6^1\times 3^4\times2^4}\rangle$ \\\cline{2-3}
&10& $|\phi_{12^8\times 4^1\times 3^1}\rangle$, $|\phi_{4^1\times 2^9}\rangle$, $|\phi_{8^3\times 2^7}\rangle$, $|\phi_{9^2\times 3^8}\rangle$, $|\phi_{12^1\times 6^3\times2^6}\rangle$ \\\cline{2-3}
&11& $|\phi_{12^9\times 4^1\times 3^1}\rangle$, $|\phi_{4^2\times 2^9}\rangle$ \\\cline{2-3}
&12& $|\phi_{12^{10}\times 4^1\times 3^1}\rangle$, $|\phi_{8^2\times 4^2\times2^8}\rangle$, $|\phi_{12^2\times 3^2\times2^8}\rangle$ \\\cline{2-3}
&13& $|\phi_{12^{11}\times 4^1\times 3^1}\rangle$, $|\phi_{4^1\times 2^{12}}\rangle$, $|\phi_{9^1\times 3^{12}}\rangle$ \\\cline{2-3}
&14& $|\phi_{12^{12}\times 4^1\times 3^1}\rangle$, $|\phi_{12^3\times 2^{11}}\rangle$  \\\cline{2-3}
&15& $|\phi_{12^{13}\times 4^1\times 3^1}\rangle$, $|\phi_{6^1\times3^7\times 2^7}\rangle$, $|\phi_{10^1\times5^7\times 2^7}\rangle$ \\\cline{2-3}
&16& $|\phi_{12^{14}\times 4^1\times 3^1}\rangle$, $|\phi_{8^2\times 2^{14}}\rangle$, $|\phi_{8^1\times 4^3\times2^{12}}\rangle$, $|\phi_{12^1\times 3^3\times2^{12}}\rangle$ \\\cline{2-3}
&17& $|\phi_{12^{15}\times 4^1\times 3^1}\rangle$, $|\phi_{6^1\times3^8\times 2^8}\rangle$, $|\phi_{10^1\times5^8\times 2^8}\rangle$ \\\cline{2-3}
&18& $|\phi_{12^{16}\times 4^1\times 3^1}\rangle$, $|\phi_{3^9\times 2^9}\rangle$, $|\phi_{5^9\times 2^9}\rangle$ \\\cline{2-3}
&19& $|\phi_{12^{17}\times 4^1\times 3^1}\rangle$, $|\phi_{6^1\times3^9\times 2^9}\rangle$, $|\phi_{10^1\times5^9\times 2^9}\rangle$ \\\cline{2-3}
&20& $|\phi_{12^{18}\times 4^1\times 3^1}\rangle$, $|\phi_{3^{10}\times 2^{10}}\rangle$, $|\phi_{5^{10}\times 2^{10}}\rangle$ \\\cline{2-3}
&21& $|\phi_{12^{19}\times 4^1\times 3^1}\rangle$, $|\phi_{6^1\times3^{10}\times 2^{10}}\rangle$, $|\phi_{10^1\times5^{10}\times 2^{10}}\rangle$ \\\cline{2-3}
&22& $|\phi_{12^{20}\times 4^1\times 3^1}\rangle$, $|\phi_{8^1\times 2^{21}}\rangle$, $|\phi_{3^{11}\times 2^{11}}\rangle$, $|\phi_{5^{11}\times 2^{11}}\rangle$ \\\hline
\end{tabular}}
\end{center}
\end{tab}

\vskip 0.6cm

\begin{tab}\label{24runs}
An {\rm IrMOA}$(24,a+b+c+d+e,12^a6^b4^c3^d2^e,2)$ and the corresponding two-uniform states. Take the ${\rm MOA}(24,21,4^12^{20},2)$, ${\rm MOA}(24,13,6^14^12^{11},2))$, and ${\rm MOA}(24,15,4^13^12^{13},2)$ in \cite{warren} whose MDs are 11, 6, and 6, respectively. From Lemma \ref{Mdel}, we can obtain an {\rm IrMOA}$(24,1+e,4^12^e,2)$ for $e=12,\ldots,20$, an {\rm IrMOA}$(24,2+e,6^14^12^e,2)$ for $e=8,9,10,11$, and an {\rm IrMOA}$(24,15,4^13^12^{13},2)$. Moreover, in Theorem \ref{czgouzao}, let $A={\rm MOA}(24,13,12^12^{12},2)=[A_1,A_2]=[(12)\oplus\bm{0}_2, D(12,12,2)\oplus \bm{(2)}]$, where {\rm MD}$(A_2)=6$ and $D(12,12,2)$ is from \cite{warren}. Then, we can obtain the other IrMOAs including a special {\rm IrMOA}$(24,9,3^12^8,2)$ by replacing the 12-level column by an {\rm MOA}$(12,5,3^12^4,2)$, an {\rm MOA}$(12,2,4^13^1,2)$, and an {\rm MOA}$(12,3,6^12^2,2)$, respectively.
\begin{center}
{\begin{tabular}{|l|l|l|l|l|l|}\hline
\multicolumn{5}{|c|}{${\rm IrMOA}(24,12^a6^b4^c3^d2^e,2)$} & Two-uniform states\\\hline
$a$&$b$&$c$&$d$&$e$& $|\phi_{12^a6^b4^c3^d2^e}\rangle$\\\hline
1& & & &8, 9, 10, 11, 12 & $|\phi_{12^12^8}\rangle$, $|\phi_{12^12^9}\rangle$, $|\phi_{12^12^{10}}\rangle$, $|\phi_{12^12^{11}}\rangle$, $|\phi_{12^12^{12}}\rangle$\\\hline
 &1& & &9,10,11,12,13,14 & $|\phi_{6^12^9}\rangle$, $|\phi_{6^12^{10}}\rangle$, $|\phi_{6^12^{11}}\rangle$, $|\phi_{6^12^{12}}\rangle$, $|\phi_{6^12^{13}}\rangle$, $|\phi_{6^12^{14}}\rangle$\\\hline
 & &1& &9,10,11,\ldots,20 & $|\phi_{4^12^9}\rangle$, $|\phi_{4^12^{10}}\rangle$, $\ldots$, $|\phi_{4^12^{20}}\rangle$\\\hline
 & & &1&8,9,10,\ldots,16 & $|\phi_{3^12^8}\rangle$, $|\phi_{3^12^9}\rangle$, $|\phi_{3^12^{10}}\rangle$, $\ldots$, $|\phi_{3^12^{16}}\rangle$\\\hline
 &1&1& &8,9,10,11 & $|\phi_{6^14^12^8}\rangle$, $|\phi_{6^14^12^{9}}\rangle$, $|\phi_{6^14^12^{10}}\rangle$, $|\phi_{6^14^12^{11}}\rangle$\\\hline
 & &1&1&9,10,11,12,13 & $|\phi_{4^13^12^9}\rangle$, $|\phi_{4^13^12^{10}}\rangle$, $|\phi_{4^13^12^{11}}\rangle$, $|\phi_{4^13^12^{12}}\rangle$, $|\phi_{4^13^12^{13}}\rangle$\\\hline
\end{tabular}}
\end{center}
\end{tab}

\vskip 0.6cm

\begin{tab}\label{36runs}
Selective {\rm IrMOA}$(36,12^a9^b6^c4^d3^e2^f,2)$ and corresponding two-uniform states. By using $D(12,$ $12,3)$ in \cite{warren}, we have an {\rm MOA}$(36,12^13^{12},2)=[\bm{(12)}\oplus\bm{0}_3, D(12,12,3)\oplus\bm{(3)}]$, where {\rm MD}$(D(12,12,3)\oplus \bm{(3)})=8$. Then, by Theorem \ref{czgouzao}, we can obtain many IrMOAs by using any subarray of {\rm OA}$(12,2^{11},2)$, {\rm MOA}$(12,3^12^4,2)$, {\rm MOA}$(12,4^13^1,2)$, and {\rm MOA}$(12,6^12^2,2)$ to replace the 12-level column. By Theorem \ref{tihuan}, we can construct an {\rm IrMOA}$(36,3^52^m,2)$ for $m=8,9,10,11$ and an {\rm IrMOA}$(36,3^62^m,2)$ for $m=8,9,10,11$ from the ${\rm MOA}(36,2^93^46^2,2)$ in \cite{warren}. By Lemma \ref{Mdel} and the known MOAs of size 36 in \cite{warren}, we can obtain an {\rm IrMOA}$(36,9^12^m,2)$ for $m=13,14,15,16$ from the ${\rm MOA}(36,9^12^{16},2)$; an {\rm IrMOA}$(36,3^12^m,2)$ for $m=17,\ldots,27$ from the ${\rm MOA}(36,3^12^{27},2)$; an {\rm IrMOA}$(36,3^22^m,2)$ for $m=15,\ldots,20$ from the ${\rm MOA}(36,3^22^{20},2)$; an {\rm IrMOA}$(36,6^13^12^m,2)$ for $m=13,\ldots,18$ and an {\rm IrMOA}$(36,6^12^m,$ $2)$ for $m=14,15,\ldots,18$ from the ${\rm MOA}(36,$ $6^13^12^{18},2)$; an {\rm IrMOA}$(36,3^42^m,2)$ for $m=11,12,\ldots,16$, an {\rm IrMOA}$(36,3^32^m,2)$ for $m=12,13,\ldots,16$, an {\rm IrMOA}$(36,3^22^m,2)$ for $m=13,14,\ldots,16$, and an {\rm IrMOA}$(36,3^12^m,2)$ for $m=14,15,16$ from the ${\rm MOA}(36,$ $3^42^{16},2)$.
\begin{center}
{\begin{tabular}{|l|l|l|l|l|l|l|}\hline
\multicolumn{6}{|c|}{${\rm IrMOA}(36,12^a9^b6^c4^d3^e2^f,2)$} & Two-uniform states \\\hline
$a$&$b$&$c$&$d$&$e$&$f$& $|\phi_{12^a9^b6^c4^d3^e2^f}\rangle$\\\hline
1& & & &7,8,\ldots,12& & $|\phi_{12^13^7}\rangle$, $|\phi_{12^13^8}\rangle$, $\ldots$, $|\phi_{12^13^{12}}\rangle$ \\\hline
 &1& & & &13,14,15,16 & $|\phi_{9^12^{13}}\rangle$, $|\phi_{9^12^{14}}\rangle$, $|\phi_{9^12^{15}}\rangle$, $|\phi_{9^12^{16}}\rangle$ \\\hline
 &&1& & &14,15,\ldots,18 & $|\phi_{6^12^{14}}\rangle$, $|\phi_{6^12^{15}}\rangle$, $\ldots$, $|\phi_{6^12^{18}}\rangle$ \\\hline
 &&1& &1&13,14,\ldots,18 & $|\phi_{6^13^12^{13}}\rangle$, $|\phi_{6^13^12^{14}}\rangle$, $\ldots$, $|\phi_{6^13^12^{18}}\rangle$ \\\hline
 & &1& &7,8,\ldots,12& & $|\phi_{6^13^7}\rangle$, $|\phi_{6^13^8}\rangle$, $\ldots$, $|\phi_{6^13^{12}}\rangle$ \\\hline
 & &1& &7,8,\ldots,12&1 & $|\phi_{6^13^72^1}\rangle$, $|\phi_{6^13^82^1}\rangle$, $\ldots$, $|\phi_{6^13^{12}2^1}\rangle$ \\\hline
 & &1& &7,8,\ldots,12&2 & $|\phi_{6^13^72^2}\rangle$, $|\phi_{6^13^82^2}\rangle$, $\ldots$, $|\phi_{6^13^{12}2^2}\rangle$ \\\hline
 & & &1&7,8,\ldots,13&  & $|\phi_{4^12^7}\rangle$, $|\phi_{4^12^8}\rangle$, $\ldots$, $|\phi_{4^12^{13}}\rangle$ \\\hline
 & & & &1&14,15,\ldots,27 & $|\phi_{3^12^{14}}\rangle$, $|\phi_{3^12^{15}}\rangle$, $\ldots$, $|\phi_{3^12^{27}}\rangle$ \\\hline
 & & & &2&13,14,\ldots,20 & $|\phi_{3^22^{13}}\rangle$, $|\phi_{3^22^{14}}\rangle$, $\ldots$, $|\phi_{3^22^{20}}\rangle$ \\\hline
 & & & &3&12,13,\ldots,16 & $|\phi_{3^32^{12}}\rangle$, $|\phi_{3^32^{13}}\rangle$, $\ldots$, $|\phi_{3^32^{16}}\rangle$ \\\hline
 & & & &4&11,12,\ldots,16 & $|\phi_{3^42^{11}}\rangle$, $|\phi_{3^42^{12}}\rangle$, $\ldots$, $|\phi_{3^42^{16}}\rangle$ \\\hline
 & & & &5&8,9,10,11      & $|\phi_{3^52^{8}}\rangle$, $|\phi_{3^52^9}\rangle$, $|\phi_{3^52^{10}}\rangle$, $|\phi_{3^52^{11}}\rangle$ \\\hline
 & & & &6&8,9,10,11      & $|\phi_{3^62^{8}}\rangle$, $|\phi_{3^62^9}\rangle$, $|\phi_{3^62^{10}}\rangle$, $|\phi_{3^62^{11}}\rangle$\\\hline
 & & & &7&1,2,\ldots,11  & $|\phi_{3^72^{1}}\rangle$, $|\phi_{3^72^2}\rangle$, $\ldots$, $|\phi_{3^72^{11}}\rangle$ \\\hline
 & & & &8&1,2,\ldots,11  & $|\phi_{3^82^{1}}\rangle$, $|\phi_{3^82^2}\rangle$, $\ldots$, $|\phi_{3^82^{11}}\rangle$ \\\hline
 & & & &9&1,2,\ldots,11  & $|\phi_{3^92^{1}}\rangle$, $|\phi_{3^92^2}\rangle$, $\ldots$, $|\phi_{3^92^{11}}\rangle$ \\\hline
 & & & &10&1,2,\ldots,11 & $|\phi_{3^{10}2^{1}}\rangle$, $|\phi_{3^{10}2^2}\rangle$, $\ldots$, $|\phi_{3^{10}2^{11}}\rangle$ \\\hline
 & & & &11&1,2,\ldots,11 & $|\phi_{3^{11}2^{1}}\rangle$, $|\phi_{3^{11}2^2}\rangle$, $\ldots$, $|\phi_{3^{11}2^{11}}\rangle$ \\\hline
 & & & &12&1,2,\ldots,11 & $|\phi_{3^{12}2^{1}}\rangle$, $|\phi_{3^{12}2^2}\rangle$, $\ldots$, $|\phi_{3^{12}2^{11}}\rangle$ \\\hline
 & & & &13&1,2,3,4       & $|\phi_{3^{13}2^{1}}\rangle$, $|\phi_{3^{13}2^2}\rangle$, $|\phi_{3^{13}2^3}\rangle$, $|\phi_{3^{13}2^4}\rangle$ \\\hline
\end{tabular}}
\end{center}
\end{tab}

{\bf Note:} We can further obtain a large number of IrMOAs from the other known MOAs in \cite{warren}.

\newpage

\begin{tab}\label{72runs}
{\small \begin{center}
{\rm IrMOA}$(72,12^a6^b4^c3^d2^e,2)$ and corresponding two-uniform states\\
{\begin{tabular}{|l|l|l|l|l|l|}\hline
\multicolumn{5}{|c|}{${\rm IrMOA}(72,12^a6^b4^c3^d2^e,2)$} & Two-uniform states \\\hline
$a$ &$b$ &$c$ &$d$ &$e$ & $|\phi_{12^a6^b4^c3^d2^e}\rangle$\\\hline
1&6& & & & $|\phi_{12^16^6}\rangle$ \\\hline
1&5& &1&0,1 & $|\phi_{12^16^53^1}\rangle$, $|\phi_{12^16^53^12^1}\rangle$ \\\hline
1&5& & &0,1& $|\phi_{12^16^5}\rangle$, $|\phi_{12^16^52^1}\rangle$ \\\hline
1&4& &2&0,1,2 & $|\phi_{12^16^43^2}\rangle$, $|\phi_{12^16^43^22^1}\rangle$, $|\phi_{12^16^43^22^2}\rangle$ \\\hline
1&4& &1&0,1,2& $|\phi_{12^16^43^1}\rangle$, $|\phi_{12^16^43^12^1}\rangle$, $|\phi_{12^16^43^12^2}\rangle$ \\\hline
1&4& & &0,1,2& $|\phi_{12^16^4}\rangle$, $|\phi_{12^16^42^1}\rangle$, $|\phi_{12^16^42^2}\rangle$ \\\hline
1&3& &3&1,2,3& $|\phi_{12^16^33^32^1}\rangle$, $|\phi_{12^16^33^32^2}\rangle$, $|\phi_{12^16^33^32^3}\rangle$ \\\hline
1&3& &2&1,2,3& $|\phi_{12^16^33^22^1}\rangle$, $|\phi_{12^16^33^22^2}\rangle$, $|\phi_{12^16^33^22^3}\rangle$ \\\hline
1&3& &1&1,2,3& $|\phi_{12^16^33^12^1}\rangle$, $|\phi_{12^16^33^12^2}\rangle$, $|\phi_{12^16^33^12^3}\rangle$ \\\hline
1&2& &4&2,3,4& $|\phi_{12^16^23^42^2}\rangle$, $|\phi_{12^16^23^42^3}\rangle$, $|\phi_{12^16^23^42^4}\rangle$ \\\hline
1&2& &3&2,3,4& $|\phi_{12^16^23^32^2}\rangle$, $|\phi_{12^16^23^32^3}\rangle$, $|\phi_{12^16^23^32^4}\rangle$ \\\hline
1&2& &2&2,3,4& $|\phi_{12^16^23^22^2}\rangle$, $|\phi_{12^16^23^22^3}\rangle$, $|\phi_{12^16^23^22^4}\rangle$ \\\hline
1&1& &5&3,4,5& $|\phi_{12^16^13^52^3}\rangle$, $|\phi_{12^16^13^52^4}\rangle$, $|\phi_{12^16^13^52^5}\rangle$ \\\hline
1&1& &4&3,4,5& $|\phi_{12^16^13^42^3}\rangle$, $|\phi_{12^16^13^42^4}\rangle$, $|\phi_{12^16^13^42^5}\rangle$ \\\hline
1&1& &3&3,4,5& $|\phi_{12^16^13^32^3}\rangle$, $|\phi_{12^16^13^32^4}\rangle$, $|\phi_{12^16^13^32^5}\rangle$ \\\hline
1& & &6&4,5,6& $|\phi_{12^13^62^4}\rangle$, $|\phi_{12^13^62^5}\rangle$, $|\phi_{12^13^62^6}\rangle$ \\\hline
1& & &5&4,5,6& $|\phi_{12^13^52^4}\rangle$, $|\phi_{12^13^52^5}\rangle$, $|\phi_{12^13^52^6}\rangle$ \\\hline
1& & &4&4,5,6& $|\phi_{12^13^42^4}\rangle$, $|\phi_{12^13^42^5}\rangle$, $|\phi_{12^13^42^6}\rangle$ \\\hline
 &7& & &1,2& $|\phi_{6^72^1}\rangle$, $|\phi_{6^72^2}\rangle$ \\\hline
 &6&1&& & $|\phi_{6^64^1}\rangle$ \\\hline
 &6&1&1& & $|\phi_{6^64^13^1}\rangle$ \\\hline
 &6& &1&0,1,\ldots,4& $|\phi_{6^63^1}\rangle$, $|\phi_{6^63^12^1}\rangle$, $\ldots$, $|\phi_{6^63^12^4}\rangle$ \\\hline
 &6& & &1,2,\ldots,11& $|\phi_{6^62^1}\rangle$, $|\phi_{6^62^2}\rangle$, $\ldots$, $|\phi_{6^62^{11}}\rangle$\\\hline
 &5&1&2&0,1& $|\phi_{6^54^13^2}\rangle$, $|\phi_{6^54^13^22^1}\rangle$ \\\hline
 &5&1&1&0,1& $|\phi_{6^54^13^1}\rangle$, $|\phi_{6^54^13^12^1}\rangle$ \\\hline
 &5&1& &0,1& $|\phi_{6^54^1}\rangle$, $|\phi_{6^54^12^1}\rangle$ \\\hline
 &5& &2&0,1,2,3,4,5& $|\phi_{6^53^2}\rangle$, $|\phi_{6^53^22^1}\rangle$, $\ldots$, $|\phi_{6^53^22^5}\rangle$ \\\hline
 &5& &1&0,1,\ldots,12& $|\phi_{6^53^1}\rangle$, $|\phi_{6^53^12^1}\rangle$, $\ldots$, $|\phi_{6^53^12^{12}}\rangle$ \\\hline
 &5& & &1,2,\ldots,12& $|\phi_{6^52^1}\rangle$, $|\phi_{6^52^2}\rangle$, $\ldots$, $|\phi_{6^52^{12}}\rangle$ \\\hline
 &4&1&3&0,1,2& $|\phi_{6^44^13^3}\rangle$, $|\phi_{6^44^13^32^1}\rangle$, $|\phi_{6^44^13^32^2}\rangle$ \\\hline
 &4&1&2&0,1,2& $|\phi_{6^44^13^2}\rangle$, $|\phi_{6^44^13^22^1}\rangle$, $|\phi_{6^44^13^22^2}\rangle$ \\\hline
 &4&1&1&0,1,2& $|\phi_{6^44^13^1}\rangle$, $|\phi_{6^44^13^12^1}\rangle$, $|\phi_{6^44^13^12^2}\rangle$ \\\hline
 &4& &3&1,2,3,4,5,6& $|\phi_{6^43^32^1}\rangle$, $|\phi_{6^43^32^2}\rangle$, $\ldots$, $|\phi_{6^43^32^6}\rangle$ \\\hline
 &4& &2&1,2,\ldots,13& $|\phi_{6^43^22^1}\rangle$, $|\phi_{6^43^22^2}\rangle$, $\ldots$, $|\phi_{6^43^22^{13}}\rangle$ \\\hline
 &4& &1&1,2,\ldots,13& $|\phi_{6^43^12^1}\rangle$, $|\phi_{6^43^12^2}\rangle$, $\ldots$, $|\phi_{6^43^12^{13}}\rangle$ \\\hline
 &4& & &6,7,\ldots,13& $|\phi_{6^42^6}\rangle$, $|\phi_{6^42^7}\rangle$, $\ldots$, $|\phi_{6^42^{13}}\rangle$ \\\hline
 &3&1&4&1,2,3& $|\phi_{6^34^13^42^1}\rangle$, $|\phi_{6^34^13^42^2}\rangle$, $|\phi_{6^34^13^42^3}\rangle$ \\\hline
 &3&1&3&1,2,3& $|\phi_{6^34^13^32^1}\rangle$, $|\phi_{6^34^13^32^2}\rangle$, $|\phi_{6^34^13^32^3}\rangle$ \\\hline
 &3&1&2&1,2,3& $|\phi_{6^34^13^22^1}\rangle$, $|\phi_{6^34^13^22^2}\rangle$, $|\phi_{6^34^13^22^3}\rangle$ \\\hline
 &3& &4&2,3,4,5,6,7& $|\phi_{6^33^42^2}\rangle$, $|\phi_{6^33^42^3}\rangle$, $\ldots$, $|\phi_{6^33^42^7}\rangle$ \\\hline
 &3& &3&2,3,\ldots,14& $|\phi_{6^33^32^2}\rangle$, $|\phi_{6^33^32^3}\rangle$, $\ldots$, $|\phi_{6^33^32^{14}}\rangle$ \\\hline
 &3& &2&2,3,\ldots,14& $|\phi_{6^33^22^2}\rangle$, $|\phi_{6^33^22^3}\rangle$, $\ldots$, $|\phi_{6^33^22^{14}}\rangle$ \\\hline
 &3& &1&7,8,\ldots,14& $|\phi_{6^33^12^7}\rangle$, $|\phi_{6^33^12^8}\rangle$, $\ldots$, $|\phi_{6^33^12^{14}}\rangle$ \\\hline
 &2&1&5&2,3,4& $|\phi_{6^24^13^52^2}\rangle$, $|\phi_{6^24^13^52^3}\rangle$, $|\phi_{6^24^13^52^4}\rangle$ \\\hline
 &2&1&4&2,3,4& $|\phi_{6^24^13^42^2}\rangle$, $|\phi_{6^24^13^42^3}\rangle$, $|\phi_{6^24^13^42^4}\rangle$ \\\hline
 &2&1&3&2,3,4& $|\phi_{6^24^13^32^2}\rangle$, $|\phi_{6^24^13^32^3}\rangle$, $|\phi_{6^24^13^32^4}\rangle$ \\\hline
 &2& &5&3,4,5,6,7,8& $|\phi_{6^23^52^3}\rangle$, $|\phi_{6^23^52^4}\rangle$, $\ldots$, $|\phi_{6^23^52^8}\rangle$ \\\hline
 &2& &4&3,4,\ldots,15& $|\phi_{6^23^42^3}\rangle$, $|\phi_{6^23^42^4}\rangle$, $\ldots$, $|\phi_{6^23^42^{15}}\rangle$ \\\hline
 &2& &3&3,4,\ldots,15& $|\phi_{6^23^32^3}\rangle$, $|\phi_{6^23^32^4}\rangle$, $\ldots$, $|\phi_{6^23^32^{15}}\rangle$ \\\hline
 &2& &2&8,9,\ldots,15& $|\phi_{6^23^22^8}\rangle$, $|\phi_{6^23^22^9}\rangle$, $\ldots$, $|\phi_{6^23^22^{15}}\rangle$ \\\hline
 &1&1&6&3,4,5& $|\phi_{6^14^13^62^3}\rangle$, $|\phi_{6^14^13^62^4}\rangle$, $|\phi_{6^14^13^62^5}\rangle$ \\\hline
 &1&1&5&3,4,5& $|\phi_{6^14^13^52^3}\rangle$, $|\phi_{6^14^13^52^4}\rangle$, $|\phi_{6^14^13^52^5}\rangle$ \\\hline
 &1&1&4&3,4,5& $|\phi_{6^14^13^42^3}\rangle$, $|\phi_{6^14^13^42^4}\rangle$, $|\phi_{6^14^13^42^5}\rangle$ \\\hline
 &1& &6&4,5,6,7,8,9& $|\phi_{6^13^62^4}\rangle$, $|\phi_{6^13^62^5}\rangle$, $\ldots$, $|\phi_{6^13^62^9}\rangle$ \\\hline
 &1& &5&4,5,\ldots,16& $|\phi_{6^13^52^4}\rangle$, $|\phi_{6^13^52^5}\rangle$, $\ldots$, $|\phi_{6^13^52^{16}}\rangle$ \\\hline
 &1& &4&4,5,\ldots,16& $|\phi_{6^13^42^4}\rangle$, $|\phi_{6^13^42^5}\rangle$, $\ldots$, $|\phi_{6^13^42^{16}}\rangle$ \\\hline
 &1& &3&9,10,\ldots,16& $|\phi_{6^13^32^9}\rangle$, $|\phi_{6^13^32^{10}}\rangle$, $\ldots$, $|\phi_{6^13^32^{16}}\rangle$ \\\hline
 & &1&7&4,5,6& $|\phi_{4^13^72^4}\rangle$, $|\phi_{4^13^72^5}\rangle$, $|\phi_{4^13^72^6}\rangle$ \\\hline
 & &1&6&4,5,6& $|\phi_{4^13^62^4}\rangle$, $|\phi_{4^13^62^5}\rangle$, $|\phi_{4^13^62^6}\rangle$ \\\hline
 & &1&5&4,5,6& $|\phi_{4^13^52^4}\rangle$, $|\phi_{4^13^52^5}\rangle$, $|\phi_{4^13^52^6}\rangle$ \\\hline
 & & &7&5,6,7,8,9,10& $|\phi_{3^72^5}\rangle$, $|\phi_{3^72^6}\rangle$, $\ldots$, $|\phi_{3^72^{10}}\rangle$ \\\hline
 & & &6&5,6,\ldots,17& $|\phi_{3^62^5}\rangle$, $|\phi_{3^62^6}\rangle$, $\ldots$, $|\phi_{3^62^{17}}\rangle$ \\\hline
 & & &5&5,6,\ldots,17& $|\phi_{3^52^5}\rangle$, $|\phi_{3^52^6}\rangle$, $\ldots$, $|\phi_{3^52^{17}}\rangle$ \\\hline
 & & &4&10,11,\ldots,17& $|\phi_{3^42^{10}}\rangle$, $|\phi_{3^42^{11}}\rangle$, $\ldots$, $|\phi_{3^42^{17}}\rangle$ \\\hline
\end{tabular}}
\end{center}}
\end{tab}

The IrMOAs in Table \ref{72runs} are constructed as follows.

We can obtain $A_2={\rm OA}(72,6,6,2)=D(12,6,6)\oplus\bm{(6)}$ with MD$(A_2)=4\geq3$, where $D(12,6,6)$ is given in \cite{warren} and $\oplus$ is the Kronecker product with
multiplication replaced by the summation on the group $\mathbb{Z}_6$. Let $A={\rm MOA}(72,7,12^16^6,2)=[\bm{(12)}\oplus \bm{0}_6, D(12,6,6)\oplus\bm{(6)}]$. We have MD$(A)=5$. So it is an IrMOA$(72,7,12^16^6,2)$.

Let $s=2$ and $B_2=B_3=B_4=B_5=B_6=B_7={\rm MOA}(6,2,3^12^1,2)$. By Theorem \ref{tihuan}, we have the following four cases.

(1) Take $B_1={\rm OA}(12,11,2,2)$. Then, we have an IrMOA$(72,5+\sum_{w=3}^7u_w+v_1+v_{12}+v_{22},$ $6^{5-\sum_{w=3}^7u_w}2^{v_1}3^{v_{12}}2^{v_{22}}3^{\sum_{w=3}^7u_w}2^{\sum_{w=3}^7u_w},2)$ for $0\leq v_1\leq11$, $0\leq v_{12},v_{22},u_3,u_4,\ldots, u_7\leq1$. More precisely, we have an IrMOA$(72,6+e,6^53^12^e,2)$ for $e=0,1,\ldots,12$ and an IrMOA$(72,5+e,6^52^e,2)$ for $e=1,2,\ldots,12$; an IrMOA$(72,6+e,6^43^22^e,2)$ and an IrMOA$(72,5+e,6^43^12^e,2)$ for $e=1,2,\ldots,13$; an IrMOA$(72,6+e,6^33^32^e,2)$ and an IrMOA$(72,5+e,6^33^22^e,2)$ for $e=2,3,\ldots,14$; an IrMOA$(72,6+e,6^23^42^e,2)$ and an IrMOA$(72,5+e,6^23^32^e,2)$ for $e=3,4,\ldots,15$; an IrMOA$(72,6+e,6^13^52^e,2)$ and an IrMOA$(72,5+e,6^13^42^e,2)$ for $e=4,5,\ldots,16$; an IrMOA$(72,6+e,3^62^e,2)$ and an IrMOA$(72,5+e,3^52^e,2)$ for $e=5,6,\ldots,17$.

(2) Take $B_1={\rm MOA}(12,5,3^12^4,2)$. Then, we have an IrMOA$(72,5+\sum_{w=3}^7u_w+v_{11}+v_{21}+v_{12}+v_{22},6^{5-\sum_{w=3}^7u_w}3^{v_{11}+v_{12}}2^{v_{21}+v_{22}}3^{\sum_{w=3}^7u_w}2^{\sum_{w=3}^7u_w},2)$
for $0\leq v_{11}\leq1$, $0\leq v_{21}\leq4$ and $0\leq v_{12},v_{22},u_3,u_4,$ $\ldots, u_7\leq1$. For details, we have an IrMOA$(72,7+e,6^53^22^e,2)$ for $e=0,1,\ldots,5$, an IrMOA$(72,7+e,6^43^32^e,2)$ for $e=1,2,\ldots,6$, an IrMOA$(72,7+e,6^33^42^e,2)$ for $e=2,3,\ldots,7$, an IrMOA$(72,7+e,6^23^52^e,2)$ for $e=3,4,\ldots,8$, an IrMOA$(72,7+e,6^13^62^e,2)$ for $e=4,5,\ldots,9$, and an IrMOA$(72,7+e,3^72^e,2)$ for $e=5,6,\ldots,10$.

(3) Let $B_1={\rm MOA}(12,2,4^13^1,2)$. Consequently, we have an IrMOA$(72,5+\sum_{w=3}^7u_w+v_{11}+v_{21}+v_{12}+v_{22},6^{5-\sum_{w=3}^7u_w}4^{v_{11}}3^{v_{21}+v_{12}}$ $2^{v_{22}}3^{\sum_{w=3}^7u_w}2^{\sum_{w=3}^7u_w},2)$
for $0\leq v_{11},v_{21},v_{12},v_{22},u_3,u_4,\ldots, u_7\leq1$. Then, the arrays IrMOA$(72,8+e,6^54^13^22^e,2)$, IrMOA$(72,7+e,6^54^13^12^e,2)$, and IrMOA$(72,6+e,6^54^12^e,2)$ can be obtained for $e=0,1$.

(4) Let $B_1=B_2={\rm MOA}(6,2,3^12^1,2)$. Then, we have an IrMOA$(72,5+\sum_{w=4}^7u_w+v_{11}+v_{21}+v_{12}+v_{22},12^16^{4-\sum_{w=4}^7u_w}3^{v_{11}+v_{12}}$ $2^{v_{21}+v_{22}}3^{\sum_{w=4}^7u_w}2^{\sum_{w=4}^7u_w},2)$
for $0\leq v_{11},v_{21},v_{12},v_{22},u_3,u_4,\ldots, u_7\leq1$. Thus, we have an IrMOA$(72,7+e,12^16^43^22^e,2)$, an IrMOA$(72,6+e,12^16^43^12^e,2)$, an IrMOA$(72,5+e,12^16^42^e,2)$ for $e=0,1,2$; an IrMOA$(72,7+e,12^16^33^32^e,2)$, an IrMOA$(72,6+e,12^16^33^22^e,2)$, an IrMOA$(72,5+e,12^16^33^12^e,2)$ for $e=1,2,3$; an IrMOA$(72,7+e,12^16^23^42^e,2)$, an IrMOA$(72,6+e,12^16^23^32^e,2)$, an IrMOA$(72,5+e,12^16^23^22^e,2)$ for $e=2,3,4$; an IrMOA$(72,7+e,12^16^13^52^e,2)$, an IrMOA$(72,6+e,12^16^13^42^e,2)$, an IrMOA$(72,5+e,12^16^13^32^e,2)$ for $e=3,4,5$; an IrMOA$(72,7+e,12^13^62^e,2)$, an IrMOA$(72,6+e,12^13^52^e,2)$, an IrMOA$(72,5+e,12^13^42^e,2)$ for $e=4,5,6$.

Replacing the $12^1$ by $4^13^1$, we can obtain an IrMOA$(72,6+\sum_{w=4}^7u_w+v_{11}+v_{21}+v_{12}+v_{22},$ $4^13^16^{4-\sum_{w=4}^7u_w}3^{v_{11}+v_{12}} 2^{v_{21}+v_{22}}3^{\sum_{w=4}^7u_w}$ $2^{\sum_{w=4}^7u_w},2)$ for $0\leq v_{11},v_{21},v_{12},v_{22},u_3,u_4,\ldots, u_7\leq1$. Then, we have an IrMOA$(72,8+e,6^44^13^32^e,2)$, an IrMOA$(72,7+e,6^44^13^22^e,2)$ and an IrMOA$(72,6+e,6^44^13^12^e,2)$ for $e=0,1,2$; an IrMOA$(72,8+e,6^34^13^42^e,2)$, an IrMOA$(72,7+e,6^34^13^32^e,2)$ and an IrMOA$(72,6+e,6^34^13^22^e,2)$ for $e=1,2,3$; an IrMOA$(72,8+e,6^24^13^52^e,2)$, an IrMOA$(72,7+e,6^24^13^42^e,2)$ and an IrMOA$(72,6+e,6^24^13^32^e,2)$ for $e=2,3,4$; an IrMOA$(72,8+e,6^14^13^62^e,2)$, an IrMOA$(72,7+e,6^14^13^52^e,2)$ and an IrMOA$(72,6+e,6^14^13^42^e,2)$ for $e=3,4,5$; an IrMOA$(72,8+e,4^13^72^e,2)$, an IrMOA$(72,7+e,4^13^62^e,2)$ and an IrMOA$(72,6+e,4^13^52^e,2)$ for $e=4,5,6$.

Similarly, replacing the $12^1$ by $2^u$ for $6\leq u\leq11$, we can obtain an IrMOA$(72,4+u+\sum_{w=4}^7u_w+v_{11}+v_{21}+v_{12}+v_{22},2^{u}6^{4-\sum_{w=4}^7u_w}3^{v_{11}+v_{12}}2^{v_{21}+v_{22}}3^{\sum_{w=4}^7u_w}$ $2^{\sum_{w=4}^7u_w},2)$ for $6\leq u\leq11$, $0\leq v_{11},v_{21},v_{12},v_{22},$ $u_3,u_4,\ldots, u_7\leq1$. Then, we have an IrMOA$(72,4+e,6^42^e,2)$ for $e=6,7,\ldots,13$; an IrMOA$(72,4+e,6^33^12^e,2)$ for $e=7,8,\ldots,14$; an IrMOA$(72,4+e,6^23^22^e,2)$ for $e=8,9,\ldots,15$; an IrMOA$(72,4+e,6^13^32^e,2)$ for $e=9,10,\ldots,16$; an IrMOA$(72,4+e,3^42^e,2)$ for $e=10,11,\ldots,17$.

Let $s=1$. We have the following two cases.

(1) Take $B_1={\rm OA}(12,11,2,2)$, ${\rm MOA}(12,5,3^12^4,2)$, ${\rm MOA}(12,2,4^13^1,2)$ and ${\rm MOA}(12,3,6^12^2,2)$, respectively. We have an IrMOA$(72,6+e,6^62^e,2)$ for $e=1,2,\ldots,11$; an IrMOA$(72,7+e,6^63^12^e,2)$ for $e=0,1,2,3,4$; an IrMOA$(72,7+e,6^64^13^e,2)$ for $e=0,1$; and an IrMOA$(72,7+e,6^72^e,2)$ for $e=1,2$.

(2) By taking $B_1$ to be an {\rm MOA}$(6,2,3^12^1,2)$ we find an IrMOA$(72,7+e,12^16^53^12^e,2)$ and an IrMOA$(72,6+e,12^16^52^e,$ $2)$ for $e=0,1$.  \hfill $\blacksquare$


\begin{thebibliography}{99}
\bibitem{goy18} D. Goyeneche, Z. Raissi, S. D. Martino, and K. \.{Z}yczkowski, Entanglement and quantum combinatorial designs 2018 Phys. Rev. A {\bf 97} 062326
\bibitem{fei19}M. Li, Z. Wang, J. Wang, S. Shen, and S. Fei, The norms of Bloch vectors and classification of four-qudits quantum states 2019 Euro. Phys. Lett. {\bf 129} 20006
\bibitem{wangyl15}Y. Wang, M. Li, Z. Zheng, and S. Fei, Nonlocality of orthogonal product-basis quantum states 2015 Phys. Rev. A {\bf 92} 032313
\bibitem{wangyl14}Y. Wang, M. Li, Z. Zheng, and S. Fei, Unextendible maximally entangled bases in $\mathbb{C}^d\otimes \mathbb{C}^d$ 2014 Phys. Rev. A {\bf 90} 034301
\bibitem{goy14} D. Goyeneche and K. \.{Z}yczkowski, Genuinely multipartite entangled states and orthogonal arrays 2014 Phys. Rev. A {\bf 90} 022316
\bibitem{arn13} L. Arnaud and N. J. Cerf, Exploring pure quantum states with maximally mixed reductions 2013 Phys. Rev. A {\bf 87} 012319
\bibitem{sco04} A. J. Scott, Multipartite entanglement, quantum-error-correcting codes, and entangling power of quantum evolutions 2004 Phys. Rev. A {\bf 69} 052330
\bibitem{bou97} D. Bouwmeester, J. Pan, K. Mattle, M. Eibl, H. Weinfurter, and A. Zeilinger, Experimental quantum teleportation 1997 Nature {\bf 390} 575-579
\bibitem{zhao04} Z. Zhao, Y. Chen, A. Zhang, T. Yang, H. Briegel, and J. Pan, Experimental demonstration of five-photon entanglement and open-destination teleportation 2004 Nature {\bf 430} 54-58
\bibitem{rie04} M. Riebe, H. H\"affner, C. F. Roos, W. H\"ansel, J. Benhelm, G. P. T. Lancaster, T. W. K\"orber, C. Becher, F. Schmidt-Kaler, D. F. V. James, and R. Blatt, Deterministic quantum teleportation with atoms 2004 Nature {\bf 429} 734-737
\bibitem{roos04} C. F. Roos, M. Riebe, H. H\"affner, W. H\"ansel1, J. Benhelm, G. P. T. Lancaster, C. Becher, F. Schmidt-Kaler, and R. Blatt, Control and measurement of three-qubit entangled states 2004 Science {\bf 304} 1478-1480
\bibitem{lo12} H. K. Lo, M. Curty, and B. Qi, Measurement-device-independent quantum key distribution 2012 Phys. Rev. Lett. {\bf 108} 130503
\bibitem{jozsa03} R. Jozsa and N. Linden, On the role of entanglement in quantumcomputational speed-up 2003 Proc. R. Soc. A {\bf 459} 2011-2032
\bibitem{hel12} W. Helwig, W. Cui, J. I. Latorre, A. Riera, and H. K. Lo, Absolute maximal entanglement and quantum secret sharing 2012 Phys. Rev. A {\bf 86} 052335
\bibitem{gao18} J. Gao, L. Qiao, Z. Jiao, Y. Ma, C. Hu, R. Ren, A. Yang, H. Tang, M. Yung, and X. Jin, Experimental machine learning of quantum states 2018 Phys. Rev. Lett. {\bf 120} 240501
\bibitem{song17} C. Song, K. Xu, W. Liu, C. Yang, S. Zheng, H. Deng, Q. Xie, K. Huang, Q. Guo, L. Zhang et al., 10-qubit entanglement and parallel logic operations with a superconducting circuit 2017 Phys. Rev. Lett. {\bf 119} 180511
\bibitem{huber17} F. Huber, O. G\"uhne, and J. Siewert, Absolutely maximally entangled states of seven qubits do not exist 2017 Phys. Rev. Lett. {\bf 118} 200502
\bibitem{wang18} X. Wang, Y. Luo, H. Huang, M. Chen, Z. Su, C. Liu, C. Chen, W. Li, Y. Fang, X. Jiang et al., 18-qubit entanglement with six photons' three degrees of freedom 2018 Phys. Rev. Lett. {\bf 120} 260502
\bibitem{zha2013} X. Zha, C. Yuan, and Y. Zhang, Generalized criterion for a maximally multi-qubit entangled state 2013 Laser Phys. Lett. {\bf 10} 045201
\bibitem{goy16} D. Goyeneche, J. Bielawski, and K. \.{Z}yczkowski, Multipartite entanglement in heterogeneous systems 2016 Phys. Rev. A {\bf 94} 012346
\bibitem{goy15} D. Goyeneche, D. Alsina, J. Latorre, A. Riera, and K. \.{Z}yczkowski, Absolutely maximally entangled states, combinatorial designs, and multiunitary matrices 2015 Phys. Rev. A {\bf 92} 032316
\bibitem{npj} S. Pang, X. Zhang, X. Lin, and Q. Zhang, Two and three-uniform states from irredundant orthogonal arrays 2019 npj Quantum Inf. {\bf 5} 52
\bibitem{feng17} K. Feng, L. Jin, C. Xing, and C. Yuan, Multipartite entangled states, symmetric matrices, and error-correcting codes 2017 IEEE Trans. Inf. Theory {\bf 63} 5618-5627
\bibitem{limao19} M. Li and Y. Wang, $k$-uniform quantum states arising from orthogonal arrays 2019 Phys. Rev. A {\bf 99} 042332
\bibitem{chen13} G. Chen, L. Ji and J. Lei, The existence of mixed orthogonal arrays with four and five factors of strength two 2014 J. Combin. Des. {\bf 22} 323-342
\bibitem{hss} A. S. Hedayat, N. J. A. Sloane, and J. Stufken, Orthogonal Arrays: Theory and Applications (Springer-Verlag, New York, 1999)
\bibitem{jpa} S. Pang, X. Zhang, J. Du and T. Wang, Multipartite entanglement states of higher uniformity 2021 J. Phys. A: Math. Theor. {\bf 54} 015305
\bibitem{helwig} W. Helwig, Absolutely maximally entangled qudit graph states,
arXiv:1306.2879
\bibitem{yucs07} C. Yu, H. Song, and Y. Wang, Genuine tripartite entanglement semi-monotone for $2\times2\times n$-dimensional systems 2007 Quantum Inf. Comput. {\bf 7} 584-593
\bibitem{yucs08} C. Yu, L. Zhou, and H. Song, Genuine tripartite entanglement monotone of $2\times2\times n$-dimensional systems 2008 Phys. Rev. A {\bf 77} 022313
\bibitem{miya04} A. Miyake and F. Verstraete, Multipartite entanglement in $2\times2\times N$ quantum systems 2004 Phys. Rev. A {\bf 69} 012101
\bibitem{chen06} L. Chen and Y. Chen, Range criterion and classification of true entanglement in a $2\times M \times N$ system 2006 Phys. Rev. A {\bf 73} 052310
\bibitem{chenl06} L. Chen, Y. Chen, and Y. Mei, Classification of multipartite entanglement containing infinitely many kinds of states 2006 Phys. Rev. A {\bf 74} 052331
\bibitem{wangs13} S. Wang, Y. Lu, and G. Long, Entanglement classification of $2\times2\times2\times d$ quantum systems via the ranks of the multiple coefficient matrices 2013 Phys. Rev. A {\bf 87} 062305
\bibitem{cerf02} N. J. Cerf, M. Bourennane, A. Karlsson, and N. Gisin, Security
of quantum key distribution using $d$-level systems 2002 Phys. Rev. Lett. {\bf 88} 127902
\bibitem{fuji03} M. Fujiwara, M. Takeoka, J. Mizuno, and M. Sasaki, Exceeding
the classical capacity limit in a quantum optical channel 2003 Phys. Rev. Lett. {\bf 90} 167906
\bibitem{ralph07} T. C. Ralph, K. J. Resch, and A. Gilchrist, Efficient toffoli gates using qudits 2007 Phys. Rev. A {\bf 75} 022313
\bibitem{malik16} M. Malik, M. Erhard, M. Huber, M. Krenn, R. Fickler, and A. Zeilinger, Multi-photon entanglement in high dimensions 2016 Nat. Photonics {\bf 10} 248-252
\bibitem{xiao13} X. Xiao and Y. Li, Protecting qutrit-qutrit entanglement by weak measurement and reversal 2013 Eur. Phys. J. D {\bf 67} 1-7
\bibitem{xiao14} X. Xiao, Protecting qubit-qutrit entanglement from amplitude damping decoherence via weak measurement and reversal 2014 Phys. Scr. {\bf 89} 065102
\bibitem{wer01} R. F. Werner, All teleportation and dense coding schemes 2001 J. Phys. A {\bf 34} 7081-7094
\bibitem{it06} M. Rotteler and P. Wocjan, Equivalence of decoupling schemes and orthogonal arrays 2006 IEEE Trans. Inform. Theory {\bf 52} 4171-4181
\bibitem{pang18is} S. Pang, X. Wang, J. Wang, J. Du, and M. Feng, Construction and count of 1-resilient rotation symmetric Boolean functions 2018 Inf. Sci. {\bf 450} 336-342
\bibitem{h97} A. S. Hedayat, E. Seiden, and J. Stufken, On the maximal number of factors and the enumeration of 3-symbol orthogonal arrays of strength 3 and index 2 1997 J. Stat. Plann. Inference {\bf 58} 43-63
\bibitem{ss} Y. Zhang, Y. Lu, and S. Pang, Orthogonal arrays obtained by orthogonal decomposition of projection matrices 1999 Stat. Sin. {\bf 9} 595-604
\bibitem{zhang01} Y. Zhang, S. Pang, and Y. Wang, Orthogonal arrays obtained by the generalized Hadamard product 2001 Discrete Math. {\bf 238} 151-170
\bibitem{pang17} S. Pang, R. Yan, and S. Li, Schematic saturated orthogonal arrays obtained by using the contractive replacement method 2017 Comm. Stat. Theory Methods {\bf 46} 8913-8924
\bibitem{yin10} L. Ji and J. Yin, Constructions of new orthogonal arrays and covering arrays of strength three 2010 J. Combin. Theory Ser. A {\bf 117} 236-247
\bibitem{yin11} J. Yin, J. Wang, L. Ji, and Y. Li, On the existence of orthogonal arrays $OA(3, 5, 4n + 2)$ 2011 J. Combin. Theory Ser. A {\bf 118} 270-276
\bibitem{pang15} S. Pang, Y. Zhu, and Y. Wang, A class of mixed orthogonal arrays obtained from projection matrix inequalities 2015 J. Inequal. Appl. {\bf 2015} 1-9
\bibitem{pang17amas} S. Pang and L. Chen, Generalized Latin matrix and construction of orthogonal arrays 2017 Acta Math. Appl. Sin. {\bf 33} 1083-1092
\bibitem{pang18l} S. Pang, X. Lin, and J. Wang, Construction of asymmetric orthogonal arrays of strength $t$ from orthogonal partition of small orthogonal arrays 2018 IEICE Trans. Fundam. {\bf E101-A} 1267-1272
\bibitem{pang17xu} S. Pang, W. Xu, G. Chen, and Y. Wang, Construction of symmetric and asymmetric orthogonal arrays of strength $t$ from orthogonal partition 2018 Indian J. Pure Appl. Math. {\bf 49} 663-669
\bibitem{cstm} S. Pang, X. Zhang, and Q. Zhang, The Hamming distances of saturated asymmetrical orthogonal arrays with strength 2 2019 Comm. Statist. Theory Methods {\bf 49} 3895-3910
\bibitem{warren} W. F. Kuhfeld, Orthogonal arrays. http://support.sas.com/techsup/technote/ts723.html. Accessed 10 June 2019.
\bibitem{sloane} N. J. A. Sloane, A Library of Orthogonal Arrays. http://neilsloane.com/oadir/index.html. Accessed 16 June 2019.
\bibitem{h4} R. Horodecki, P. Horodecki, M. Horodecki, K. Horodecki, Quantum entanglement 2009 Rev. Mod. Phys. {\bf 81} 865
\bibitem{pasta} F. Pastawski, B. Yoshida, D. Harlow, and J. Preskill, Holographic quantum error-correcting codes: Toy models for the bulk/boundary correspondence 2015 J. High Energy Phys. {\bf 6} 149
\bibitem{zyl09} Y. Zhang, On schematic orthogonal arrays of strength two 2009 Ars. Combin. {\bf 91} 147-163
\bibitem{hast90} J. Hastad, Tensor rank is NP-complete 1990 J. Algorithms {\bf 11} 644-654
\bibitem{hm01} M. Horodecki, Entanglement measures 2001 Quantum Inf. Comput. {\bf 1} 3-26
\bibitem{wangz13}Z. Wang, S. Yu, H. Fan, and C. H. Oh, Quantum error-correcting codes over mixed alphabets 2013 Phys. Rev. A {\bf 88} 022328
\bibitem{hjg} A. S. Hedayat, J. Stufken, and G. Su, On difference schemes and orthogonal arrays of strength $t$ 1996 J. Stat. Plann. Inference {\bf 56} 307-324
\bibitem{chen17} G. Chen and J. Lei, Constructions of mixed orthogonal arrays of strength three 2017 Sci. Sin. Math. {\bf 47} 545-564 (In Chinese).
\bibitem{bushb} K. A. Bush, Orthogonal arrays of index unity 1952 Ann. Math. Stat. {\bf 23} 426-434
\end{thebibliography}
\end{document}


\begin{center}
{\Large \bf Supplementary information for ``Quantum $k$-uniform states for heterogeneous systems from irredundant mixed orthogonal arrays''}
\end{center}

IrMOAs obtained in Example 6.

{\small \begin{multicols}{3}
\noindent IrMOA$(56^4,   9,  56^7 28^1             2^1,  4)$,\\
IrMOA$(56^4,  10,  56^7      14^1                  2^2,  4)$,\\
IrMOA$(56^4,   9,  56^7      14^1            4^1      ,  4)$,\\
IrMOA$(56^4,  11,  56^7                 7^1        2^3,  4)$,\\
IrMOA$(56^4,  10,  56^7                 7^1  4^1   2^1,  4)$,\\
IrMOA$(56^4,   9,  56^7            8^1  7^1           ,  4)$,\\
IrMOA$(56^4,  10,  56^6 28^2                       2^2,  4)$,\\
IrMOA$(56^4,  11,  56^6 28^1 14^1                  2^3,  4)$,\\
IrMOA$(56^4,  12,  56^6      14^2                  2^4,  4)$,\\
IrMOA$(56^4,  10,  56^6 28^1 14^1            4^1   2^1,  4)$,\\
IrMOA$(56^4,  11,  56^6      14^2            4^1   2^2,  4)$,\\
IrMOA$(56^4,  10,  56^6      14^2            4^2      ,  4)$,\\
IrMOA$(56^4,  12,  56^6 28^1            7^1        2^4,  4)$,\\
IrMOA$(56^4,  13,  56^6      14^1       7^1        2^5,  4)$,\\
IrMOA$(56^4,  12,  56^6      14^1       7^1  4^1   2^3,  4)$,\\
IrMOA$(56^4,  14,  56^6                 7^2        2^6,  4)$,\\
IrMOA$(56^4,  11,  56^6 28^1            7^1  4^1   2^2,  4)$,\\
IrMOA$(56^4,  12,  56^6      14^1       7^1  4^1   2^3,  4)$,\\
IrMOA$(56^4,  11,  56^6      14^1       7^1  4^2   2^1,  4)$,\\
IrMOA$(56^4,  13,  56^6                 7^2  4^1   2^4,  4)$,\\
IrMOA$(56^4,  12,  56^6                 7^2  4^2   2^2,  4)$,\\
IrMOA$(56^4,  10,  56^6 28^1       8^1  7^1        2^1,  4)$,\\
IrMOA$(56^4,  11,  56^6      14^1  8^1  7^1        2^2,  4)$,\\
IrMOA$(56^4,  10,  56^6      14^1  8^1  7^1  4^1      ,  4)$,\\
IrMOA$(56^4,  12,  56^6            8^1  7^2        2^3,  4)$,\\
IrMOA$(56^4,  11,  56^6            8^1  7^2  4^1   2^1,  4)$,\\
IrMOA$(56^4,  10,  56^6            8^2  7^2           ,  4)$,\\
IrMOA$(56^4,  11,  56^5 28^3                       2^3,  4)$,\\
IrMOA$(56^4,  12,  56^5 28^2 14^1                  2^4,  4)$,\\
IrMOA$(56^4,  13,  56^5 28^1 14^2                  2^5,  4)$,\\
IrMOA$(56^4,  14,  56^5      14^3                  2^6,  4)$,\\
IrMOA$(56^4,  11,  56^5 28^2 14^1            4^1   2^2,  4)$,\\
IrMOA$(56^4,  12,  56^5 28^1 14^2            4^1   2^3,  4)$,\\
IrMOA$(56^4,  13,  56^5      14^3            4^1   2^4,  4)$,\\
IrMOA$(56^4,  11,  56^5 28^1 14^2            4^2   2^1,  4)$,\\
IrMOA$(56^4,  12,  56^5      14^3            4^2   2^2,  4)$,\\
IrMOA$(56^4,  11,  56^5      14^3            4^3      ,  4)$,\\
IrMOA$(56^4,  13,  56^5 28^2            7^1        2^5,  4)$,\\
IrMOA$(56^4,  14,  56^5 28^1 14^1       7^1        2^6,  4)$,\\
IrMOA$(56^4,  15,  56^5      14^2       7^1        2^7,  4)$,\\
IrMOA$(56^4,  13,  56^5 28^1 14^1       7^1  4^1   2^4,  4)$,\\
IrMOA$(56^4,  14,  56^5      14^2       7^1  4^1   2^5,  4)$,\\
IrMOA$(56^4,  13,  56^5      14^2       7^1  4^2   2^3,  4)$,\\
IrMOA$(56^4,  15,  56^5 28^1            7^2        2^7,  4)$,\\
IrMOA$(56^4,  16,  56^5      14^1       7^2        2^8,  4)$,\\
IrMOA$(56^4,  15,  56^5      14^1       7^2  4^1   2^6,  4)$,\\
IrMOA$(56^4,  17,  56^5                 7^3        2^9,  4)$,\\
IrMOA$(56^4,  12,  56^5 28^2            7^1  4^1   2^3,  4)$,\\
IrMOA$(56^4,  13,  56^5 28^1 14^1       7^1  4^1   2^4,  4)$,\\
IrMOA$(56^4,  14,  56^5      14^2       7^1  4^1   2^5,  4)$,\\
IrMOA$(56^4,  12,  56^5 28^1 14^1       7^1  4^2   2^2,  4)$,\\
IrMOA$(56^4,  13,  56^5      14^2       7^1  4^2   2^3,  4)$,\\
IrMOA$(56^4,  12,  56^5      14^2       7^1  4^3   2^1,  4)$,\\
IrMOA$(56^4,  14,  56^5 28^1            7^2  4^1   2^5,  4)$,\\
IrMOA$(56^4,  15,  56^5      14^1       7^2  4^1   2^6,  4)$,\\
IrMOA$(56^4,  14,  56^5      14^1       7^2  4^2   2^4,  4)$,\\
IrMOA$(56^4,  16,  56^5                 7^3  4^1   2^7,  4)$,\\
IrMOA$(56^4,  13,  56^5 28^1            7^2  4^2   2^3,  4)$,\\
IrMOA$(56^4,  14,  56^5      14^1       7^2  4^2   2^4,  4)$,\\
IrMOA$(56^4,  13,  56^5      14^1       7^2  4^3   2^2,  4)$,\\
IrMOA$(56^4,  15,  56^5                 7^3  4^2   2^5,  4)$,\\
IrMOA$(56^4,  14,  56^5                 7^3  4^3   2^3,  4)$,\\
IrMOA$(56^4,  11,  56^5 28^2       8^1  7^1        2^2,  4)$,\\
IrMOA$(56^4,  12,  56^5 28^1 14^1  8^1  7^1        2^3,  4)$,\\
IrMOA$(56^4,  13,  56^5      14^2  8^1  7^1        2^4,  4)$,\\
IrMOA$(56^4,  11,  56^5 28^1 14^1  8^1  7^1  4^1   2^1,  4)$,\\
IrMOA$(56^4,  12,  56^5      14^2  8^1  7^1  4^1   2^2,  4)$,\\
IrMOA$(56^4,  11,  56^5      14^2  8^1  7^1  4^2      ,  4)$,\\
IrMOA$(56^4,  13,  56^5 28^1       8^1  7^2        2^4,  4)$,\\
IrMOA$(56^4,  14,  56^5      14^1  8^1  7^2        2^5,  4)$,\\
IrMOA$(56^4,  13,  56^5      14^1  8^1  7^2  4^1   2^3,  4)$,\\
IrMOA$(56^4,  15,  56^5            8^1  7^3        2^6,  4)$,\\
IrMOA$(56^4,  12,  56^5 28^1       8^1  7^2  4^1   2^2,  4)$,\\
IrMOA$(56^4,  13,  56^5      14^1  8^1  7^2  4^1   2^3,  4)$,\\
IrMOA$(56^4,  12,  56^5      14^1  8^1  7^2  4^2   2^1,  4)$,\\
IrMOA$(56^4,  14,  56^5            8^1  7^3  4^1   2^4,  4)$,\\
IrMOA$(56^4,  13,  56^5            8^1  7^3  4^2   2^2,  4)$,\\
IrMOA$(56^4,  11,  56^5 28^1       8^2  7^2        2^1,  4)$,\\
IrMOA$(56^4,  12,  56^5      14^1  8^2  7^2        2^2,  4)$,\\
IrMOA$(56^4,  11,  56^5      14^1  8^2  7^2  4^1      ,  4)$,\\
IrMOA$(56^4,  13,  56^5            8^2  7^3        2^3,  4)$,\\
IrMOA$(56^4,  12,  56^5            8^2  7^3  4^1   2^1,  4)$,\\
IrMOA$(56^4,  11,  56^5            8^3  7^3           ,  4)$,\\
IrMOA$(56^4,  12,  56^4 28^4                       2^4,  4)$,\\
IrMOA$(56^4,  13,  56^4 28^3 14^1                  2^5,  4)$,\\
IrMOA$(56^4,  14,  56^4 28^2 14^2                  2^6,  4)$,\\
IrMOA$(56^4,  15,  56^4 28^1 14^3                  2^7,  4)$,\\
IrMOA$(56^4,  16,  56^4      14^4                  2^8,  4)$,\\
IrMOA$(56^4,  12,  56^4 28^3 14^1            4^1   2^3,  4)$,\\
IrMOA$(56^4,  13,  56^4 28^2 14^2            4^1   2^4,  4)$,\\
IrMOA$(56^4,  14,  56^4 28^1 14^3            4^1   2^5,  4)$,\\
IrMOA$(56^4,  15,  56^4      14^4            4^1   2^6,  4)$,\\
IrMOA$(56^4,  12,  56^4 28^2 14^2            4^2   2^2,  4)$,\\
IrMOA$(56^4,  13,  56^4 28^1 14^3            4^2   2^3,  4)$,\\
IrMOA$(56^4,  14,  56^4      14^4            4^2   2^4,  4)$,\\
IrMOA$(56^4,  12,  56^4 28^1 14^3            4^3   2^1,  4)$,\\
IrMOA$(56^4,  13,  56^4      14^4            4^3   2^2,  4)$,\\
IrMOA$(56^4,  12,  56^4      14^4            4^4      ,  4)$,\\
IrMOA$(56^4,  14,  56^4 28^3            7^1        2^6,  4)$,\\
IrMOA$(56^4,  15,  56^4 28^2 14^1       7^1        2^7,  4)$,\\
IrMOA$(56^4,  16,  56^4 28^1 14^2       7^1        2^8,  4)$,\\
IrMOA$(56^4,  17,  56^4      14^3       7^1        2^9,  4)$,\\
IrMOA$(56^4,  14,  56^4 28^2 14^1       7^1  4^1   2^5,  4)$,\\
IrMOA$(56^4,  15,  56^4 28^1 14^2       7^1  4^1   2^6,  4)$,\\
IrMOA$(56^4,  16,  56^4      14^3       7^1  4^1   2^7,  4)$,\\
IrMOA$(56^4,  14,  56^4 28^1 14^2       7^1  4^2   2^4,  4)$,\\
IrMOA$(56^4,  15,  56^4      14^3       7^1  4^2   2^5,  4)$,\\
IrMOA$(56^4,  14,  56^4      14^3       7^1  4^3   2^3,  4)$,\\
IrMOA$(56^4,  16,  56^4 28^2            7^2        2^8,  4)$,\\
IrMOA$(56^4,  17,  56^4 28^1 14^1       7^2        2^9,  4)$,\\
IrMOA$(56^4,  18,  56^4      14^2       7^2     2^{10},  4)$,\\
IrMOA$(56^4,  16,  56^4 28^1 14^1       7^2  4^1   2^7,  4)$,\\
IrMOA$(56^4,  17,  56^4      14^2       7^2  4^1   2^8,  4)$,\\
IrMOA$(56^4,  16,  56^4      14^2       7^2  4^2   2^6,  4)$,\\
IrMOA$(56^4,  18,  56^4 28^1            7^3     2^{10},  4)$,\\
IrMOA$(56^4,  19,  56^4      14^1       7^3   2^{11},  4)$,\\
IrMOA$(56^4,  18,  56^4      14^1       7^3  4^1   2^9,  4)$,\\
IrMOA$(56^4,  20,  56^4                 7^4   2^{12},  4)$,\\
IrMOA$(56^4,  13,  56^4 28^3            7^1  4^1   2^4,  4)$,\\
IrMOA$(56^4,  14,  56^4 28^2 14^1       7^1  4^1   2^5,  4)$,\\
IrMOA$(56^4,  15,  56^4 28^1 14^2       7^1  4^1   2^6,  4)$,\\
IrMOA$(56^4,  16,  56^4      14^3       7^1  4^1   2^7,  4)$,\\
IrMOA$(56^4,  13,  56^4 28^2 14^1       7^1  4^2   2^3,  4)$,\\
IrMOA$(56^4,  14,  56^4 28^1 14^2       7^1  4^2   2^4,  4)$,\\
IrMOA$(56^4,  15,  56^4      14^3       7^1  4^2   2^5,  4)$,\\
IrMOA$(56^4,  13,  56^4 28^1 14^2       7^1  4^3   2^2,  4)$,\\
IrMOA$(56^4,  14,  56^4      14^3       7^1  4^3   2^3,  4)$,\\
IrMOA$(56^4,  13,  56^4      14^3       7^1  4^4   2^1,  4)$,\\
IrMOA$(56^4,  15,  56^4 28^2            7^2  4^1   2^6,  4)$,\\
IrMOA$(56^4,  16,  56^4 28^1 14^1       7^2  4^1   2^7,  4)$,\\
IrMOA$(56^4,  17,  56^4      14^2       7^2  4^1   2^8,  4)$,\\
IrMOA$(56^4,  15,  56^4 28^1 14^1       7^2  4^2   2^5,  4)$,\\
IrMOA$(56^4,  16,  56^4      14^2       7^2  4^2   2^6,  4)$,\\
IrMOA$(56^4,  15,  56^4      14^2       7^2  4^3   2^4,  4)$,\\
IrMOA$(56^4,  17,  56^4 28^1            7^3  4^1   2^8,  4)$,\\
IrMOA$(56^4,  18,  56^4      14^1       7^3  4^1   2^9,  4)$,\\
IrMOA$(56^4,  17,  56^4      14^1       7^3  4^2   2^7,  4)$,\\
IrMOA$(56^4,  19,  56^4                 7^4  4^12^{10},  4)$,\\
IrMOA$(56^4,  14,  56^4 28^2            7^2  4^2   2^4,  4)$,\\
IrMOA$(56^4,  15,  56^4 28^1 14^1       7^2  4^2   2^5,  4)$,\\
IrMOA$(56^4,  16,  56^4      14^2       7^2  4^2   2^6,  4)$,\\
IrMOA$(56^4,  14,  56^4 28^1 14^1       7^2  4^3   2^3,  4)$,\\
IrMOA$(56^4,  15,  56^4      14^2       7^2  4^3   2^4,  4)$,\\
IrMOA$(56^4,  14,  56^4      14^2       7^2  4^4   2^2,  4)$,\\
IrMOA$(56^4,  16,  56^4 28^1            7^3  4^2   2^6,  4)$,\\
IrMOA$(56^4,  17,  56^4      14^1       7^3  4^2   2^7,  4)$,\\
IrMOA$(56^4,  16,  56^4      14^1       7^3  4^3   2^5,  4)$,\\
IrMOA$(56^4,  18,  56^4                 7^4  4^2   2^8,  4)$,\\
IrMOA$(56^4,  15,  56^4 28^1            7^3  4^3   2^4,  4)$,\\
IrMOA$(56^4,  16,  56^4      14^1       7^3  4^3   2^5,  4)$,\\
IrMOA$(56^4,  15,  56^4      14^1       7^3  4^4   2^3,  4)$,\\
IrMOA$(56^4,  17,  56^4                 7^4  4^3   2^6,  4)$,\\
IrMOA$(56^4,  16,  56^4                 7^4  4^4   2^4,  4)$,\\
IrMOA$(56^4,  12,  56^4 28^3       8^1  7^1        2^3,  4)$,\\
IrMOA$(56^4,  13,  56^4 28^2 14^1  8^1  7^1        2^4,  4)$,\\
IrMOA$(56^4,  14,  56^4 28^1 14^2  8^1  7^1        2^5,  4)$,\\
IrMOA$(56^4,  15,  56^4      14^3  8^1  7^1        2^6,  4)$,\\
IrMOA$(56^4,  12,  56^4 28^2 14^1  8^1  7^1  4^1   2^2,  4)$,\\
IrMOA$(56^4,  13,  56^4 28^1 14^2  8^1  7^1  4^1   2^3,  4)$,\\
IrMOA$(56^4,  14,  56^4      14^3  8^1  7^1  4^1   2^4,  4)$,\\
IrMOA$(56^4,  12,  56^4 28^1 14^2  8^1  7^1  4^2   2^1,  4)$,\\
IrMOA$(56^4,  13,  56^4      14^3  8^1  7^1  4^2   2^2,  4)$,\\
IrMOA$(56^4,  12,  56^4      14^3  8^1  7^1  4^3      ,  4)$,\\
IrMOA$(56^4,  14,  56^4 28^2       8^1  7^2        2^5,  4)$,\\
IrMOA$(56^4,  15,  56^4 28^1 14^1  8^1  7^2        2^6,  4)$,\\
IrMOA$(56^4,  16,  56^4      14^2  8^1  7^2        2^7,  4)$,\\
IrMOA$(56^4,  14,  56^4 28^1 14^1  8^1  7^2  4^1   2^4,  4)$,\\
IrMOA$(56^4,  15,  56^4      14^2  8^1  7^2  4^1   2^5,  4)$,\\
IrMOA$(56^4,  14,  56^4      14^2  8^1  7^2  4^2   2^3,  4)$,\\
IrMOA$(56^4,  16,  56^4 28^1       8^1  7^3        2^7,  4)$,\\
IrMOA$(56^4,  17,  56^4      14^1  8^1  7^3        2^8,  4)$,\\
IrMOA$(56^4,  16,  56^4      14^1  8^1  7^3  4^1   2^6,  4)$,\\
IrMOA$(56^4,  18,  56^4            8^1  7^4        2^9,  4)$,\\
IrMOA$(56^4,  13,  56^4 28^2       8^1  7^2  4^1   2^3,  4)$,\\
IrMOA$(56^4,  14,  56^4 28^1 14^1  8^1  7^2  4^1   2^4,  4)$,\\
IrMOA$(56^4,  15,  56^4      14^2  8^1  7^2  4^1   2^5,  4)$,\\
IrMOA$(56^4,  13,  56^4 28^1 14^1  8^1  7^2  4^2   2^2,  4)$,\\
IrMOA$(56^4,  14,  56^4      14^2  8^1  7^2  4^2   2^3,  4)$,\\
IrMOA$(56^4,  13,  56^4      14^2  8^1  7^2  4^3   2^1,  4)$,\\
IrMOA$(56^4,  15,  56^4 28^1       8^1  7^3  4^1   2^5,  4)$,\\
IrMOA$(56^4,  16,  56^4      14^1  8^1  7^3  4^1   2^6,  4)$,\\
IrMOA$(56^4,  15,  56^4      14^1  8^1  7^3  4^2   2^4,  4)$,\\
IrMOA$(56^4,  17,  56^4            8^1  7^4  4^1   2^7,  4)$,\\
IrMOA$(56^4,  14,  56^4 28^1       8^1  7^3  4^2   2^3,  4)$,\\
IrMOA$(56^4,  15,  56^4      14^1  8^1  7^3  4^2   2^4,  4)$,\\
IrMOA$(56^4,  14,  56^4      14^1  8^1  7^3  4^3   2^2,  4)$,\\
IrMOA$(56^4,  16,  56^4            8^1  7^4  4^2   2^5,  4)$,\\
IrMOA$(56^4,  15,  56^4            8^1  7^4  4^3   2^3,  4)$,\\
IrMOA$(56^4,  12,  56^4 28^2       8^2  7^2        2^2,  4)$,\\
IrMOA$(56^4,  13,  56^4 28^1 14^1  8^2  7^2        2^3,  4)$,\\
IrMOA$(56^4,  14,  56^4      14^2  8^2  7^2        2^4,  4)$,\\
IrMOA$(56^4,  12,  56^4 28^1 14^1  8^2  7^2  4^1   2^1,  4)$,\\
IrMOA$(56^4,  13,  56^4      14^2  8^2  7^2  4^1   2^2,  4)$,\\
IrMOA$(56^4,  12,  56^4      14^2  8^2  7^2  4^2      ,  4)$,\\
IrMOA$(56^4,  14,  56^4 28^1       8^2  7^3        2^4,  4)$,\\
IrMOA$(56^4,  15,  56^4      14^1  8^2  7^3        2^5,  4)$,\\
IrMOA$(56^4,  14,  56^4      14^1  8^2  7^3  4^1   2^3,  4)$,\\
IrMOA$(56^4,  16,  56^4            8^2  7^4        2^6,  4)$,\\
IrMOA$(56^4,  13,  56^4 28^1       8^2  7^3  4^1   2^2,  4)$,\\
IrMOA$(56^4,  14,  56^4      14^1  8^2  7^3  4^1   2^3,  4)$,\\
IrMOA$(56^4,  13,  56^4      14^1  8^2  7^3  4^2   2^1,  4)$,\\
IrMOA$(56^4,  15,  56^4            8^2  7^4  4^1   2^4,  4)$,\\
IrMOA$(56^4,  14,  56^4            8^2  7^4  4^2   2^2,  4)$,\\
IrMOA$(56^4,  12,  56^4 28^1       8^3  7^3        2^1,  4)$,\\
IrMOA$(56^4,  13,  56^4      14^1  8^3  7^3        2^2,  4)$,\\
IrMOA$(56^4,  12,  56^4      14^1  8^3  7^3  4^1      ,  4)$,\\
IrMOA$(56^4,  14,  56^4            8^3  7^4        2^3,  4)$,\\
IrMOA$(56^4,  13,  56^4            8^3  7^4  4^1   2^1,  4)$,\\
IrMOA$(56^4,  12,  56^4            8^4  7^4           ,  4)$,\\
IrMOA$(56^4,  13,  56^3 28^5                       2^5,  4)$,\\
IrMOA$(56^4,  14,  56^3 28^4 14^1                  2^6,  4)$,\\
IrMOA$(56^4,  15,  56^3 28^3 14^2                  2^7,  4)$,\\
IrMOA$(56^4,  16,  56^3 28^2 14^3                  2^8,  4)$,\\
IrMOA$(56^4,  17,  56^3 28^1 14^4                  2^9,  4)$,\\
IrMOA$(56^4,  18,  56^3      14^5               2^{10},  4)$,\\
IrMOA$(56^4,  13,  56^3 28^4 14^1            4^1   2^4,  4)$,\\
IrMOA$(56^4,  14,  56^3 28^3 14^2            4^1   2^5,  4)$,\\
IrMOA$(56^4,  15,  56^3 28^2 14^3            4^1   2^6,  4)$,\\
IrMOA$(56^4,  16,  56^3 28^1 14^4            4^1   2^7,  4)$,\\
IrMOA$(56^4,  17,  56^3      14^5            4^1   2^8,  4)$,\\
IrMOA$(56^4,  13,  56^3 28^3 14^2            4^2   2^3,  4)$,\\
IrMOA$(56^4,  14,  56^3 28^2 14^3            4^2   2^4,  4)$,\\
IrMOA$(56^4,  15,  56^3 28^1 14^4            4^2   2^5,  4)$,\\
IrMOA$(56^4,  16,  56^3      14^5            4^2   2^6,  4)$,\\
IrMOA$(56^4,  13,  56^3 28^2 14^3            4^3   2^2,  4)$,\\
IrMOA$(56^4,  14,  56^3 28^1 14^4            4^3   2^3,  4)$,\\
IrMOA$(56^4,  15,  56^3      14^5            4^3   2^4,  4)$,\\
IrMOA$(56^4,  13,  56^3 28^1 14^4            4^4   2^1,  4)$,\\
IrMOA$(56^4,  14,  56^3      14^5            4^4   2^2,  4)$,\\
IrMOA$(56^4,  13,  56^3      14^5            4^5      ,  4)$,\\
IrMOA$(56^4,  15,  56^3 28^4            7^1        2^7,  4)$,\\
IrMOA$(56^4,  16,  56^3 28^3 14^1       7^1        2^8,  4)$,\\
IrMOA$(56^4,  17,  56^3 28^2 14^2       7^1        2^9,  4)$,\\
IrMOA$(56^4,  18,  56^3 28^1 14^3       7^1     2^{10},  4)$,\\
IrMOA$(56^4,  19,  56^3      14^4       7^1   2^{11},  4)$,\\
IrMOA$(56^4,  15,  56^3 28^3 14^1       7^1  4^1   2^6,  4)$,\\
IrMOA$(56^4,  16,  56^3 28^2 14^2       7^1  4^1   2^7,  4)$,\\
IrMOA$(56^4,  17,  56^3 28^1 14^3       7^1  4^1   2^8,  4)$,\\
IrMOA$(56^4,  18,  56^3      14^4       7^1  4^1   2^9,  4)$,\\
IrMOA$(56^4,  15,  56^3 28^2 14^2       7^1  4^2   2^5,  4)$,\\
IrMOA$(56^4,  16,  56^3 28^1 14^3       7^1  4^2   2^6,  4)$,\\
IrMOA$(56^4,  17,  56^3      14^4       7^1  4^2   2^7,  4)$,\\
IrMOA$(56^4,  15,  56^3 28^1 14^3       7^1  4^3   2^4,  4)$,\\
IrMOA$(56^4,  16,  56^3      14^4       7^1  4^3   2^5,  4)$,\\
IrMOA$(56^4,  15,  56^3      14^4       7^1  4^4   2^3,  4)$,\\
IrMOA$(56^4,  17,  56^3 28^3            7^2        2^9,  4)$,\\
IrMOA$(56^4,  18,  56^3 28^2 14^1       7^2     2^{10},  4)$,\\
IrMOA$(56^4,  19,  56^3 28^1 14^2       7^2   2^{11},  4)$,\\
IrMOA$(56^4,  20,  56^3      14^3       7^2   2^{12},  4)$,\\
IrMOA$(56^4,  17,  56^3 28^2 14^1       7^2  4^1   2^8,  4)$,\\
IrMOA$(56^4,  18,  56^3 28^1 14^2       7^2  4^1   2^9,  4)$,\\
IrMOA$(56^4,  19,  56^3      14^3       7^2  4^12^{10},  4)$,\\
IrMOA$(56^4,  17,  56^3 28^1 14^2       7^2  4^2   2^7,  4)$,\\
IrMOA$(56^4,  18,  56^3      14^3       7^2  4^2   2^8,  4)$,\\
IrMOA$(56^4,  17,  56^3      14^3       7^2  4^3   2^6,  4)$,\\
IrMOA$(56^4,  19,  56^3 28^2            7^3   2^{11},  4)$,\\
IrMOA$(56^4,  20,  56^3 28^1 14^1       7^3   2^{12},  4)$,\\
IrMOA$(56^4,  21,  56^3      14^2       7^3   2^{13},  4)$,\\
IrMOA$(56^4,  19,  56^3 28^1 14^1       7^3  4^12^{10},  4)$,\\
IrMOA$(56^4,  20,  56^3      14^2       7^3  4^12^{11},  4)$,\\
IrMOA$(56^4,  19,  56^3      14^2       7^3  4^2   2^9,  4)$,\\
IrMOA$(56^4,  21,  56^3 28^1            7^4   2^{13},  4)$,\\
IrMOA$(56^4,  22,  56^3      14^1       7^4   2^{14},  4)$,\\
IrMOA$(56^4,  21,  56^3      14^1       7^4  4^12^{12},  4)$,\\
IrMOA$(56^4,  23,  56^3                 7^5   2^{15},  4)$,\\
IrMOA$(56^4,  14,  56^3 28^4            7^1  4^1   2^5,  4)$,\\
IrMOA$(56^4,  15,  56^3 28^3 14^1       7^1  4^1   2^6,  4)$,\\
IrMOA$(56^4,  16,  56^3 28^2 14^2       7^1  4^1   2^7,  4)$,\\
IrMOA$(56^4,  17,  56^3 28^1 14^3       7^1  4^1   2^8,  4)$,\\
IrMOA$(56^4,  18,  56^3      14^4       7^1  4^1   2^9,  4)$,\\
IrMOA$(56^4,  14,  56^3 28^3 14^1       7^1  4^2   2^4,  4)$,\\
IrMOA$(56^4,  15,  56^3 28^2 14^2       7^1  4^2   2^5,  4)$,\\
IrMOA$(56^4,  16,  56^3 28^1 14^3       7^1  4^2   2^6,  4)$,\\
IrMOA$(56^4,  17,  56^3      14^4       7^1  4^2   2^7,  4)$,\\
IrMOA$(56^4,  14,  56^3 28^2 14^2       7^1  4^3   2^3,  4)$,\\
IrMOA$(56^4,  15,  56^3 28^1 14^3       7^1  4^3   2^4,  4)$,\\
IrMOA$(56^4,  16,  56^3      14^4       7^1  4^3   2^5,  4)$,\\
IrMOA$(56^4,  14,  56^3 28^1 14^3       7^1  4^4   2^2,  4)$,\\
IrMOA$(56^4,  15,  56^3      14^4       7^1  4^4   2^3,  4)$,\\
IrMOA$(56^4,  14,  56^3      14^4       7^1  4^5   2^1,  4)$,\\
IrMOA$(56^4,  16,  56^3 28^3            7^2  4^1   2^7,  4)$,\\
IrMOA$(56^4,  17,  56^3 28^2 14^1       7^2  4^1   2^8,  4)$,\\
IrMOA$(56^4,  18,  56^3 28^1 14^2       7^2  4^1   2^9,  4)$,\\
IrMOA$(56^4,  19,  56^3      14^3       7^2  4^12^{10},  4)$,\\
IrMOA$(56^4,  16,  56^3 28^2 14^1       7^2  4^2   2^6,  4)$,\\
IrMOA$(56^4,  17,  56^3 28^1 14^2       7^2  4^2   2^7,  4)$,\\
IrMOA$(56^4,  18,  56^3      14^3       7^2  4^2   2^8,  4)$,\\
IrMOA$(56^4,  16,  56^3 28^1 14^2       7^2  4^3   2^5,  4)$,\\
IrMOA$(56^4,  17,  56^3      14^3       7^2  4^3   2^6,  4)$,\\
IrMOA$(56^4,  16,  56^3      14^3       7^2  4^4   2^4,  4)$,\\
IrMOA$(56^4,  18,  56^3 28^2            7^3  4^1   2^9,  4)$,\\
IrMOA$(56^4,  19,  56^3 28^1 14^1       7^3  4^12^{10},  4)$,\\
IrMOA$(56^4,  20,  56^3      14^2       7^3  4^12^{11},  4)$,\\
IrMOA$(56^4,  18,  56^3 28^1 14^1       7^3  4^2   2^8,  4)$,\\
IrMOA$(56^4,  19,  56^3      14^2       7^3  4^2   2^9,  4)$,\\
IrMOA$(56^4,  18,  56^3      14^2       7^3  4^3   2^7,  4)$,\\
IrMOA$(56^4,  20,  56^3 28^1            7^4  4^12^{11},  4)$,\\
IrMOA$(56^4,  21,  56^3      14^1       7^4  4^12^{12},  4)$,\\
IrMOA$(56^4,  20,  56^3      14^1       7^4  4^22^{10},  4)$,\\
IrMOA$(56^4,  22,  56^3                 7^5  4^12^{13},  4)$,\\
IrMOA$(56^4,  15,  56^3 28^3            7^2  4^2   2^5,  4)$,\\
IrMOA$(56^4,  16,  56^3 28^2 14^1       7^2  4^2   2^6,  4)$,\\
IrMOA$(56^4,  17,  56^3 28^1 14^2       7^2  4^2   2^7,  4)$,\\
IrMOA$(56^4,  18,  56^3      14^3       7^2  4^2   2^8,  4)$,\\
IrMOA$(56^4,  15,  56^3 28^2 14^1       7^2  4^3   2^4,  4)$,\\
IrMOA$(56^4,  16,  56^3 28^1 14^2       7^2  4^3   2^5,  4)$,\\
IrMOA$(56^4,  17,  56^3      14^3       7^2  4^3   2^6,  4)$,\\
IrMOA$(56^4,  15,  56^3 28^1 14^2       7^2  4^4   2^3,  4)$,\\
IrMOA$(56^4,  16,  56^3      14^3       7^2  4^4   2^4,  4)$,\\
IrMOA$(56^4,  15,  56^3      14^3       7^2  4^5   2^2,  4)$,\\
IrMOA$(56^4,  17,  56^3 28^2            7^3  4^2   2^7,  4)$,\\
IrMOA$(56^4,  18,  56^3 28^1 14^1       7^3  4^2   2^8,  4)$,\\
IrMOA$(56^4,  19,  56^3      14^2       7^3  4^2   2^9,  4)$,\\
IrMOA$(56^4,  17,  56^3 28^1 14^1       7^3  4^3   2^6,  4)$,\\
IrMOA$(56^4,  18,  56^3      14^2       7^3  4^3   2^7,  4)$,\\
IrMOA$(56^4,  17,  56^3      14^2       7^3  4^4   2^5,  4)$,\\
IrMOA$(56^4,  19,  56^3 28^1            7^4  4^2   2^9,  4)$,\\
IrMOA$(56^4,  20,  56^3      14^1       7^4  4^22^{10},  4)$,\\
IrMOA$(56^4,  19,  56^3      14^1       7^4  4^3   2^8,  4)$,\\
IrMOA$(56^4,  21,  56^3                 7^5  4^22^{11},  4)$,\\
IrMOA$(56^4,  16,  56^3 28^2            7^3  4^3   2^5,  4)$,\\
IrMOA$(56^4,  17,  56^3 28^1 14^1       7^3  4^3   2^6,  4)$,\\
IrMOA$(56^4,  18,  56^3      14^2       7^3  4^3   2^7,  4)$,\\
IrMOA$(56^4,  16,  56^3 28^1 14^1       7^3  4^4   2^4,  4)$,\\
IrMOA$(56^4,  17,  56^3      14^2       7^3  4^4   2^5,  4)$,\\
IrMOA$(56^4,  16,  56^3      14^2       7^3  4^5   2^3,  4)$,\\
IrMOA$(56^4,  18,  56^3 28^1            7^4  4^3   2^7,  4)$,\\
IrMOA$(56^4,  19,  56^3      14^1       7^4  4^3   2^8,  4)$,\\
IrMOA$(56^4,  18,  56^3      14^1       7^4  4^4   2^6,  4)$,\\
IrMOA$(56^4,  20,  56^3                 7^5  4^3   2^9,  4)$,\\
IrMOA$(56^4,  17,  56^3 28^1            7^4  4^4   2^5,  4)$,\\
IrMOA$(56^4,  18,  56^3      14^1       7^4  4^4   2^6,  4)$,\\
IrMOA$(56^4,  17,  56^3      14^1       7^4  4^5   2^4,  4)$,\\
IrMOA$(56^4,  19,  56^3                 7^5  4^4   2^7,  4)$,\\
IrMOA$(56^4,  18,  56^3                 7^5  4^5   2^5,  4)$,\\
IrMOA$(56^4,  13,  56^3 28^4       8^1  7^1        2^4,  4)$,\\
IrMOA$(56^4,  14,  56^3 28^3 14^1  8^1  7^1        2^5,  4)$,\\
IrMOA$(56^4,  15,  56^3 28^2 14^2  8^1  7^1        2^6,  4)$,\\
IrMOA$(56^4,  16,  56^3 28^1 14^3  8^1  7^1        2^7,  4)$,\\
IrMOA$(56^4,  17,  56^3      14^4  8^1  7^1        2^8,  4)$,\\
IrMOA$(56^4,  13,  56^3 28^3 14^1  8^1  7^1  4^1   2^3,  4)$,\\
IrMOA$(56^4,  14,  56^3 28^2 14^2  8^1  7^1  4^1   2^4,  4)$,\\
IrMOA$(56^4,  15,  56^3 28^1 14^3  8^1  7^1  4^1   2^5,  4)$,\\
IrMOA$(56^4,  16,  56^3      14^4  8^1  7^1  4^1   2^6,  4)$,\\
IrMOA$(56^4,  13,  56^3 28^2 14^2  8^1  7^1  4^2   2^2,  4)$,\\
IrMOA$(56^4,  14,  56^3 28^1 14^3  8^1  7^1  4^2   2^3,  4)$,\\
IrMOA$(56^4,  15,  56^3      14^4  8^1  7^1  4^2   2^4,  4)$,\\
IrMOA$(56^4,  13,  56^3 28^1 14^3  8^1  7^1  4^3   2^1,  4)$,\\
IrMOA$(56^4,  14,  56^3      14^4  8^1  7^1  4^3   2^2,  4)$,\\
IrMOA$(56^4,  13,  56^3      14^4  8^1  7^1  4^4      ,  4)$,\\
IrMOA$(56^4,  15,  56^3 28^3       8^1  7^2        2^6,  4)$,\\
IrMOA$(56^4,  16,  56^3 28^2 14^1  8^1  7^2        2^7,  4)$,\\
IrMOA$(56^4,  17,  56^3 28^1 14^2  8^1  7^2        2^8,  4)$,\\
IrMOA$(56^4,  18,  56^3      14^3  8^1  7^2        2^9,  4)$,\\
IrMOA$(56^4,  15,  56^3 28^2 14^1  8^1  7^2  4^1   2^5,  4)$,\\
IrMOA$(56^4,  16,  56^3 28^1 14^2  8^1  7^2  4^1   2^6,  4)$,\\
IrMOA$(56^4,  17,  56^3      14^3  8^1  7^2  4^1   2^7,  4)$,\\
IrMOA$(56^4,  15,  56^3 28^1 14^2  8^1  7^2  4^2   2^4,  4)$,\\
IrMOA$(56^4,  16,  56^3      14^3  8^1  7^2  4^2   2^5,  4)$,\\
IrMOA$(56^4,  15,  56^3      14^3  8^1  7^2  4^3   2^3,  4)$,\\
IrMOA$(56^4,  17,  56^3 28^2       8^1  7^3        2^8,  4)$,\\
IrMOA$(56^4,  18,  56^3 28^1 14^1  8^1  7^3        2^9,  4)$,\\
IrMOA$(56^4,  19,  56^3      14^2  8^1  7^3     2^{10},  4)$,\\
IrMOA$(56^4,  17,  56^3 28^1 14^1  8^1  7^3  4^1   2^7,  4)$,\\
IrMOA$(56^4,  18,  56^3      14^2  8^1  7^3  4^1   2^8,  4)$,\\
IrMOA$(56^4,  17,  56^3      14^2  8^1  7^3  4^2   2^6,  4)$,\\
IrMOA$(56^4,  19,  56^3 28^1       8^1  7^4     2^{10},  4)$,\\
IrMOA$(56^4,  20,  56^3      14^1  8^1  7^4   2^{11},  4)$,\\
IrMOA$(56^4,  19,  56^3      14^1  8^1  7^4  4^1   2^9,  4)$,\\
IrMOA$(56^4,  21,  56^3            8^1  7^5     2^{12},  4)$,\\
IrMOA$(56^4,  14,  56^3 28^3       8^1  7^2  4^1   2^4,  4)$,\\
IrMOA$(56^4,  15,  56^3 28^2 14^1  8^1  7^2  4^1   2^5,  4)$,\\
IrMOA$(56^4,  16,  56^3 28^1 14^2  8^1  7^2  4^1   2^6,  4)$,\\
IrMOA$(56^4,  17,  56^3      14^3  8^1  7^2  4^1   2^7,  4)$,\\
IrMOA$(56^4,  14,  56^3 28^2 14^1  8^1  7^2  4^2   2^3,  4)$,\\
IrMOA$(56^4,  15,  56^3 28^1 14^2  8^1  7^2  4^2   2^4,  4)$,\\
IrMOA$(56^4,  16,  56^3      14^3  8^1  7^2  4^2   2^5,  4)$,\\
IrMOA$(56^4,  14,  56^3 28^1 14^2  8^1  7^2  4^3   2^2,  4)$,\\
IrMOA$(56^4,  15,  56^3      14^3  8^1  7^2  4^3   2^3,  4)$,\\
IrMOA$(56^4,  14,  56^3      14^3  8^1  7^2  4^4   2^1,  4)$,\\
IrMOA$(56^4,  16,  56^3 28^2       8^1  7^3  4^1   2^6,  4)$,\\
IrMOA$(56^4,  17,  56^3 28^1 14^1  8^1  7^3  4^1   2^7,  4)$,\\
IrMOA$(56^4,  18,  56^3      14^2  8^1  7^3  4^1   2^8,  4)$,\\
IrMOA$(56^4,  16,  56^3 28^1 14^1  8^1  7^3  4^2   2^5,  4)$,\\
IrMOA$(56^4,  17,  56^3      14^2  8^1  7^3  4^2   2^6,  4)$,\\
IrMOA$(56^4,  16,  56^3      14^2  8^1  7^3  4^3   2^4,  4)$,\\
IrMOA$(56^4,  18,  56^3 28^1       8^1  7^4  4^1   2^8,  4)$,\\
IrMOA$(56^4,  19,  56^3      14^1  8^1  7^4  4^1   2^9,  4)$,\\
IrMOA$(56^4,  18,  56^3      14^1  8^1  7^4  4^2   2^7,  4)$,\\
IrMOA$(56^4,  20,  56^3            8^1  7^5  4^12^{10},  4)$,\\
IrMOA$(56^4,  15,  56^3 28^2       8^1  7^3  4^2   2^4,  4)$,\\
IrMOA$(56^4,  16,  56^3 28^1 14^1  8^1  7^3  4^2   2^5,  4)$,\\
IrMOA$(56^4,  17,  56^3      14^2  8^1  7^3  4^2   2^6,  4)$,\\
IrMOA$(56^4,  15,  56^3 28^1 14^1  8^1  7^3  4^3   2^3,  4)$,\\
IrMOA$(56^4,  16,  56^3      14^2  8^1  7^3  4^3   2^4,  4)$,\\
IrMOA$(56^4,  15,  56^3      14^2  8^1  7^3  4^4   2^2,  4)$,\\
IrMOA$(56^4,  17,  56^3 28^1       8^1  7^4  4^2   2^6,  4)$,\\
IrMOA$(56^4,  18,  56^3      14^1  8^1  7^4  4^2   2^7,  4)$,\\
IrMOA$(56^4,  17,  56^3      14^1  8^1  7^4  4^3   2^5,  4)$,\\
IrMOA$(56^4,  19,  56^3            8^1  7^5  4^2   2^8,  4)$,\\
IrMOA$(56^4,  16,  56^3 28^1       8^1  7^4  4^3   2^4,  4)$,\\
IrMOA$(56^4,  17,  56^3      14^1  8^1  7^4  4^3   2^5,  4)$,\\
IrMOA$(56^4,  16,  56^3      14^1  8^1  7^4  4^4   2^3,  4)$,\\
IrMOA$(56^4,  18,  56^3            8^1  7^5  4^3   2^6,  4)$,\\
IrMOA$(56^4,  17,  56^3            8^1  7^5  4^4   2^4,  4)$,\\
IrMOA$(56^4,  13,  56^3 28^3       8^2  7^2        2^3,  4)$,\\
IrMOA$(56^4,  14,  56^3 28^2 14^1  8^2  7^2        2^4,  4)$,\\
IrMOA$(56^4,  15,  56^3 28^1 14^2  8^2  7^2        2^5,  4)$,\\
IrMOA$(56^4,  16,  56^3      14^3  8^2  7^2        2^6,  4)$,\\
IrMOA$(56^4,  13,  56^3 28^2 14^1  8^2  7^2  4^1   2^2,  4)$,\\
IrMOA$(56^4,  14,  56^3 28^1 14^2  8^2  7^2  4^1   2^3,  4)$,\\
IrMOA$(56^4,  15,  56^3      14^3  8^2  7^2  4^1   2^4,  4)$,\\
IrMOA$(56^4,  13,  56^3 28^1 14^2  8^2  7^2  4^2   2^1,  4)$,\\
IrMOA$(56^4,  14,  56^3      14^3  8^2  7^2  4^2   2^2,  4)$,\\
IrMOA$(56^4,  13,  56^3      14^3  8^2  7^2  4^3      ,  4)$,\\
IrMOA$(56^4,  15,  56^3 28^2       8^2  7^3        2^5,  4)$,\\
IrMOA$(56^4,  16,  56^3 28^1 14^1  8^2  7^3        2^6,  4)$,\\
IrMOA$(56^4,  17,  56^3      14^2  8^2  7^3        2^7,  4)$,\\
IrMOA$(56^4,  15,  56^3 28^1 14^1  8^2  7^3  4^1   2^4,  4)$,\\
IrMOA$(56^4,  16,  56^3      14^2  8^2  7^3  4^1   2^5,  4)$,\\
IrMOA$(56^4,  15,  56^3      14^2  8^2  7^3  4^2   2^3,  4)$,\\
IrMOA$(56^4,  17,  56^3 28^1       8^2  7^4        2^7,  4)$,\\
IrMOA$(56^4,  18,  56^3      14^1  8^2  7^4        2^8,  4)$,\\
IrMOA$(56^4,  17,  56^3      14^1  8^2  7^4  4^1   2^6,  4)$,\\
IrMOA$(56^4,  19,  56^3            8^2  7^5        2^9,  4)$,\\
IrMOA$(56^4,  14,  56^3 28^2       8^2  7^3  4^1   2^3,  4)$,\\
IrMOA$(56^4,  15,  56^3 28^1 14^1  8^2  7^3  4^1   2^4,  4)$,\\
IrMOA$(56^4,  16,  56^3      14^2  8^2  7^3  4^1   2^5,  4)$,\\
IrMOA$(56^4,  14,  56^3 28^1 14^1  8^2  7^3  4^2   2^2,  4)$,\\
IrMOA$(56^4,  15,  56^3      14^2  8^2  7^3  4^2   2^3,  4)$,\\
IrMOA$(56^4,  14,  56^3      14^2  8^2  7^3  4^3   2^1,  4)$,\\
IrMOA$(56^4,  16,  56^3 28^1       8^2  7^4  4^1   2^5,  4)$,\\
IrMOA$(56^4,  17,  56^3      14^1  8^2  7^4  4^1   2^6,  4)$,\\
IrMOA$(56^4,  16,  56^3      14^1  8^2  7^4  4^2   2^4,  4)$,\\
IrMOA$(56^4,  18,  56^3            8^2  7^5  4^1   2^7,  4)$,\\
IrMOA$(56^4,  15,  56^3 28^1       8^2  7^4  4^2   2^3,  4)$,\\
IrMOA$(56^4,  16,  56^3      14^1  8^2  7^4  4^2   2^4,  4)$,\\
IrMOA$(56^4,  15,  56^3      14^1  8^2  7^4  4^3   2^2,  4)$,\\
IrMOA$(56^4,  17,  56^3            8^2  7^5  4^2   2^5,  4)$,\\
IrMOA$(56^4,  16,  56^3            8^2  7^5  4^3   2^3,  4)$,\\
IrMOA$(56^4,  13,  56^3 28^2       8^3  7^3        2^2,  4)$,\\
IrMOA$(56^4,  14,  56^3 28^1 14^1  8^3  7^3        2^3,  4)$,\\
IrMOA$(56^4,  15,  56^3      14^2  8^3  7^3        2^4,  4)$,\\
IrMOA$(56^4,  13,  56^3 28^1 14^1  8^3  7^3  4^1   2^1,  4)$,\\
IrMOA$(56^4,  14,  56^3      14^2  8^3  7^3  4^1   2^2,  4)$,\\
IrMOA$(56^4,  13,  56^3      14^2  8^3  7^3  4^2      ,  4)$,\\
IrMOA$(56^4,  15,  56^3 28^1       8^3  7^4        2^4,  4)$,\\
IrMOA$(56^4,  16,  56^3      14^1  8^3  7^4        2^5,  4)$,\\
IrMOA$(56^4,  15,  56^3      14^1  8^3  7^4  4^1   2^3,  4)$,\\
IrMOA$(56^4,  17,  56^3            8^3  7^5        2^6,  4)$,\\
IrMOA$(56^4,  14,  56^3 28^1       8^3  7^4  4^1   2^2,  4)$,\\
IrMOA$(56^4,  15,  56^3      14^1  8^3  7^4  4^1   2^3,  4)$,\\
IrMOA$(56^4,  14,  56^3      14^1  8^3  7^4  4^2   2^1,  4)$,\\
IrMOA$(56^4,  16,  56^3            8^3  7^5  4^1   2^4,  4)$,\\
IrMOA$(56^4,  15,  56^3            8^3  7^5  4^2   2^2,  4)$,\\
IrMOA$(56^4,  13,  56^3 28^1       8^4  7^4        2^1,  4)$,\\
IrMOA$(56^4,  14,  56^3      14^1  8^4  7^4        2^2,  4)$,\\
IrMOA$(56^4,  13,  56^3      14^1  8^4  7^4  4^1      ,  4)$,\\
IrMOA$(56^4,  15,  56^3            8^4  7^5        2^3,  4)$,\\
IrMOA$(56^4,  14,  56^3            8^4  7^5  4^1   2^1,  4)$,\\
IrMOA$(56^4,  13,  56^3            8^5  7^5           ,  4)$,\\
IrMOA$(56^4,  14,  56^2 28^6                       2^6,  4)$,\\
IrMOA$(56^4,  15,  56^2 28^5 14^1                  2^7,  4)$,\\
IrMOA$(56^4,  16,  56^2 28^4 14^2                  2^8,  4)$,\\
IrMOA$(56^4,  17,  56^2 28^3 14^3                  2^9,  4)$,\\
IrMOA$(56^4,  18,  56^2 28^2 14^4               2^{10},  4)$,\\
IrMOA$(56^4,  19,  56^2 28^1 14^5               2^{11},  4)$,\\
IrMOA$(56^4,  20,  56^2      14^6               2^{12},  4)$,\\
IrMOA$(56^4,  14,  56^2 28^5 14^1            4^1   2^5,  4)$,\\
IrMOA$(56^4,  15,  56^2 28^4 14^2            4^1   2^6,  4)$,\\
IrMOA$(56^4,  16,  56^2 28^3 14^3            4^1   2^7,  4)$,\\
IrMOA$(56^4,  17,  56^2 28^2 14^4            4^1   2^8,  4)$,\\
IrMOA$(56^4,  18,  56^2 28^1 14^5            4^1   2^9,  4)$,\\
IrMOA$(56^4,  19,  56^2      14^6            4^12^{10},  4)$,\\
IrMOA$(56^4,  14,  56^2 28^4 14^2            4^2   2^4,  4)$,\\
IrMOA$(56^4,  15,  56^2 28^3 14^3            4^2   2^5,  4)$,\\
IrMOA$(56^4,  16,  56^2 28^2 14^4            4^2   2^6,  4)$,\\
IrMOA$(56^4,  17,  56^2 28^1 14^5            4^2   2^7,  4)$,\\
IrMOA$(56^4,  18,  56^2      14^6            4^2   2^8,  4)$,\\
IrMOA$(56^4,  14,  56^2 28^3 14^3            4^3   2^3,  4)$,\\
IrMOA$(56^4,  15,  56^2 28^2 14^4            4^3   2^4,  4)$,\\
IrMOA$(56^4,  16,  56^2 28^1 14^5            4^3   2^5,  4)$,\\
IrMOA$(56^4,  17,  56^2      14^6            4^3   2^6,  4)$,\\
IrMOA$(56^4,  14,  56^2 28^2 14^4            4^4   2^2,  4)$,\\
IrMOA$(56^4,  15,  56^2 28^1 14^5            4^4   2^3,  4)$,\\
IrMOA$(56^4,  16,  56^2      14^6            4^4   2^4,  4)$,\\
IrMOA$(56^4,  14,  56^2 28^1 14^5            4^5   2^1,  4)$,\\
IrMOA$(56^4,  15,  56^2      14^6            4^5   2^2,  4)$,\\
IrMOA$(56^4,  14,  56^2      14^6            4^6      ,  4)$,\\
IrMOA$(56^4,  16,  56^2 28^5            7^1        2^8,  4)$,\\
IrMOA$(56^4,  17,  56^2 28^4 14^1       7^1        2^9,  4)$,\\
IrMOA$(56^4,  18,  56^2 28^3 14^2       7^1     2^{10},  4)$,\\
IrMOA$(56^4,  19,  56^2 28^2 14^3       7^1     2^{11},  4)$,\\
IrMOA$(56^4,  20,  56^2 28^1 14^4       7^1     2^{12},  4)$,\\
IrMOA$(56^4,  21,  56^2      14^5       7^1     2^{13},  4)$,\\
IrMOA$(56^4,  16,  56^2 28^4 14^1       7^1  4^1   2^7,  4)$,\\
IrMOA$(56^4,  17,  56^2 28^3 14^2       7^1  4^1   2^8,  4)$,\\
IrMOA$(56^4,  18,  56^2 28^2 14^3       7^1  4^1   2^9,  4)$,\\
IrMOA$(56^4,  19,  56^2 28^1 14^4       7^1  4^12^{10},  4)$,\\
IrMOA$(56^4,  20,  56^2      14^5       7^1  4^12^{11},  4)$,\\
IrMOA$(56^4,  16,  56^2 28^3 14^2       7^1  4^2   2^6,  4)$,\\
IrMOA$(56^4,  17,  56^2 28^2 14^3       7^1  4^2   2^7,  4)$,\\
IrMOA$(56^4,  18,  56^2 28^1 14^4       7^1  4^2   2^8,  4)$,\\
IrMOA$(56^4,  19,  56^2      14^5       7^1  4^2   2^9,  4)$,\\
IrMOA$(56^4,  16,  56^2 28^2 14^3       7^1  4^3   2^5,  4)$,\\
IrMOA$(56^4,  17,  56^2 28^1 14^4       7^1  4^3   2^6,  4)$,\\
IrMOA$(56^4,  18,  56^2      14^5       7^1  4^3   2^7,  4)$,\\
IrMOA$(56^4,  16,  56^2 28^1 14^4       7^1  4^4   2^4,  4)$,\\
IrMOA$(56^4,  17,  56^2      14^5       7^1  4^4   2^5,  4)$,\\
IrMOA$(56^4,  16,  56^2      14^5       7^1  4^5   2^3,  4)$,\\
IrMOA$(56^4,  18,  56^2 28^4            7^2     2^{10},  4)$,\\
IrMOA$(56^4,  19,  56^2 28^3 14^1       7^2     2^{11},  4)$,\\
IrMOA$(56^4,  20,  56^2 28^2 14^2       7^2     2^{12},  4)$,\\
IrMOA$(56^4,  21,  56^2 28^1 14^3       7^2     2^{13},  4)$,\\
IrMOA$(56^4,  22,  56^2      14^4       7^2     2^{14},  4)$,\\
IrMOA$(56^4,  18,  56^2 28^3 14^1       7^2  4^1   2^9,  4)$,\\
IrMOA$(56^4,  19,  56^2 28^2 14^2       7^2  4^12^{10},  4)$,\\
IrMOA$(56^4,  20,  56^2 28^1 14^3       7^2  4^12^{11},  4)$,\\
IrMOA$(56^4,  21,  56^2      14^4       7^2  4^12^{12},  4)$,\\
IrMOA$(56^4,  18,  56^2 28^2 14^2       7^2  4^2   2^8,  4)$,\\
IrMOA$(56^4,  19,  56^2 28^1 14^3       7^2  4^2   2^9,  4)$,\\
IrMOA$(56^4,  20,  56^2      14^4       7^2  4^22^{10},  4)$,\\
IrMOA$(56^4,  18,  56^2 28^1 14^3       7^2  4^3   2^7,  4)$,\\
IrMOA$(56^4,  19,  56^2      14^4       7^2  4^3   2^8,  4)$,\\
IrMOA$(56^4,  18,  56^2      14^4       7^2  4^4   2^6,  4)$,\\
IrMOA$(56^4,  20,  56^2 28^3            7^3     2^{12},  4)$,\\
IrMOA$(56^4,  21,  56^2 28^2 14^1       7^3     2^{13},  4)$,\\
IrMOA$(56^4,  22,  56^2 28^1 14^2       7^3     2^{14},  4)$,\\
IrMOA$(56^4,  23,  56^2      14^3       7^3     2^{15},  4)$,\\
IrMOA$(56^4,  20,  56^2 28^2 14^1       7^3  4^12^{11},  4)$,\\
IrMOA$(56^4,  21,  56^2 28^1 14^2       7^3  4^12^{12},  4)$,\\
IrMOA$(56^4,  22,  56^2      14^3       7^3  4^12^{13},  4)$,\\
IrMOA$(56^4,  20,  56^2 28^1 14^2       7^3  4^22^{10},  4)$,\\
IrMOA$(56^4,  21,  56^2      14^3       7^3  4^22^{11},  4)$,\\
IrMOA$(56^4,  20,  56^2      14^3       7^3  4^3   2^9,  4)$,\\
IrMOA$(56^4,  22,  56^2 28^2            7^4     2^{14},  4)$,\\
IrMOA$(56^4,  23,  56^2 28^1 14^1       7^4     2^{15},  4)$,\\
IrMOA$(56^4,  24,  56^2      14^2       7^4     2^{16},  4)$,\\
IrMOA$(56^4,  22,  56^2 28^1 14^1       7^4  4^12^{13},  4)$,\\
IrMOA$(56^4,  23,  56^2      14^2       7^4  4^12^{14},  4)$,\\
IrMOA$(56^4,  22,  56^2      14^2       7^4  4^22^{12},  4)$,\\
IrMOA$(56^4,  24,  56^2 28^1            7^5     2^{16},  4)$,\\
IrMOA$(56^4,  25,  56^2      14^1       7^5     2^{17},  4)$,\\
IrMOA$(56^4,  24,  56^2      14^1       7^5  4^12^{15},  4)$,\\
IrMOA$(56^4,  26,  56^2                 7^6     2^{18},  4)$,\\
IrMOA$(56^4,  15,  56^2 28^5            7^1  4^1   2^6,  4)$,\\
IrMOA$(56^4,  16,  56^2 28^4 14^1       7^1  4^1   2^7,  4)$,\\
IrMOA$(56^4,  17,  56^2 28^3 14^2       7^1  4^1   2^8,  4)$,\\
IrMOA$(56^4,  18,  56^2 28^2 14^3       7^1  4^1   2^9,  4)$,\\
IrMOA$(56^4,  19,  56^2 28^1 14^4       7^1  4^12^{10},  4)$,\\
IrMOA$(56^4,  20,  56^2      14^5       7^1  4^12^{11},  4)$,\\
IrMOA$(56^4,  15,  56^2 28^4 14^1       7^1  4^2   2^5,  4)$,\\
IrMOA$(56^4,  16,  56^2 28^3 14^2       7^1  4^2   2^6,  4)$,\\
IrMOA$(56^4,  17,  56^2 28^2 14^3       7^1  4^2   2^7,  4)$,\\
IrMOA$(56^4,  18,  56^2 28^1 14^4       7^1  4^2   2^8,  4)$,\\
IrMOA$(56^4,  19,  56^2      14^5       7^1  4^2   2^9,  4)$,\\
IrMOA$(56^4,  15,  56^2 28^3 14^2       7^1  4^3   2^4,  4)$,\\
IrMOA$(56^4,  16,  56^2 28^2 14^3       7^1  4^3   2^5,  4)$,\\
IrMOA$(56^4,  17,  56^2 28^1 14^4       7^1  4^3   2^6,  4)$,\\
IrMOA$(56^4,  18,  56^2      14^5       7^1  4^3   2^7,  4)$,\\
IrMOA$(56^4,  15,  56^2 28^2 14^3       7^1  4^4   2^3,  4)$,\\
IrMOA$(56^4,  16,  56^2 28^1 14^4       7^1  4^4   2^4,  4)$,\\
IrMOA$(56^4,  17,  56^2      14^5       7^1  4^4   2^5,  4)$,\\
IrMOA$(56^4,  15,  56^2 28^1 14^4       7^1  4^5   2^2,  4)$,\\
IrMOA$(56^4,  16,  56^2      14^5       7^1  4^5   2^3,  4)$,\\
IrMOA$(56^4,  15,  56^2      14^5       7^1  4^6   2^1,  4)$,\\
IrMOA$(56^4,  17,  56^2 28^4            7^2  4^1   2^8,  4)$,\\
IrMOA$(56^4,  18,  56^2 28^3 14^1       7^2  4^1   2^9,  4)$,\\
IrMOA$(56^4,  19,  56^2 28^2 14^2       7^2  4^12^{10},  4)$,\\
IrMOA$(56^4,  20,  56^2 28^1 14^3       7^2  4^12^{11},  4)$,\\
IrMOA$(56^4,  21,  56^2      14^4       7^2  4^12^{12},  4)$,\\
IrMOA$(56^4,  17,  56^2 28^3 14^1       7^2  4^2   2^7,  4)$,\\
IrMOA$(56^4,  18,  56^2 28^2 14^2       7^2  4^2   2^8,  4)$,\\
IrMOA$(56^4,  19,  56^2 28^1 14^3       7^2  4^2   2^9,  4)$,\\
IrMOA$(56^4,  20,  56^2      14^4       7^2  4^22^{10},  4)$,\\
IrMOA$(56^4,  17,  56^2 28^2 14^2       7^2  4^3   2^6,  4)$,\\
IrMOA$(56^4,  18,  56^2 28^1 14^3       7^2  4^3   2^7,  4)$,\\
IrMOA$(56^4,  19,  56^2      14^4       7^2  4^3   2^8,  4)$,\\
IrMOA$(56^4,  17,  56^2 28^1 14^3       7^2  4^4   2^5,  4)$,\\
IrMOA$(56^4,  18,  56^2      14^4       7^2  4^4   2^6,  4)$,\\
IrMOA$(56^4,  17,  56^2      14^4       7^2  4^5   2^4,  4)$,\\
IrMOA$(56^4,  19,  56^2 28^3            7^3  4^12^{10},  4)$,\\
IrMOA$(56^4,  20,  56^2 28^2 14^1       7^3  4^12^{11},  4)$,\\
IrMOA$(56^4,  21,  56^2 28^1 14^2       7^3  4^12^{12},  4)$,\\
IrMOA$(56^4,  22,  56^2      14^3       7^3  4^12^{13},  4)$,\\
IrMOA$(56^4,  19,  56^2 28^2 14^1       7^3  4^2   2^9,  4)$,\\
IrMOA$(56^4,  20,  56^2 28^1 14^2       7^3  4^22^{10},  4)$,\\
IrMOA$(56^4,  21,  56^2      14^3       7^3  4^22^{11},  4)$,\\
IrMOA$(56^4,  19,  56^2 28^1 14^2       7^3  4^3   2^8,  4)$,\\
IrMOA$(56^4,  20,  56^2      14^3       7^3  4^3   2^9,  4)$,\\
IrMOA$(56^4,  19,  56^2      14^3       7^3  4^4   2^7,  4)$,\\
IrMOA$(56^4,  21,  56^2 28^2            7^4  4^12^{12},  4)$,\\
IrMOA$(56^4,  22,  56^2 28^1 14^1       7^4  4^12^{13},  4)$,\\
IrMOA$(56^4,  23,  56^2      14^2       7^4  4^12^{14},  4)$,\\
IrMOA$(56^4,  21,  56^2 28^1 14^1       7^4  4^22^{11},  4)$,\\
IrMOA$(56^4,  22,  56^2      14^2       7^4  4^22^{12},  4)$,\\
IrMOA$(56^4,  21,  56^2      14^2       7^4  4^32^{10},  4)$,\\
IrMOA$(56^4,  23,  56^2 28^1            7^5  4^12^{14},  4)$,\\
IrMOA$(56^4,  24,  56^2      14^1       7^5  4^12^{15},  4)$,\\
IrMOA$(56^4,  23,  56^2      14^1       7^5  4^22^{13},  4)$,\\
IrMOA$(56^4,  25,  56^2                 7^6  4^12^{16},  4)$,\\
IrMOA$(56^4,  16,  56^2 28^4            7^2  4^2   2^6,  4)$,\\
IrMOA$(56^4,  17,  56^2 28^3 14^1       7^2  4^2   2^7,  4)$,\\
IrMOA$(56^4,  18,  56^2 28^2 14^2       7^2  4^2   2^8,  4)$,\\
IrMOA$(56^4,  19,  56^2 28^1 14^3       7^2  4^2   2^9,  4)$,\\
IrMOA$(56^4,  20,  56^2      14^4       7^2  4^22^{10},  4)$,\\
IrMOA$(56^4,  16,  56^2 28^3 14^1       7^2  4^3   2^5,  4)$,\\
IrMOA$(56^4,  17,  56^2 28^2 14^2       7^2  4^3   2^6,  4)$,\\
IrMOA$(56^4,  18,  56^2 28^1 14^3       7^2  4^3   2^7,  4)$,\\
IrMOA$(56^4,  19,  56^2      14^4       7^2  4^3   2^8,  4)$,\\
IrMOA$(56^4,  16,  56^2 28^2 14^2       7^2  4^4   2^4,  4)$,\\
IrMOA$(56^4,  17,  56^2 28^1 14^3       7^2  4^4   2^5,  4)$,\\
IrMOA$(56^4,  18,  56^2      14^4       7^2  4^4   2^6,  4)$,\\
IrMOA$(56^4,  16,  56^2 28^1 14^3       7^2  4^5   2^3,  4)$,\\
IrMOA$(56^4,  17,  56^2      14^4       7^2  4^5   2^4,  4)$,\\
IrMOA$(56^4,  16,  56^2      14^4       7^2  4^6   2^2,  4)$,\\
IrMOA$(56^4,  18,  56^2 28^3            7^3  4^2   2^8,  4)$,\\
IrMOA$(56^4,  19,  56^2 28^2 14^1       7^3  4^2   2^9,  4)$,\\
IrMOA$(56^4,  20,  56^2 28^1 14^2       7^3  4^22^{10},  4)$,\\
IrMOA$(56^4,  21,  56^2      14^3       7^3  4^22^{11},  4)$,\\
IrMOA$(56^4,  18,  56^2 28^2 14^1       7^3  4^3   2^7,  4)$,\\
IrMOA$(56^4,  19,  56^2 28^1 14^2       7^3  4^3   2^8,  4)$,\\
IrMOA$(56^4,  20,  56^2      14^3       7^3  4^3   2^9,  4)$,\\
IrMOA$(56^4,  18,  56^2 28^1 14^2       7^3  4^4   2^6,  4)$,\\
IrMOA$(56^4,  19,  56^2      14^3       7^3  4^4   2^7,  4)$,\\
IrMOA$(56^4,  18,  56^2      14^3       7^3  4^5   2^5,  4)$,\\
IrMOA$(56^4,  20,  56^2 28^2            7^4  4^22^{10},  4)$,\\
IrMOA$(56^4,  21,  56^2 28^1 14^1       7^4  4^22^{11},  4)$,\\
IrMOA$(56^4,  22,  56^2      14^2       7^4  4^22^{12},  4)$,\\
IrMOA$(56^4,  20,  56^2 28^1 14^1       7^4  4^3   2^9,  4)$,\\
IrMOA$(56^4,  21,  56^2      14^2       7^4  4^32^{10},  4)$,\\
IrMOA$(56^4,  20,  56^2      14^2       7^4  4^4   2^8,  4)$,\\
IrMOA$(56^4,  22,  56^2 28^1            7^5  4^22^{12},  4)$,\\
IrMOA$(56^4,  23,  56^2      14^1       7^5  4^22^{13},  4)$,\\
IrMOA$(56^4,  22,  56^2      14^1       7^5  4^32^{11},  4)$,\\
IrMOA$(56^4,  24,  56^2                 7^6  4^22^{14},  4)$,\\
IrMOA$(56^4,  17,  56^2 28^3            7^3  4^3   2^6,  4)$,\\
IrMOA$(56^4,  18,  56^2 28^2 14^1       7^3  4^3   2^7,  4)$,\\
IrMOA$(56^4,  19,  56^2 28^1 14^2       7^3  4^3   2^8,  4)$,\\
IrMOA$(56^4,  20,  56^2      14^3       7^3  4^3   2^9,  4)$,\\
IrMOA$(56^4,  17,  56^2 28^2 14^1       7^3  4^4   2^5,  4)$,\\
IrMOA$(56^4,  18,  56^2 28^1 14^2       7^3  4^4   2^6,  4)$,\\
IrMOA$(56^4,  19,  56^2      14^3       7^3  4^4   2^7,  4)$,\\
IrMOA$(56^4,  17,  56^2 28^1 14^2       7^3  4^5   2^4,  4)$,\\
IrMOA$(56^4,  18,  56^2      14^3       7^3  4^5   2^5,  4)$,\\
IrMOA$(56^4,  17,  56^2      14^3       7^3  4^6   2^3,  4)$,\\
IrMOA$(56^4,  19,  56^2 28^2            7^4  4^3   2^8,  4)$,\\
IrMOA$(56^4,  20,  56^2 28^1 14^1       7^4  4^3   2^9,  4)$,\\
IrMOA$(56^4,  21,  56^2      14^2       7^4  4^32^{10},  4)$,\\
IrMOA$(56^4,  19,  56^2 28^1 14^1       7^4  4^4   2^7,  4)$,\\
IrMOA$(56^4,  20,  56^2      14^2       7^4  4^4   2^8,  4)$,\\
IrMOA$(56^4,  19,  56^2      14^2       7^4  4^5   2^6,  4)$,\\
IrMOA$(56^4,  21,  56^2 28^1            7^5  4^32^{10},  4)$,\\
IrMOA$(56^4,  22,  56^2      14^1       7^5  4^32^{11},  4)$,\\
IrMOA$(56^4,  21,  56^2      14^1       7^5  4^4   2^9,  4)$,\\
IrMOA$(56^4,  23,  56^2                 7^6  4^32^{12},  4)$,\\
IrMOA$(56^4,  18,  56^2 28^2            7^4  4^4   2^6,  4)$,\\
IrMOA$(56^4,  19,  56^2 28^1 14^1       7^4  4^4   2^7,  4)$,\\
IrMOA$(56^4,  20,  56^2      14^2       7^4  4^4   2^8,  4)$,\\
IrMOA$(56^4,  18,  56^2 28^1 14^1       7^4  4^5   2^5,  4)$,\\
IrMOA$(56^4,  19,  56^2      14^2       7^4  4^5   2^6,  4)$,\\
IrMOA$(56^4,  18,  56^2      14^2       7^4  4^6   2^4,  4)$,\\
IrMOA$(56^4,  20,  56^2 28^1            7^5  4^4   2^8,  4)$,\\
IrMOA$(56^4,  21,  56^2      14^1       7^5  4^4   2^9,  4)$,\\
IrMOA$(56^4,  20,  56^2      14^1       7^5  4^5   2^7,  4)$,\\
IrMOA$(56^4,  22,  56^2                 7^6  4^42^{10},  4)$,\\
IrMOA$(56^4,  19,  56^2 28^1            7^5  4^5   2^6,  4)$,\\
IrMOA$(56^4,  20,  56^2      14^1       7^5  4^5   2^7,  4)$,\\
IrMOA$(56^4,  19,  56^2      14^1       7^5  4^6   2^5,  4)$,\\
IrMOA$(56^4,  21,  56^2                 7^6  4^5   2^8,  4)$,\\
IrMOA$(56^4,  20,  56^2                 7^6  4^6   2^6,  4)$,\\
IrMOA$(56^4,  14,  56^2 28^5       8^1  7^1        2^5,  4)$,\\
IrMOA$(56^4,  15,  56^2 28^4 14^1  8^1  7^1        2^6,  4)$,\\
IrMOA$(56^4,  16,  56^2 28^3 14^2  8^1  7^1        2^7,  4)$,\\
IrMOA$(56^4,  17,  56^2 28^2 14^3  8^1  7^1        2^8,  4)$,\\
IrMOA$(56^4,  18,  56^2 28^1 14^4  8^1  7^1        2^9,  4)$,\\
IrMOA$(56^4,  19,  56^2      14^5  8^1  7^1     2^{10},  4)$,\\
IrMOA$(56^4,  14,  56^2 28^4 14^1  8^1  7^1  4^1   2^4,  4)$,\\
IrMOA$(56^4,  15,  56^2 28^3 14^2  8^1  7^1  4^1   2^5,  4)$,\\
IrMOA$(56^4,  16,  56^2 28^2 14^3  8^1  7^1  4^1   2^6,  4)$,\\
IrMOA$(56^4,  17,  56^2 28^1 14^4  8^1  7^1  4^1   2^7,  4)$,\\
IrMOA$(56^4,  18,  56^2      14^5  8^1  7^1  4^1   2^8,  4)$,\\
IrMOA$(56^4,  14,  56^2 28^3 14^2  8^1  7^1  4^2   2^3,  4)$,\\
IrMOA$(56^4,  15,  56^2 28^2 14^3  8^1  7^1  4^2   2^4,  4)$,\\
IrMOA$(56^4,  16,  56^2 28^1 14^4  8^1  7^1  4^2   2^5,  4)$,\\
IrMOA$(56^4,  17,  56^2      14^5  8^1  7^1  4^2   2^6,  4)$,\\
IrMOA$(56^4,  14,  56^2 28^2 14^3  8^1  7^1  4^3   2^2,  4)$,\\
IrMOA$(56^4,  15,  56^2 28^1 14^4  8^1  7^1  4^3   2^3,  4)$,\\
IrMOA$(56^4,  16,  56^2      14^5  8^1  7^1  4^3   2^4,  4)$,\\
IrMOA$(56^4,  14,  56^2 28^1 14^4  8^1  7^1  4^4   2^1,  4)$,\\
IrMOA$(56^4,  15,  56^2      14^5  8^1  7^1  4^4   2^2,  4)$,\\
IrMOA$(56^4,  14,  56^2      14^5  8^1  7^1  4^5      ,  4)$,\\
IrMOA$(56^4,  16,  56^2 28^4       8^1  7^2        2^7,  4)$,\\
IrMOA$(56^4,  17,  56^2 28^3 14^1  8^1  7^2        2^8,  4)$,\\
IrMOA$(56^4,  18,  56^2 28^2 14^2  8^1  7^2        2^9,  4)$,\\
IrMOA$(56^4,  19,  56^2 28^1 14^3  8^1  7^2     2^{10},  4)$,\\
IrMOA$(56^4,  20,  56^2      14^4  8^1  7^2     2^{11},  4)$,\\
IrMOA$(56^4,  16,  56^2 28^3 14^1  8^1  7^2  4^1   2^6,  4)$,\\
IrMOA$(56^4,  17,  56^2 28^2 14^2  8^1  7^2  4^1   2^7,  4)$,\\
IrMOA$(56^4,  18,  56^2 28^1 14^3  8^1  7^2  4^1   2^8,  4)$,\\
IrMOA$(56^4,  19,  56^2      14^4  8^1  7^2  4^1   2^9,  4)$,\\
IrMOA$(56^4,  16,  56^2 28^2 14^2  8^1  7^2  4^2   2^5,  4)$,\\
IrMOA$(56^4,  17,  56^2 28^1 14^3  8^1  7^2  4^2   2^6,  4)$,\\
IrMOA$(56^4,  18,  56^2      14^4  8^1  7^2  4^2   2^7,  4)$,\\
IrMOA$(56^4,  16,  56^2 28^1 14^3  8^1  7^2  4^3   2^4,  4)$,\\
IrMOA$(56^4,  17,  56^2      14^4  8^1  7^2  4^3   2^5,  4)$,\\
IrMOA$(56^4,  16,  56^2      14^4  8^1  7^2  4^4   2^3,  4)$,\\
IrMOA$(56^4,  18,  56^2 28^3       8^1  7^3        2^9,  4)$,\\
IrMOA$(56^4,  19,  56^2 28^2 14^1  8^1  7^3     2^{10},  4)$,\\
IrMOA$(56^4,  20,  56^2 28^1 14^2  8^1  7^3     2^{11},  4)$,\\
IrMOA$(56^4,  21,  56^2      14^3  8^1  7^3     2^{12},  4)$,\\
IrMOA$(56^4,  18,  56^2 28^2 14^1  8^1  7^3  4^1   2^8,  4)$,\\
IrMOA$(56^4,  19,  56^2 28^1 14^2  8^1  7^3  4^1   2^9,  4)$,\\
IrMOA$(56^4,  20,  56^2      14^3  8^1  7^3  4^12^{10},  4)$,\\
IrMOA$(56^4,  18,  56^2 28^1 14^2  8^1  7^3  4^2   2^7,  4)$,\\
IrMOA$(56^4,  19,  56^2      14^3  8^1  7^3  4^2   2^8,  4)$,\\
IrMOA$(56^4,  18,  56^2      14^3  8^1  7^3  4^3   2^6,  4)$,\\
IrMOA$(56^4,  20,  56^2 28^2       8^1  7^4     2^{11},  4)$,\\
IrMOA$(56^4,  21,  56^2 28^1 14^1  8^1  7^4     2^{12},  4)$,\\
IrMOA$(56^4,  22,  56^2      14^2  8^1  7^4     2^{13},  4)$,\\
IrMOA$(56^4,  20,  56^2 28^1 14^1  8^1  7^4  4^12^{10},  4)$,\\
IrMOA$(56^4,  21,  56^2      14^2  8^1  7^4  4^12^{11},  4)$,\\
IrMOA$(56^4,  20,  56^2      14^2  8^1  7^4  4^2   2^9,  4)$,\\
IrMOA$(56^4,  22,  56^2 28^1       8^1  7^5     2^{13},  4)$,\\
IrMOA$(56^4,  23,  56^2      14^1  8^1  7^5     2^{14},  4)$,\\
IrMOA$(56^4,  22,  56^2      14^1  8^1  7^5  4^12^{12},  4)$,\\
IrMOA$(56^4,  24,  56^2            8^1  7^6     2^{15},  4)$,\\
IrMOA$(56^4,  15,  56^2 28^4       8^1  7^2  4^1   2^5,  4)$,\\
IrMOA$(56^4,  16,  56^2 28^3 14^1  8^1  7^2  4^1   2^6,  4)$,\\
IrMOA$(56^4,  17,  56^2 28^2 14^2  8^1  7^2  4^1   2^7,  4)$,\\
IrMOA$(56^4,  18,  56^2 28^1 14^3  8^1  7^2  4^1   2^8,  4)$,\\
IrMOA$(56^4,  19,  56^2      14^4  8^1  7^2  4^1   2^9,  4)$,\\
IrMOA$(56^4,  15,  56^2 28^3 14^1  8^1  7^2  4^2   2^4,  4)$,\\
IrMOA$(56^4,  16,  56^2 28^2 14^2  8^1  7^2  4^2   2^5,  4)$,\\
IrMOA$(56^4,  17,  56^2 28^1 14^3  8^1  7^2  4^2   2^6,  4)$,\\
IrMOA$(56^4,  18,  56^2      14^4  8^1  7^2  4^2   2^7,  4)$,\\
IrMOA$(56^4,  15,  56^2 28^2 14^2  8^1  7^2  4^3   2^3,  4)$,\\
IrMOA$(56^4,  16,  56^2 28^1 14^3  8^1  7^2  4^3   2^4,  4)$,\\
IrMOA$(56^4,  17,  56^2      14^4  8^1  7^2  4^3   2^5,  4)$,\\
IrMOA$(56^4,  15,  56^2 28^1 14^3  8^1  7^2  4^4   2^2,  4)$,\\
IrMOA$(56^4,  16,  56^2      14^4  8^1  7^2  4^4   2^3,  4)$,\\
IrMOA$(56^4,  15,  56^2      14^4  8^1  7^2  4^5   2^1,  4)$,\\
IrMOA$(56^4,  17,  56^2 28^3       8^1  7^3  4^1   2^7,  4)$,\\
IrMOA$(56^4,  18,  56^2 28^2 14^1  8^1  7^3  4^1   2^8,  4)$,\\
IrMOA$(56^4,  19,  56^2 28^1 14^2  8^1  7^3  4^1   2^9,  4)$,\\
IrMOA$(56^4,  20,  56^2      14^3  8^1  7^3  4^12^{10},  4)$,\\
IrMOA$(56^4,  17,  56^2 28^2 14^1  8^1  7^3  4^2   2^6,  4)$,\\
IrMOA$(56^4,  18,  56^2 28^1 14^2  8^1  7^3  4^2   2^7,  4)$,\\
IrMOA$(56^4,  19,  56^2      14^3  8^1  7^3  4^2   2^8,  4)$,\\
IrMOA$(56^4,  17,  56^2 28^1 14^2  8^1  7^3  4^3   2^5,  4)$,\\
IrMOA$(56^4,  18,  56^2      14^3  8^1  7^3  4^3   2^6,  4)$,\\
IrMOA$(56^4,  17,  56^2      14^3  8^1  7^3  4^4   2^4,  4)$,\\
IrMOA$(56^4,  19,  56^2 28^2       8^1  7^4  4^1   2^9,  4)$,\\
IrMOA$(56^4,  20,  56^2 28^1 14^1  8^1  7^4  4^12^{10},  4)$,\\
IrMOA$(56^4,  21,  56^2      14^2  8^1  7^4  4^12^{11},  4)$,\\
IrMOA$(56^4,  19,  56^2 28^1 14^1  8^1  7^4  4^2   2^8,  4)$,\\
IrMOA$(56^4,  20,  56^2      14^2  8^1  7^4  4^2   2^9,  4)$,\\
IrMOA$(56^4,  19,  56^2      14^2  8^1  7^4  4^3   2^7,  4)$,\\
IrMOA$(56^4,  21,  56^2 28^1       8^1  7^5  4^12^{11},  4)$,\\
IrMOA$(56^4,  22,  56^2      14^1  8^1  7^5  4^12^{12},  4)$,\\
IrMOA$(56^4,  21,  56^2      14^1  8^1  7^5  4^22^{10},  4)$,\\
IrMOA$(56^4,  23,  56^2            8^1  7^6  4^12^{13},  4)$,\\
IrMOA$(56^4,  16,  56^2 28^3       8^1  7^3  4^2   2^5,  4)$,\\
IrMOA$(56^4,  17,  56^2 28^2 14^1  8^1  7^3  4^2   2^6,  4)$,\\
IrMOA$(56^4,  18,  56^2 28^1 14^2  8^1  7^3  4^2   2^7,  4)$,\\
IrMOA$(56^4,  19,  56^2      14^3  8^1  7^3  4^2   2^8,  4)$,\\
IrMOA$(56^4,  16,  56^2 28^2 14^1  8^1  7^3  4^3   2^4,  4)$,\\
IrMOA$(56^4,  17,  56^2 28^1 14^2  8^1  7^3  4^3   2^5,  4)$,\\
IrMOA$(56^4,  18,  56^2      14^3  8^1  7^3  4^3   2^6,  4)$,\\
IrMOA$(56^4,  16,  56^2 28^1 14^2  8^1  7^3  4^4   2^3,  4)$,\\
IrMOA$(56^4,  17,  56^2      14^3  8^1  7^3  4^4   2^4,  4)$,\\
IrMOA$(56^4,  16,  56^2      14^3  8^1  7^3  4^5   2^2,  4)$,\\
IrMOA$(56^4,  18,  56^2 28^2       8^1  7^4  4^2   2^7,  4)$,\\
IrMOA$(56^4,  19,  56^2 28^1 14^1  8^1  7^4  4^2   2^8,  4)$,\\
IrMOA$(56^4,  20,  56^2      14^2  8^1  7^4  4^2   2^9,  4)$,\\
IrMOA$(56^4,  18,  56^2 28^1 14^1  8^1  7^4  4^3   2^6,  4)$,\\
IrMOA$(56^4,  19,  56^2      14^2  8^1  7^4  4^3   2^7,  4)$,\\
IrMOA$(56^4,  18,  56^2      14^2  8^1  7^4  4^4   2^5,  4)$,\\
IrMOA$(56^4,  20,  56^2 28^1       8^1  7^5  4^2   2^9,  4)$,\\
IrMOA$(56^4,  21,  56^2      14^1  8^1  7^5  4^22^{10},  4)$,\\
IrMOA$(56^4,  20,  56^2      14^1  8^1  7^5  4^3   2^8,  4)$,\\
IrMOA$(56^4,  22,  56^2            8^1  7^6  4^22^{11},  4)$,\\
IrMOA$(56^4,  17,  56^2 28^2       8^1  7^4  4^3   2^5,  4)$,\\
IrMOA$(56^4,  18,  56^2 28^1 14^1  8^1  7^4  4^3   2^6,  4)$,\\
IrMOA$(56^4,  19,  56^2      14^2  8^1  7^4  4^3   2^7,  4)$,\\
IrMOA$(56^4,  17,  56^2 28^1 14^1  8^1  7^4  4^4   2^4,  4)$,\\
IrMOA$(56^4,  18,  56^2      14^2  8^1  7^4  4^4   2^5,  4)$,\\
IrMOA$(56^4,  17,  56^2      14^2  8^1  7^4  4^5   2^3,  4)$,\\
IrMOA$(56^4,  19,  56^2 28^1       8^1  7^5  4^3   2^7,  4)$,\\
IrMOA$(56^4,  20,  56^2      14^1  8^1  7^5  4^3   2^8,  4)$,\\
IrMOA$(56^4,  19,  56^2      14^1  8^1  7^5  4^4   2^6,  4)$,\\
IrMOA$(56^4,  21,  56^2            8^1  7^6  4^3   2^9,  4)$,\\
IrMOA$(56^4,  18,  56^2 28^1       8^1  7^5  4^4   2^5,  4)$,\\
IrMOA$(56^4,  19,  56^2      14^1  8^1  7^5  4^4   2^6,  4)$,\\
IrMOA$(56^4,  18,  56^2      14^1  8^1  7^5  4^5   2^4,  4)$,\\
IrMOA$(56^4,  20,  56^2            8^1  7^6  4^4   2^7,  4)$,\\
IrMOA$(56^4,  19,  56^2            8^1  7^6  4^5   2^5,  4)$,\\
IrMOA$(56^4,  14,  56^2 28^4       8^2  7^2        2^4,  4)$,\\
IrMOA$(56^4,  15,  56^2 28^3 14^1  8^2  7^2        2^5,  4)$,\\
IrMOA$(56^4,  16,  56^2 28^2 14^2  8^2  7^2        2^6,  4)$,\\
IrMOA$(56^4,  17,  56^2 28^1 14^3  8^2  7^2        2^7,  4)$,\\
IrMOA$(56^4,  18,  56^2      14^4  8^2  7^2        2^8,  4)$,\\
IrMOA$(56^4,  14,  56^2 28^3 14^1  8^2  7^2  4^1   2^3,  4)$,\\
IrMOA$(56^4,  15,  56^2 28^2 14^2  8^2  7^2  4^1   2^4,  4)$,\\
IrMOA$(56^4,  16,  56^2 28^1 14^3  8^2  7^2  4^1   2^5,  4)$,\\
IrMOA$(56^4,  17,  56^2      14^4  8^2  7^2  4^1   2^6,  4)$,\\
IrMOA$(56^4,  14,  56^2 28^2 14^2  8^2  7^2  4^2   2^2,  4)$,\\
IrMOA$(56^4,  15,  56^2 28^1 14^3  8^2  7^2  4^2   2^3,  4)$,\\
IrMOA$(56^4,  16,  56^2      14^4  8^2  7^2  4^2   2^4,  4)$,\\
IrMOA$(56^4,  14,  56^2 28^1 14^3  8^2  7^2  4^3   2^1,  4)$,\\
IrMOA$(56^4,  15,  56^2      14^4  8^2  7^2  4^3   2^2,  4)$,\\
IrMOA$(56^4,  14,  56^2      14^4  8^2  7^2  4^4      ,  4)$,\\
IrMOA$(56^4,  16,  56^2 28^3       8^2  7^3        2^6,  4)$,\\
IrMOA$(56^4,  17,  56^2 28^2 14^1  8^2  7^3        2^7,  4)$,\\
IrMOA$(56^4,  18,  56^2 28^1 14^2  8^2  7^3        2^8,  4)$,\\
IrMOA$(56^4,  19,  56^2      14^3  8^2  7^3        2^9,  4)$,\\
IrMOA$(56^4,  16,  56^2 28^2 14^1  8^2  7^3  4^1   2^5,  4)$,\\
IrMOA$(56^4,  17,  56^2 28^1 14^2  8^2  7^3  4^1   2^6,  4)$,\\
IrMOA$(56^4,  18,  56^2      14^3  8^2  7^3  4^1   2^7,  4)$,\\
IrMOA$(56^4,  16,  56^2 28^1 14^2  8^2  7^3  4^2   2^4,  4)$,\\
IrMOA$(56^4,  17,  56^2      14^3  8^2  7^3  4^2   2^5,  4)$,\\
IrMOA$(56^4,  16,  56^2      14^3  8^2  7^3  4^3   2^3,  4)$,\\
IrMOA$(56^4,  18,  56^2 28^2       8^2  7^4        2^8,  4)$,\\
IrMOA$(56^4,  19,  56^2 28^1 14^1  8^2  7^4        2^9,  4)$,\\
IrMOA$(56^4,  20,  56^2      14^2  8^2  7^4     2^{10},  4)$,\\
IrMOA$(56^4,  18,  56^2 28^1 14^1  8^2  7^4  4^1   2^7,  4)$,\\
IrMOA$(56^4,  19,  56^2      14^2  8^2  7^4  4^1   2^8,  4)$,\\
IrMOA$(56^4,  18,  56^2      14^2  8^2  7^4  4^2   2^6,  4)$,\\
IrMOA$(56^4,  20,  56^2 28^1       8^2  7^5     2^{10},  4)$,\\
IrMOA$(56^4,  21,  56^2      14^1  8^2  7^5     2^{11},  4)$,\\
IrMOA$(56^4,  20,  56^2      14^1  8^2  7^5  4^1   2^9,  4)$,\\
IrMOA$(56^4,  22,  56^2            8^2  7^6     2^{12},  4)$,\\
IrMOA$(56^4,  15,  56^2 28^3       8^2  7^3  4^1   2^4,  4)$,\\
IrMOA$(56^4,  16,  56^2 28^2 14^1  8^2  7^3  4^1   2^5,  4)$,\\
IrMOA$(56^4,  17,  56^2 28^1 14^2  8^2  7^3  4^1   2^6,  4)$,\\
IrMOA$(56^4,  18,  56^2      14^3  8^2  7^3  4^1   2^7,  4)$,\\
IrMOA$(56^4,  15,  56^2 28^2 14^1  8^2  7^3  4^2   2^3,  4)$,\\
IrMOA$(56^4,  16,  56^2 28^1 14^2  8^2  7^3  4^2   2^4,  4)$,\\
IrMOA$(56^4,  17,  56^2      14^3  8^2  7^3  4^2   2^5,  4)$,\\
IrMOA$(56^4,  15,  56^2 28^1 14^2  8^2  7^3  4^3   2^2,  4)$,\\
IrMOA$(56^4,  16,  56^2      14^3  8^2  7^3  4^3   2^3,  4)$,\\
IrMOA$(56^4,  15,  56^2      14^3  8^2  7^3  4^4   2^1,  4)$,\\
IrMOA$(56^4,  17,  56^2 28^2       8^2  7^4  4^1   2^6,  4)$,\\
IrMOA$(56^4,  18,  56^2 28^1 14^1  8^2  7^4  4^1   2^7,  4)$,\\
IrMOA$(56^4,  19,  56^2      14^2  8^2  7^4  4^1   2^8,  4)$,\\
IrMOA$(56^4,  17,  56^2 28^1 14^1  8^2  7^4  4^2   2^5,  4)$,\\
IrMOA$(56^4,  18,  56^2      14^2  8^2  7^4  4^2   2^6,  4)$,\\
IrMOA$(56^4,  17,  56^2      14^2  8^2  7^4  4^3   2^4,  4)$,\\
IrMOA$(56^4,  19,  56^2 28^1       8^2  7^5  4^1   2^8,  4)$,\\
IrMOA$(56^4,  20,  56^2      14^1  8^2  7^5  4^1   2^9,  4)$,\\
IrMOA$(56^4,  19,  56^2      14^1  8^2  7^5  4^2   2^7,  4)$,\\
IrMOA$(56^4,  21,  56^2            8^2  7^6  4^12^{10},  4)$,\\
IrMOA$(56^4,  16,  56^2 28^2       8^2  7^4  4^2   2^4,  4)$,\\
IrMOA$(56^4,  17,  56^2 28^1 14^1  8^2  7^4  4^2   2^5,  4)$,\\
IrMOA$(56^4,  18,  56^2      14^2  8^2  7^4  4^2   2^6,  4)$,\\
IrMOA$(56^4,  16,  56^2 28^1 14^1  8^2  7^4  4^3   2^3,  4)$,\\
IrMOA$(56^4,  17,  56^2      14^2  8^2  7^4  4^3   2^4,  4)$,\\
IrMOA$(56^4,  16,  56^2      14^2  8^2  7^4  4^4   2^2,  4)$,\\
IrMOA$(56^4,  18,  56^2 28^1       8^2  7^5  4^2   2^6,  4)$,\\
IrMOA$(56^4,  19,  56^2      14^1  8^2  7^5  4^2   2^7,  4)$,\\
IrMOA$(56^4,  18,  56^2      14^1  8^2  7^5  4^3   2^5,  4)$,\\
IrMOA$(56^4,  20,  56^2            8^2  7^6  4^2   2^8,  4)$,\\
IrMOA$(56^4,  17,  56^2 28^1       8^2  7^5  4^3   2^4,  4)$,\\
IrMOA$(56^4,  18,  56^2      14^1  8^2  7^5  4^3   2^5,  4)$,\\
IrMOA$(56^4,  17,  56^2      14^1  8^2  7^5  4^4   2^3,  4)$,\\
IrMOA$(56^4,  19,  56^2            8^2  7^6  4^3   2^6,  4)$,\\
IrMOA$(56^4,  18,  56^2            8^2  7^6  4^4   2^4,  4)$,\\
IrMOA$(56^4,  14,  56^2 28^3       8^3  7^3        2^3,  4)$,\\
IrMOA$(56^4,  15,  56^2 28^2 14^1  8^3  7^3        2^4,  4)$,\\
IrMOA$(56^4,  16,  56^2 28^1 14^2  8^3  7^3        2^5,  4)$,\\
IrMOA$(56^4,  17,  56^2      14^3  8^3  7^3        2^6,  4)$,\\
IrMOA$(56^4,  14,  56^2 28^2 14^1  8^3  7^3  4^1   2^2,  4)$,\\
IrMOA$(56^4,  15,  56^2 28^1 14^2  8^3  7^3  4^1   2^3,  4)$,\\
IrMOA$(56^4,  16,  56^2      14^3  8^3  7^3  4^1   2^4,  4)$,\\
IrMOA$(56^4,  14,  56^2 28^1 14^2  8^3  7^3  4^2   2^1,  4)$,\\
IrMOA$(56^4,  15,  56^2      14^3  8^3  7^3  4^2   2^2,  4)$,\\
IrMOA$(56^4,  14,  56^2      14^3  8^3  7^3  4^3      ,  4)$,\\
IrMOA$(56^4,  16,  56^2 28^2       8^3  7^4        2^5,  4)$,\\
IrMOA$(56^4,  17,  56^2 28^1 14^1  8^3  7^4        2^6,  4)$,\\
IrMOA$(56^4,  18,  56^2      14^2  8^3  7^4        2^7,  4)$,\\
IrMOA$(56^4,  16,  56^2 28^1 14^1  8^3  7^4  4^1   2^4,  4)$,\\
IrMOA$(56^4,  17,  56^2      14^2  8^3  7^4  4^1   2^5,  4)$,\\
IrMOA$(56^4,  16,  56^2      14^2  8^3  7^4  4^2   2^3,  4)$,\\
IrMOA$(56^4,  18,  56^2 28^1       8^3  7^5        2^7,  4)$,\\
IrMOA$(56^4,  19,  56^2      14^1  8^3  7^5        2^8,  4)$,\\
IrMOA$(56^4,  18,  56^2      14^1  8^3  7^5  4^1   2^6,  4)$,\\
IrMOA$(56^4,  20,  56^2            8^3  7^6        2^9,  4)$,\\
IrMOA$(56^4,  15,  56^2 28^2       8^3  7^4  4^1   2^3,  4)$,\\
IrMOA$(56^4,  16,  56^2 28^1 14^1  8^3  7^4  4^1   2^4,  4)$,\\
IrMOA$(56^4,  17,  56^2      14^2  8^3  7^4  4^1   2^5,  4)$,\\
IrMOA$(56^4,  15,  56^2 28^1 14^1  8^3  7^4  4^2   2^2,  4)$,\\
IrMOA$(56^4,  16,  56^2      14^2  8^3  7^4  4^2   2^3,  4)$,\\
IrMOA$(56^4,  15,  56^2      14^2  8^3  7^4  4^3   2^1,  4)$,\\
IrMOA$(56^4,  17,  56^2 28^1       8^3  7^5  4^1   2^5,  4)$,\\
IrMOA$(56^4,  18,  56^2      14^1  8^3  7^5  4^1   2^6,  4)$,\\
IrMOA$(56^4,  17,  56^2      14^1  8^3  7^5  4^2   2^4,  4)$,\\
IrMOA$(56^4,  19,  56^2            8^3  7^6  4^1   2^7,  4)$,\\
IrMOA$(56^4,  16,  56^2 28^1       8^3  7^5  4^2   2^3,  4)$,\\
IrMOA$(56^4,  17,  56^2      14^1  8^3  7^5  4^2   2^4,  4)$,\\
IrMOA$(56^4,  16,  56^2      14^1  8^3  7^5  4^3   2^2,  4)$,\\
IrMOA$(56^4,  18,  56^2            8^3  7^6  4^2   2^5,  4)$,\\
IrMOA$(56^4,  17,  56^2            8^3  7^6  4^3   2^3,  4)$,\\
IrMOA$(56^4,  14,  56^2 28^2       8^4  7^4        2^2,  4)$,\\
IrMOA$(56^4,  15,  56^2 28^1 14^1  8^4  7^4        2^3,  4)$,\\
IrMOA$(56^4,  16,  56^2      14^2  8^4  7^4        2^4,  4)$,\\
IrMOA$(56^4,  14,  56^2 28^1 14^1  8^4  7^4  4^1   2^1,  4)$,\\
IrMOA$(56^4,  15,  56^2      14^2  8^4  7^4  4^1   2^2,  4)$,\\
IrMOA$(56^4,  14,  56^2      14^2  8^4  7^4  4^2      ,  4)$,\\
IrMOA$(56^4,  16,  56^2 28^1       8^4  7^5        2^4,  4)$,\\
IrMOA$(56^4,  17,  56^2      14^1  8^4  7^5        2^5,  4)$,\\
IrMOA$(56^4,  16,  56^2      14^1  8^4  7^5  4^1   2^3,  4)$,\\
IrMOA$(56^4,  18,  56^2            8^4  7^6        2^6,  4)$,\\
IrMOA$(56^4,  15,  56^2 28^1       8^4  7^5  4^1   2^2,  4)$,\\
IrMOA$(56^4,  16,  56^2      14^1  8^4  7^5  4^1   2^3,  4)$,\\
IrMOA$(56^4,  15,  56^2      14^1  8^4  7^5  4^2   2^1,  4)$,\\
IrMOA$(56^4,  17,  56^2            8^4  7^6  4^1   2^4,  4)$,\\
IrMOA$(56^4,  16,  56^2            8^4  7^6  4^2   2^2,  4)$,\\
IrMOA$(56^4,  14,  56^2 28^1       8^5  7^5        2^1,  4)$,\\
IrMOA$(56^4,  15,  56^2      14^1  8^5  7^5        2^2,  4)$,\\
IrMOA$(56^4,  14,  56^2      14^1  8^5  7^5  4^1      ,  4)$,\\
IrMOA$(56^4,  16,  56^2            8^5  7^6        2^3,  4)$,\\
IrMOA$(56^4,  15,  56^2            8^5  7^6  4^1   2^1,  4)$,\\
IrMOA$(56^4,  14,  56^2            8^6  7^6           ,  4)$,\\
IrMOA$(56^4,  15,  56^1 28^7                       2^7,  4)$,\\
IrMOA$(56^4,  16,  56^1 28^6 14^1                  2^8,  4)$,\\
IrMOA$(56^4,  17,  56^1 28^5 14^2                  2^9,  4)$,\\
IrMOA$(56^4,  18,  56^1 28^4 14^3               2^{10},  4)$,\\
IrMOA$(56^4,  19,  56^1 28^3 14^4               2^{11},  4)$,\\
IrMOA$(56^4,  20,  56^1 28^2 14^5               2^{12},  4)$,\\
IrMOA$(56^4,  21,  56^1 28^1 14^6               2^{13},  4)$,\\
IrMOA$(56^4,  22,  56^1      14^7               2^{14},  4)$,\\
IrMOA$(56^4,  15,  56^1 28^6 14^1            4^1   2^6,  4)$,\\
IrMOA$(56^4,  16,  56^1 28^5 14^2            4^1   2^7,  4)$,\\
IrMOA$(56^4,  17,  56^1 28^4 14^3            4^1   2^8,  4)$,\\
IrMOA$(56^4,  18,  56^1 28^3 14^4            4^1   2^9,  4)$,\\
IrMOA$(56^4,  19,  56^1 28^2 14^5            4^12^{10},  4)$,\\
IrMOA$(56^4,  20,  56^1 28^1 14^6            4^12^{11},  4)$,\\
IrMOA$(56^4,  21,  56^1      14^7            4^12^{12},  4)$,\\
IrMOA$(56^4,  15,  56^1 28^5 14^2            4^2   2^5,  4)$,\\
IrMOA$(56^4,  16,  56^1 28^4 14^3            4^2   2^6,  4)$,\\
IrMOA$(56^4,  17,  56^1 28^3 14^4            4^2   2^7,  4)$,\\
IrMOA$(56^4,  18,  56^1 28^2 14^5            4^2   2^8,  4)$,\\
IrMOA$(56^4,  19,  56^1 28^1 14^6            4^2   2^9,  4)$,\\
IrMOA$(56^4,  20,  56^1      14^7            4^22^{10},  4)$,\\
IrMOA$(56^4,  15,  56^1 28^4 14^3            4^3   2^4,  4)$,\\
IrMOA$(56^4,  16,  56^1 28^3 14^4            4^3   2^5,  4)$,\\
IrMOA$(56^4,  17,  56^1 28^2 14^5            4^3   2^6,  4)$,\\
IrMOA$(56^4,  18,  56^1 28^1 14^6            4^3   2^7,  4)$,\\
IrMOA$(56^4,  19,  56^1      14^7            4^3   2^8,  4)$,\\
IrMOA$(56^4,  15,  56^1 28^3 14^4            4^4   2^3,  4)$,\\
IrMOA$(56^4,  16,  56^1 28^2 14^5            4^4   2^4,  4)$,\\
IrMOA$(56^4,  17,  56^1 28^1 14^6            4^4   2^5,  4)$,\\
IrMOA$(56^4,  18,  56^1      14^7            4^4   2^6,  4)$,\\
IrMOA$(56^4,  15,  56^1 28^2 14^5            4^5   2^2,  4)$,\\
IrMOA$(56^4,  16,  56^1 28^1 14^6            4^5   2^3,  4)$,\\
IrMOA$(56^4,  17,  56^1      14^7            4^5   2^4,  4)$,\\
IrMOA$(56^4,  15,  56^1 28^1 14^6            4^6   2^1,  4)$,\\
IrMOA$(56^4,  16,  56^1      14^7            4^6   2^2,  4)$,\\
IrMOA$(56^4,  15,  56^1      14^7            4^7      ,  4)$,\\
IrMOA$(56^4,  17,  56^1 28^6            7^1        2^9,  4)$,\\
IrMOA$(56^4,  18,  56^1 28^5 14^1       7^1     2^{10},  4)$,\\
IrMOA$(56^4,  19,  56^1 28^4 14^2       7^1     2^{11},  4)$,\\
IrMOA$(56^4,  20,  56^1 28^3 14^3       7^1     2^{12},  4)$,\\
IrMOA$(56^4,  21,  56^1 28^2 14^4       7^1     2^{13},  4)$,\\
IrMOA$(56^4,  22,  56^1 28^1 14^5       7^1     2^{14},  4)$,\\
IrMOA$(56^4,  23,  56^1      14^6       7^1     2^{15},  4)$,\\
IrMOA$(56^4,  17,  56^1 28^5 14^1       7^1  4^1   2^8,  4)$,\\
IrMOA$(56^4,  18,  56^1 28^4 14^2       7^1  4^1   2^9,  4)$,\\
IrMOA$(56^4,  19,  56^1 28^3 14^3       7^1  4^12^{10},  4)$,\\
IrMOA$(56^4,  20,  56^1 28^2 14^4       7^1  4^12^{11},  4)$,\\
IrMOA$(56^4,  21,  56^1 28^1 14^5       7^1  4^12^{12},  4)$,\\
IrMOA$(56^4,  22,  56^1      14^6       7^1  4^12^{13},  4)$,\\
IrMOA$(56^4,  17,  56^1 28^4 14^2       7^1  4^2   2^7,  4)$,\\
IrMOA$(56^4,  18,  56^1 28^3 14^3       7^1  4^2   2^8,  4)$,\\
IrMOA$(56^4,  19,  56^1 28^2 14^4       7^1  4^2   2^9,  4)$,\\
IrMOA$(56^4,  20,  56^1 28^1 14^5       7^1  4^22^{10},  4)$,\\
IrMOA$(56^4,  21,  56^1      14^6       7^1  4^22^{11},  4)$,\\
IrMOA$(56^4,  17,  56^1 28^3 14^3       7^1  4^3   2^6,  4)$,\\
IrMOA$(56^4,  18,  56^1 28^2 14^4       7^1  4^3   2^7,  4)$,\\
IrMOA$(56^4,  19,  56^1 28^1 14^5       7^1  4^3   2^8,  4)$,\\
IrMOA$(56^4,  20,  56^1      14^6       7^1  4^3   2^9,  4)$,\\
IrMOA$(56^4,  17,  56^1 28^2 14^4       7^1  4^4   2^5,  4)$,\\
IrMOA$(56^4,  18,  56^1 28^1 14^5       7^1  4^4   2^6,  4)$,\\
IrMOA$(56^4,  19,  56^1      14^6       7^1  4^4   2^7,  4)$,\\
IrMOA$(56^4,  17,  56^1 28^1 14^5       7^1  4^5   2^4,  4)$,\\
IrMOA$(56^4,  18,  56^1      14^6       7^1  4^5   2^5,  4)$,\\
IrMOA$(56^4,  17,  56^1      14^6       7^1  4^6   2^3,  4)$,\\
IrMOA$(56^4,  19,  56^1 28^5            7^2     2^{11},  4)$,\\
IrMOA$(56^4,  20,  56^1 28^4 14^1       7^2     2^{12},  4)$,\\
IrMOA$(56^4,  21,  56^1 28^3 14^2       7^2     2^{13},  4)$,\\
IrMOA$(56^4,  22,  56^1 28^2 14^3       7^2     2^{14},  4)$,\\
IrMOA$(56^4,  23,  56^1 28^1 14^4       7^2     2^{15},  4)$,\\
IrMOA$(56^4,  24,  56^1      14^5       7^2     2^{16},  4)$,\\
IrMOA$(56^4,  19,  56^1 28^4 14^1       7^2  4^12^{10},  4)$,\\
IrMOA$(56^4,  20,  56^1 28^3 14^2       7^2  4^12^{11},  4)$,\\
IrMOA$(56^4,  21,  56^1 28^2 14^3       7^2  4^12^{12},  4)$,\\
IrMOA$(56^4,  22,  56^1 28^1 14^4       7^2  4^12^{13},  4)$,\\
IrMOA$(56^4,  23,  56^1      14^5       7^2  4^12^{14},  4)$,\\
IrMOA$(56^4,  19,  56^1 28^3 14^2       7^2  4^2   2^9,  4)$,\\
IrMOA$(56^4,  20,  56^1 28^2 14^3       7^2  4^22^{10},  4)$,\\
IrMOA$(56^4,  21,  56^1 28^1 14^4       7^2  4^22^{11},  4)$,\\
IrMOA$(56^4,  22,  56^1      14^5       7^2  4^22^{12},  4)$,\\
IrMOA$(56^4,  19,  56^1 28^2 14^3       7^2  4^3   2^8,  4)$,\\
IrMOA$(56^4,  20,  56^1 28^1 14^4       7^2  4^3   2^9,  4)$,\\
IrMOA$(56^4,  21,  56^1      14^5       7^2  4^32^{10},  4)$,\\
IrMOA$(56^4,  19,  56^1 28^1 14^4       7^2  4^4   2^7,  4)$,\\
IrMOA$(56^4,  20,  56^1      14^5       7^2  4^4   2^8,  4)$,\\
IrMOA$(56^4,  19,  56^1      14^5       7^2  4^5   2^6,  4)$,\\
IrMOA$(56^4,  21,  56^1 28^4            7^3     2^{13},  4)$,\\
IrMOA$(56^4,  22,  56^1 28^3 14^1       7^3     2^{14},  4)$,\\
IrMOA$(56^4,  23,  56^1 28^2 14^2       7^3     2^{15},  4)$,\\
IrMOA$(56^4,  24,  56^1 28^1 14^3       7^3     2^{16},  4)$,\\
IrMOA$(56^4,  25,  56^1      14^4       7^3     2^{17},  4)$,\\
IrMOA$(56^4,  21,  56^1 28^3 14^1       7^3  4^12^{12},  4)$,\\
IrMOA$(56^4,  22,  56^1 28^2 14^2       7^3  4^12^{13},  4)$,\\
IrMOA$(56^4,  23,  56^1 28^1 14^3       7^3  4^12^{14},  4)$,\\
IrMOA$(56^4,  24,  56^1      14^4       7^3  4^12^{15},  4)$,\\
IrMOA$(56^4,  21,  56^1 28^2 14^2       7^3  4^22^{11},  4)$,\\
IrMOA$(56^4,  22,  56^1 28^1 14^3       7^3  4^22^{12},  4)$,\\
IrMOA$(56^4,  23,  56^1      14^4       7^3  4^22^{13},  4)$,\\
IrMOA$(56^4,  21,  56^1 28^1 14^3       7^3  4^32^{10},  4)$,\\
IrMOA$(56^4,  22,  56^1      14^4       7^3  4^32^{11},  4)$,\\
IrMOA$(56^4,  21,  56^1      14^4       7^3  4^4   2^9,  4)$,\\
IrMOA$(56^4,  23,  56^1 28^3            7^4     2^{15},  4)$,\\
IrMOA$(56^4,  24,  56^1 28^2 14^1       7^4     2^{16},  4)$,\\
IrMOA$(56^4,  25,  56^1 28^1 14^2       7^4     2^{17},  4)$,\\
IrMOA$(56^4,  26,  56^1      14^3       7^4     2^{18},  4)$,\\
IrMOA$(56^4,  23,  56^1 28^2 14^1       7^4  4^12^{14},  4)$,\\
IrMOA$(56^4,  24,  56^1 28^1 14^2       7^4  4^12^{15},  4)$,\\
IrMOA$(56^4,  25,  56^1      14^3       7^4  4^12^{16},  4)$,\\
IrMOA$(56^4,  23,  56^1 28^1 14^2       7^4  4^22^{13},  4)$,\\
IrMOA$(56^4,  24,  56^1      14^3       7^4  4^22^{14},  4)$,\\
IrMOA$(56^4,  23,  56^1      14^3       7^4  4^32^{12},  4)$,\\
IrMOA$(56^4,  25,  56^1 28^2            7^5     2^{17},  4)$,\\
IrMOA$(56^4,  26,  56^1 28^1 14^1       7^5     2^{18},  4)$,\\
IrMOA$(56^4,  27,  56^1      14^2       7^5     2^{19},  4)$,\\
IrMOA$(56^4,  25,  56^1 28^1 14^1       7^5  4^12^{16},  4)$,\\
IrMOA$(56^4,  26,  56^1      14^2       7^5  4^12^{17},  4)$,\\
IrMOA$(56^4,  25,  56^1      14^2       7^5  4^22^{15},  4)$,\\
IrMOA$(56^4,  27,  56^1 28^1            7^6     2^{19},  4)$,\\
IrMOA$(56^4,  28,  56^1      14^1       7^6     2^{20},  4)$,\\
IrMOA$(56^4,  27,  56^1      14^1       7^6  4^12^{18},  4)$,\\
IrMOA$(56^4,  29,  56^1                 7^7     2^{21},  4)$,\\
IrMOA$(56^4,  16,  56^1 28^6            7^1  4^1   2^7,  4)$,\\
IrMOA$(56^4,  17,  56^1 28^5 14^1       7^1  4^1   2^8,  4)$,\\
IrMOA$(56^4,  18,  56^1 28^4 14^2       7^1  4^1   2^9,  4)$,\\
IrMOA$(56^4,  19,  56^1 28^3 14^3       7^1  4^12^{10},  4)$,\\
IrMOA$(56^4,  20,  56^1 28^2 14^4       7^1  4^12^{11},  4)$,\\
IrMOA$(56^4,  21,  56^1 28^1 14^5       7^1  4^12^{12},  4)$,\\
IrMOA$(56^4,  22,  56^1      14^6       7^1  4^12^{13},  4)$,\\
IrMOA$(56^4,  16,  56^1 28^5 14^1       7^1  4^2   2^6,  4)$,\\
IrMOA$(56^4,  17,  56^1 28^4 14^2       7^1  4^2   2^7,  4)$,\\
IrMOA$(56^4,  18,  56^1 28^3 14^3       7^1  4^2   2^8,  4)$,\\
IrMOA$(56^4,  19,  56^1 28^2 14^4       7^1  4^2   2^9,  4)$,\\
IrMOA$(56^4,  20,  56^1 28^1 14^5       7^1  4^22^{10},  4)$,\\
IrMOA$(56^4,  21,  56^1      14^6       7^1  4^22^{11},  4)$,\\
IrMOA$(56^4,  16,  56^1 28^4 14^2       7^1  4^3   2^5,  4)$,\\
IrMOA$(56^4,  17,  56^1 28^3 14^3       7^1  4^3   2^6,  4)$,\\
IrMOA$(56^4,  18,  56^1 28^2 14^4       7^1  4^3   2^7,  4)$,\\
IrMOA$(56^4,  19,  56^1 28^1 14^5       7^1  4^3   2^8,  4)$,\\
IrMOA$(56^4,  20,  56^1      14^6       7^1  4^3   2^9,  4)$,\\
IrMOA$(56^4,  16,  56^1 28^3 14^3       7^1  4^4   2^4,  4)$,\\
IrMOA$(56^4,  17,  56^1 28^2 14^4       7^1  4^4   2^5,  4)$,\\
IrMOA$(56^4,  18,  56^1 28^1 14^5       7^1  4^4   2^6,  4)$,\\
IrMOA$(56^4,  19,  56^1      14^6       7^1  4^4   2^7,  4)$,\\
IrMOA$(56^4,  16,  56^1 28^2 14^4       7^1  4^5   2^3,  4)$,\\
IrMOA$(56^4,  17,  56^1 28^1 14^5       7^1  4^5   2^4,  4)$,\\
IrMOA$(56^4,  18,  56^1      14^6       7^1  4^5   2^5,  4)$,\\
IrMOA$(56^4,  16,  56^1 28^1 14^5       7^1  4^6   2^2,  4)$,\\
IrMOA$(56^4,  17,  56^1      14^6       7^1  4^6   2^3,  4)$,\\
IrMOA$(56^4,  16,  56^1      14^6       7^1  4^7   2^1,  4)$,\\
IrMOA$(56^4,  18,  56^1 28^5            7^2  4^1   2^9,  4)$,\\
IrMOA$(56^4,  19,  56^1 28^4 14^1       7^2  4^12^{10},  4)$,\\
IrMOA$(56^4,  20,  56^1 28^3 14^2       7^2  4^12^{11},  4)$,\\
IrMOA$(56^4,  21,  56^1 28^2 14^3       7^2  4^12^{12},  4)$,\\
IrMOA$(56^4,  22,  56^1 28^1 14^4       7^2  4^12^{13},  4)$,\\
IrMOA$(56^4,  23,  56^1      14^5       7^2  4^12^{14},  4)$,\\
IrMOA$(56^4,  18,  56^1 28^4 14^1       7^2  4^2   2^8,  4)$,\\
IrMOA$(56^4,  19,  56^1 28^3 14^2       7^2  4^2   2^9,  4)$,\\
IrMOA$(56^4,  20,  56^1 28^2 14^3       7^2  4^22^{10},  4)$,\\
IrMOA$(56^4,  21,  56^1 28^1 14^4       7^2  4^22^{11},  4)$,\\
IrMOA$(56^4,  22,  56^1      14^5       7^2  4^22^{12},  4)$,\\
IrMOA$(56^4,  18,  56^1 28^3 14^2       7^2  4^3   2^7,  4)$,\\
IrMOA$(56^4,  19,  56^1 28^2 14^3       7^2  4^3   2^8,  4)$,\\
IrMOA$(56^4,  20,  56^1 28^1 14^4       7^2  4^3   2^9,  4)$,\\
IrMOA$(56^4,  21,  56^1      14^5       7^2  4^32^{10},  4)$,\\
IrMOA$(56^4,  18,  56^1 28^2 14^3       7^2  4^4   2^6,  4)$,\\
IrMOA$(56^4,  19,  56^1 28^1 14^4       7^2  4^4   2^7,  4)$,\\
IrMOA$(56^4,  20,  56^1      14^5       7^2  4^4   2^8,  4)$,\\
IrMOA$(56^4,  18,  56^1 28^1 14^4       7^2  4^5   2^5,  4)$,\\
IrMOA$(56^4,  19,  56^1      14^5       7^2  4^5   2^6,  4)$,\\
IrMOA$(56^4,  18,  56^1      14^5       7^2  4^6   2^4,  4)$,\\
IrMOA$(56^4,  20,  56^1 28^4            7^3  4^12^{11},  4)$,\\
IrMOA$(56^4,  21,  56^1 28^3 14^1       7^3  4^12^{12},  4)$,\\
IrMOA$(56^4,  22,  56^1 28^2 14^2       7^3  4^12^{13},  4)$,\\
IrMOA$(56^4,  23,  56^1 28^1 14^3       7^3  4^12^{14},  4)$,\\
IrMOA$(56^4,  24,  56^1      14^4       7^3  4^12^{15},  4)$,\\
IrMOA$(56^4,  20,  56^1 28^3 14^1       7^3  4^22^{10},  4)$,\\
IrMOA$(56^4,  21,  56^1 28^2 14^2       7^3  4^22^{11},  4)$,\\
IrMOA$(56^4,  22,  56^1 28^1 14^3       7^3  4^22^{12},  4)$,\\
IrMOA$(56^4,  23,  56^1      14^4       7^3  4^22^{13},  4)$,\\
IrMOA$(56^4,  20,  56^1 28^2 14^2       7^3  4^3   2^9,  4)$,\\
IrMOA$(56^4,  21,  56^1 28^1 14^3       7^3  4^32^{10},  4)$,\\
IrMOA$(56^4,  22,  56^1      14^4       7^3  4^32^{11},  4)$,\\
IrMOA$(56^4,  20,  56^1 28^1 14^3       7^3  4^4   2^8,  4)$,\\
IrMOA$(56^4,  21,  56^1      14^4       7^3  4^4   2^9,  4)$,\\
IrMOA$(56^4,  20,  56^1      14^4       7^3  4^5   2^7,  4)$,\\
IrMOA$(56^4,  22,  56^1 28^3            7^4  4^12^{13},  4)$,\\
IrMOA$(56^4,  23,  56^1 28^2 14^1       7^4  4^12^{14},  4)$,\\
IrMOA$(56^4,  24,  56^1 28^1 14^2       7^4  4^12^{15},  4)$,\\
IrMOA$(56^4,  25,  56^1      14^3       7^4  4^12^{16},  4)$,\\
IrMOA$(56^4,  22,  56^1 28^2 14^1       7^4  4^22^{12},  4)$,\\
IrMOA$(56^4,  23,  56^1 28^1 14^2       7^4  4^22^{13},  4)$,\\
IrMOA$(56^4,  24,  56^1      14^3       7^4  4^22^{14},  4)$,\\
IrMOA$(56^4,  22,  56^1 28^1 14^2       7^4  4^32^{11},  4)$,\\
IrMOA$(56^4,  23,  56^1      14^3       7^4  4^32^{12},  4)$,\\
IrMOA$(56^4,  22,  56^1      14^3       7^4  4^42^{10},  4)$,\\
IrMOA$(56^4,  24,  56^1 28^2            7^5  4^12^{15},  4)$,\\
IrMOA$(56^4,  25,  56^1 28^1 14^1       7^5  4^12^{16},  4)$,\\
IrMOA$(56^4,  26,  56^1      14^2       7^5  4^12^{17},  4)$,\\
IrMOA$(56^4,  24,  56^1 28^1 14^1       7^5  4^22^{14},  4)$,\\
IrMOA$(56^4,  25,  56^1      14^2       7^5  4^22^{15},  4)$,\\
IrMOA$(56^4,  24,  56^1      14^2       7^5  4^32^{13},  4)$,\\
IrMOA$(56^4,  26,  56^1 28^1            7^6  4^12^{17},  4)$,\\
IrMOA$(56^4,  27,  56^1      14^1       7^6  4^12^{18},  4)$,\\
IrMOA$(56^4,  26,  56^1      14^1       7^6  4^22^{16},  4)$,\\
IrMOA$(56^4,  28,  56^1                 7^7  4^12^{19},  4)$,\\
IrMOA$(56^4,  17,  56^1 28^5            7^2  4^2   2^7,  4)$,\\
IrMOA$(56^4,  18,  56^1 28^4 14^1       7^2  4^2   2^8,  4)$,\\
IrMOA$(56^4,  19,  56^1 28^3 14^2       7^2  4^2   2^9,  4)$,\\
IrMOA$(56^4,  20,  56^1 28^2 14^3       7^2  4^22^{10},  4)$,\\
IrMOA$(56^4,  21,  56^1 28^1 14^4       7^2  4^22^{11},  4)$,\\
IrMOA$(56^4,  22,  56^1      14^5       7^2  4^22^{12},  4)$,\\
IrMOA$(56^4,  17,  56^1 28^4 14^1       7^2  4^3   2^6,  4)$,\\
IrMOA$(56^4,  18,  56^1 28^3 14^2       7^2  4^3   2^7,  4)$,\\
IrMOA$(56^4,  19,  56^1 28^2 14^3       7^2  4^3   2^8,  4)$,\\
IrMOA$(56^4,  20,  56^1 28^1 14^4       7^2  4^3   2^9,  4)$,\\
IrMOA$(56^4,  21,  56^1      14^5       7^2  4^32^{10},  4)$,\\
IrMOA$(56^4,  17,  56^1 28^3 14^2       7^2  4^4   2^5,  4)$,\\
IrMOA$(56^4,  18,  56^1 28^2 14^3       7^2  4^4   2^6,  4)$,\\
IrMOA$(56^4,  19,  56^1 28^1 14^4       7^2  4^4   2^7,  4)$,\\
IrMOA$(56^4,  20,  56^1      14^5       7^2  4^4   2^8,  4)$,\\
IrMOA$(56^4,  17,  56^1 28^2 14^3       7^2  4^5   2^4,  4)$,\\
IrMOA$(56^4,  18,  56^1 28^1 14^4       7^2  4^5   2^5,  4)$,\\
IrMOA$(56^4,  19,  56^1      14^5       7^2  4^5   2^6,  4)$,\\
IrMOA$(56^4,  17,  56^1 28^1 14^4       7^2  4^6   2^3,  4)$,\\
IrMOA$(56^4,  18,  56^1      14^5       7^2  4^6   2^4,  4)$,\\
IrMOA$(56^4,  17,  56^1      14^5       7^2  4^7   2^2,  4)$,\\
IrMOA$(56^4,  19,  56^1 28^4            7^3  4^2   2^9,  4)$,\\
IrMOA$(56^4,  20,  56^1 28^3 14^1       7^3  4^22^{10},  4)$,\\
IrMOA$(56^4,  21,  56^1 28^2 14^2       7^3  4^22^{11},  4)$,\\
IrMOA$(56^4,  22,  56^1 28^1 14^3       7^3  4^22^{12},  4)$,\\
IrMOA$(56^4,  23,  56^1      14^4       7^3  4^22^{13},  4)$,\\
IrMOA$(56^4,  19,  56^1 28^3 14^1       7^3  4^3   2^8,  4)$,\\
IrMOA$(56^4,  20,  56^1 28^2 14^2       7^3  4^3   2^9,  4)$,\\
IrMOA$(56^4,  21,  56^1 28^1 14^3       7^3  4^32^{10},  4)$,\\
IrMOA$(56^4,  22,  56^1      14^4       7^3  4^32^{11},  4)$,\\
IrMOA$(56^4,  19,  56^1 28^2 14^2       7^3  4^4   2^7,  4)$,\\
IrMOA$(56^4,  20,  56^1 28^1 14^3       7^3  4^4   2^8,  4)$,\\
IrMOA$(56^4,  21,  56^1      14^4       7^3  4^4   2^9,  4)$,\\
IrMOA$(56^4,  19,  56^1 28^1 14^3       7^3  4^5   2^6,  4)$,\\
IrMOA$(56^4,  20,  56^1      14^4       7^3  4^5   2^7,  4)$,\\
IrMOA$(56^4,  19,  56^1      14^4       7^3  4^6   2^5,  4)$,\\
IrMOA$(56^4,  21,  56^1 28^3            7^4  4^22^{11},  4)$,\\
IrMOA$(56^4,  22,  56^1 28^2 14^1       7^4  4^22^{12},  4)$,\\
IrMOA$(56^4,  23,  56^1 28^1 14^2       7^4  4^22^{13},  4)$,\\
IrMOA$(56^4,  24,  56^1      14^3       7^4  4^22^{14},  4)$,\\
IrMOA$(56^4,  21,  56^1 28^2 14^1       7^4  4^32^{10},  4)$,\\
IrMOA$(56^4,  22,  56^1 28^1 14^2       7^4  4^32^{11},  4)$,\\
IrMOA$(56^4,  23,  56^1      14^3       7^4  4^32^{12},  4)$,\\
IrMOA$(56^4,  21,  56^1 28^1 14^2       7^4  4^4   2^9,  4)$,\\
IrMOA$(56^4,  22,  56^1      14^3       7^4  4^42^{10},  4)$,\\
IrMOA$(56^4,  21,  56^1      14^3       7^4  4^5   2^8,  4)$,\\
IrMOA$(56^4,  23,  56^1 28^2            7^5  4^22^{13},  4)$,\\
IrMOA$(56^4,  24,  56^1 28^1 14^1       7^5  4^22^{14},  4)$,\\
IrMOA$(56^4,  25,  56^1      14^2       7^5  4^22^{15},  4)$,\\
IrMOA$(56^4,  23,  56^1 28^1 14^1       7^5  4^32^{12},  4)$,\\
IrMOA$(56^4,  24,  56^1      14^2       7^5  4^32^{13},  4)$,\\
IrMOA$(56^4,  23,  56^1      14^2       7^5  4^42^{11},  4)$,\\
IrMOA$(56^4,  25,  56^1 28^1            7^6  4^22^{15},  4)$,\\
IrMOA$(56^4,  26,  56^1      14^1       7^6  4^22^{16},  4)$,\\
IrMOA$(56^4,  25,  56^1      14^1       7^6  4^32^{14},  4)$,\\
IrMOA$(56^4,  27,  56^1                 7^7  4^22^{17},  4)$,\\
IrMOA$(56^4,  18,  56^1 28^4            7^3  4^3   2^7,  4)$,\\
IrMOA$(56^4,  19,  56^1 28^3 14^1       7^3  4^3   2^8,  4)$,\\
IrMOA$(56^4,  20,  56^1 28^2 14^2       7^3  4^3   2^9,  4)$,\\
IrMOA$(56^4,  21,  56^1 28^1 14^3       7^3  4^32^{10},  4)$,\\
IrMOA$(56^4,  22,  56^1      14^4       7^3  4^32^{11},  4)$,\\
IrMOA$(56^4,  18,  56^1 28^3 14^1       7^3  4^4   2^6,  4)$,\\
IrMOA$(56^4,  19,  56^1 28^2 14^2       7^3  4^4   2^7,  4)$,\\
IrMOA$(56^4,  20,  56^1 28^1 14^3       7^3  4^4   2^8,  4)$,\\
IrMOA$(56^4,  21,  56^1      14^4       7^3  4^4   2^9,  4)$,\\
IrMOA$(56^4,  18,  56^1 28^2 14^2       7^3  4^5   2^5,  4)$,\\
IrMOA$(56^4,  19,  56^1 28^1 14^3       7^3  4^5   2^6,  4)$,\\
IrMOA$(56^4,  20,  56^1      14^4       7^3  4^5   2^7,  4)$,\\
IrMOA$(56^4,  18,  56^1 28^1 14^3       7^3  4^6   2^4,  4)$,\\
IrMOA$(56^4,  19,  56^1      14^4       7^3  4^6   2^5,  4)$,\\
IrMOA$(56^4,  18,  56^1      14^4       7^3  4^7   2^3,  4)$,\\
IrMOA$(56^4,  20,  56^1 28^3            7^4  4^3   2^9,  4)$,\\
IrMOA$(56^4,  21,  56^1 28^2 14^1       7^4  4^32^{10},  4)$,\\
IrMOA$(56^4,  22,  56^1 28^1 14^2       7^4  4^32^{11},  4)$,\\
IrMOA$(56^4,  23,  56^1      14^3       7^4  4^32^{12},  4)$,\\
IrMOA$(56^4,  20,  56^1 28^2 14^1       7^4  4^4   2^8,  4)$,\\
IrMOA$(56^4,  21,  56^1 28^1 14^2       7^4  4^4   2^9,  4)$,\\
IrMOA$(56^4,  22,  56^1      14^3       7^4  4^42^{10},  4)$,\\
IrMOA$(56^4,  20,  56^1 28^1 14^2       7^4  4^5   2^7,  4)$,\\
IrMOA$(56^4,  21,  56^1      14^3       7^4  4^5   2^8,  4)$,\\
IrMOA$(56^4,  20,  56^1      14^3       7^4  4^6   2^6,  4)$,\\
IrMOA$(56^4,  22,  56^1 28^2            7^5  4^32^{11},  4)$,\\
IrMOA$(56^4,  23,  56^1 28^1 14^1       7^5  4^32^{12},  4)$,\\
IrMOA$(56^4,  24,  56^1      14^2       7^5  4^32^{13},  4)$,\\
IrMOA$(56^4,  22,  56^1 28^1 14^1       7^5  4^42^{10},  4)$,\\
IrMOA$(56^4,  23,  56^1      14^2       7^5  4^42^{11},  4)$,\\
IrMOA$(56^4,  22,  56^1      14^2       7^5  4^5   2^9,  4)$,\\
IrMOA$(56^4,  24,  56^1 28^1            7^6  4^32^{13},  4)$,\\
IrMOA$(56^4,  25,  56^1      14^1       7^6  4^32^{14},  4)$,\\
IrMOA$(56^4,  24,  56^1      14^1       7^6  4^42^{12},  4)$,\\
IrMOA$(56^4,  26,  56^1                 7^7  4^32^{15},  4)$,\\
IrMOA$(56^4,  19,  56^1 28^3            7^4  4^4   2^7,  4)$,\\
IrMOA$(56^4,  20,  56^1 28^2 14^1       7^4  4^4   2^8,  4)$,\\
IrMOA$(56^4,  21,  56^1 28^1 14^2       7^4  4^4   2^9,  4)$,\\
IrMOA$(56^4,  22,  56^1      14^3       7^4  4^42^{10},  4)$,\\
IrMOA$(56^4,  19,  56^1 28^2 14^1       7^4  4^5   2^6,  4)$,\\
IrMOA$(56^4,  20,  56^1 28^1 14^2       7^4  4^5   2^7,  4)$,\\
IrMOA$(56^4,  21,  56^1      14^3       7^4  4^5   2^8,  4)$,\\
IrMOA$(56^4,  19,  56^1 28^1 14^2       7^4  4^6   2^5,  4)$,\\
IrMOA$(56^4,  20,  56^1      14^3       7^4  4^6   2^6,  4)$,\\
IrMOA$(56^4,  19,  56^1      14^3       7^4  4^7   2^4,  4)$,\\
IrMOA$(56^4,  21,  56^1 28^2            7^5  4^4   2^9,  4)$,\\
IrMOA$(56^4,  22,  56^1 28^1 14^1       7^5  4^42^{10},  4)$,\\
IrMOA$(56^4,  23,  56^1      14^2       7^5  4^42^{11},  4)$,\\
IrMOA$(56^4,  21,  56^1 28^1 14^1       7^5  4^5   2^8,  4)$,\\
IrMOA$(56^4,  22,  56^1      14^2       7^5  4^5   2^9,  4)$,\\
IrMOA$(56^4,  21,  56^1      14^2       7^5  4^6   2^7,  4)$,\\
IrMOA$(56^4,  23,  56^1 28^1            7^6  4^42^{11},  4)$,\\
IrMOA$(56^4,  24,  56^1      14^1       7^6  4^42^{12},  4)$,\\
IrMOA$(56^4,  23,  56^1      14^1       7^6  4^52^{10},  4)$,\\
IrMOA$(56^4,  25,  56^1                 7^7  4^42^{13},  4)$,\\
IrMOA$(56^4,  20,  56^1 28^2            7^5  4^5   2^7,  4)$,\\
IrMOA$(56^4,  21,  56^1 28^1 14^1       7^5  4^5   2^8,  4)$,\\
IrMOA$(56^4,  22,  56^1      14^2       7^5  4^5   2^9,  4)$,\\
IrMOA$(56^4,  20,  56^1 28^1 14^1       7^5  4^6   2^6,  4)$,\\
IrMOA$(56^4,  21,  56^1      14^2       7^5  4^6   2^7,  4)$,\\
IrMOA$(56^4,  20,  56^1      14^2       7^5  4^7   2^5,  4)$,\\
IrMOA$(56^4,  22,  56^1 28^1            7^6  4^5   2^9,  4)$,\\
IrMOA$(56^4,  23,  56^1      14^1       7^6  4^52^{10},  4)$,\\
IrMOA$(56^4,  22,  56^1      14^1       7^6  4^6   2^8,  4)$,\\
IrMOA$(56^4,  24,  56^1                 7^7  4^52^{11},  4)$,\\
IrMOA$(56^4,  21,  56^1 28^1            7^6  4^6   2^7,  4)$,\\
IrMOA$(56^4,  22,  56^1      14^1       7^6  4^6   2^8,  4)$,\\
IrMOA$(56^4,  21,  56^1      14^1       7^6  4^7   2^6,  4)$,\\
IrMOA$(56^4,  23,  56^1                 7^7  4^6   2^9,  4)$,\\
IrMOA$(56^4,  22,  56^1                 7^7  4^7   2^7,  4)$,\\
IrMOA$(56^4,  15,  56^1 28^6       8^1  7^1        2^6,  4)$,\\
IrMOA$(56^4,  16,  56^1 28^5 14^1  8^1  7^1        2^7,  4)$,\\
IrMOA$(56^4,  17,  56^1 28^4 14^2  8^1  7^1        2^8,  4)$,\\
IrMOA$(56^4,  18,  56^1 28^3 14^3  8^1  7^1        2^9,  4)$,\\
IrMOA$(56^4,  19,  56^1 28^2 14^4  8^1  7^1     2^{10},  4)$,\\
IrMOA$(56^4,  20,  56^1 28^1 14^5  8^1  7^1     2^{11},  4)$,\\
IrMOA$(56^4,  21,  56^1      14^6  8^1  7^1     2^{12},  4)$,\\
IrMOA$(56^4,  15,  56^1 28^5 14^1  8^1  7^1  4^1   2^5,  4)$,\\
IrMOA$(56^4,  16,  56^1 28^4 14^2  8^1  7^1  4^1   2^6,  4)$,\\
IrMOA$(56^4,  17,  56^1 28^3 14^3  8^1  7^1  4^1   2^7,  4)$,\\
IrMOA$(56^4,  18,  56^1 28^2 14^4  8^1  7^1  4^1   2^8,  4)$,\\
IrMOA$(56^4,  19,  56^1 28^1 14^5  8^1  7^1  4^1   2^9,  4)$,\\
IrMOA$(56^4,  20,  56^1      14^6  8^1  7^1  4^12^{10},  4)$,\\
IrMOA$(56^4,  15,  56^1 28^4 14^2  8^1  7^1  4^2   2^4,  4)$,\\
IrMOA$(56^4,  16,  56^1 28^3 14^3  8^1  7^1  4^2   2^5,  4)$,\\
IrMOA$(56^4,  17,  56^1 28^2 14^4  8^1  7^1  4^2   2^6,  4)$,\\
IrMOA$(56^4,  18,  56^1 28^1 14^5  8^1  7^1  4^2   2^7,  4)$,\\
IrMOA$(56^4,  19,  56^1      14^6  8^1  7^1  4^2   2^8,  4)$,\\
IrMOA$(56^4,  15,  56^1 28^3 14^3  8^1  7^1  4^3   2^3,  4)$,\\
IrMOA$(56^4,  16,  56^1 28^2 14^4  8^1  7^1  4^3   2^4,  4)$,\\
IrMOA$(56^4,  17,  56^1 28^1 14^5  8^1  7^1  4^3   2^5,  4)$,\\
IrMOA$(56^4,  18,  56^1      14^6  8^1  7^1  4^3   2^6,  4)$,\\
IrMOA$(56^4,  15,  56^1 28^2 14^4  8^1  7^1  4^4   2^2,  4)$,\\
IrMOA$(56^4,  16,  56^1 28^1 14^5  8^1  7^1  4^4   2^3,  4)$,\\
IrMOA$(56^4,  17,  56^1      14^6  8^1  7^1  4^4   2^4,  4)$,\\
IrMOA$(56^4,  15,  56^1 28^1 14^5  8^1  7^1  4^5   2^1,  4)$,\\
IrMOA$(56^4,  16,  56^1      14^6  8^1  7^1  4^5   2^2,  4)$,\\
IrMOA$(56^4,  15,  56^1      14^6  8^1  7^1  4^6      ,  4)$,\\
IrMOA$(56^4,  17,  56^1 28^5       8^1  7^2        2^8,  4)$,\\
IrMOA$(56^4,  18,  56^1 28^4 14^1  8^1  7^2        2^9,  4)$,\\
IrMOA$(56^4,  19,  56^1 28^3 14^2  8^1  7^2     2^{10},  4)$,\\
IrMOA$(56^4,  20,  56^1 28^2 14^3  8^1  7^2     2^{11},  4)$,\\
IrMOA$(56^4,  21,  56^1 28^1 14^4  8^1  7^2     2^{12},  4)$,\\
IrMOA$(56^4,  22,  56^1      14^5  8^1  7^2     2^{13},  4)$,\\
IrMOA$(56^4,  17,  56^1 28^4 14^1  8^1  7^2  4^1   2^7,  4)$,\\
IrMOA$(56^4,  18,  56^1 28^3 14^2  8^1  7^2  4^1   2^8,  4)$,\\
IrMOA$(56^4,  19,  56^1 28^2 14^3  8^1  7^2  4^1   2^9,  4)$,\\
IrMOA$(56^4,  20,  56^1 28^1 14^4  8^1  7^2  4^12^{10},  4)$,\\
IrMOA$(56^4,  21,  56^1      14^5  8^1  7^2  4^12^{11},  4)$,\\
IrMOA$(56^4,  17,  56^1 28^3 14^2  8^1  7^2  4^2   2^6,  4)$,\\
IrMOA$(56^4,  18,  56^1 28^2 14^3  8^1  7^2  4^2   2^7,  4)$,\\
IrMOA$(56^4,  19,  56^1 28^1 14^4  8^1  7^2  4^2   2^8,  4)$,\\
IrMOA$(56^4,  20,  56^1      14^5  8^1  7^2  4^2   2^9,  4)$,\\
IrMOA$(56^4,  17,  56^1 28^2 14^3  8^1  7^2  4^3   2^5,  4)$,\\
IrMOA$(56^4,  18,  56^1 28^1 14^4  8^1  7^2  4^3   2^6,  4)$,\\
IrMOA$(56^4,  19,  56^1      14^5  8^1  7^2  4^3   2^7,  4)$,\\
IrMOA$(56^4,  17,  56^1 28^1 14^4  8^1  7^2  4^4   2^4,  4)$,\\
IrMOA$(56^4,  18,  56^1      14^5  8^1  7^2  4^4   2^5,  4)$,\\
IrMOA$(56^4,  17,  56^1      14^5  8^1  7^2  4^5   2^3,  4)$,\\
IrMOA$(56^4,  19,  56^1 28^4       8^1  7^3     2^{10},  4)$,\\
IrMOA$(56^4,  20,  56^1 28^3 14^1  8^1  7^3     2^{11},  4)$,\\
IrMOA$(56^4,  21,  56^1 28^2 14^2  8^1  7^3     2^{12},  4)$,\\
IrMOA$(56^4,  22,  56^1 28^1 14^3  8^1  7^3     2^{13},  4)$,\\
IrMOA$(56^4,  23,  56^1      14^4  8^1  7^3     2^{14},  4)$,\\
IrMOA$(56^4,  19,  56^1 28^3 14^1  8^1  7^3  4^1   2^9,  4)$,\\
IrMOA$(56^4,  20,  56^1 28^2 14^2  8^1  7^3  4^12^{10},  4)$,\\
IrMOA$(56^4,  21,  56^1 28^1 14^3  8^1  7^3  4^12^{11},  4)$,\\
IrMOA$(56^4,  22,  56^1      14^4  8^1  7^3  4^12^{12},  4)$,\\
IrMOA$(56^4,  19,  56^1 28^2 14^2  8^1  7^3  4^2   2^8,  4)$,\\
IrMOA$(56^4,  20,  56^1 28^1 14^3  8^1  7^3  4^2   2^9,  4)$,\\
IrMOA$(56^4,  21,  56^1      14^4  8^1  7^3  4^22^{10},  4)$,\\
IrMOA$(56^4,  19,  56^1 28^1 14^3  8^1  7^3  4^3   2^7,  4)$,\\
IrMOA$(56^4,  20,  56^1      14^4  8^1  7^3  4^3   2^8,  4)$,\\
IrMOA$(56^4,  19,  56^1      14^4  8^1  7^3  4^4   2^6,  4)$,\\
IrMOA$(56^4,  21,  56^1 28^3       8^1  7^4     2^{12},  4)$,\\
IrMOA$(56^4,  22,  56^1 28^2 14^1  8^1  7^4     2^{13},  4)$,\\
IrMOA$(56^4,  23,  56^1 28^1 14^2  8^1  7^4     2^{14},  4)$,\\
IrMOA$(56^4,  24,  56^1      14^3  8^1  7^4     2^{15},  4)$,\\
IrMOA$(56^4,  21,  56^1 28^2 14^1  8^1  7^4  4^12^{11},  4)$,\\
IrMOA$(56^4,  22,  56^1 28^1 14^2  8^1  7^4  4^12^{12},  4)$,\\
IrMOA$(56^4,  23,  56^1      14^3  8^1  7^4  4^12^{13},  4)$,\\
IrMOA$(56^4,  21,  56^1 28^1 14^2  8^1  7^4  4^22^{10},  4)$,\\
IrMOA$(56^4,  22,  56^1      14^3  8^1  7^4  4^22^{11},  4)$,\\
IrMOA$(56^4,  21,  56^1      14^3  8^1  7^4  4^3   2^9,  4)$,\\
IrMOA$(56^4,  23,  56^1 28^2       8^1  7^5     2^{14},  4)$,\\
IrMOA$(56^4,  24,  56^1 28^1 14^1  8^1  7^5     2^{15},  4)$,\\
IrMOA$(56^4,  25,  56^1      14^2  8^1  7^5     2^{16},  4)$,\\
IrMOA$(56^4,  23,  56^1 28^1 14^1  8^1  7^5  4^12^{13},  4)$,\\
IrMOA$(56^4,  24,  56^1      14^2  8^1  7^5  4^12^{14},  4)$,\\
IrMOA$(56^4,  23,  56^1      14^2  8^1  7^5  4^22^{12},  4)$,\\
IrMOA$(56^4,  25,  56^1 28^1       8^1  7^6     2^{16},  4)$,\\
IrMOA$(56^4,  26,  56^1      14^1  8^1  7^6     2^{17},  4)$,\\
IrMOA$(56^4,  25,  56^1      14^1  8^1  7^6  4^12^{15},  4)$,\\
IrMOA$(56^4,  27,  56^1            8^1  7^7     2^{18},  4)$,\\
IrMOA$(56^4,  16,  56^1 28^5       8^1  7^2  4^1   2^6,  4)$,\\
IrMOA$(56^4,  17,  56^1 28^4 14^1  8^1  7^2  4^1   2^7,  4)$,\\
IrMOA$(56^4,  18,  56^1 28^3 14^2  8^1  7^2  4^1   2^8,  4)$,\\
IrMOA$(56^4,  19,  56^1 28^2 14^3  8^1  7^2  4^1   2^9,  4)$,\\
IrMOA$(56^4,  20,  56^1 28^1 14^4  8^1  7^2  4^12^{10},  4)$,\\
IrMOA$(56^4,  21,  56^1      14^5  8^1  7^2  4^12^{11},  4)$,\\
IrMOA$(56^4,  16,  56^1 28^4 14^1  8^1  7^2  4^2   2^5,  4)$,\\
IrMOA$(56^4,  17,  56^1 28^3 14^2  8^1  7^2  4^2   2^6,  4)$,\\
IrMOA$(56^4,  18,  56^1 28^2 14^3  8^1  7^2  4^2   2^7,  4)$,\\
IrMOA$(56^4,  19,  56^1 28^1 14^4  8^1  7^2  4^2   2^8,  4)$,\\
IrMOA$(56^4,  20,  56^1      14^5  8^1  7^2  4^2   2^9,  4)$,\\
IrMOA$(56^4,  16,  56^1 28^3 14^2  8^1  7^2  4^3   2^4,  4)$,\\
IrMOA$(56^4,  17,  56^1 28^2 14^3  8^1  7^2  4^3   2^5,  4)$,\\
IrMOA$(56^4,  18,  56^1 28^1 14^4  8^1  7^2  4^3   2^6,  4)$,\\
IrMOA$(56^4,  19,  56^1      14^5  8^1  7^2  4^3   2^7,  4)$,\\
IrMOA$(56^4,  16,  56^1 28^2 14^3  8^1  7^2  4^4   2^3,  4)$,\\
IrMOA$(56^4,  17,  56^1 28^1 14^4  8^1  7^2  4^4   2^4,  4)$,\\
IrMOA$(56^4,  18,  56^1      14^5  8^1  7^2  4^4   2^5,  4)$,\\
IrMOA$(56^4,  16,  56^1 28^1 14^4  8^1  7^2  4^5   2^2,  4)$,\\
IrMOA$(56^4,  17,  56^1      14^5  8^1  7^2  4^5   2^3,  4)$,\\
IrMOA$(56^4,  16,  56^1      14^5  8^1  7^2  4^6   2^1,  4)$,\\
IrMOA$(56^4,  18,  56^1 28^4       8^1  7^3  4^1   2^8,  4)$,\\
IrMOA$(56^4,  19,  56^1 28^3 14^1  8^1  7^3  4^1   2^9,  4)$,\\
IrMOA$(56^4,  20,  56^1 28^2 14^2  8^1  7^3  4^12^{10},  4)$,\\
IrMOA$(56^4,  21,  56^1 28^1 14^3  8^1  7^3  4^12^{11},  4)$,\\
IrMOA$(56^4,  22,  56^1      14^4  8^1  7^3  4^12^{12},  4)$,\\
IrMOA$(56^4,  18,  56^1 28^3 14^1  8^1  7^3  4^2   2^7,  4)$,\\
IrMOA$(56^4,  19,  56^1 28^2 14^2  8^1  7^3  4^2   2^8,  4)$,\\
IrMOA$(56^4,  20,  56^1 28^1 14^3  8^1  7^3  4^2   2^9,  4)$,\\
IrMOA$(56^4,  21,  56^1      14^4  8^1  7^3  4^22^{10},  4)$,\\
IrMOA$(56^4,  18,  56^1 28^2 14^2  8^1  7^3  4^3   2^6,  4)$,\\
IrMOA$(56^4,  19,  56^1 28^1 14^3  8^1  7^3  4^3   2^7,  4)$,\\
IrMOA$(56^4,  20,  56^1      14^4  8^1  7^3  4^3   2^8,  4)$,\\
IrMOA$(56^4,  18,  56^1 28^1 14^3  8^1  7^3  4^4   2^5,  4)$,\\
IrMOA$(56^4,  19,  56^1      14^4  8^1  7^3  4^4   2^6,  4)$,\\
IrMOA$(56^4,  18,  56^1      14^4  8^1  7^3  4^5   2^4,  4)$,\\
IrMOA$(56^4,  20,  56^1 28^3       8^1  7^4  4^12^{10},  4)$,\\
IrMOA$(56^4,  21,  56^1 28^2 14^1  8^1  7^4  4^12^{11},  4)$,\\
IrMOA$(56^4,  22,  56^1 28^1 14^2  8^1  7^4  4^12^{12},  4)$,\\
IrMOA$(56^4,  23,  56^1      14^3  8^1  7^4  4^12^{13},  4)$,\\
IrMOA$(56^4,  20,  56^1 28^2 14^1  8^1  7^4  4^2   2^9,  4)$,\\
IrMOA$(56^4,  21,  56^1 28^1 14^2  8^1  7^4  4^22^{10},  4)$,\\
IrMOA$(56^4,  22,  56^1      14^3  8^1  7^4  4^22^{11},  4)$,\\
IrMOA$(56^4,  20,  56^1 28^1 14^2  8^1  7^4  4^3   2^8,  4)$,\\
IrMOA$(56^4,  21,  56^1      14^3  8^1  7^4  4^3   2^9,  4)$,\\
IrMOA$(56^4,  20,  56^1      14^3  8^1  7^4  4^4   2^7,  4)$,\\
IrMOA$(56^4,  22,  56^1 28^2       8^1  7^5  4^12^{12},  4)$,\\
IrMOA$(56^4,  23,  56^1 28^1 14^1  8^1  7^5  4^12^{13},  4)$,\\
IrMOA$(56^4,  24,  56^1      14^2  8^1  7^5  4^12^{14},  4)$,\\
IrMOA$(56^4,  22,  56^1 28^1 14^1  8^1  7^5  4^22^{11},  4)$,\\
IrMOA$(56^4,  23,  56^1      14^2  8^1  7^5  4^22^{12},  4)$,\\
IrMOA$(56^4,  22,  56^1      14^2  8^1  7^5  4^32^{10},  4)$,\\
IrMOA$(56^4,  24,  56^1 28^1       8^1  7^6  4^12^{14},  4)$,\\
IrMOA$(56^4,  25,  56^1      14^1  8^1  7^6  4^12^{15},  4)$,\\
IrMOA$(56^4,  24,  56^1      14^1  8^1  7^6  4^22^{13},  4)$,\\
IrMOA$(56^4,  26,  56^1            8^1  7^7  4^12^{16},  4)$,\\
IrMOA$(56^4,  17,  56^1 28^4       8^1  7^3  4^2   2^6,  4)$,\\
IrMOA$(56^4,  18,  56^1 28^3 14^1  8^1  7^3  4^2   2^7,  4)$,\\
IrMOA$(56^4,  19,  56^1 28^2 14^2  8^1  7^3  4^2   2^8,  4)$,\\
IrMOA$(56^4,  20,  56^1 28^1 14^3  8^1  7^3  4^2   2^9,  4)$,\\
IrMOA$(56^4,  21,  56^1      14^4  8^1  7^3  4^22^{10},  4)$,\\
IrMOA$(56^4,  17,  56^1 28^3 14^1  8^1  7^3  4^3   2^5,  4)$,\\
IrMOA$(56^4,  18,  56^1 28^2 14^2  8^1  7^3  4^3   2^6,  4)$,\\
IrMOA$(56^4,  19,  56^1 28^1 14^3  8^1  7^3  4^3   2^7,  4)$,\\
IrMOA$(56^4,  20,  56^1      14^4  8^1  7^3  4^3   2^8,  4)$,\\
IrMOA$(56^4,  17,  56^1 28^2 14^2  8^1  7^3  4^4   2^4,  4)$,\\
IrMOA$(56^4,  18,  56^1 28^1 14^3  8^1  7^3  4^4   2^5,  4)$,\\
IrMOA$(56^4,  19,  56^1      14^4  8^1  7^3  4^4   2^6,  4)$,\\
IrMOA$(56^4,  17,  56^1 28^1 14^3  8^1  7^3  4^5   2^3,  4)$,\\
IrMOA$(56^4,  18,  56^1      14^4  8^1  7^3  4^5   2^4,  4)$,\\
IrMOA$(56^4,  17,  56^1      14^4  8^1  7^3  4^6   2^2,  4)$,\\
IrMOA$(56^4,  19,  56^1 28^3       8^1  7^4  4^2   2^8,  4)$,\\
IrMOA$(56^4,  20,  56^1 28^2 14^1  8^1  7^4  4^2   2^9,  4)$,\\
IrMOA$(56^4,  21,  56^1 28^1 14^2  8^1  7^4  4^22^{10},  4)$,\\
IrMOA$(56^4,  22,  56^1      14^3  8^1  7^4  4^22^{11},  4)$,\\
IrMOA$(56^4,  19,  56^1 28^2 14^1  8^1  7^4  4^3   2^7,  4)$,\\
IrMOA$(56^4,  20,  56^1 28^1 14^2  8^1  7^4  4^3   2^8,  4)$,\\
IrMOA$(56^4,  21,  56^1      14^3  8^1  7^4  4^3   2^9,  4)$,\\
IrMOA$(56^4,  19,  56^1 28^1 14^2  8^1  7^4  4^4   2^6,  4)$,\\
IrMOA$(56^4,  20,  56^1      14^3  8^1  7^4  4^4   2^7,  4)$,\\
IrMOA$(56^4,  19,  56^1      14^3  8^1  7^4  4^5   2^5,  4)$,\\
IrMOA$(56^4,  21,  56^1 28^2       8^1  7^5  4^22^{10},  4)$,\\
IrMOA$(56^4,  22,  56^1 28^1 14^1  8^1  7^5  4^22^{11},  4)$,\\
IrMOA$(56^4,  23,  56^1      14^2  8^1  7^5  4^22^{12},  4)$,\\
IrMOA$(56^4,  21,  56^1 28^1 14^1  8^1  7^5  4^3   2^9,  4)$,\\
IrMOA$(56^4,  22,  56^1      14^2  8^1  7^5  4^32^{10},  4)$,\\
IrMOA$(56^4,  21,  56^1      14^2  8^1  7^5  4^4   2^8,  4)$,\\
IrMOA$(56^4,  23,  56^1 28^1       8^1  7^6  4^22^{12},  4)$,\\
IrMOA$(56^4,  24,  56^1      14^1  8^1  7^6  4^22^{13},  4)$,\\
IrMOA$(56^4,  23,  56^1      14^1  8^1  7^6  4^32^{11},  4)$,\\
IrMOA$(56^4,  25,  56^1            8^1  7^7  4^22^{14},  4)$,\\
IrMOA$(56^4,  18,  56^1 28^3       8^1  7^4  4^3   2^6,  4)$,\\
IrMOA$(56^4,  19,  56^1 28^2 14^1  8^1  7^4  4^3   2^7,  4)$,\\
IrMOA$(56^4,  20,  56^1 28^1 14^2  8^1  7^4  4^3   2^8,  4)$,\\
IrMOA$(56^4,  21,  56^1      14^3  8^1  7^4  4^3   2^9,  4)$,\\
IrMOA$(56^4,  18,  56^1 28^2 14^1  8^1  7^4  4^4   2^5,  4)$,\\
IrMOA$(56^4,  19,  56^1 28^1 14^2  8^1  7^4  4^4   2^6,  4)$,\\
IrMOA$(56^4,  20,  56^1      14^3  8^1  7^4  4^4   2^7,  4)$,\\
IrMOA$(56^4,  18,  56^1 28^1 14^2  8^1  7^4  4^5   2^4,  4)$,\\
IrMOA$(56^4,  19,  56^1      14^3  8^1  7^4  4^5   2^5,  4)$,\\
IrMOA$(56^4,  18,  56^1      14^3  8^1  7^4  4^6   2^3,  4)$,\\
IrMOA$(56^4,  20,  56^1 28^2       8^1  7^5  4^3   2^8,  4)$,\\
IrMOA$(56^4,  21,  56^1 28^1 14^1  8^1  7^5  4^3   2^9,  4)$,\\
IrMOA$(56^4,  22,  56^1      14^2  8^1  7^5  4^32^{10},  4)$,\\
IrMOA$(56^4,  20,  56^1 28^1 14^1  8^1  7^5  4^4   2^7,  4)$,\\
IrMOA$(56^4,  21,  56^1      14^2  8^1  7^5  4^4   2^8,  4)$,\\
IrMOA$(56^4,  20,  56^1      14^2  8^1  7^5  4^5   2^6,  4)$,\\
IrMOA$(56^4,  22,  56^1 28^1       8^1  7^6  4^32^{10},  4)$,\\
IrMOA$(56^4,  23,  56^1      14^1  8^1  7^6  4^32^{11},  4)$,\\
IrMOA$(56^4,  22,  56^1      14^1  8^1  7^6  4^4   2^9,  4)$,\\
IrMOA$(56^4,  24,  56^1            8^1  7^7  4^32^{12},  4)$,\\
IrMOA$(56^4,  19,  56^1 28^2       8^1  7^5  4^4   2^6,  4)$,\\
IrMOA$(56^4,  20,  56^1 28^1 14^1  8^1  7^5  4^4   2^7,  4)$,\\
IrMOA$(56^4,  21,  56^1      14^2  8^1  7^5  4^4   2^8,  4)$,\\
IrMOA$(56^4,  19,  56^1 28^1 14^1  8^1  7^5  4^5   2^5,  4)$,\\
IrMOA$(56^4,  20,  56^1      14^2  8^1  7^5  4^5   2^6,  4)$,\\
IrMOA$(56^4,  19,  56^1      14^2  8^1  7^5  4^6   2^4,  4)$,\\
IrMOA$(56^4,  21,  56^1 28^1       8^1  7^6  4^4   2^8,  4)$,\\
IrMOA$(56^4,  22,  56^1      14^1  8^1  7^6  4^4   2^9,  4)$,\\
IrMOA$(56^4,  21,  56^1      14^1  8^1  7^6  4^5   2^7,  4)$,\\
IrMOA$(56^4,  23,  56^1            8^1  7^7  4^42^{10},  4)$,\\
IrMOA$(56^4,  20,  56^1 28^1       8^1  7^6  4^5   2^6,  4)$,\\
IrMOA$(56^4,  21,  56^1      14^1  8^1  7^6  4^5   2^7,  4)$,\\
IrMOA$(56^4,  20,  56^1      14^1  8^1  7^6  4^6   2^5,  4)$,\\
IrMOA$(56^4,  22,  56^1            8^1  7^7  4^5   2^8,  4)$,\\
IrMOA$(56^4,  21,  56^1            8^1  7^7  4^6   2^6,  4)$,\\
IrMOA$(56^4,  15,  56^1 28^5       8^2  7^2        2^5,  4)$,\\
IrMOA$(56^4,  16,  56^1 28^4 14^1  8^2  7^2        2^6,  4)$,\\
IrMOA$(56^4,  17,  56^1 28^3 14^2  8^2  7^2        2^7,  4)$,\\
IrMOA$(56^4,  18,  56^1 28^2 14^3  8^2  7^2        2^8,  4)$,\\
IrMOA$(56^4,  19,  56^1 28^1 14^4  8^2  7^2        2^9,  4)$,\\
IrMOA$(56^4,  20,  56^1      14^5  8^2  7^2     2^{10},  4)$,\\
IrMOA$(56^4,  15,  56^1 28^4 14^1  8^2  7^2  4^1   2^4,  4)$,\\
IrMOA$(56^4,  16,  56^1 28^3 14^2  8^2  7^2  4^1   2^5,  4)$,\\
IrMOA$(56^4,  17,  56^1 28^2 14^3  8^2  7^2  4^1   2^6,  4)$,\\
IrMOA$(56^4,  18,  56^1 28^1 14^4  8^2  7^2  4^1   2^7,  4)$,\\
IrMOA$(56^4,  19,  56^1      14^5  8^2  7^2  4^1   2^8,  4)$,\\
IrMOA$(56^4,  15,  56^1 28^3 14^2  8^2  7^2  4^2   2^3,  4)$,\\
IrMOA$(56^4,  16,  56^1 28^2 14^3  8^2  7^2  4^2   2^4,  4)$,\\
IrMOA$(56^4,  17,  56^1 28^1 14^4  8^2  7^2  4^2   2^5,  4)$,\\
IrMOA$(56^4,  18,  56^1      14^5  8^2  7^2  4^2   2^6,  4)$,\\
IrMOA$(56^4,  15,  56^1 28^2 14^3  8^2  7^2  4^3   2^2,  4)$,\\
IrMOA$(56^4,  16,  56^1 28^1 14^4  8^2  7^2  4^3   2^3,  4)$,\\
IrMOA$(56^4,  17,  56^1      14^5  8^2  7^2  4^3   2^4,  4)$,\\
IrMOA$(56^4,  15,  56^1 28^1 14^4  8^2  7^2  4^4   2^1,  4)$,\\
IrMOA$(56^4,  16,  56^1      14^5  8^2  7^2  4^4   2^2,  4)$,\\
IrMOA$(56^4,  15,  56^1      14^5  8^2  7^2  4^5      ,  4)$,\\
IrMOA$(56^4,  17,  56^1 28^4       8^2  7^3        2^7,  4)$,\\
IrMOA$(56^4,  18,  56^1 28^3 14^1  8^2  7^3        2^8,  4)$,\\
IrMOA$(56^4,  19,  56^1 28^2 14^2  8^2  7^3        2^9,  4)$,\\
IrMOA$(56^4,  20,  56^1 28^1 14^3  8^2  7^3     2^{10},  4)$,\\
IrMOA$(56^4,  21,  56^1      14^4  8^2  7^3     2^{11},  4)$,\\
IrMOA$(56^4,  17,  56^1 28^3 14^1  8^2  7^3  4^1   2^6,  4)$,\\
IrMOA$(56^4,  18,  56^1 28^2 14^2  8^2  7^3  4^1   2^7,  4)$,\\
IrMOA$(56^4,  19,  56^1 28^1 14^3  8^2  7^3  4^1   2^8,  4)$,\\
IrMOA$(56^4,  20,  56^1      14^4  8^2  7^3  4^1   2^9,  4)$,\\
IrMOA$(56^4,  17,  56^1 28^2 14^2  8^2  7^3  4^2   2^5,  4)$,\\
IrMOA$(56^4,  18,  56^1 28^1 14^3  8^2  7^3  4^2   2^6,  4)$,\\
IrMOA$(56^4,  19,  56^1      14^4  8^2  7^3  4^2   2^7,  4)$,\\
IrMOA$(56^4,  17,  56^1 28^1 14^3  8^2  7^3  4^3   2^4,  4)$,\\
IrMOA$(56^4,  18,  56^1      14^4  8^2  7^3  4^3   2^5,  4)$,\\
IrMOA$(56^4,  17,  56^1      14^4  8^2  7^3  4^4   2^3,  4)$,\\
IrMOA$(56^4,  19,  56^1 28^3       8^2  7^4        2^9,  4)$,\\
IrMOA$(56^4,  20,  56^1 28^2 14^1  8^2  7^4     2^{10},  4)$,\\
IrMOA$(56^4,  21,  56^1 28^1 14^2  8^2  7^4     2^{11},  4)$,\\
IrMOA$(56^4,  22,  56^1      14^3  8^2  7^4     2^{12},  4)$,\\
IrMOA$(56^4,  19,  56^1 28^2 14^1  8^2  7^4  4^1   2^8,  4)$,\\
IrMOA$(56^4,  20,  56^1 28^1 14^2  8^2  7^4  4^1   2^9,  4)$,\\
IrMOA$(56^4,  21,  56^1      14^3  8^2  7^4  4^12^{10},  4)$,\\
IrMOA$(56^4,  19,  56^1 28^1 14^2  8^2  7^4  4^2   2^7,  4)$,\\
IrMOA$(56^4,  20,  56^1      14^3  8^2  7^4  4^2   2^8,  4)$,\\
IrMOA$(56^4,  19,  56^1      14^3  8^2  7^4  4^3   2^6,  4)$,\\
IrMOA$(56^4,  21,  56^1 28^2       8^2  7^5     2^{11},  4)$,\\
IrMOA$(56^4,  22,  56^1 28^1 14^1  8^2  7^5     2^{12},  4)$,\\
IrMOA$(56^4,  23,  56^1      14^2  8^2  7^5     2^{13},  4)$,\\
IrMOA$(56^4,  21,  56^1 28^1 14^1  8^2  7^5  4^12^{10},  4)$,\\
IrMOA$(56^4,  22,  56^1      14^2  8^2  7^5  4^12^{11},  4)$,\\
IrMOA$(56^4,  21,  56^1      14^2  8^2  7^5  4^2   2^9,  4)$,\\
IrMOA$(56^4,  23,  56^1 28^1       8^2  7^6     2^{13},  4)$,\\
IrMOA$(56^4,  24,  56^1      14^1  8^2  7^6     2^{14},  4)$,\\
IrMOA$(56^4,  23,  56^1      14^1  8^2  7^6  4^12^{12},  4)$,\\
IrMOA$(56^4,  25,  56^1            8^2  7^7     2^{15},  4)$,\\
IrMOA$(56^4,  16,  56^1 28^4       8^2  7^3  4^1   2^5,  4)$,\\
IrMOA$(56^4,  17,  56^1 28^3 14^1  8^2  7^3  4^1   2^6,  4)$,\\
IrMOA$(56^4,  18,  56^1 28^2 14^2  8^2  7^3  4^1   2^7,  4)$,\\
IrMOA$(56^4,  19,  56^1 28^1 14^3  8^2  7^3  4^1   2^8,  4)$,\\
IrMOA$(56^4,  20,  56^1      14^4  8^2  7^3  4^1   2^9,  4)$,\\
IrMOA$(56^4,  16,  56^1 28^3 14^1  8^2  7^3  4^2   2^4,  4)$,\\
IrMOA$(56^4,  17,  56^1 28^2 14^2  8^2  7^3  4^2   2^5,  4)$,\\
IrMOA$(56^4,  18,  56^1 28^1 14^3  8^2  7^3  4^2   2^6,  4)$,\\
IrMOA$(56^4,  19,  56^1      14^4  8^2  7^3  4^2   2^7,  4)$,\\
IrMOA$(56^4,  16,  56^1 28^2 14^2  8^2  7^3  4^3   2^3,  4)$,\\
IrMOA$(56^4,  17,  56^1 28^1 14^3  8^2  7^3  4^3   2^4,  4)$,\\
IrMOA$(56^4,  18,  56^1      14^4  8^2  7^3  4^3   2^5,  4)$,\\
IrMOA$(56^4,  16,  56^1 28^1 14^3  8^2  7^3  4^4   2^2,  4)$,\\
IrMOA$(56^4,  17,  56^1      14^4  8^2  7^3  4^4   2^3,  4)$,\\
IrMOA$(56^4,  16,  56^1      14^4  8^2  7^3  4^5   2^1,  4)$,\\
IrMOA$(56^4,  18,  56^1 28^3       8^2  7^4  4^1   2^7,  4)$,\\
IrMOA$(56^4,  19,  56^1 28^2 14^1  8^2  7^4  4^1   2^8,  4)$,\\
IrMOA$(56^4,  20,  56^1 28^1 14^2  8^2  7^4  4^1   2^9,  4)$,\\
IrMOA$(56^4,  21,  56^1      14^3  8^2  7^4  4^12^{10},  4)$,\\
IrMOA$(56^4,  18,  56^1 28^2 14^1  8^2  7^4  4^2   2^6,  4)$,\\
IrMOA$(56^4,  19,  56^1 28^1 14^2  8^2  7^4  4^2   2^7,  4)$,\\
IrMOA$(56^4,  20,  56^1      14^3  8^2  7^4  4^2   2^8,  4)$,\\
IrMOA$(56^4,  18,  56^1 28^1 14^2  8^2  7^4  4^3   2^5,  4)$,\\
IrMOA$(56^4,  19,  56^1      14^3  8^2  7^4  4^3   2^6,  4)$,\\
IrMOA$(56^4,  18,  56^1      14^3  8^2  7^4  4^4   2^4,  4)$,\\
IrMOA$(56^4,  20,  56^1 28^2       8^2  7^5  4^1   2^9,  4)$,\\
IrMOA$(56^4,  21,  56^1 28^1 14^1  8^2  7^5  4^12^{10},  4)$,\\
IrMOA$(56^4,  22,  56^1      14^2  8^2  7^5  4^12^{11},  4)$,\\
IrMOA$(56^4,  20,  56^1 28^1 14^1  8^2  7^5  4^2   2^8,  4)$,\\
IrMOA$(56^4,  21,  56^1      14^2  8^2  7^5  4^2   2^9,  4)$,\\
IrMOA$(56^4,  20,  56^1      14^2  8^2  7^5  4^3   2^7,  4)$,\\
IrMOA$(56^4,  22,  56^1 28^1       8^2  7^6  4^12^{11},  4)$,\\
IrMOA$(56^4,  23,  56^1      14^1  8^2  7^6  4^12^{12},  4)$,\\
IrMOA$(56^4,  22,  56^1      14^1  8^2  7^6  4^22^{10},  4)$,\\
IrMOA$(56^4,  24,  56^1            8^2  7^7  4^12^{13},  4)$,\\
IrMOA$(56^4,  17,  56^1 28^3       8^2  7^4  4^2   2^5,  4)$,\\
IrMOA$(56^4,  18,  56^1 28^2 14^1  8^2  7^4  4^2   2^6,  4)$,\\
IrMOA$(56^4,  19,  56^1 28^1 14^2  8^2  7^4  4^2   2^7,  4)$,\\
IrMOA$(56^4,  20,  56^1      14^3  8^2  7^4  4^2   2^8,  4)$,\\
IrMOA$(56^4,  17,  56^1 28^2 14^1  8^2  7^4  4^3   2^4,  4)$,\\
IrMOA$(56^4,  18,  56^1 28^1 14^2  8^2  7^4  4^3   2^5,  4)$,\\
IrMOA$(56^4,  19,  56^1      14^3  8^2  7^4  4^3   2^6,  4)$,\\
IrMOA$(56^4,  17,  56^1 28^1 14^2  8^2  7^4  4^4   2^3,  4)$,\\
IrMOA$(56^4,  18,  56^1      14^3  8^2  7^4  4^4   2^4,  4)$,\\
IrMOA$(56^4,  17,  56^1      14^3  8^2  7^4  4^5   2^2,  4)$,\\
IrMOA$(56^4,  19,  56^1 28^2       8^2  7^5  4^2   2^7,  4)$,\\
IrMOA$(56^4,  20,  56^1 28^1 14^1  8^2  7^5  4^2   2^8,  4)$,\\
IrMOA$(56^4,  21,  56^1      14^2  8^2  7^5  4^2   2^9,  4)$,\\
IrMOA$(56^4,  19,  56^1 28^1 14^1  8^2  7^5  4^3   2^6,  4)$,\\
IrMOA$(56^4,  20,  56^1      14^2  8^2  7^5  4^3   2^7,  4)$,\\
IrMOA$(56^4,  19,  56^1      14^2  8^2  7^5  4^4   2^5,  4)$,\\
IrMOA$(56^4,  21,  56^1 28^1       8^2  7^6  4^2   2^9,  4)$,\\
IrMOA$(56^4,  22,  56^1      14^1  8^2  7^6  4^22^{10},  4)$,\\
IrMOA$(56^4,  21,  56^1      14^1  8^2  7^6  4^3   2^8,  4)$,\\
IrMOA$(56^4,  23,  56^1            8^2  7^7  4^22^{11},  4)$,\\
IrMOA$(56^4,  18,  56^1 28^2       8^2  7^5  4^3   2^5,  4)$,\\
IrMOA$(56^4,  19,  56^1 28^1 14^1  8^2  7^5  4^3   2^6,  4)$,\\
IrMOA$(56^4,  20,  56^1      14^2  8^2  7^5  4^3   2^7,  4)$,\\
IrMOA$(56^4,  18,  56^1 28^1 14^1  8^2  7^5  4^4   2^4,  4)$,\\
IrMOA$(56^4,  19,  56^1      14^2  8^2  7^5  4^4   2^5,  4)$,\\
IrMOA$(56^4,  18,  56^1      14^2  8^2  7^5  4^5   2^3,  4)$,\\
IrMOA$(56^4,  20,  56^1 28^1       8^2  7^6  4^3   2^7,  4)$,\\
IrMOA$(56^4,  21,  56^1      14^1  8^2  7^6  4^3   2^8,  4)$,\\
IrMOA$(56^4,  20,  56^1      14^1  8^2  7^6  4^4   2^6,  4)$,\\
IrMOA$(56^4,  22,  56^1            8^2  7^7  4^3   2^9,  4)$,\\
IrMOA$(56^4,  19,  56^1 28^1       8^2  7^6  4^4   2^5,  4)$,\\
IrMOA$(56^4,  20,  56^1      14^1  8^2  7^6  4^4   2^6,  4)$,\\
IrMOA$(56^4,  19,  56^1      14^1  8^2  7^6  4^5   2^4,  4)$,\\
IrMOA$(56^4,  21,  56^1            8^2  7^7  4^4   2^7,  4)$,\\
IrMOA$(56^4,  20,  56^1            8^2  7^7  4^5   2^5,  4)$,\\
IrMOA$(56^4,  15,  56^1 28^4       8^3  7^3        2^4,  4)$,\\
IrMOA$(56^4,  16,  56^1 28^3 14^1  8^3  7^3        2^5,  4)$,\\
IrMOA$(56^4,  17,  56^1 28^2 14^2  8^3  7^3        2^6,  4)$,\\
IrMOA$(56^4,  18,  56^1 28^1 14^3  8^3  7^3        2^7,  4)$,\\
IrMOA$(56^4,  19,  56^1      14^4  8^3  7^3        2^8,  4)$,\\
IrMOA$(56^4,  15,  56^1 28^3 14^1  8^3  7^3  4^1   2^3,  4)$,\\
IrMOA$(56^4,  16,  56^1 28^2 14^2  8^3  7^3  4^1   2^4,  4)$,\\
IrMOA$(56^4,  17,  56^1 28^1 14^3  8^3  7^3  4^1   2^5,  4)$,\\
IrMOA$(56^4,  18,  56^1      14^4  8^3  7^3  4^1   2^6,  4)$,\\
IrMOA$(56^4,  15,  56^1 28^2 14^2  8^3  7^3  4^2   2^2,  4)$,\\
IrMOA$(56^4,  16,  56^1 28^1 14^3  8^3  7^3  4^2   2^3,  4)$,\\
IrMOA$(56^4,  17,  56^1      14^4  8^3  7^3  4^2   2^4,  4)$,\\
IrMOA$(56^4,  15,  56^1 28^1 14^3  8^3  7^3  4^3   2^1,  4)$,\\
IrMOA$(56^4,  16,  56^1      14^4  8^3  7^3  4^3   2^2,  4)$,\\
IrMOA$(56^4,  15,  56^1      14^4  8^3  7^3  4^4      ,  4)$,\\
IrMOA$(56^4,  17,  56^1 28^3       8^3  7^4        2^6,  4)$,\\
IrMOA$(56^4,  18,  56^1 28^2 14^1  8^3  7^4        2^7,  4)$,\\
IrMOA$(56^4,  19,  56^1 28^1 14^2  8^3  7^4        2^8,  4)$,\\
IrMOA$(56^4,  20,  56^1      14^3  8^3  7^4        2^9,  4)$,\\
IrMOA$(56^4,  17,  56^1 28^2 14^1  8^3  7^4  4^1   2^5,  4)$,\\
IrMOA$(56^4,  18,  56^1 28^1 14^2  8^3  7^4  4^1   2^6,  4)$,\\
IrMOA$(56^4,  19,  56^1      14^3  8^3  7^4  4^1   2^7,  4)$,\\
IrMOA$(56^4,  17,  56^1 28^1 14^2  8^3  7^4  4^2   2^4,  4)$,\\
IrMOA$(56^4,  18,  56^1      14^3  8^3  7^4  4^2   2^5,  4)$,\\
IrMOA$(56^4,  17,  56^1      14^3  8^3  7^4  4^3   2^3,  4)$,\\
IrMOA$(56^4,  19,  56^1 28^2       8^3  7^5        2^8,  4)$,\\
IrMOA$(56^4,  20,  56^1 28^1 14^1  8^3  7^5        2^9,  4)$,\\
IrMOA$(56^4,  21,  56^1      14^2  8^3  7^5     2^{10},  4)$,\\
IrMOA$(56^4,  19,  56^1 28^1 14^1  8^3  7^5  4^1   2^7,  4)$,\\
IrMOA$(56^4,  20,  56^1      14^2  8^3  7^5  4^1   2^8,  4)$,\\
IrMOA$(56^4,  19,  56^1      14^2  8^3  7^5  4^2   2^6,  4)$,\\
IrMOA$(56^4,  21,  56^1 28^1       8^3  7^6     2^{10},  4)$,\\
IrMOA$(56^4,  22,  56^1      14^1  8^3  7^6     2^{11},  4)$,\\
IrMOA$(56^4,  21,  56^1      14^1  8^3  7^6  4^1   2^9,  4)$,\\
IrMOA$(56^4,  23,  56^1            8^3  7^7     2^{12},  4)$,\\
IrMOA$(56^4,  16,  56^1 28^3       8^3  7^4  4^1   2^4,  4)$,\\
IrMOA$(56^4,  17,  56^1 28^2 14^1  8^3  7^4  4^1   2^5,  4)$,\\
IrMOA$(56^4,  18,  56^1 28^1 14^2  8^3  7^4  4^1   2^6,  4)$,\\
IrMOA$(56^4,  19,  56^1      14^3  8^3  7^4  4^1   2^7,  4)$,\\
IrMOA$(56^4,  16,  56^1 28^2 14^1  8^3  7^4  4^2   2^3,  4)$,\\
IrMOA$(56^4,  17,  56^1 28^1 14^2  8^3  7^4  4^2   2^4,  4)$,\\
IrMOA$(56^4,  18,  56^1      14^3  8^3  7^4  4^2   2^5,  4)$,\\
IrMOA$(56^4,  16,  56^1 28^1 14^2  8^3  7^4  4^3   2^2,  4)$,\\
IrMOA$(56^4,  17,  56^1      14^3  8^3  7^4  4^3   2^3,  4)$,\\
IrMOA$(56^4,  16,  56^1      14^3  8^3  7^4  4^4   2^1,  4)$,\\
IrMOA$(56^4,  18,  56^1 28^2       8^3  7^5  4^1   2^6,  4)$,\\
IrMOA$(56^4,  19,  56^1 28^1 14^1  8^3  7^5  4^1   2^7,  4)$,\\
IrMOA$(56^4,  20,  56^1      14^2  8^3  7^5  4^1   2^8,  4)$,\\
IrMOA$(56^4,  18,  56^1 28^1 14^1  8^3  7^5  4^2   2^5,  4)$,\\
IrMOA$(56^4,  19,  56^1      14^2  8^3  7^5  4^2   2^6,  4)$,\\
IrMOA$(56^4,  18,  56^1      14^2  8^3  7^5  4^3   2^4,  4)$,\\
IrMOA$(56^4,  20,  56^1 28^1       8^3  7^6  4^1   2^8,  4)$,\\
IrMOA$(56^4,  21,  56^1      14^1  8^3  7^6  4^1   2^9,  4)$,\\
IrMOA$(56^4,  20,  56^1      14^1  8^3  7^6  4^2   2^7,  4)$,\\
IrMOA$(56^4,  22,  56^1            8^3  7^7  4^12^{10},  4)$,\\
IrMOA$(56^4,  17,  56^1 28^2       8^3  7^5  4^2   2^4,  4)$,\\
IrMOA$(56^4,  18,  56^1 28^1 14^1  8^3  7^5  4^2   2^5,  4)$,\\
IrMOA$(56^4,  19,  56^1      14^2  8^3  7^5  4^2   2^6,  4)$,\\
IrMOA$(56^4,  17,  56^1 28^1 14^1  8^3  7^5  4^3   2^3,  4)$,\\
IrMOA$(56^4,  18,  56^1      14^2  8^3  7^5  4^3   2^4,  4)$,\\
IrMOA$(56^4,  17,  56^1      14^2  8^3  7^5  4^4   2^2,  4)$,\\
IrMOA$(56^4,  19,  56^1 28^1       8^3  7^6  4^2   2^6,  4)$,\\
IrMOA$(56^4,  20,  56^1      14^1  8^3  7^6  4^2   2^7,  4)$,\\
IrMOA$(56^4,  19,  56^1      14^1  8^3  7^6  4^3   2^5,  4)$,\\
IrMOA$(56^4,  21,  56^1            8^3  7^7  4^2   2^8,  4)$,\\
IrMOA$(56^4,  18,  56^1 28^1       8^3  7^6  4^3   2^4,  4)$,\\
IrMOA$(56^4,  19,  56^1      14^1  8^3  7^6  4^3   2^5,  4)$,\\
IrMOA$(56^4,  18,  56^1      14^1  8^3  7^6  4^4   2^3,  4)$,\\
IrMOA$(56^4,  20,  56^1            8^3  7^7  4^3   2^6,  4)$,\\
IrMOA$(56^4,  19,  56^1            8^3  7^7  4^4   2^4,  4)$,\\
IrMOA$(56^4,  15,  56^1 28^3       8^4  7^4        2^3,  4)$,\\
IrMOA$(56^4,  16,  56^1 28^2 14^1  8^4  7^4        2^4,  4)$,\\
IrMOA$(56^4,  17,  56^1 28^1 14^2  8^4  7^4        2^5,  4)$,\\
IrMOA$(56^4,  18,  56^1      14^3  8^4  7^4        2^6,  4)$,\\
IrMOA$(56^4,  15,  56^1 28^2 14^1  8^4  7^4  4^1   2^2,  4)$,\\
IrMOA$(56^4,  16,  56^1 28^1 14^2  8^4  7^4  4^1   2^3,  4)$,\\
IrMOA$(56^4,  17,  56^1      14^3  8^4  7^4  4^1   2^4,  4)$,\\
IrMOA$(56^4,  15,  56^1 28^1 14^2  8^4  7^4  4^2   2^1,  4)$,\\
IrMOA$(56^4,  16,  56^1      14^3  8^4  7^4  4^2   2^2,  4)$,\\
IrMOA$(56^4,  15,  56^1      14^3  8^4  7^4  4^3      ,  4)$,\\
IrMOA$(56^4,  17,  56^1 28^2       8^4  7^5        2^5,  4)$,\\
IrMOA$(56^4,  18,  56^1 28^1 14^1  8^4  7^5        2^6,  4)$,\\
IrMOA$(56^4,  19,  56^1      14^2  8^4  7^5        2^7,  4)$,\\
IrMOA$(56^4,  17,  56^1 28^1 14^1  8^4  7^5  4^1   2^4,  4)$,\\
IrMOA$(56^4,  18,  56^1      14^2  8^4  7^5  4^1   2^5,  4)$,\\
IrMOA$(56^4,  17,  56^1      14^2  8^4  7^5  4^2   2^3,  4)$,\\
IrMOA$(56^4,  19,  56^1 28^1       8^4  7^6        2^7,  4)$,\\
IrMOA$(56^4,  20,  56^1      14^1  8^4  7^6        2^8,  4)$,\\
IrMOA$(56^4,  19,  56^1      14^1  8^4  7^6  4^1   2^6,  4)$,\\
IrMOA$(56^4,  21,  56^1            8^4  7^7        2^9,  4)$,\\
IrMOA$(56^4,  16,  56^1 28^2       8^4  7^5  4^1   2^3,  4)$,\\
IrMOA$(56^4,  17,  56^1 28^1 14^1  8^4  7^5  4^1   2^4,  4)$,\\
IrMOA$(56^4,  18,  56^1      14^2  8^4  7^5  4^1   2^5,  4)$,\\
IrMOA$(56^4,  16,  56^1 28^1 14^1  8^4  7^5  4^2   2^2,  4)$,\\
IrMOA$(56^4,  17,  56^1      14^2  8^4  7^5  4^2   2^3,  4)$,\\
IrMOA$(56^4,  16,  56^1      14^2  8^4  7^5  4^3   2^1,  4)$,\\
IrMOA$(56^4,  18,  56^1 28^1       8^4  7^6  4^1   2^5,  4)$,\\
IrMOA$(56^4,  19,  56^1      14^1  8^4  7^6  4^1   2^6,  4)$,\\
IrMOA$(56^4,  18,  56^1      14^1  8^4  7^6  4^2   2^4,  4)$,\\
IrMOA$(56^4,  20,  56^1            8^4  7^7  4^1   2^7,  4)$,\\
IrMOA$(56^4,  17,  56^1 28^1       8^4  7^6  4^2   2^3,  4)$,\\
IrMOA$(56^4,  18,  56^1      14^1  8^4  7^6  4^2   2^4,  4)$,\\
IrMOA$(56^4,  17,  56^1      14^1  8^4  7^6  4^3   2^2,  4)$,\\
IrMOA$(56^4,  19,  56^1            8^4  7^7  4^2   2^5,  4)$,\\
IrMOA$(56^4,  18,  56^1            8^4  7^7  4^3   2^3,  4)$,\\
IrMOA$(56^4,  15,  56^1 28^2       8^5  7^5        2^2,  4)$,\\
IrMOA$(56^4,  16,  56^1 28^1 14^1  8^5  7^5        2^3,  4)$,\\
IrMOA$(56^4,  17,  56^1      14^2  8^5  7^5        2^4,  4)$,\\
IrMOA$(56^4,  15,  56^1 28^1 14^1  8^5  7^5  4^1   2^1,  4)$,\\
IrMOA$(56^4,  16,  56^1      14^2  8^5  7^5  4^1   2^2,  4)$,\\
IrMOA$(56^4,  15,  56^1      14^2  8^5  7^5  4^2      ,  4)$,\\
IrMOA$(56^4,  17,  56^1 28^1       8^5  7^6        2^4,  4)$,\\
IrMOA$(56^4,  18,  56^1      14^1  8^5  7^6        2^5,  4)$,\\
IrMOA$(56^4,  17,  56^1      14^1  8^5  7^6  4^1   2^3,  4)$,\\
IrMOA$(56^4,  19,  56^1            8^5  7^7        2^6,  4)$,\\
IrMOA$(56^4,  16,  56^1 28^1       8^5  7^6  4^1   2^2,  4)$,\\
IrMOA$(56^4,  17,  56^1      14^1  8^5  7^6  4^1   2^3,  4)$,\\
IrMOA$(56^4,  16,  56^1      14^1  8^5  7^6  4^2   2^1,  4)$,\\
IrMOA$(56^4,  18,  56^1            8^5  7^7  4^1   2^4,  4)$,\\
IrMOA$(56^4,  17,  56^1            8^5  7^7  4^2   2^2,  4)$,\\
IrMOA$(56^4,  15,  56^1 28^1       8^6  7^6        2^1,  4)$,\\
IrMOA$(56^4,  16,  56^1      14^1  8^6  7^6        2^2,  4)$,\\
IrMOA$(56^4,  15,  56^1      14^1  8^6  7^6  4^1      ,  4)$,\\
IrMOA$(56^4,  17,  56^1            8^6  7^7        2^3,  4)$,\\
IrMOA$(56^4,  16,  56^1            8^6  7^7  4^1   2^1,  4)$,\\
IrMOA$(56^4,  15,  56^1            8^7  7^7           ,  4)$,\\
IrMOA$(56^4,  16,       28^8                       2^8,  4)$,\\
IrMOA$(56^4,  17,       28^7 14^1                  2^9,  4)$,\\
IrMOA$(56^4,  18,       28^6 14^2               2^{10},  4)$,\\
IrMOA$(56^4,  19,       28^5 14^3               2^{11},  4)$,\\
IrMOA$(56^4,  20,       28^4 14^4               2^{12},  4)$,\\
IrMOA$(56^4,  21,       28^3 14^5               2^{13},  4)$,\\
IrMOA$(56^4,  22,       28^2 14^6               2^{14},  4)$,\\
IrMOA$(56^4,  23,       28^1 14^7               2^{15},  4)$,\\
IrMOA$(56^4,  24,            14^8               2^{16},  4)$,\\
IrMOA$(56^4,  16,       28^7 14^1            4^1   2^7,  4)$,\\
IrMOA$(56^4,  17,       28^6 14^2            4^1   2^8,  4)$,\\
IrMOA$(56^4,  18,       28^5 14^3            4^1   2^9,  4)$,\\
IrMOA$(56^4,  19,       28^4 14^4            4^12^{10},  4)$,\\
IrMOA$(56^4,  20,       28^3 14^5            4^12^{11},  4)$,\\
IrMOA$(56^4,  21,       28^2 14^6            4^12^{12},  4)$,\\
IrMOA$(56^4,  22,       28^1 14^7            4^12^{13},  4)$,\\
IrMOA$(56^4,  23,            14^8            4^12^{14},  4)$,\\
IrMOA$(56^4,  16,       28^6 14^2            4^2   2^6,  4)$,\\
IrMOA$(56^4,  17,       28^5 14^3            4^2   2^7,  4)$,\\
IrMOA$(56^4,  18,       28^4 14^4            4^2   2^8,  4)$,\\
IrMOA$(56^4,  19,       28^3 14^5            4^2   2^9,  4)$,\\
IrMOA$(56^4,  20,       28^2 14^6            4^22^{10},  4)$,\\
IrMOA$(56^4,  21,       28^1 14^7            4^22^{11},  4)$,\\
IrMOA$(56^4,  22,            14^8            4^22^{12},  4)$,\\
IrMOA$(56^4,  16,       28^5 14^3            4^3   2^5,  4)$,\\
IrMOA$(56^4,  17,       28^4 14^4            4^3   2^6,  4)$,\\
IrMOA$(56^4,  18,       28^3 14^5            4^3   2^7,  4)$,\\
IrMOA$(56^4,  19,       28^2 14^6            4^3   2^8,  4)$,\\
IrMOA$(56^4,  20,       28^1 14^7            4^3   2^9,  4)$,\\
IrMOA$(56^4,  21,            14^8            4^32^{10},  4)$,\\
IrMOA$(56^4,  16,       28^4 14^4            4^4   2^4,  4)$,\\
IrMOA$(56^4,  17,       28^3 14^5            4^4   2^5,  4)$,\\
IrMOA$(56^4,  18,       28^2 14^6            4^4   2^6,  4)$,\\
IrMOA$(56^4,  19,       28^1 14^7            4^4   2^7,  4)$,\\
IrMOA$(56^4,  20,            14^8            4^4   2^8,  4)$,\\
IrMOA$(56^4,  16,       28^3 14^5            4^5   2^3,  4)$,\\
IrMOA$(56^4,  17,       28^2 14^6            4^5   2^4,  4)$,\\
IrMOA$(56^4,  18,       28^1 14^7            4^5   2^5,  4)$,\\
IrMOA$(56^4,  19,            14^8            4^5   2^6,  4)$,\\
IrMOA$(56^4,  16,       28^2 14^6            4^6   2^2,  4)$,\\
IrMOA$(56^4,  17,       28^1 14^7            4^6   2^3,  4)$,\\
IrMOA$(56^4,  18,            14^8            4^6   2^4,  4)$,\\
IrMOA$(56^4,  16,       28^1 14^7            4^7   2^1,  4)$,\\
IrMOA$(56^4,  17,            14^8            4^7   2^2,  4)$,\\
IrMOA$(56^4,  16,            14^8            4^8      ,  4)$,\\
IrMOA$(56^4,  18,       28^7            7^1     2^{10},  4)$,\\
IrMOA$(56^4,  19,       28^6 14^1       7^1     2^{11},  4)$,\\
IrMOA$(56^4,  20,       28^5 14^2       7^1     2^{12},  4)$,\\
IrMOA$(56^4,  21,       28^4 14^3       7^1     2^{13},  4)$,\\
IrMOA$(56^4,  22,       28^3 14^4       7^1     2^{14},  4)$,\\
IrMOA$(56^4,  23,       28^2 14^5       7^1     2^{15},  4)$,\\
IrMOA$(56^4,  24,       28^1 14^6       7^1     2^{16},  4)$,\\
IrMOA$(56^4,  25,            14^7       7^1     2^{17},  4)$,\\
IrMOA$(56^4,  18,       28^6 14^1       7^1  4^1   2^9,  4)$,\\
IrMOA$(56^4,  19,       28^5 14^2       7^1  4^12^{10},  4)$,\\
IrMOA$(56^4,  20,       28^4 14^3       7^1  4^12^{11},  4)$,\\
IrMOA$(56^4,  21,       28^3 14^4       7^1  4^12^{12},  4)$,\\
IrMOA$(56^4,  22,       28^2 14^5       7^1  4^12^{13},  4)$,\\
IrMOA$(56^4,  23,       28^1 14^6       7^1  4^12^{14},  4)$,\\
IrMOA$(56^4,  24,            14^7       7^1  4^12^{15},  4)$,\\
IrMOA$(56^4,  18,       28^5 14^2       7^1  4^2   2^8,  4)$,\\
IrMOA$(56^4,  19,       28^4 14^3       7^1  4^2   2^9,  4)$,\\
IrMOA$(56^4,  20,       28^3 14^4       7^1  4^22^{10},  4)$,\\
IrMOA$(56^4,  21,       28^2 14^5       7^1  4^22^{11},  4)$,\\
IrMOA$(56^4,  22,       28^1 14^6       7^1  4^22^{12},  4)$,\\
IrMOA$(56^4,  23,            14^7       7^1  4^22^{13},  4)$,\\
IrMOA$(56^4,  18,       28^4 14^3       7^1  4^3   2^7,  4)$,\\
IrMOA$(56^4,  19,       28^3 14^4       7^1  4^3   2^8,  4)$,\\
IrMOA$(56^4,  20,       28^2 14^5       7^1  4^3   2^9,  4)$,\\
IrMOA$(56^4,  21,       28^1 14^6       7^1  4^32^{10},  4)$,\\
IrMOA$(56^4,  22,            14^7       7^1  4^32^{11},  4)$,\\
IrMOA$(56^4,  18,       28^3 14^4       7^1  4^4   2^6,  4)$,\\
IrMOA$(56^4,  19,       28^2 14^5       7^1  4^4   2^7,  4)$,\\
IrMOA$(56^4,  20,       28^1 14^6       7^1  4^4   2^8,  4)$,\\
IrMOA$(56^4,  21,            14^7       7^1  4^4   2^9,  4)$,\\
IrMOA$(56^4,  18,       28^2 14^5       7^1  4^5   2^5,  4)$,\\
IrMOA$(56^4,  19,       28^1 14^6       7^1  4^5   2^6,  4)$,\\
IrMOA$(56^4,  20,            14^7       7^1  4^5   2^7,  4)$,\\
IrMOA$(56^4,  18,       28^1 14^6       7^1  4^6   2^4,  4)$,\\
IrMOA$(56^4,  19,            14^7       7^1  4^6   2^5,  4)$,\\
IrMOA$(56^4,  18,            14^7       7^1  4^7   2^3,  4)$,\\
IrMOA$(56^4,  20,       28^6            7^2     2^{12},  4)$,\\
IrMOA$(56^4,  21,       28^5 14^1       7^2     2^{13},  4)$,\\
IrMOA$(56^4,  22,       28^4 14^2       7^2     2^{14},  4)$,\\
IrMOA$(56^4,  23,       28^3 14^3       7^2     2^{15},  4)$,\\
IrMOA$(56^4,  24,       28^2 14^4       7^2     2^{16},  4)$,\\
IrMOA$(56^4,  25,       28^1 14^5       7^2     2^{17},  4)$,\\
IrMOA$(56^4,  26,            14^6       7^2     2^{18},  4)$,\\
IrMOA$(56^4,  20,       28^5 14^1       7^2  4^12^{11},  4)$,\\
IrMOA$(56^4,  21,       28^4 14^2       7^2  4^12^{12},  4)$,\\
IrMOA$(56^4,  22,       28^3 14^3       7^2  4^12^{13},  4)$,\\
IrMOA$(56^4,  23,       28^2 14^4       7^2  4^12^{14},  4)$,\\
IrMOA$(56^4,  24,       28^1 14^5       7^2  4^12^{15},  4)$,\\
IrMOA$(56^4,  25,            14^6       7^2  4^12^{16},  4)$,\\
IrMOA$(56^4,  20,       28^4 14^2       7^2  4^22^{10},  4)$,\\
IrMOA$(56^4,  21,       28^3 14^3       7^2  4^22^{11},  4)$,\\
IrMOA$(56^4,  22,       28^2 14^4       7^2  4^22^{12},  4)$,\\
IrMOA$(56^4,  23,       28^1 14^5       7^2  4^22^{13},  4)$,\\
IrMOA$(56^4,  24,            14^6       7^2  4^22^{14},  4)$,\\
IrMOA$(56^4,  20,       28^3 14^3       7^2  4^3   2^9,  4)$,\\
IrMOA$(56^4,  21,       28^2 14^4       7^2  4^32^{10},  4)$,\\
IrMOA$(56^4,  22,       28^1 14^5       7^2  4^32^{11},  4)$,\\
IrMOA$(56^4,  23,            14^6       7^2  4^32^{12},  4)$,\\
IrMOA$(56^4,  20,       28^2 14^4       7^2  4^4   2^8,  4)$,\\
IrMOA$(56^4,  21,       28^1 14^5       7^2  4^4   2^9,  4)$,\\
IrMOA$(56^4,  22,            14^6       7^2  4^42^{10},  4)$,\\
IrMOA$(56^4,  20,       28^1 14^5       7^2  4^5   2^7,  4)$,\\
IrMOA$(56^4,  21,            14^6       7^2  4^5   2^8,  4)$,\\
IrMOA$(56^4,  20,            14^6       7^2  4^6   2^6,  4)$,\\
IrMOA$(56^4,  22,       28^5            7^3     2^{14},  4)$,\\
IrMOA$(56^4,  23,       28^4 14^1       7^3     2^{15},  4)$,\\
IrMOA$(56^4,  24,       28^3 14^2       7^3     2^{16},  4)$,\\
IrMOA$(56^4,  25,       28^2 14^3       7^3     2^{17},  4)$,\\
IrMOA$(56^4,  26,       28^1 14^4       7^3     2^{18},  4)$,\\
IrMOA$(56^4,  27,            14^5       7^3     2^{19},  4)$,\\
IrMOA$(56^4,  22,       28^4 14^1       7^3  4^12^{13},  4)$,\\
IrMOA$(56^4,  23,       28^3 14^2       7^3  4^12^{14},  4)$,\\
IrMOA$(56^4,  24,       28^2 14^3       7^3  4^12^{15},  4)$,\\
IrMOA$(56^4,  25,       28^1 14^4       7^3  4^12^{16},  4)$,\\
IrMOA$(56^4,  26,            14^5       7^3  4^12^{17},  4)$,\\
IrMOA$(56^4,  22,       28^3 14^2       7^3  4^22^{12},  4)$,\\
IrMOA$(56^4,  23,       28^2 14^3       7^3  4^22^{13},  4)$,\\
IrMOA$(56^4,  24,       28^1 14^4       7^3  4^22^{14},  4)$,\\
IrMOA$(56^4,  25,            14^5       7^3  4^22^{15},  4)$,\\
IrMOA$(56^4,  22,       28^2 14^3       7^3  4^32^{11},  4)$,\\
IrMOA$(56^4,  23,       28^1 14^4       7^3  4^32^{12},  4)$,\\
IrMOA$(56^4,  24,            14^5       7^3  4^32^{13},  4)$,\\
IrMOA$(56^4,  22,       28^1 14^4       7^3  4^42^{10},  4)$,\\
IrMOA$(56^4,  23,            14^5       7^3  4^42^{11},  4)$,\\
IrMOA$(56^4,  22,            14^5       7^3  4^5   2^9,  4)$,\\
IrMOA$(56^4,  24,       28^4            7^4     2^{16},  4)$,\\
IrMOA$(56^4,  25,       28^3 14^1       7^4     2^{17},  4)$,\\
IrMOA$(56^4,  26,       28^2 14^2       7^4     2^{18},  4)$,\\
IrMOA$(56^4,  27,       28^1 14^3       7^4     2^{19},  4)$,\\
IrMOA$(56^4,  28,            14^4       7^4     2^{20},  4)$,\\
IrMOA$(56^4,  24,       28^3 14^1       7^4  4^12^{15},  4)$,\\
IrMOA$(56^4,  25,       28^2 14^2       7^4  4^12^{16},  4)$,\\
IrMOA$(56^4,  26,       28^1 14^3       7^4  4^12^{17},  4)$,\\
IrMOA$(56^4,  27,            14^4       7^4  4^12^{18},  4)$,\\
IrMOA$(56^4,  24,       28^2 14^2       7^4  4^22^{14},  4)$,\\
IrMOA$(56^4,  25,       28^1 14^3       7^4  4^22^{15},  4)$,\\
IrMOA$(56^4,  26,            14^4       7^4  4^22^{16},  4)$,\\
IrMOA$(56^4,  24,       28^1 14^3       7^4  4^32^{13},  4)$,\\
IrMOA$(56^4,  25,            14^4       7^4  4^32^{14},  4)$,\\
IrMOA$(56^4,  24,            14^4       7^4  4^42^{12},  4)$,\\
IrMOA$(56^4,  26,       28^3            7^5     2^{18},  4)$,\\
IrMOA$(56^4,  27,       28^2 14^1       7^5     2^{19},  4)$,\\
IrMOA$(56^4,  28,       28^1 14^2       7^5     2^{20},  4)$,\\
IrMOA$(56^4,  29,            14^3       7^5     2^{21},  4)$,\\
IrMOA$(56^4,  26,       28^2 14^1       7^5  4^12^{17},  4)$,\\
IrMOA$(56^4,  27,       28^1 14^2       7^5  4^12^{18},  4)$,\\
IrMOA$(56^4,  28,            14^3       7^5  4^12^{19},  4)$,\\
IrMOA$(56^4,  26,       28^1 14^2       7^5  4^22^{16},  4)$,\\
IrMOA$(56^4,  27,            14^3       7^5  4^22^{17},  4)$,\\
IrMOA$(56^4,  26,            14^3       7^5  4^32^{15},  4)$,\\
IrMOA$(56^4,  28,       28^2            7^6     2^{20},  4)$,\\
IrMOA$(56^4,  29,       28^1 14^1       7^6     2^{21},  4)$,\\
IrMOA$(56^4,  30,            14^2       7^6     2^{22},  4)$,\\
IrMOA$(56^4,  28,       28^1 14^1       7^6  4^12^{19},  4)$,\\
IrMOA$(56^4,  29,            14^2       7^6  4^12^{20},  4)$,\\
IrMOA$(56^4,  28,            14^2       7^6  4^22^{18},  4)$,\\
IrMOA$(56^4,  30,       28^1            7^7     2^{22},  4)$,\\
IrMOA$(56^4,  31,            14^1       7^7     2^{23},  4)$,\\
IrMOA$(56^4,  30,            14^1       7^7  4^12^{21},  4)$,\\
IrMOA$(56^4,  32,                       7^8     2^{24},  4)$,\\
IrMOA$(56^4,  17,       28^7            7^1  4^1   2^8,  4)$,\\
IrMOA$(56^4,  18,       28^6 14^1       7^1  4^1   2^9,  4)$,\\
IrMOA$(56^4,  19,       28^5 14^2       7^1  4^12^{10},  4)$,\\
IrMOA$(56^4,  20,       28^4 14^3       7^1  4^12^{11},  4)$,\\
IrMOA$(56^4,  21,       28^3 14^4       7^1  4^12^{12},  4)$,\\
IrMOA$(56^4,  22,       28^2 14^5       7^1  4^12^{13},  4)$,\\
IrMOA$(56^4,  23,       28^1 14^6       7^1  4^12^{14},  4)$,\\
IrMOA$(56^4,  24,            14^7       7^1  4^12^{15},  4)$,\\
IrMOA$(56^4,  17,       28^6 14^1       7^1  4^2   2^7,  4)$,\\
IrMOA$(56^4,  18,       28^5 14^2       7^1  4^2   2^8,  4)$,\\
IrMOA$(56^4,  19,       28^4 14^3       7^1  4^2   2^9,  4)$,\\
IrMOA$(56^4,  20,       28^3 14^4       7^1  4^22^{10},  4)$,\\
IrMOA$(56^4,  21,       28^2 14^5       7^1  4^22^{11},  4)$,\\
IrMOA$(56^4,  22,       28^1 14^6       7^1  4^22^{12},  4)$,\\
IrMOA$(56^4,  23,            14^7       7^1  4^22^{13},  4)$,\\
IrMOA$(56^4,  17,       28^5 14^2       7^1  4^3   2^6,  4)$,\\
IrMOA$(56^4,  18,       28^4 14^3       7^1  4^3   2^7,  4)$,\\
IrMOA$(56^4,  19,       28^3 14^4       7^1  4^3   2^8,  4)$,\\
IrMOA$(56^4,  20,       28^2 14^5       7^1  4^3   2^9,  4)$,\\
IrMOA$(56^4,  21,       28^1 14^6       7^1  4^32^{10},  4)$,\\
IrMOA$(56^4,  22,            14^7       7^1  4^32^{11},  4)$,\\
IrMOA$(56^4,  17,       28^4 14^3       7^1  4^4   2^5,  4)$,\\
IrMOA$(56^4,  18,       28^3 14^4       7^1  4^4   2^6,  4)$,\\
IrMOA$(56^4,  19,       28^2 14^5       7^1  4^4   2^7,  4)$,\\
IrMOA$(56^4,  20,       28^1 14^6       7^1  4^4   2^8,  4)$,\\
IrMOA$(56^4,  21,            14^7       7^1  4^4   2^9,  4)$,\\
IrMOA$(56^4,  17,       28^3 14^4       7^1  4^5   2^4,  4)$,\\
IrMOA$(56^4,  18,       28^2 14^5       7^1  4^5   2^5,  4)$,\\
IrMOA$(56^4,  19,       28^1 14^6       7^1  4^5   2^6,  4)$,\\
IrMOA$(56^4,  20,            14^7       7^1  4^5   2^7,  4)$,\\
IrMOA$(56^4,  17,       28^2 14^5       7^1  4^6   2^3,  4)$,\\
IrMOA$(56^4,  18,       28^1 14^6       7^1  4^6   2^4,  4)$,\\
IrMOA$(56^4,  19,            14^7       7^1  4^6   2^5,  4)$,\\
IrMOA$(56^4,  17,       28^1 14^6       7^1  4^7   2^2,  4)$,\\
IrMOA$(56^4,  18,            14^7       7^1  4^7   2^3,  4)$,\\
IrMOA$(56^4,  17,            14^7       7^1  4^8   2^1,  4)$,\\
IrMOA$(56^4,  19,       28^6            7^2  4^12^{10},  4)$,\\
IrMOA$(56^4,  20,       28^5 14^1       7^2  4^12^{11},  4)$,\\
IrMOA$(56^4,  21,       28^4 14^2       7^2  4^12^{12},  4)$,\\
IrMOA$(56^4,  22,       28^3 14^3       7^2  4^12^{13},  4)$,\\
IrMOA$(56^4,  23,       28^2 14^4       7^2  4^12^{14},  4)$,\\
IrMOA$(56^4,  24,       28^1 14^5       7^2  4^12^{15},  4)$,\\
IrMOA$(56^4,  25,            14^6       7^2  4^12^{16},  4)$,\\
IrMOA$(56^4,  19,       28^5 14^1       7^2  4^2   2^9,  4)$,\\
IrMOA$(56^4,  20,       28^4 14^2       7^2  4^22^{10},  4)$,\\
IrMOA$(56^4,  21,       28^3 14^3       7^2  4^22^{11},  4)$,\\
IrMOA$(56^4,  22,       28^2 14^4       7^2  4^22^{12},  4)$,\\
IrMOA$(56^4,  23,       28^1 14^5       7^2  4^22^{13},  4)$,\\
IrMOA$(56^4,  24,            14^6       7^2  4^22^{14},  4)$,\\
IrMOA$(56^4,  19,       28^4 14^2       7^2  4^3   2^8,  4)$,\\
IrMOA$(56^4,  20,       28^3 14^3       7^2  4^3   2^9,  4)$,\\
IrMOA$(56^4,  21,       28^2 14^4       7^2  4^32^{10},  4)$,\\
IrMOA$(56^4,  22,       28^1 14^5       7^2  4^32^{11},  4)$,\\
IrMOA$(56^4,  23,            14^6       7^2  4^32^{12},  4)$,\\
IrMOA$(56^4,  19,       28^3 14^3       7^2  4^4   2^7,  4)$,\\
IrMOA$(56^4,  20,       28^2 14^4       7^2  4^4   2^8,  4)$,\\
IrMOA$(56^4,  21,       28^1 14^5       7^2  4^4   2^9,  4)$,\\
IrMOA$(56^4,  22,            14^6       7^2  4^42^{10},  4)$,\\
IrMOA$(56^4,  19,       28^2 14^4       7^2  4^5   2^6,  4)$,\\
IrMOA$(56^4,  20,       28^1 14^5       7^2  4^5   2^7,  4)$,\\
IrMOA$(56^4,  21,            14^6       7^2  4^5   2^8,  4)$,\\
IrMOA$(56^4,  19,       28^1 14^5       7^2  4^6   2^5,  4)$,\\
IrMOA$(56^4,  20,            14^6       7^2  4^6   2^6,  4)$,\\
IrMOA$(56^4,  19,            14^6       7^2  4^7   2^4,  4)$,\\
IrMOA$(56^4,  21,       28^5            7^3  4^12^{12},  4)$,\\
IrMOA$(56^4,  22,       28^4 14^1       7^3  4^12^{13},  4)$,\\
IrMOA$(56^4,  23,       28^3 14^2       7^3  4^12^{14},  4)$,\\
IrMOA$(56^4,  24,       28^2 14^3       7^3  4^12^{15},  4)$,\\
IrMOA$(56^4,  25,       28^1 14^4       7^3  4^12^{16},  4)$,\\
IrMOA$(56^4,  26,            14^5       7^3  4^12^{17},  4)$,\\
IrMOA$(56^4,  21,       28^4 14^1       7^3  4^22^{11},  4)$,\\
IrMOA$(56^4,  22,       28^3 14^2       7^3  4^22^{12},  4)$,\\
IrMOA$(56^4,  23,       28^2 14^3       7^3  4^22^{13},  4)$,\\
IrMOA$(56^4,  24,       28^1 14^4       7^3  4^22^{14},  4)$,\\
IrMOA$(56^4,  25,            14^5       7^3  4^22^{15},  4)$,\\
IrMOA$(56^4,  21,       28^3 14^2       7^3  4^32^{10},  4)$,\\
IrMOA$(56^4,  22,       28^2 14^3       7^3  4^32^{11},  4)$,\\
IrMOA$(56^4,  23,       28^1 14^4       7^3  4^32^{12},  4)$,\\
IrMOA$(56^4,  24,            14^5       7^3  4^32^{13},  4)$,\\
IrMOA$(56^4,  21,       28^2 14^3       7^3  4^4   2^9,  4)$,\\
IrMOA$(56^4,  22,       28^1 14^4       7^3  4^42^{10},  4)$,\\
IrMOA$(56^4,  23,            14^5       7^3  4^42^{11},  4)$,\\
IrMOA$(56^4,  21,       28^1 14^4       7^3  4^5   2^8,  4)$,\\
IrMOA$(56^4,  22,            14^5       7^3  4^5   2^9,  4)$,\\
IrMOA$(56^4,  21,            14^5       7^3  4^6   2^7,  4)$,\\
IrMOA$(56^4,  23,       28^4            7^4  4^12^{14},  4)$,\\
IrMOA$(56^4,  24,       28^3 14^1       7^4  4^12^{15},  4)$,\\
IrMOA$(56^4,  25,       28^2 14^2       7^4  4^12^{16},  4)$,\\
IrMOA$(56^4,  26,       28^1 14^3       7^4  4^12^{17},  4)$,\\
IrMOA$(56^4,  27,            14^4       7^4  4^12^{18},  4)$,\\
IrMOA$(56^4,  23,       28^3 14^1       7^4  4^22^{13},  4)$,\\
IrMOA$(56^4,  24,       28^2 14^2       7^4  4^22^{14},  4)$,\\
IrMOA$(56^4,  25,       28^1 14^3       7^4  4^22^{15},  4)$,\\
IrMOA$(56^4,  26,            14^4       7^4  4^22^{16},  4)$,\\
IrMOA$(56^4,  23,       28^2 14^2       7^4  4^32^{12},  4)$,\\
IrMOA$(56^4,  24,       28^1 14^3       7^4  4^32^{13},  4)$,\\
IrMOA$(56^4,  25,            14^4       7^4  4^32^{14},  4)$,\\
IrMOA$(56^4,  23,       28^1 14^3       7^4  4^42^{11},  4)$,\\
IrMOA$(56^4,  24,            14^4       7^4  4^42^{12},  4)$,\\
IrMOA$(56^4,  23,            14^4       7^4  4^52^{10},  4)$,\\
IrMOA$(56^4,  25,       28^3            7^5  4^12^{16},  4)$,\\
IrMOA$(56^4,  26,       28^2 14^1       7^5  4^12^{17},  4)$,\\
IrMOA$(56^4,  27,       28^1 14^2       7^5  4^12^{18},  4)$,\\
IrMOA$(56^4,  28,            14^3       7^5  4^12^{19},  4)$,\\
IrMOA$(56^4,  25,       28^2 14^1       7^5  4^22^{15},  4)$,\\
IrMOA$(56^4,  26,       28^1 14^2       7^5  4^22^{16},  4)$,\\
IrMOA$(56^4,  27,            14^3       7^5  4^22^{17},  4)$,\\
IrMOA$(56^4,  25,       28^1 14^2       7^5  4^32^{14},  4)$,\\
IrMOA$(56^4,  26,            14^3       7^5  4^32^{15},  4)$,\\
IrMOA$(56^4,  25,            14^3       7^5  4^42^{13},  4)$,\\
IrMOA$(56^4,  27,       28^2            7^6  4^12^{18},  4)$,\\
IrMOA$(56^4,  28,       28^1 14^1       7^6  4^12^{19},  4)$,\\
IrMOA$(56^4,  29,            14^2       7^6  4^12^{20},  4)$,\\
IrMOA$(56^4,  27,       28^1 14^1       7^6  4^22^{17},  4)$,\\
IrMOA$(56^4,  28,            14^2       7^6  4^22^{18},  4)$,\\
IrMOA$(56^4,  27,            14^2       7^6  4^32^{16},  4)$,\\
IrMOA$(56^4,  29,       28^1            7^7  4^12^{20},  4)$,\\
IrMOA$(56^4,  30,            14^1       7^7  4^12^{21},  4)$,\\
IrMOA$(56^4,  29,            14^1       7^7  4^22^{19},  4)$,\\
IrMOA$(56^4,  31,                       7^8  4^12^{22},  4)$,\\
IrMOA$(56^4,  18,       28^6            7^2  4^2   2^8,  4)$,\\
IrMOA$(56^4,  19,       28^5 14^1       7^2  4^2   2^9,  4)$,\\
IrMOA$(56^4,  20,       28^4 14^2       7^2  4^22^{10},  4)$,\\
IrMOA$(56^4,  21,       28^3 14^3       7^2  4^22^{11},  4)$,\\
IrMOA$(56^4,  22,       28^2 14^4       7^2  4^22^{12},  4)$,\\
IrMOA$(56^4,  23,       28^1 14^5       7^2  4^22^{13},  4)$,\\
IrMOA$(56^4,  24,            14^6       7^2  4^22^{14},  4)$,\\
IrMOA$(56^4,  18,       28^5 14^1       7^2  4^3   2^7,  4)$,\\
IrMOA$(56^4,  19,       28^4 14^2       7^2  4^3   2^8,  4)$,\\
IrMOA$(56^4,  20,       28^3 14^3       7^2  4^3   2^9,  4)$,\\
IrMOA$(56^4,  21,       28^2 14^4       7^2  4^32^{10},  4)$,\\
IrMOA$(56^4,  22,       28^1 14^5       7^2  4^32^{11},  4)$,\\
IrMOA$(56^4,  23,            14^6       7^2  4^32^{12},  4)$,\\
IrMOA$(56^4,  18,       28^4 14^2       7^2  4^4   2^6,  4)$,\\
IrMOA$(56^4,  19,       28^3 14^3       7^2  4^4   2^7,  4)$,\\
IrMOA$(56^4,  20,       28^2 14^4       7^2  4^4   2^8,  4)$,\\
IrMOA$(56^4,  21,       28^1 14^5       7^2  4^4   2^9,  4)$,\\
IrMOA$(56^4,  22,            14^6       7^2  4^42^{10},  4)$,\\
IrMOA$(56^4,  18,       28^3 14^3       7^2  4^5   2^5,  4)$,\\
IrMOA$(56^4,  19,       28^2 14^4       7^2  4^5   2^6,  4)$,\\
IrMOA$(56^4,  20,       28^1 14^5       7^2  4^5   2^7,  4)$,\\
IrMOA$(56^4,  21,            14^6       7^2  4^5   2^8,  4)$,\\
IrMOA$(56^4,  18,       28^2 14^4       7^2  4^6   2^4,  4)$,\\
IrMOA$(56^4,  19,       28^1 14^5       7^2  4^6   2^5,  4)$,\\
IrMOA$(56^4,  20,            14^6       7^2  4^6   2^6,  4)$,\\
IrMOA$(56^4,  18,       28^1 14^5       7^2  4^7   2^3,  4)$,\\
IrMOA$(56^4,  19,            14^6       7^2  4^7   2^4,  4)$,\\
IrMOA$(56^4,  18,            14^6       7^2  4^8   2^2,  4)$,\\
IrMOA$(56^4,  20,       28^5            7^3  4^22^{10},  4)$,\\
IrMOA$(56^4,  21,       28^4 14^1       7^3  4^22^{11},  4)$,\\
IrMOA$(56^4,  22,       28^3 14^2       7^3  4^22^{12},  4)$,\\
IrMOA$(56^4,  23,       28^2 14^3       7^3  4^22^{13},  4)$,\\
IrMOA$(56^4,  24,       28^1 14^4       7^3  4^22^{14},  4)$,\\
IrMOA$(56^4,  25,            14^5       7^3  4^22^{15},  4)$,\\
IrMOA$(56^4,  20,       28^4 14^1       7^3  4^3   2^9,  4)$,\\
IrMOA$(56^4,  21,       28^3 14^2       7^3  4^32^{10},  4)$,\\
IrMOA$(56^4,  22,       28^2 14^3       7^3  4^32^{11},  4)$,\\
IrMOA$(56^4,  23,       28^1 14^4       7^3  4^32^{12},  4)$,\\
IrMOA$(56^4,  24,            14^5       7^3  4^32^{13},  4)$,\\
IrMOA$(56^4,  20,       28^3 14^2       7^3  4^4   2^8,  4)$,\\
IrMOA$(56^4,  21,       28^2 14^3       7^3  4^4   2^9,  4)$,\\
IrMOA$(56^4,  22,       28^1 14^4       7^3  4^42^{10},  4)$,\\
IrMOA$(56^4,  23,            14^5       7^3  4^42^{11},  4)$,\\
IrMOA$(56^4,  20,       28^2 14^3       7^3  4^5   2^7,  4)$,\\
IrMOA$(56^4,  21,       28^1 14^4       7^3  4^5   2^8,  4)$,\\
IrMOA$(56^4,  22,            14^5       7^3  4^5   2^9,  4)$,\\
IrMOA$(56^4,  20,       28^1 14^4       7^3  4^6   2^6,  4)$,\\
IrMOA$(56^4,  21,            14^5       7^3  4^6   2^7,  4)$,\\
IrMOA$(56^4,  20,            14^5       7^3  4^7   2^5,  4)$,\\
IrMOA$(56^4,  22,       28^4            7^4  4^22^{12},  4)$,\\
IrMOA$(56^4,  23,       28^3 14^1       7^4  4^22^{13},  4)$,\\
IrMOA$(56^4,  24,       28^2 14^2       7^4  4^22^{14},  4)$,\\
IrMOA$(56^4,  25,       28^1 14^3       7^4  4^22^{15},  4)$,\\
IrMOA$(56^4,  26,            14^4       7^4  4^22^{16},  4)$,\\
IrMOA$(56^4,  22,       28^3 14^1       7^4  4^32^{11},  4)$,\\
IrMOA$(56^4,  23,       28^2 14^2       7^4  4^32^{12},  4)$,\\
IrMOA$(56^4,  24,       28^1 14^3       7^4  4^32^{13},  4)$,\\
IrMOA$(56^4,  25,            14^4       7^4  4^32^{14},  4)$,\\
IrMOA$(56^4,  22,       28^2 14^2       7^4  4^42^{10},  4)$,\\
IrMOA$(56^4,  23,       28^1 14^3       7^4  4^42^{11},  4)$,\\
IrMOA$(56^4,  24,            14^4       7^4  4^42^{12},  4)$,\\
IrMOA$(56^4,  22,       28^1 14^3       7^4  4^5   2^9,  4)$,\\
IrMOA$(56^4,  23,            14^4       7^4  4^52^{10},  4)$,\\
IrMOA$(56^4,  22,            14^4       7^4  4^6   2^8,  4)$,\\
IrMOA$(56^4,  24,       28^3            7^5  4^22^{14},  4)$,\\
IrMOA$(56^4,  25,       28^2 14^1       7^5  4^22^{15},  4)$,\\
IrMOA$(56^4,  26,       28^1 14^2       7^5  4^22^{16},  4)$,\\
IrMOA$(56^4,  27,            14^3       7^5  4^22^{17},  4)$,\\
IrMOA$(56^4,  24,       28^2 14^1       7^5  4^32^{13},  4)$,\\
IrMOA$(56^4,  25,       28^1 14^2       7^5  4^32^{14},  4)$,\\
IrMOA$(56^4,  26,            14^3       7^5  4^32^{15},  4)$,\\
IrMOA$(56^4,  24,       28^1 14^2       7^5  4^42^{12},  4)$,\\
IrMOA$(56^4,  25,            14^3       7^5  4^42^{13},  4)$,\\
IrMOA$(56^4,  24,            14^3       7^5  4^52^{11},  4)$,\\
IrMOA$(56^4,  26,       28^2            7^6  4^22^{16},  4)$,\\
IrMOA$(56^4,  27,       28^1 14^1       7^6  4^22^{17},  4)$,\\
IrMOA$(56^4,  28,            14^2       7^6  4^22^{18},  4)$,\\
IrMOA$(56^4,  26,       28^1 14^1       7^6  4^32^{15},  4)$,\\
IrMOA$(56^4,  27,            14^2       7^6  4^32^{16},  4)$,\\
IrMOA$(56^4,  26,            14^2       7^6  4^42^{14},  4)$,\\
IrMOA$(56^4,  28,       28^1            7^7  4^22^{18},  4)$,\\
IrMOA$(56^4,  29,            14^1       7^7  4^22^{19},  4)$,\\
IrMOA$(56^4,  28,            14^1       7^7  4^32^{17},  4)$,\\
IrMOA$(56^4,  30,                       7^8  4^22^{20},  4)$,\\
IrMOA$(56^4,  19,       28^5            7^3  4^3   2^8,  4)$,\\
IrMOA$(56^4,  20,       28^4 14^1       7^3  4^3   2^9,  4)$,\\
IrMOA$(56^4,  21,       28^3 14^2       7^3  4^32^{10},  4)$,\\
IrMOA$(56^4,  22,       28^2 14^3       7^3  4^32^{11},  4)$,\\
IrMOA$(56^4,  23,       28^1 14^4       7^3  4^32^{12},  4)$,\\
IrMOA$(56^4,  24,            14^5       7^3  4^32^{13},  4)$,\\
IrMOA$(56^4,  19,       28^4 14^1       7^3  4^4   2^7,  4)$,\\
IrMOA$(56^4,  20,       28^3 14^2       7^3  4^4   2^8,  4)$,\\
IrMOA$(56^4,  21,       28^2 14^3       7^3  4^4   2^9,  4)$,\\
IrMOA$(56^4,  22,       28^1 14^4       7^3  4^42^{10},  4)$,\\
IrMOA$(56^4,  23,            14^5       7^3  4^42^{11},  4)$,\\
IrMOA$(56^4,  19,       28^3 14^2       7^3  4^5   2^6,  4)$,\\
IrMOA$(56^4,  20,       28^2 14^3       7^3  4^5   2^7,  4)$,\\
IrMOA$(56^4,  21,       28^1 14^4       7^3  4^5   2^8,  4)$,\\
IrMOA$(56^4,  22,            14^5       7^3  4^5   2^9,  4)$,\\
IrMOA$(56^4,  19,       28^2 14^3       7^3  4^6   2^5,  4)$,\\
IrMOA$(56^4,  20,       28^1 14^4       7^3  4^6   2^6,  4)$,\\
IrMOA$(56^4,  21,            14^5       7^3  4^6   2^7,  4)$,\\
IrMOA$(56^4,  19,       28^1 14^4       7^3  4^7   2^4,  4)$,\\
IrMOA$(56^4,  20,            14^5       7^3  4^7   2^5,  4)$,\\
IrMOA$(56^4,  19,            14^5       7^3  4^8   2^3,  4)$,\\
IrMOA$(56^4,  21,       28^4            7^4  4^32^{10},  4)$,\\
IrMOA$(56^4,  22,       28^3 14^1       7^4  4^32^{11},  4)$,\\
IrMOA$(56^4,  23,       28^2 14^2       7^4  4^32^{12},  4)$,\\
IrMOA$(56^4,  24,       28^1 14^3       7^4  4^32^{13},  4)$,\\
IrMOA$(56^4,  25,            14^4       7^4  4^32^{14},  4)$,\\
IrMOA$(56^4,  21,       28^3 14^1       7^4  4^4   2^9,  4)$,\\
IrMOA$(56^4,  22,       28^2 14^2       7^4  4^42^{10},  4)$,\\
IrMOA$(56^4,  23,       28^1 14^3       7^4  4^42^{11},  4)$,\\
IrMOA$(56^4,  24,            14^4       7^4  4^42^{12},  4)$,\\
IrMOA$(56^4,  21,       28^2 14^2       7^4  4^5   2^8,  4)$,\\
IrMOA$(56^4,  22,       28^1 14^3       7^4  4^5   2^9,  4)$,\\
IrMOA$(56^4,  23,            14^4       7^4  4^52^{10},  4)$,\\
IrMOA$(56^4,  21,       28^1 14^3       7^4  4^6   2^7,  4)$,\\
IrMOA$(56^4,  22,            14^4       7^4  4^6   2^8,  4)$,\\
IrMOA$(56^4,  21,            14^4       7^4  4^7   2^6,  4)$,\\
IrMOA$(56^4,  23,       28^3            7^5  4^32^{12},  4)$,\\
IrMOA$(56^4,  24,       28^2 14^1       7^5  4^32^{13},  4)$,\\
IrMOA$(56^4,  25,       28^1 14^2       7^5  4^32^{14},  4)$,\\
IrMOA$(56^4,  26,            14^3       7^5  4^32^{15},  4)$,\\
IrMOA$(56^4,  23,       28^2 14^1       7^5  4^42^{11},  4)$,\\
IrMOA$(56^4,  24,       28^1 14^2       7^5  4^42^{12},  4)$,\\
IrMOA$(56^4,  25,            14^3       7^5  4^42^{13},  4)$,\\
IrMOA$(56^4,  23,       28^1 14^2       7^5  4^52^{10},  4)$,\\
IrMOA$(56^4,  24,            14^3       7^5  4^52^{11},  4)$,\\
IrMOA$(56^4,  23,            14^3       7^5  4^6   2^9,  4)$,\\
IrMOA$(56^4,  25,       28^2            7^6  4^32^{14},  4)$,\\
IrMOA$(56^4,  26,       28^1 14^1       7^6  4^32^{15},  4)$,\\
IrMOA$(56^4,  27,            14^2       7^6  4^32^{16},  4)$,\\
IrMOA$(56^4,  25,       28^1 14^1       7^6  4^42^{13},  4)$,\\
IrMOA$(56^4,  26,            14^2       7^6  4^42^{14},  4)$,\\
IrMOA$(56^4,  25,            14^2       7^6  4^52^{12},  4)$,\\
IrMOA$(56^4,  27,       28^1            7^7  4^32^{16},  4)$,\\
IrMOA$(56^4,  28,            14^1       7^7  4^32^{17},  4)$,\\
IrMOA$(56^4,  27,            14^1       7^7  4^42^{15},  4)$,\\
IrMOA$(56^4,  29,                       7^8  4^32^{18},  4)$,\\
IrMOA$(56^4,  20,       28^4            7^4  4^4   2^8,  4)$,\\
IrMOA$(56^4,  21,       28^3 14^1       7^4  4^4   2^9,  4)$,\\
IrMOA$(56^4,  22,       28^2 14^2       7^4  4^42^{10},  4)$,\\
IrMOA$(56^4,  23,       28^1 14^3       7^4  4^42^{11},  4)$,\\
IrMOA$(56^4,  24,            14^4       7^4  4^42^{12},  4)$,\\
IrMOA$(56^4,  20,       28^3 14^1       7^4  4^5   2^7,  4)$,\\
IrMOA$(56^4,  21,       28^2 14^2       7^4  4^5   2^8,  4)$,\\
IrMOA$(56^4,  22,       28^1 14^3       7^4  4^5   2^9,  4)$,\\
IrMOA$(56^4,  23,            14^4       7^4  4^52^{10},  4)$,\\
IrMOA$(56^4,  20,       28^2 14^2       7^4  4^6   2^6,  4)$,\\
IrMOA$(56^4,  21,       28^1 14^3       7^4  4^6   2^7,  4)$,\\
IrMOA$(56^4,  22,            14^4       7^4  4^6   2^8,  4)$,\\
IrMOA$(56^4,  20,       28^1 14^3       7^4  4^7   2^5,  4)$,\\
IrMOA$(56^4,  21,            14^4       7^4  4^7   2^6,  4)$,\\
IrMOA$(56^4,  20,            14^4       7^4  4^8   2^4,  4)$,\\
IrMOA$(56^4,  22,       28^3            7^5  4^42^{10},  4)$,\\
IrMOA$(56^4,  23,       28^2 14^1       7^5  4^42^{11},  4)$,\\
IrMOA$(56^4,  24,       28^1 14^2       7^5  4^42^{12},  4)$,\\
IrMOA$(56^4,  25,            14^3       7^5  4^42^{13},  4)$,\\
IrMOA$(56^4,  22,       28^2 14^1       7^5  4^5   2^9,  4)$,\\
IrMOA$(56^4,  23,       28^1 14^2       7^5  4^52^{10},  4)$,\\
IrMOA$(56^4,  24,            14^3       7^5  4^52^{11},  4)$,\\
IrMOA$(56^4,  22,       28^1 14^2       7^5  4^6   2^8,  4)$,\\
IrMOA$(56^4,  23,            14^3       7^5  4^6   2^9,  4)$,\\
IrMOA$(56^4,  22,            14^3       7^5  4^7   2^7,  4)$,\\
IrMOA$(56^4,  24,       28^2            7^6  4^42^{12},  4)$,\\
IrMOA$(56^4,  25,       28^1 14^1       7^6  4^42^{13},  4)$,\\
IrMOA$(56^4,  26,            14^2       7^6  4^42^{14},  4)$,\\
IrMOA$(56^4,  24,       28^1 14^1       7^6  4^52^{11},  4)$,\\
IrMOA$(56^4,  25,            14^2       7^6  4^52^{12},  4)$,\\
IrMOA$(56^4,  24,            14^2       7^6  4^62^{10},  4)$,\\
IrMOA$(56^4,  26,       28^1            7^7  4^42^{14},  4)$,\\
IrMOA$(56^4,  27,            14^1       7^7  4^42^{15},  4)$,\\
IrMOA$(56^4,  26,            14^1       7^7  4^52^{13},  4)$,\\
IrMOA$(56^4,  28,                       7^8  4^42^{16},  4)$,\\
IrMOA$(56^4,  21,       28^3            7^5  4^5   2^8,  4)$,\\
IrMOA$(56^4,  22,       28^2 14^1       7^5  4^5   2^9,  4)$,\\
IrMOA$(56^4,  23,       28^1 14^2       7^5  4^52^{10},  4)$,\\
IrMOA$(56^4,  24,            14^3       7^5  4^52^{11},  4)$,\\
IrMOA$(56^4,  21,       28^2 14^1       7^5  4^6   2^7,  4)$,\\
IrMOA$(56^4,  22,       28^1 14^2       7^5  4^6   2^8,  4)$,\\
IrMOA$(56^4,  23,            14^3       7^5  4^6   2^9,  4)$,\\
IrMOA$(56^4,  21,       28^1 14^2       7^5  4^7   2^6,  4)$,\\
IrMOA$(56^4,  22,            14^3       7^5  4^7   2^7,  4)$,\\
IrMOA$(56^4,  21,            14^3       7^5  4^8   2^5,  4)$,\\
IrMOA$(56^4,  23,       28^2            7^6  4^52^{10},  4)$,\\
IrMOA$(56^4,  24,       28^1 14^1       7^6  4^52^{11},  4)$,\\
IrMOA$(56^4,  25,            14^2       7^6  4^52^{12},  4)$,\\
IrMOA$(56^4,  23,       28^1 14^1       7^6  4^6   2^9,  4)$,\\
IrMOA$(56^4,  24,            14^2       7^6  4^62^{10},  4)$,\\
IrMOA$(56^4,  23,            14^2       7^6  4^7   2^8,  4)$,\\
IrMOA$(56^4,  25,       28^1            7^7  4^52^{12},  4)$,\\
IrMOA$(56^4,  26,            14^1       7^7  4^52^{13},  4)$,\\
IrMOA$(56^4,  25,            14^1       7^7  4^62^{11},  4)$,\\
IrMOA$(56^4,  27,                       7^8  4^52^{14},  4)$,\\
IrMOA$(56^4,  22,       28^2            7^6  4^6   2^8,  4)$,\\
IrMOA$(56^4,  23,       28^1 14^1       7^6  4^6   2^9,  4)$,\\
IrMOA$(56^4,  24,            14^2       7^6  4^62^{10},  4)$,\\
IrMOA$(56^4,  22,       28^1 14^1       7^6  4^7   2^7,  4)$,\\
IrMOA$(56^4,  23,            14^2       7^6  4^7   2^8,  4)$,\\
IrMOA$(56^4,  22,            14^2       7^6  4^8   2^6,  4)$,\\
IrMOA$(56^4,  24,       28^1            7^7  4^62^{10},  4)$,\\
IrMOA$(56^4,  25,            14^1       7^7  4^62^{11},  4)$,\\
IrMOA$(56^4,  24,            14^1       7^7  4^7   2^9,  4)$,\\
IrMOA$(56^4,  26,                       7^8  4^62^{12},  4)$,\\
IrMOA$(56^4,  23,       28^1            7^7  4^7   2^8,  4)$,\\
IrMOA$(56^4,  24,            14^1       7^7  4^7   2^9,  4)$,\\
IrMOA$(56^4,  23,            14^1       7^7  4^8   2^7,  4)$,\\
IrMOA$(56^4,  25,                       7^8  4^72^{10},  4)$,\\
IrMOA$(56^4,  24,                       7^8  4^8   2^8,  4)$,\\
IrMOA$(56^4,  16,       28^7       8^1  7^1        2^7,  4)$,\\
IrMOA$(56^4,  17,       28^6 14^1  8^1  7^1        2^8,  4)$,\\
IrMOA$(56^4,  18,       28^5 14^2  8^1  7^1        2^9,  4)$,\\
IrMOA$(56^4,  19,       28^4 14^3  8^1  7^1     2^{10},  4)$,\\
IrMOA$(56^4,  20,       28^3 14^4  8^1  7^1     2^{11},  4)$,\\
IrMOA$(56^4,  21,       28^2 14^5  8^1  7^1     2^{12},  4)$,\\
IrMOA$(56^4,  22,       28^1 14^6  8^1  7^1     2^{13},  4)$,\\
IrMOA$(56^4,  23,            14^7  8^1  7^1     2^{14},  4)$,\\
IrMOA$(56^4,  16,       28^6 14^1  8^1  7^1  4^1   2^6,  4)$,\\
IrMOA$(56^4,  17,       28^5 14^2  8^1  7^1  4^1   2^7,  4)$,\\
IrMOA$(56^4,  18,       28^4 14^3  8^1  7^1  4^1   2^8,  4)$,\\
IrMOA$(56^4,  19,       28^3 14^4  8^1  7^1  4^1   2^9,  4)$,\\
IrMOA$(56^4,  20,       28^2 14^5  8^1  7^1  4^12^{10},  4)$,\\
IrMOA$(56^4,  21,       28^1 14^6  8^1  7^1  4^12^{11},  4)$,\\
IrMOA$(56^4,  22,            14^7  8^1  7^1  4^12^{12},  4)$,\\
IrMOA$(56^4,  16,       28^5 14^2  8^1  7^1  4^2   2^5,  4)$,\\
IrMOA$(56^4,  17,       28^4 14^3  8^1  7^1  4^2   2^6,  4)$,\\
IrMOA$(56^4,  18,       28^3 14^4  8^1  7^1  4^2   2^7,  4)$,\\
IrMOA$(56^4,  19,       28^2 14^5  8^1  7^1  4^2   2^8,  4)$,\\
IrMOA$(56^4,  20,       28^1 14^6  8^1  7^1  4^2   2^9,  4)$,\\
IrMOA$(56^4,  21,            14^7  8^1  7^1  4^22^{10},  4)$,\\
IrMOA$(56^4,  16,       28^4 14^3  8^1  7^1  4^3   2^4,  4)$,\\
IrMOA$(56^4,  17,       28^3 14^4  8^1  7^1  4^3   2^5,  4)$,\\
IrMOA$(56^4,  18,       28^2 14^5  8^1  7^1  4^3   2^6,  4)$,\\
IrMOA$(56^4,  19,       28^1 14^6  8^1  7^1  4^3   2^7,  4)$,\\
IrMOA$(56^4,  20,            14^7  8^1  7^1  4^3   2^8,  4)$,\\
IrMOA$(56^4,  16,       28^3 14^4  8^1  7^1  4^4   2^3,  4)$,\\
IrMOA$(56^4,  17,       28^2 14^5  8^1  7^1  4^4   2^4,  4)$,\\
IrMOA$(56^4,  18,       28^1 14^6  8^1  7^1  4^4   2^5,  4)$,\\
IrMOA$(56^4,  19,            14^7  8^1  7^1  4^4   2^6,  4)$,\\
IrMOA$(56^4,  16,       28^2 14^5  8^1  7^1  4^5   2^2,  4)$,\\
IrMOA$(56^4,  17,       28^1 14^6  8^1  7^1  4^5   2^3,  4)$,\\
IrMOA$(56^4,  18,            14^7  8^1  7^1  4^5   2^4,  4)$,\\
IrMOA$(56^4,  16,       28^1 14^6  8^1  7^1  4^6   2^1,  4)$,\\
IrMOA$(56^4,  17,            14^7  8^1  7^1  4^6   2^2,  4)$,\\
IrMOA$(56^4,  16,            14^7  8^1  7^1  4^7      ,  4)$,\\
IrMOA$(56^4,  18,       28^6       8^1  7^2        2^9,  4)$,\\
IrMOA$(56^4,  19,       28^5 14^1  8^1  7^2     2^{10},  4)$,\\
IrMOA$(56^4,  20,       28^4 14^2  8^1  7^2     2^{11},  4)$,\\
IrMOA$(56^4,  21,       28^3 14^3  8^1  7^2     2^{12},  4)$,\\
IrMOA$(56^4,  22,       28^2 14^4  8^1  7^2     2^{13},  4)$,\\
IrMOA$(56^4,  23,       28^1 14^5  8^1  7^2     2^{14},  4)$,\\
IrMOA$(56^4,  24,            14^6  8^1  7^2     2^{15},  4)$,\\
IrMOA$(56^4,  18,       28^5 14^1  8^1  7^2  4^1   2^8,  4)$,\\
IrMOA$(56^4,  19,       28^4 14^2  8^1  7^2  4^1   2^9,  4)$,\\
IrMOA$(56^4,  20,       28^3 14^3  8^1  7^2  4^12^{10},  4)$,\\
IrMOA$(56^4,  21,       28^2 14^4  8^1  7^2  4^12^{11},  4)$,\\
IrMOA$(56^4,  22,       28^1 14^5  8^1  7^2  4^12^{12},  4)$,\\
IrMOA$(56^4,  23,            14^6  8^1  7^2  4^12^{13},  4)$,\\
IrMOA$(56^4,  18,       28^4 14^2  8^1  7^2  4^2   2^7,  4)$,\\
IrMOA$(56^4,  19,       28^3 14^3  8^1  7^2  4^2   2^8,  4)$,\\
IrMOA$(56^4,  20,       28^2 14^4  8^1  7^2  4^2   2^9,  4)$,\\
IrMOA$(56^4,  21,       28^1 14^5  8^1  7^2  4^22^{10},  4)$,\\
IrMOA$(56^4,  22,            14^6  8^1  7^2  4^22^{11},  4)$,\\
IrMOA$(56^4,  18,       28^3 14^3  8^1  7^2  4^3   2^6,  4)$,\\
IrMOA$(56^4,  19,       28^2 14^4  8^1  7^2  4^3   2^7,  4)$,\\
IrMOA$(56^4,  20,       28^1 14^5  8^1  7^2  4^3   2^8,  4)$,\\
IrMOA$(56^4,  21,            14^6  8^1  7^2  4^3   2^9,  4)$,\\
IrMOA$(56^4,  18,       28^2 14^4  8^1  7^2  4^4   2^5,  4)$,\\
IrMOA$(56^4,  19,       28^1 14^5  8^1  7^2  4^4   2^6,  4)$,\\
IrMOA$(56^4,  20,            14^6  8^1  7^2  4^4   2^7,  4)$,\\
IrMOA$(56^4,  18,       28^1 14^5  8^1  7^2  4^5   2^4,  4)$,\\
IrMOA$(56^4,  19,            14^6  8^1  7^2  4^5   2^5,  4)$,\\
IrMOA$(56^4,  18,            14^6  8^1  7^2  4^6   2^3,  4)$,\\
IrMOA$(56^4,  20,       28^5       8^1  7^3     2^{11},  4)$,\\
IrMOA$(56^4,  21,       28^4 14^1  8^1  7^3     2^{12},  4)$,\\
IrMOA$(56^4,  22,       28^3 14^2  8^1  7^3     2^{13},  4)$,\\
IrMOA$(56^4,  23,       28^2 14^3  8^1  7^3     2^{14},  4)$,\\
IrMOA$(56^4,  24,       28^1 14^4  8^1  7^3     2^{15},  4)$,\\
IrMOA$(56^4,  25,            14^5  8^1  7^3     2^{16},  4)$,\\
IrMOA$(56^4,  20,       28^4 14^1  8^1  7^3  4^12^{10},  4)$,\\
IrMOA$(56^4,  21,       28^3 14^2  8^1  7^3  4^12^{11},  4)$,\\
IrMOA$(56^4,  22,       28^2 14^3  8^1  7^3  4^12^{12},  4)$,\\
IrMOA$(56^4,  23,       28^1 14^4  8^1  7^3  4^12^{13},  4)$,\\
IrMOA$(56^4,  24,            14^5  8^1  7^3  4^12^{14},  4)$,\\
IrMOA$(56^4,  20,       28^3 14^2  8^1  7^3  4^2   2^9,  4)$,\\
IrMOA$(56^4,  21,       28^2 14^3  8^1  7^3  4^22^{10},  4)$,\\
IrMOA$(56^4,  22,       28^1 14^4  8^1  7^3  4^22^{11},  4)$,\\
IrMOA$(56^4,  23,            14^5  8^1  7^3  4^22^{12},  4)$,\\
IrMOA$(56^4,  20,       28^2 14^3  8^1  7^3  4^3   2^8,  4)$,\\
IrMOA$(56^4,  21,       28^1 14^4  8^1  7^3  4^3   2^9,  4)$,\\
IrMOA$(56^4,  22,            14^5  8^1  7^3  4^32^{10},  4)$,\\
IrMOA$(56^4,  20,       28^1 14^4  8^1  7^3  4^4   2^7,  4)$,\\
IrMOA$(56^4,  21,            14^5  8^1  7^3  4^4   2^8,  4)$,\\
IrMOA$(56^4,  20,            14^5  8^1  7^3  4^5   2^6,  4)$,\\
IrMOA$(56^4,  22,       28^4       8^1  7^4     2^{13},  4)$,\\
IrMOA$(56^4,  23,       28^3 14^1  8^1  7^4     2^{14},  4)$,\\
IrMOA$(56^4,  24,       28^2 14^2  8^1  7^4     2^{15},  4)$,\\
IrMOA$(56^4,  25,       28^1 14^3  8^1  7^4     2^{16},  4)$,\\
IrMOA$(56^4,  26,            14^4  8^1  7^4     2^{17},  4)$,\\
IrMOA$(56^4,  22,       28^3 14^1  8^1  7^4  4^12^{12},  4)$,\\
IrMOA$(56^4,  23,       28^2 14^2  8^1  7^4  4^12^{13},  4)$,\\
IrMOA$(56^4,  24,       28^1 14^3  8^1  7^4  4^12^{14},  4)$,\\
IrMOA$(56^4,  25,            14^4  8^1  7^4  4^12^{15},  4)$,\\
IrMOA$(56^4,  22,       28^2 14^2  8^1  7^4  4^22^{11},  4)$,\\
IrMOA$(56^4,  23,       28^1 14^3  8^1  7^4  4^22^{12},  4)$,\\
IrMOA$(56^4,  24,            14^4  8^1  7^4  4^22^{13},  4)$,\\
IrMOA$(56^4,  22,       28^1 14^3  8^1  7^4  4^32^{10},  4)$,\\
IrMOA$(56^4,  23,            14^4  8^1  7^4  4^32^{11},  4)$,\\
IrMOA$(56^4,  22,            14^4  8^1  7^4  4^4   2^9,  4)$,\\
IrMOA$(56^4,  24,       28^3       8^1  7^5     2^{15},  4)$,\\
IrMOA$(56^4,  25,       28^2 14^1  8^1  7^5     2^{16},  4)$,\\
IrMOA$(56^4,  26,       28^1 14^2  8^1  7^5     2^{17},  4)$,\\
IrMOA$(56^4,  27,            14^3  8^1  7^5     2^{18},  4)$,\\
IrMOA$(56^4,  24,       28^2 14^1  8^1  7^5  4^12^{14},  4)$,\\
IrMOA$(56^4,  25,       28^1 14^2  8^1  7^5  4^12^{15},  4)$,\\
IrMOA$(56^4,  26,            14^3  8^1  7^5  4^12^{16},  4)$,\\
IrMOA$(56^4,  24,       28^1 14^2  8^1  7^5  4^22^{13},  4)$,\\
IrMOA$(56^4,  25,            14^3  8^1  7^5  4^22^{14},  4)$,\\
IrMOA$(56^4,  24,            14^3  8^1  7^5  4^32^{12},  4)$,\\
IrMOA$(56^4,  26,       28^2       8^1  7^6     2^{17},  4)$,\\
IrMOA$(56^4,  27,       28^1 14^1  8^1  7^6     2^{18},  4)$,\\
IrMOA$(56^4,  28,            14^2  8^1  7^6     2^{19},  4)$,\\
IrMOA$(56^4,  26,       28^1 14^1  8^1  7^6  4^12^{16},  4)$,\\
IrMOA$(56^4,  27,            14^2  8^1  7^6  4^12^{17},  4)$,\\
IrMOA$(56^4,  26,            14^2  8^1  7^6  4^22^{15},  4)$,\\
IrMOA$(56^4,  28,       28^1       8^1  7^7     2^{19},  4)$,\\
IrMOA$(56^4,  29,            14^1  8^1  7^7     2^{20},  4)$,\\
IrMOA$(56^4,  28,            14^1  8^1  7^7  4^12^{18},  4)$,\\
IrMOA$(56^4,  30,                  8^1  7^8     2^{21},  4)$,\\
IrMOA$(56^4,  17,       28^6       8^1  7^2  4^1   2^7,  4)$,\\
IrMOA$(56^4,  18,       28^5 14^1  8^1  7^2  4^1   2^8,  4)$,\\
IrMOA$(56^4,  19,       28^4 14^2  8^1  7^2  4^1   2^9,  4)$,\\
IrMOA$(56^4,  20,       28^3 14^3  8^1  7^2  4^12^{10},  4)$,\\
IrMOA$(56^4,  21,       28^2 14^4  8^1  7^2  4^12^{11},  4)$,\\
IrMOA$(56^4,  22,       28^1 14^5  8^1  7^2  4^12^{12},  4)$,\\
IrMOA$(56^4,  23,            14^6  8^1  7^2  4^12^{13},  4)$,\\
IrMOA$(56^4,  17,       28^5 14^1  8^1  7^2  4^2   2^6,  4)$,\\
IrMOA$(56^4,  18,       28^4 14^2  8^1  7^2  4^2   2^7,  4)$,\\
IrMOA$(56^4,  19,       28^3 14^3  8^1  7^2  4^2   2^8,  4)$,\\
IrMOA$(56^4,  20,       28^2 14^4  8^1  7^2  4^2   2^9,  4)$,\\
IrMOA$(56^4,  21,       28^1 14^5  8^1  7^2  4^22^{10},  4)$,\\
IrMOA$(56^4,  22,            14^6  8^1  7^2  4^22^{11},  4)$,\\
IrMOA$(56^4,  17,       28^4 14^2  8^1  7^2  4^3   2^5,  4)$,\\
IrMOA$(56^4,  18,       28^3 14^3  8^1  7^2  4^3   2^6,  4)$,\\
IrMOA$(56^4,  19,       28^2 14^4  8^1  7^2  4^3   2^7,  4)$,\\
IrMOA$(56^4,  20,       28^1 14^5  8^1  7^2  4^3   2^8,  4)$,\\
IrMOA$(56^4,  21,            14^6  8^1  7^2  4^3   2^9,  4)$,\\
IrMOA$(56^4,  17,       28^3 14^3  8^1  7^2  4^4   2^4,  4)$,\\
IrMOA$(56^4,  18,       28^2 14^4  8^1  7^2  4^4   2^5,  4)$,\\
IrMOA$(56^4,  19,       28^1 14^5  8^1  7^2  4^4   2^6,  4)$,\\
IrMOA$(56^4,  20,            14^6  8^1  7^2  4^4   2^7,  4)$,\\
IrMOA$(56^4,  17,       28^2 14^4  8^1  7^2  4^5   2^3,  4)$,\\
IrMOA$(56^4,  18,       28^1 14^5  8^1  7^2  4^5   2^4,  4)$,\\
IrMOA$(56^4,  19,            14^6  8^1  7^2  4^5   2^5,  4)$,\\
IrMOA$(56^4,  17,       28^1 14^5  8^1  7^2  4^6   2^2,  4)$,\\
IrMOA$(56^4,  18,            14^6  8^1  7^2  4^6   2^3,  4)$,\\
IrMOA$(56^4,  17,            14^6  8^1  7^2  4^7   2^1,  4)$,\\
IrMOA$(56^4,  19,       28^5       8^1  7^3  4^1   2^9,  4)$,\\
IrMOA$(56^4,  20,       28^4 14^1  8^1  7^3  4^12^{10},  4)$,\\
IrMOA$(56^4,  21,       28^3 14^2  8^1  7^3  4^12^{11},  4)$,\\
IrMOA$(56^4,  22,       28^2 14^3  8^1  7^3  4^12^{12},  4)$,\\
IrMOA$(56^4,  23,       28^1 14^4  8^1  7^3  4^12^{13},  4)$,\\
IrMOA$(56^4,  24,            14^5  8^1  7^3  4^12^{14},  4)$,\\
IrMOA$(56^4,  19,       28^4 14^1  8^1  7^3  4^2   2^8,  4)$,\\
IrMOA$(56^4,  20,       28^3 14^2  8^1  7^3  4^2   2^9,  4)$,\\
IrMOA$(56^4,  21,       28^2 14^3  8^1  7^3  4^22^{10},  4)$,\\
IrMOA$(56^4,  22,       28^1 14^4  8^1  7^3  4^22^{11},  4)$,\\
IrMOA$(56^4,  23,            14^5  8^1  7^3  4^22^{12},  4)$,\\
IrMOA$(56^4,  19,       28^3 14^2  8^1  7^3  4^3   2^7,  4)$,\\
IrMOA$(56^4,  20,       28^2 14^3  8^1  7^3  4^3   2^8,  4)$,\\
IrMOA$(56^4,  21,       28^1 14^4  8^1  7^3  4^3   2^9,  4)$,\\
IrMOA$(56^4,  22,            14^5  8^1  7^3  4^32^{10},  4)$,\\
IrMOA$(56^4,  19,       28^2 14^3  8^1  7^3  4^4   2^6,  4)$,\\
IrMOA$(56^4,  20,       28^1 14^4  8^1  7^3  4^4   2^7,  4)$,\\
IrMOA$(56^4,  21,            14^5  8^1  7^3  4^4   2^8,  4)$,\\
IrMOA$(56^4,  19,       28^1 14^4  8^1  7^3  4^5   2^5,  4)$,\\
IrMOA$(56^4,  20,            14^5  8^1  7^3  4^5   2^6,  4)$,\\
IrMOA$(56^4,  19,            14^5  8^1  7^3  4^6   2^4,  4)$,\\
IrMOA$(56^4,  21,       28^4       8^1  7^4  4^12^{11},  4)$,\\
IrMOA$(56^4,  22,       28^3 14^1  8^1  7^4  4^12^{12},  4)$,\\
IrMOA$(56^4,  23,       28^2 14^2  8^1  7^4  4^12^{13},  4)$,\\
IrMOA$(56^4,  24,       28^1 14^3  8^1  7^4  4^12^{14},  4)$,\\
IrMOA$(56^4,  25,            14^4  8^1  7^4  4^12^{15},  4)$,\\
IrMOA$(56^4,  21,       28^3 14^1  8^1  7^4  4^22^{10},  4)$,\\
IrMOA$(56^4,  22,       28^2 14^2  8^1  7^4  4^22^{11},  4)$,\\
IrMOA$(56^4,  23,       28^1 14^3  8^1  7^4  4^22^{12},  4)$,\\
IrMOA$(56^4,  24,            14^4  8^1  7^4  4^22^{13},  4)$,\\
IrMOA$(56^4,  21,       28^2 14^2  8^1  7^4  4^3   2^9,  4)$,\\
IrMOA$(56^4,  22,       28^1 14^3  8^1  7^4  4^32^{10},  4)$,\\
IrMOA$(56^4,  23,            14^4  8^1  7^4  4^32^{11},  4)$,\\
IrMOA$(56^4,  21,       28^1 14^3  8^1  7^4  4^4   2^8,  4)$,\\
IrMOA$(56^4,  22,            14^4  8^1  7^4  4^4   2^9,  4)$,\\
IrMOA$(56^4,  21,            14^4  8^1  7^4  4^5   2^7,  4)$,\\
IrMOA$(56^4,  23,       28^3       8^1  7^5  4^12^{13},  4)$,\\
IrMOA$(56^4,  24,       28^2 14^1  8^1  7^5  4^12^{14},  4)$,\\
IrMOA$(56^4,  25,       28^1 14^2  8^1  7^5  4^12^{15},  4)$,\\
IrMOA$(56^4,  26,            14^3  8^1  7^5  4^12^{16},  4)$,\\
IrMOA$(56^4,  23,       28^2 14^1  8^1  7^5  4^22^{12},  4)$,\\
IrMOA$(56^4,  24,       28^1 14^2  8^1  7^5  4^22^{13},  4)$,\\
IrMOA$(56^4,  25,            14^3  8^1  7^5  4^22^{14},  4)$,\\
IrMOA$(56^4,  23,       28^1 14^2  8^1  7^5  4^32^{11},  4)$,\\
IrMOA$(56^4,  24,            14^3  8^1  7^5  4^32^{12},  4)$,\\
IrMOA$(56^4,  23,            14^3  8^1  7^5  4^42^{10},  4)$,\\
IrMOA$(56^4,  25,       28^2       8^1  7^6  4^12^{15},  4)$,\\
IrMOA$(56^4,  26,       28^1 14^1  8^1  7^6  4^12^{16},  4)$,\\
IrMOA$(56^4,  27,            14^2  8^1  7^6  4^12^{17},  4)$,\\
IrMOA$(56^4,  25,       28^1 14^1  8^1  7^6  4^22^{14},  4)$,\\
IrMOA$(56^4,  26,            14^2  8^1  7^6  4^22^{15},  4)$,\\
IrMOA$(56^4,  25,            14^2  8^1  7^6  4^32^{13},  4)$,\\
IrMOA$(56^4,  27,       28^1       8^1  7^7  4^12^{17},  4)$,\\
IrMOA$(56^4,  28,            14^1  8^1  7^7  4^12^{18},  4)$,\\
IrMOA$(56^4,  27,            14^1  8^1  7^7  4^22^{16},  4)$,\\
IrMOA$(56^4,  29,                  8^1  7^8  4^12^{19},  4)$,\\
IrMOA$(56^4,  18,       28^5       8^1  7^3  4^2   2^7,  4)$,\\
IrMOA$(56^4,  19,       28^4 14^1  8^1  7^3  4^2   2^8,  4)$,\\
IrMOA$(56^4,  20,       28^3 14^2  8^1  7^3  4^2   2^9,  4)$,\\
IrMOA$(56^4,  21,       28^2 14^3  8^1  7^3  4^22^{10},  4)$,\\
IrMOA$(56^4,  22,       28^1 14^4  8^1  7^3  4^22^{11},  4)$,\\
IrMOA$(56^4,  23,            14^5  8^1  7^3  4^22^{12},  4)$,\\
IrMOA$(56^4,  18,       28^4 14^1  8^1  7^3  4^3   2^6,  4)$,\\
IrMOA$(56^4,  19,       28^3 14^2  8^1  7^3  4^3   2^7,  4)$,\\
IrMOA$(56^4,  20,       28^2 14^3  8^1  7^3  4^3   2^8,  4)$,\\
IrMOA$(56^4,  21,       28^1 14^4  8^1  7^3  4^3   2^9,  4)$,\\
IrMOA$(56^4,  22,            14^5  8^1  7^3  4^32^{10},  4)$,\\
IrMOA$(56^4,  18,       28^3 14^2  8^1  7^3  4^4   2^5,  4)$,\\
IrMOA$(56^4,  19,       28^2 14^3  8^1  7^3  4^4   2^6,  4)$,\\
IrMOA$(56^4,  20,       28^1 14^4  8^1  7^3  4^4   2^7,  4)$,\\
IrMOA$(56^4,  21,            14^5  8^1  7^3  4^4   2^8,  4)$,\\
IrMOA$(56^4,  18,       28^2 14^3  8^1  7^3  4^5   2^4,  4)$,\\
IrMOA$(56^4,  19,       28^1 14^4  8^1  7^3  4^5   2^5,  4)$,\\
IrMOA$(56^4,  20,            14^5  8^1  7^3  4^5   2^6,  4)$,\\
IrMOA$(56^4,  18,       28^1 14^4  8^1  7^3  4^6   2^3,  4)$,\\
IrMOA$(56^4,  19,            14^5  8^1  7^3  4^6   2^4,  4)$,\\
IrMOA$(56^4,  18,            14^5  8^1  7^3  4^7   2^2,  4)$,\\
IrMOA$(56^4,  20,       28^4       8^1  7^4  4^2   2^9,  4)$,\\
IrMOA$(56^4,  21,       28^3 14^1  8^1  7^4  4^22^{10},  4)$,\\
IrMOA$(56^4,  22,       28^2 14^2  8^1  7^4  4^22^{11},  4)$,\\
IrMOA$(56^4,  23,       28^1 14^3  8^1  7^4  4^22^{12},  4)$,\\
IrMOA$(56^4,  24,            14^4  8^1  7^4  4^22^{13},  4)$,\\
IrMOA$(56^4,  20,       28^3 14^1  8^1  7^4  4^3   2^8,  4)$,\\
IrMOA$(56^4,  21,       28^2 14^2  8^1  7^4  4^3   2^9,  4)$,\\
IrMOA$(56^4,  22,       28^1 14^3  8^1  7^4  4^32^{10},  4)$,\\
IrMOA$(56^4,  23,            14^4  8^1  7^4  4^32^{11},  4)$,\\
IrMOA$(56^4,  20,       28^2 14^2  8^1  7^4  4^4   2^7,  4)$,\\
IrMOA$(56^4,  21,       28^1 14^3  8^1  7^4  4^4   2^8,  4)$,\\
IrMOA$(56^4,  22,            14^4  8^1  7^4  4^4   2^9,  4)$,\\
IrMOA$(56^4,  20,       28^1 14^3  8^1  7^4  4^5   2^6,  4)$,\\
IrMOA$(56^4,  21,            14^4  8^1  7^4  4^5   2^7,  4)$,\\
IrMOA$(56^4,  20,            14^4  8^1  7^4  4^6   2^5,  4)$,\\
IrMOA$(56^4,  22,       28^3       8^1  7^5  4^22^{11},  4)$,\\
IrMOA$(56^4,  23,       28^2 14^1  8^1  7^5  4^22^{12},  4)$,\\
IrMOA$(56^4,  24,       28^1 14^2  8^1  7^5  4^22^{13},  4)$,\\
IrMOA$(56^4,  25,            14^3  8^1  7^5  4^22^{14},  4)$,\\
IrMOA$(56^4,  22,       28^2 14^1  8^1  7^5  4^32^{10},  4)$,\\
IrMOA$(56^4,  23,       28^1 14^2  8^1  7^5  4^32^{11},  4)$,\\
IrMOA$(56^4,  24,            14^3  8^1  7^5  4^32^{12},  4)$,\\
IrMOA$(56^4,  22,       28^1 14^2  8^1  7^5  4^4   2^9,  4)$,\\
IrMOA$(56^4,  23,            14^3  8^1  7^5  4^42^{10},  4)$,\\
IrMOA$(56^4,  22,            14^3  8^1  7^5  4^5   2^8,  4)$,\\
IrMOA$(56^4,  24,       28^2       8^1  7^6  4^22^{13},  4)$,\\
IrMOA$(56^4,  25,       28^1 14^1  8^1  7^6  4^22^{14},  4)$,\\
IrMOA$(56^4,  26,            14^2  8^1  7^6  4^22^{15},  4)$,\\
IrMOA$(56^4,  24,       28^1 14^1  8^1  7^6  4^32^{12},  4)$,\\
IrMOA$(56^4,  25,            14^2  8^1  7^6  4^32^{13},  4)$,\\
IrMOA$(56^4,  24,            14^2  8^1  7^6  4^42^{11},  4)$,\\
IrMOA$(56^4,  26,       28^1       8^1  7^7  4^22^{15},  4)$,\\
IrMOA$(56^4,  27,            14^1  8^1  7^7  4^22^{16},  4)$,\\
IrMOA$(56^4,  26,            14^1  8^1  7^7  4^32^{14},  4)$,\\
IrMOA$(56^4,  28,                  8^1  7^8  4^22^{17},  4)$,\\
IrMOA$(56^4,  19,       28^4       8^1  7^4  4^3   2^7,  4)$,\\
IrMOA$(56^4,  20,       28^3 14^1  8^1  7^4  4^3   2^8,  4)$,\\
IrMOA$(56^4,  21,       28^2 14^2  8^1  7^4  4^3   2^9,  4)$,\\
IrMOA$(56^4,  22,       28^1 14^3  8^1  7^4  4^32^{10},  4)$,\\
IrMOA$(56^4,  23,            14^4  8^1  7^4  4^32^{11},  4)$,\\
IrMOA$(56^4,  19,       28^3 14^1  8^1  7^4  4^4   2^6,  4)$,\\
IrMOA$(56^4,  20,       28^2 14^2  8^1  7^4  4^4   2^7,  4)$,\\
IrMOA$(56^4,  21,       28^1 14^3  8^1  7^4  4^4   2^8,  4)$,\\
IrMOA$(56^4,  22,            14^4  8^1  7^4  4^4   2^9,  4)$,\\
IrMOA$(56^4,  19,       28^2 14^2  8^1  7^4  4^5   2^5,  4)$,\\
IrMOA$(56^4,  20,       28^1 14^3  8^1  7^4  4^5   2^6,  4)$,\\
IrMOA$(56^4,  21,            14^4  8^1  7^4  4^5   2^7,  4)$,\\
IrMOA$(56^4,  19,       28^1 14^3  8^1  7^4  4^6   2^4,  4)$,\\
IrMOA$(56^4,  20,            14^4  8^1  7^4  4^6   2^5,  4)$,\\
IrMOA$(56^4,  19,            14^4  8^1  7^4  4^7   2^3,  4)$,\\
IrMOA$(56^4,  21,       28^3       8^1  7^5  4^3   2^9,  4)$,\\
IrMOA$(56^4,  22,       28^2 14^1  8^1  7^5  4^32^{10},  4)$,\\
IrMOA$(56^4,  23,       28^1 14^2  8^1  7^5  4^32^{11},  4)$,\\
IrMOA$(56^4,  24,            14^3  8^1  7^5  4^32^{12},  4)$,\\
IrMOA$(56^4,  21,       28^2 14^1  8^1  7^5  4^4   2^8,  4)$,\\
IrMOA$(56^4,  22,       28^1 14^2  8^1  7^5  4^4   2^9,  4)$,\\
IrMOA$(56^4,  23,            14^3  8^1  7^5  4^42^{10},  4)$,\\
IrMOA$(56^4,  21,       28^1 14^2  8^1  7^5  4^5   2^7,  4)$,\\
IrMOA$(56^4,  22,            14^3  8^1  7^5  4^5   2^8,  4)$,\\
IrMOA$(56^4,  21,            14^3  8^1  7^5  4^6   2^6,  4)$,\\
IrMOA$(56^4,  23,       28^2       8^1  7^6  4^32^{11},  4)$,\\
IrMOA$(56^4,  24,       28^1 14^1  8^1  7^6  4^32^{12},  4)$,\\
IrMOA$(56^4,  25,            14^2  8^1  7^6  4^32^{13},  4)$,\\
IrMOA$(56^4,  23,       28^1 14^1  8^1  7^6  4^42^{10},  4)$,\\
IrMOA$(56^4,  24,            14^2  8^1  7^6  4^42^{11},  4)$,\\
IrMOA$(56^4,  23,            14^2  8^1  7^6  4^5   2^9,  4)$,\\
IrMOA$(56^4,  25,       28^1       8^1  7^7  4^32^{13},  4)$,\\
IrMOA$(56^4,  26,            14^1  8^1  7^7  4^32^{14},  4)$,\\
IrMOA$(56^4,  25,            14^1  8^1  7^7  4^42^{12},  4)$,\\
IrMOA$(56^4,  27,                  8^1  7^8  4^32^{15},  4)$,\\
IrMOA$(56^4,  20,       28^3       8^1  7^5  4^4   2^7,  4)$,\\
IrMOA$(56^4,  21,       28^2 14^1  8^1  7^5  4^4   2^8,  4)$,\\
IrMOA$(56^4,  22,       28^1 14^2  8^1  7^5  4^4   2^9,  4)$,\\
IrMOA$(56^4,  23,            14^3  8^1  7^5  4^42^{10},  4)$,\\
IrMOA$(56^4,  20,       28^2 14^1  8^1  7^5  4^5   2^6,  4)$,\\
IrMOA$(56^4,  21,       28^1 14^2  8^1  7^5  4^5   2^7,  4)$,\\
IrMOA$(56^4,  22,            14^3  8^1  7^5  4^5   2^8,  4)$,\\
IrMOA$(56^4,  20,       28^1 14^2  8^1  7^5  4^6   2^5,  4)$,\\
IrMOA$(56^4,  21,            14^3  8^1  7^5  4^6   2^6,  4)$,\\
IrMOA$(56^4,  20,            14^3  8^1  7^5  4^7   2^4,  4)$,\\
IrMOA$(56^4,  22,       28^2       8^1  7^6  4^4   2^9,  4)$,\\
IrMOA$(56^4,  23,       28^1 14^1  8^1  7^6  4^42^{10},  4)$,\\
IrMOA$(56^4,  24,            14^2  8^1  7^6  4^42^{11},  4)$,\\
IrMOA$(56^4,  22,       28^1 14^1  8^1  7^6  4^5   2^8,  4)$,\\
IrMOA$(56^4,  23,            14^2  8^1  7^6  4^5   2^9,  4)$,\\
IrMOA$(56^4,  22,            14^2  8^1  7^6  4^6   2^7,  4)$,\\
IrMOA$(56^4,  24,       28^1       8^1  7^7  4^42^{11},  4)$,\\
IrMOA$(56^4,  25,            14^1  8^1  7^7  4^42^{12},  4)$,\\
IrMOA$(56^4,  24,            14^1  8^1  7^7  4^52^{10},  4)$,\\
IrMOA$(56^4,  26,                  8^1  7^8  4^42^{13},  4)$,\\
IrMOA$(56^4,  21,       28^2       8^1  7^6  4^5   2^7,  4)$,\\
IrMOA$(56^4,  22,       28^1 14^1  8^1  7^6  4^5   2^8,  4)$,\\
IrMOA$(56^4,  23,            14^2  8^1  7^6  4^5   2^9,  4)$,\\
IrMOA$(56^4,  21,       28^1 14^1  8^1  7^6  4^6   2^6,  4)$,\\
IrMOA$(56^4,  22,            14^2  8^1  7^6  4^6   2^7,  4)$,\\
IrMOA$(56^4,  21,            14^2  8^1  7^6  4^7   2^5,  4)$,\\
IrMOA$(56^4,  23,       28^1       8^1  7^7  4^5   2^9,  4)$,\\
IrMOA$(56^4,  24,            14^1  8^1  7^7  4^52^{10},  4)$,\\
IrMOA$(56^4,  23,            14^1  8^1  7^7  4^6   2^8,  4)$,\\
IrMOA$(56^4,  25,                  8^1  7^8  4^52^{11},  4)$,\\
IrMOA$(56^4,  22,       28^1       8^1  7^7  4^6   2^7,  4)$,\\
IrMOA$(56^4,  23,            14^1  8^1  7^7  4^6   2^8,  4)$,\\
IrMOA$(56^4,  22,            14^1  8^1  7^7  4^7   2^6,  4)$,\\
IrMOA$(56^4,  24,                  8^1  7^8  4^6   2^9,  4)$,\\
IrMOA$(56^4,  23,                  8^1  7^8  4^7   2^7,  4)$,\\
IrMOA$(56^4,  16,       28^6       8^2  7^2        2^6,  4)$,\\
IrMOA$(56^4,  17,       28^5 14^1  8^2  7^2        2^7,  4)$,\\
IrMOA$(56^4,  18,       28^4 14^2  8^2  7^2        2^8,  4)$,\\
IrMOA$(56^4,  19,       28^3 14^3  8^2  7^2        2^9,  4)$,\\
IrMOA$(56^4,  20,       28^2 14^4  8^2  7^2     2^{10},  4)$,\\
IrMOA$(56^4,  21,       28^1 14^5  8^2  7^2     2^{11},  4)$,\\
IrMOA$(56^4,  22,            14^6  8^2  7^2     2^{12},  4)$,\\
IrMOA$(56^4,  16,       28^5 14^1  8^2  7^2  4^1   2^5,  4)$,\\
IrMOA$(56^4,  17,       28^4 14^2  8^2  7^2  4^1   2^6,  4)$,\\
IrMOA$(56^4,  18,       28^3 14^3  8^2  7^2  4^1   2^7,  4)$,\\
IrMOA$(56^4,  19,       28^2 14^4  8^2  7^2  4^1   2^8,  4)$,\\
IrMOA$(56^4,  20,       28^1 14^5  8^2  7^2  4^1   2^9,  4)$,\\
IrMOA$(56^4,  21,            14^6  8^2  7^2  4^12^{10},  4)$,\\
IrMOA$(56^4,  16,       28^4 14^2  8^2  7^2  4^2   2^4,  4)$,\\
IrMOA$(56^4,  17,       28^3 14^3  8^2  7^2  4^2   2^5,  4)$,\\
IrMOA$(56^4,  18,       28^2 14^4  8^2  7^2  4^2   2^6,  4)$,\\
IrMOA$(56^4,  19,       28^1 14^5  8^2  7^2  4^2   2^7,  4)$,\\
IrMOA$(56^4,  20,            14^6  8^2  7^2  4^2   2^8,  4)$,\\
IrMOA$(56^4,  16,       28^3 14^3  8^2  7^2  4^3   2^3,  4)$,\\
IrMOA$(56^4,  17,       28^2 14^4  8^2  7^2  4^3   2^4,  4)$,\\
IrMOA$(56^4,  18,       28^1 14^5  8^2  7^2  4^3   2^5,  4)$,\\
IrMOA$(56^4,  19,            14^6  8^2  7^2  4^3   2^6,  4)$,\\
IrMOA$(56^4,  16,       28^2 14^4  8^2  7^2  4^4   2^2,  4)$,\\
IrMOA$(56^4,  17,       28^1 14^5  8^2  7^2  4^4   2^3,  4)$,\\
IrMOA$(56^4,  18,            14^6  8^2  7^2  4^4   2^4,  4)$,\\
IrMOA$(56^4,  16,       28^1 14^5  8^2  7^2  4^5   2^1,  4)$,\\
IrMOA$(56^4,  17,            14^6  8^2  7^2  4^5   2^2,  4)$,\\
IrMOA$(56^4,  16,            14^6  8^2  7^2  4^6      ,  4)$,\\
IrMOA$(56^4,  18,       28^5       8^2  7^3        2^8,  4)$,\\
IrMOA$(56^4,  19,       28^4 14^1  8^2  7^3        2^9,  4)$,\\
IrMOA$(56^4,  20,       28^3 14^2  8^2  7^3     2^{10},  4)$,\\
IrMOA$(56^4,  21,       28^2 14^3  8^2  7^3     2^{11},  4)$,\\
IrMOA$(56^4,  22,       28^1 14^4  8^2  7^3     2^{12},  4)$,\\
IrMOA$(56^4,  23,            14^5  8^2  7^3     2^{13},  4)$,\\
IrMOA$(56^4,  18,       28^4 14^1  8^2  7^3  4^1   2^7,  4)$,\\
IrMOA$(56^4,  19,       28^3 14^2  8^2  7^3  4^1   2^8,  4)$,\\
IrMOA$(56^4,  20,       28^2 14^3  8^2  7^3  4^1   2^9,  4)$,\\
IrMOA$(56^4,  21,       28^1 14^4  8^2  7^3  4^12^{10},  4)$,\\
IrMOA$(56^4,  22,            14^5  8^2  7^3  4^12^{11},  4)$,\\
IrMOA$(56^4,  18,       28^3 14^2  8^2  7^3  4^2   2^6,  4)$,\\
IrMOA$(56^4,  19,       28^2 14^3  8^2  7^3  4^2   2^7,  4)$,\\
IrMOA$(56^4,  20,       28^1 14^4  8^2  7^3  4^2   2^8,  4)$,\\
IrMOA$(56^4,  21,            14^5  8^2  7^3  4^2   2^9,  4)$,\\
IrMOA$(56^4,  18,       28^2 14^3  8^2  7^3  4^3   2^5,  4)$,\\
IrMOA$(56^4,  19,       28^1 14^4  8^2  7^3  4^3   2^6,  4)$,\\
IrMOA$(56^4,  20,            14^5  8^2  7^3  4^3   2^7,  4)$,\\
IrMOA$(56^4,  18,       28^1 14^4  8^2  7^3  4^4   2^4,  4)$,\\
IrMOA$(56^4,  19,            14^5  8^2  7^3  4^4   2^5,  4)$,\\
IrMOA$(56^4,  18,            14^5  8^2  7^3  4^5   2^3,  4)$,\\
IrMOA$(56^4,  20,       28^4       8^2  7^4     2^{10},  4)$,\\
IrMOA$(56^4,  21,       28^3 14^1  8^2  7^4     2^{11},  4)$,\\
IrMOA$(56^4,  22,       28^2 14^2  8^2  7^4     2^{12},  4)$,\\
IrMOA$(56^4,  23,       28^1 14^3  8^2  7^4     2^{13},  4)$,\\
IrMOA$(56^4,  24,            14^4  8^2  7^4     2^{14},  4)$,\\
IrMOA$(56^4,  20,       28^3 14^1  8^2  7^4  4^1   2^9,  4)$,\\
IrMOA$(56^4,  21,       28^2 14^2  8^2  7^4  4^12^{10},  4)$,\\
IrMOA$(56^4,  22,       28^1 14^3  8^2  7^4  4^12^{11},  4)$,\\
IrMOA$(56^4,  23,            14^4  8^2  7^4  4^12^{12},  4)$,\\
IrMOA$(56^4,  20,       28^2 14^2  8^2  7^4  4^2   2^8,  4)$,\\
IrMOA$(56^4,  21,       28^1 14^3  8^2  7^4  4^2   2^9,  4)$,\\
IrMOA$(56^4,  22,            14^4  8^2  7^4  4^22^{10},  4)$,\\
IrMOA$(56^4,  20,       28^1 14^3  8^2  7^4  4^3   2^7,  4)$,\\
IrMOA$(56^4,  21,            14^4  8^2  7^4  4^3   2^8,  4)$,\\
IrMOA$(56^4,  20,            14^4  8^2  7^4  4^4   2^6,  4)$,\\
IrMOA$(56^4,  22,       28^3       8^2  7^5     2^{12},  4)$,\\
IrMOA$(56^4,  23,       28^2 14^1  8^2  7^5     2^{13},  4)$,\\
IrMOA$(56^4,  24,       28^1 14^2  8^2  7^5     2^{14},  4)$,\\
IrMOA$(56^4,  25,            14^3  8^2  7^5     2^{15},  4)$,\\
IrMOA$(56^4,  22,       28^2 14^1  8^2  7^5  4^12^{11},  4)$,\\
IrMOA$(56^4,  23,       28^1 14^2  8^2  7^5  4^12^{12},  4)$,\\
IrMOA$(56^4,  24,            14^3  8^2  7^5  4^12^{13},  4)$,\\
IrMOA$(56^4,  22,       28^1 14^2  8^2  7^5  4^22^{10},  4)$,\\
IrMOA$(56^4,  23,            14^3  8^2  7^5  4^22^{11},  4)$,\\
IrMOA$(56^4,  22,            14^3  8^2  7^5  4^3   2^9,  4)$,\\
IrMOA$(56^4,  24,       28^2       8^2  7^6     2^{14},  4)$,\\
IrMOA$(56^4,  25,       28^1 14^1  8^2  7^6     2^{15},  4)$,\\
IrMOA$(56^4,  26,            14^2  8^2  7^6     2^{16},  4)$,\\
IrMOA$(56^4,  24,       28^1 14^1  8^2  7^6  4^12^{13},  4)$,\\
IrMOA$(56^4,  25,            14^2  8^2  7^6  4^12^{14},  4)$,\\
IrMOA$(56^4,  24,            14^2  8^2  7^6  4^22^{12},  4)$,\\
IrMOA$(56^4,  26,       28^1       8^2  7^7     2^{16},  4)$,\\
IrMOA$(56^4,  27,            14^1  8^2  7^7     2^{17},  4)$,\\
IrMOA$(56^4,  26,            14^1  8^2  7^7  4^12^{15},  4)$,\\
IrMOA$(56^4,  28,                  8^2  7^8     2^{18},  4)$,\\
IrMOA$(56^4,  17,       28^5       8^2  7^3  4^1   2^6,  4)$,\\
IrMOA$(56^4,  18,       28^4 14^1  8^2  7^3  4^1   2^7,  4)$,\\
IrMOA$(56^4,  19,       28^3 14^2  8^2  7^3  4^1   2^8,  4)$,\\
IrMOA$(56^4,  20,       28^2 14^3  8^2  7^3  4^1   2^9,  4)$,\\
IrMOA$(56^4,  21,       28^1 14^4  8^2  7^3  4^12^{10},  4)$,\\
IrMOA$(56^4,  22,            14^5  8^2  7^3  4^12^{11},  4)$,\\
IrMOA$(56^4,  17,       28^4 14^1  8^2  7^3  4^2   2^5,  4)$,\\
IrMOA$(56^4,  18,       28^3 14^2  8^2  7^3  4^2   2^6,  4)$,\\
IrMOA$(56^4,  19,       28^2 14^3  8^2  7^3  4^2   2^7,  4)$,\\
IrMOA$(56^4,  20,       28^1 14^4  8^2  7^3  4^2   2^8,  4)$,\\
IrMOA$(56^4,  21,            14^5  8^2  7^3  4^2   2^9,  4)$,\\
IrMOA$(56^4,  17,       28^3 14^2  8^2  7^3  4^3   2^4,  4)$,\\
IrMOA$(56^4,  18,       28^2 14^3  8^2  7^3  4^3   2^5,  4)$,\\
IrMOA$(56^4,  19,       28^1 14^4  8^2  7^3  4^3   2^6,  4)$,\\
IrMOA$(56^4,  20,            14^5  8^2  7^3  4^3   2^7,  4)$,\\
IrMOA$(56^4,  17,       28^2 14^3  8^2  7^3  4^4   2^3,  4)$,\\
IrMOA$(56^4,  18,       28^1 14^4  8^2  7^3  4^4   2^4,  4)$,\\
IrMOA$(56^4,  19,            14^5  8^2  7^3  4^4   2^5,  4)$,\\
IrMOA$(56^4,  17,       28^1 14^4  8^2  7^3  4^5   2^2,  4)$,\\
IrMOA$(56^4,  18,            14^5  8^2  7^3  4^5   2^3,  4)$,\\
IrMOA$(56^4,  17,            14^5  8^2  7^3  4^6   2^1,  4)$,\\
IrMOA$(56^4,  19,       28^4       8^2  7^4  4^1   2^8,  4)$,\\
IrMOA$(56^4,  20,       28^3 14^1  8^2  7^4  4^1   2^9,  4)$,\\
IrMOA$(56^4,  21,       28^2 14^2  8^2  7^4  4^12^{10},  4)$,\\
IrMOA$(56^4,  22,       28^1 14^3  8^2  7^4  4^12^{11},  4)$,\\
IrMOA$(56^4,  23,            14^4  8^2  7^4  4^12^{12},  4)$,\\
IrMOA$(56^4,  19,       28^3 14^1  8^2  7^4  4^2   2^7,  4)$,\\
IrMOA$(56^4,  20,       28^2 14^2  8^2  7^4  4^2   2^8,  4)$,\\
IrMOA$(56^4,  21,       28^1 14^3  8^2  7^4  4^2   2^9,  4)$,\\
IrMOA$(56^4,  22,            14^4  8^2  7^4  4^22^{10},  4)$,\\
IrMOA$(56^4,  19,       28^2 14^2  8^2  7^4  4^3   2^6,  4)$,\\
IrMOA$(56^4,  20,       28^1 14^3  8^2  7^4  4^3   2^7,  4)$,\\
IrMOA$(56^4,  21,            14^4  8^2  7^4  4^3   2^8,  4)$,\\
IrMOA$(56^4,  19,       28^1 14^3  8^2  7^4  4^4   2^5,  4)$,\\
IrMOA$(56^4,  20,            14^4  8^2  7^4  4^4   2^6,  4)$,\\
IrMOA$(56^4,  19,            14^4  8^2  7^4  4^5   2^4,  4)$,\\
IrMOA$(56^4,  21,       28^3       8^2  7^5  4^12^{10},  4)$,\\
IrMOA$(56^4,  22,       28^2 14^1  8^2  7^5  4^12^{11},  4)$,\\
IrMOA$(56^4,  23,       28^1 14^2  8^2  7^5  4^12^{12},  4)$,\\
IrMOA$(56^4,  24,            14^3  8^2  7^5  4^12^{13},  4)$,\\
IrMOA$(56^4,  21,       28^2 14^1  8^2  7^5  4^2   2^9,  4)$,\\
IrMOA$(56^4,  22,       28^1 14^2  8^2  7^5  4^22^{10},  4)$,\\
IrMOA$(56^4,  23,            14^3  8^2  7^5  4^22^{11},  4)$,\\
IrMOA$(56^4,  21,       28^1 14^2  8^2  7^5  4^3   2^8,  4)$,\\
IrMOA$(56^4,  22,            14^3  8^2  7^5  4^3   2^9,  4)$,\\
IrMOA$(56^4,  21,            14^3  8^2  7^5  4^4   2^7,  4)$,\\
IrMOA$(56^4,  23,       28^2       8^2  7^6  4^12^{12},  4)$,\\
IrMOA$(56^4,  24,       28^1 14^1  8^2  7^6  4^12^{13},  4)$,\\
IrMOA$(56^4,  25,            14^2  8^2  7^6  4^12^{14},  4)$,\\
IrMOA$(56^4,  23,       28^1 14^1  8^2  7^6  4^22^{11},  4)$,\\
IrMOA$(56^4,  24,            14^2  8^2  7^6  4^22^{12},  4)$,\\
IrMOA$(56^4,  23,            14^2  8^2  7^6  4^32^{10},  4)$,\\
IrMOA$(56^4,  25,       28^1       8^2  7^7  4^12^{14},  4)$,\\
IrMOA$(56^4,  26,            14^1  8^2  7^7  4^12^{15},  4)$,\\
IrMOA$(56^4,  25,            14^1  8^2  7^7  4^22^{13},  4)$,\\
IrMOA$(56^4,  27,                  8^2  7^8  4^12^{16},  4)$,\\
IrMOA$(56^4,  18,       28^4       8^2  7^4  4^2   2^6,  4)$,\\
IrMOA$(56^4,  19,       28^3 14^1  8^2  7^4  4^2   2^7,  4)$,\\
IrMOA$(56^4,  20,       28^2 14^2  8^2  7^4  4^2   2^8,  4)$,\\
IrMOA$(56^4,  21,       28^1 14^3  8^2  7^4  4^2   2^9,  4)$,\\
IrMOA$(56^4,  22,            14^4  8^2  7^4  4^22^{10},  4)$,\\
IrMOA$(56^4,  18,       28^3 14^1  8^2  7^4  4^3   2^5,  4)$,\\
IrMOA$(56^4,  19,       28^2 14^2  8^2  7^4  4^3   2^6,  4)$,\\
IrMOA$(56^4,  20,       28^1 14^3  8^2  7^4  4^3   2^7,  4)$,\\
IrMOA$(56^4,  21,            14^4  8^2  7^4  4^3   2^8,  4)$,\\
IrMOA$(56^4,  18,       28^2 14^2  8^2  7^4  4^4   2^4,  4)$,\\
IrMOA$(56^4,  19,       28^1 14^3  8^2  7^4  4^4   2^5,  4)$,\\
IrMOA$(56^4,  20,            14^4  8^2  7^4  4^4   2^6,  4)$,\\
IrMOA$(56^4,  18,       28^1 14^3  8^2  7^4  4^5   2^3,  4)$,\\
IrMOA$(56^4,  19,            14^4  8^2  7^4  4^5   2^4,  4)$,\\
IrMOA$(56^4,  18,            14^4  8^2  7^4  4^6   2^2,  4)$,\\
IrMOA$(56^4,  20,       28^3       8^2  7^5  4^2   2^8,  4)$,\\
IrMOA$(56^4,  21,       28^2 14^1  8^2  7^5  4^2   2^9,  4)$,\\
IrMOA$(56^4,  22,       28^1 14^2  8^2  7^5  4^22^{10},  4)$,\\
IrMOA$(56^4,  23,            14^3  8^2  7^5  4^22^{11},  4)$,\\
IrMOA$(56^4,  20,       28^2 14^1  8^2  7^5  4^3   2^7,  4)$,\\
IrMOA$(56^4,  21,       28^1 14^2  8^2  7^5  4^3   2^8,  4)$,\\
IrMOA$(56^4,  22,            14^3  8^2  7^5  4^3   2^9,  4)$,\\
IrMOA$(56^4,  20,       28^1 14^2  8^2  7^5  4^4   2^6,  4)$,\\
IrMOA$(56^4,  21,            14^3  8^2  7^5  4^4   2^7,  4)$,\\
IrMOA$(56^4,  20,            14^3  8^2  7^5  4^5   2^5,  4)$,\\
IrMOA$(56^4,  22,       28^2       8^2  7^6  4^22^{10},  4)$,\\
IrMOA$(56^4,  23,       28^1 14^1  8^2  7^6  4^22^{11},  4)$,\\
IrMOA$(56^4,  24,            14^2  8^2  7^6  4^22^{12},  4)$,\\
IrMOA$(56^4,  22,       28^1 14^1  8^2  7^6  4^3   2^9,  4)$,\\
IrMOA$(56^4,  23,            14^2  8^2  7^6  4^32^{10},  4)$,\\
IrMOA$(56^4,  22,            14^2  8^2  7^6  4^4   2^8,  4)$,\\
IrMOA$(56^4,  24,       28^1       8^2  7^7  4^22^{12},  4)$,\\
IrMOA$(56^4,  25,            14^1  8^2  7^7  4^22^{13},  4)$,\\
IrMOA$(56^4,  24,            14^1  8^2  7^7  4^32^{11},  4)$,\\
IrMOA$(56^4,  26,                  8^2  7^8  4^22^{14},  4)$,\\
IrMOA$(56^4,  19,       28^3       8^2  7^5  4^3   2^6,  4)$,\\
IrMOA$(56^4,  20,       28^2 14^1  8^2  7^5  4^3   2^7,  4)$,\\
IrMOA$(56^4,  21,       28^1 14^2  8^2  7^5  4^3   2^8,  4)$,\\
IrMOA$(56^4,  22,            14^3  8^2  7^5  4^3   2^9,  4)$,\\
IrMOA$(56^4,  19,       28^2 14^1  8^2  7^5  4^4   2^5,  4)$,\\
IrMOA$(56^4,  20,       28^1 14^2  8^2  7^5  4^4   2^6,  4)$,\\
IrMOA$(56^4,  21,            14^3  8^2  7^5  4^4   2^7,  4)$,\\
IrMOA$(56^4,  19,       28^1 14^2  8^2  7^5  4^5   2^4,  4)$,\\
IrMOA$(56^4,  20,            14^3  8^2  7^5  4^5   2^5,  4)$,\\
IrMOA$(56^4,  19,            14^3  8^2  7^5  4^6   2^3,  4)$,\\
IrMOA$(56^4,  21,       28^2       8^2  7^6  4^3   2^8,  4)$,\\
IrMOA$(56^4,  22,       28^1 14^1  8^2  7^6  4^3   2^9,  4)$,\\
IrMOA$(56^4,  23,            14^2  8^2  7^6  4^32^{10},  4)$,\\
IrMOA$(56^4,  21,       28^1 14^1  8^2  7^6  4^4   2^7,  4)$,\\
IrMOA$(56^4,  22,            14^2  8^2  7^6  4^4   2^8,  4)$,\\
IrMOA$(56^4,  21,            14^2  8^2  7^6  4^5   2^6,  4)$,\\
IrMOA$(56^4,  23,       28^1       8^2  7^7  4^32^{10},  4)$,\\
IrMOA$(56^4,  24,            14^1  8^2  7^7  4^32^{11},  4)$,\\
IrMOA$(56^4,  23,            14^1  8^2  7^7  4^4   2^9,  4)$,\\
IrMOA$(56^4,  25,                  8^2  7^8  4^32^{12},  4)$,\\
IrMOA$(56^4,  20,       28^2       8^2  7^6  4^4   2^6,  4)$,\\
IrMOA$(56^4,  21,       28^1 14^1  8^2  7^6  4^4   2^7,  4)$,\\
IrMOA$(56^4,  22,            14^2  8^2  7^6  4^4   2^8,  4)$,\\
IrMOA$(56^4,  20,       28^1 14^1  8^2  7^6  4^5   2^5,  4)$,\\
IrMOA$(56^4,  21,            14^2  8^2  7^6  4^5   2^6,  4)$,\\
IrMOA$(56^4,  20,            14^2  8^2  7^6  4^6   2^4,  4)$,\\
IrMOA$(56^4,  22,       28^1       8^2  7^7  4^4   2^8,  4)$,\\
IrMOA$(56^4,  23,            14^1  8^2  7^7  4^4   2^9,  4)$,\\
IrMOA$(56^4,  22,            14^1  8^2  7^7  4^5   2^7,  4)$,\\
IrMOA$(56^4,  24,                  8^2  7^8  4^42^{10},  4)$,\\
IrMOA$(56^4,  21,       28^1       8^2  7^7  4^5   2^6,  4)$,\\
IrMOA$(56^4,  22,            14^1  8^2  7^7  4^5   2^7,  4)$,\\
IrMOA$(56^4,  21,            14^1  8^2  7^7  4^6   2^5,  4)$,\\
IrMOA$(56^4,  23,                  8^2  7^8  4^5   2^8,  4)$,\\
IrMOA$(56^4,  22,                  8^2  7^8  4^6   2^6,  4)$,\\
IrMOA$(56^4,  16,       28^5       8^3  7^3        2^5,  4)$,\\
IrMOA$(56^4,  17,       28^4 14^1  8^3  7^3        2^6,  4)$,\\
IrMOA$(56^4,  18,       28^3 14^2  8^3  7^3        2^7,  4)$,\\
IrMOA$(56^4,  19,       28^2 14^3  8^3  7^3        2^8,  4)$,\\
IrMOA$(56^4,  20,       28^1 14^4  8^3  7^3        2^9,  4)$,\\
IrMOA$(56^4,  21,            14^5  8^3  7^3     2^{10},  4)$,\\
IrMOA$(56^4,  16,       28^4 14^1  8^3  7^3  4^1   2^4,  4)$,\\
IrMOA$(56^4,  17,       28^3 14^2  8^3  7^3  4^1   2^5,  4)$,\\
IrMOA$(56^4,  18,       28^2 14^3  8^3  7^3  4^1   2^6,  4)$,\\
IrMOA$(56^4,  19,       28^1 14^4  8^3  7^3  4^1   2^7,  4)$,\\
IrMOA$(56^4,  20,            14^5  8^3  7^3  4^1   2^8,  4)$,\\
IrMOA$(56^4,  16,       28^3 14^2  8^3  7^3  4^2   2^3,  4)$,\\
IrMOA$(56^4,  17,       28^2 14^3  8^3  7^3  4^2   2^4,  4)$,\\
IrMOA$(56^4,  18,       28^1 14^4  8^3  7^3  4^2   2^5,  4)$,\\
IrMOA$(56^4,  19,            14^5  8^3  7^3  4^2   2^6,  4)$,\\
IrMOA$(56^4,  16,       28^2 14^3  8^3  7^3  4^3   2^2,  4)$,\\
IrMOA$(56^4,  17,       28^1 14^4  8^3  7^3  4^3   2^3,  4)$,\\
IrMOA$(56^4,  18,            14^5  8^3  7^3  4^3   2^4,  4)$,\\
IrMOA$(56^4,  16,       28^1 14^4  8^3  7^3  4^4   2^1,  4)$,\\
IrMOA$(56^4,  17,            14^5  8^3  7^3  4^4   2^2,  4)$,\\
IrMOA$(56^4,  16,            14^5  8^3  7^3  4^5      ,  4)$,\\
IrMOA$(56^4,  18,       28^4       8^3  7^4        2^7,  4)$,\\
IrMOA$(56^4,  19,       28^3 14^1  8^3  7^4        2^8,  4)$,\\
IrMOA$(56^4,  20,       28^2 14^2  8^3  7^4        2^9,  4)$,\\
IrMOA$(56^4,  21,       28^1 14^3  8^3  7^4     2^{10},  4)$,\\
IrMOA$(56^4,  22,            14^4  8^3  7^4     2^{11},  4)$,\\
IrMOA$(56^4,  18,       28^3 14^1  8^3  7^4  4^1   2^6,  4)$,\\
IrMOA$(56^4,  19,       28^2 14^2  8^3  7^4  4^1   2^7,  4)$,\\
IrMOA$(56^4,  20,       28^1 14^3  8^3  7^4  4^1   2^8,  4)$,\\
IrMOA$(56^4,  21,            14^4  8^3  7^4  4^1   2^9,  4)$,\\
IrMOA$(56^4,  18,       28^2 14^2  8^3  7^4  4^2   2^5,  4)$,\\
IrMOA$(56^4,  19,       28^1 14^3  8^3  7^4  4^2   2^6,  4)$,\\
IrMOA$(56^4,  20,            14^4  8^3  7^4  4^2   2^7,  4)$,\\
IrMOA$(56^4,  18,       28^1 14^3  8^3  7^4  4^3   2^4,  4)$,\\
IrMOA$(56^4,  19,            14^4  8^3  7^4  4^3   2^5,  4)$,\\
IrMOA$(56^4,  18,            14^4  8^3  7^4  4^4   2^3,  4)$,\\
IrMOA$(56^4,  20,       28^3       8^3  7^5        2^9,  4)$,\\
IrMOA$(56^4,  21,       28^2 14^1  8^3  7^5     2^{10},  4)$,\\
IrMOA$(56^4,  22,       28^1 14^2  8^3  7^5     2^{11},  4)$,\\
IrMOA$(56^4,  23,            14^3  8^3  7^5     2^{12},  4)$,\\
IrMOA$(56^4,  20,       28^2 14^1  8^3  7^5  4^1   2^8,  4)$,\\
IrMOA$(56^4,  21,       28^1 14^2  8^3  7^5  4^1   2^9,  4)$,\\
IrMOA$(56^4,  22,            14^3  8^3  7^5  4^12^{10},  4)$,\\
IrMOA$(56^4,  20,       28^1 14^2  8^3  7^5  4^2   2^7,  4)$,\\
IrMOA$(56^4,  21,            14^3  8^3  7^5  4^2   2^8,  4)$,\\
IrMOA$(56^4,  20,            14^3  8^3  7^5  4^3   2^6,  4)$,\\
IrMOA$(56^4,  22,       28^2       8^3  7^6     2^{11},  4)$,\\
IrMOA$(56^4,  23,       28^1 14^1  8^3  7^6     2^{12},  4)$,\\
IrMOA$(56^4,  24,            14^2  8^3  7^6     2^{13},  4)$,\\
IrMOA$(56^4,  22,       28^1 14^1  8^3  7^6  4^12^{10},  4)$,\\
IrMOA$(56^4,  23,            14^2  8^3  7^6  4^12^{11},  4)$,\\
IrMOA$(56^4,  22,            14^2  8^3  7^6  4^2   2^9,  4)$,\\
IrMOA$(56^4,  24,       28^1       8^3  7^7     2^{13},  4)$,\\
IrMOA$(56^4,  25,            14^1  8^3  7^7     2^{14},  4)$,\\
IrMOA$(56^4,  24,            14^1  8^3  7^7  4^12^{12},  4)$,\\
IrMOA$(56^4,  26,                  8^3  7^8     2^{15},  4)$,\\
IrMOA$(56^4,  17,       28^4       8^3  7^4  4^1   2^5,  4)$,\\
IrMOA$(56^4,  18,       28^3 14^1  8^3  7^4  4^1   2^6,  4)$,\\
IrMOA$(56^4,  19,       28^2 14^2  8^3  7^4  4^1   2^7,  4)$,\\
IrMOA$(56^4,  20,       28^1 14^3  8^3  7^4  4^1   2^8,  4)$,\\
IrMOA$(56^4,  21,            14^4  8^3  7^4  4^1   2^9,  4)$,\\
IrMOA$(56^4,  17,       28^3 14^1  8^3  7^4  4^2   2^4,  4)$,\\
IrMOA$(56^4,  18,       28^2 14^2  8^3  7^4  4^2   2^5,  4)$,\\
IrMOA$(56^4,  19,       28^1 14^3  8^3  7^4  4^2   2^6,  4)$,\\
IrMOA$(56^4,  20,            14^4  8^3  7^4  4^2   2^7,  4)$,\\
IrMOA$(56^4,  17,       28^2 14^2  8^3  7^4  4^3   2^3,  4)$,\\
IrMOA$(56^4,  18,       28^1 14^3  8^3  7^4  4^3   2^4,  4)$,\\
IrMOA$(56^4,  19,            14^4  8^3  7^4  4^3   2^5,  4)$,\\
IrMOA$(56^4,  17,       28^1 14^3  8^3  7^4  4^4   2^2,  4)$,\\
IrMOA$(56^4,  18,            14^4  8^3  7^4  4^4   2^3,  4)$,\\
IrMOA$(56^4,  17,            14^4  8^3  7^4  4^5   2^1,  4)$,\\
IrMOA$(56^4,  19,       28^3       8^3  7^5  4^1   2^7,  4)$,\\
IrMOA$(56^4,  20,       28^2 14^1  8^3  7^5  4^1   2^8,  4)$,\\
IrMOA$(56^4,  21,       28^1 14^2  8^3  7^5  4^1   2^9,  4)$,\\
IrMOA$(56^4,  22,            14^3  8^3  7^5  4^12^{10},  4)$,\\
IrMOA$(56^4,  19,       28^2 14^1  8^3  7^5  4^2   2^6,  4)$,\\
IrMOA$(56^4,  20,       28^1 14^2  8^3  7^5  4^2   2^7,  4)$,\\
IrMOA$(56^4,  21,            14^3  8^3  7^5  4^2   2^8,  4)$,\\
IrMOA$(56^4,  19,       28^1 14^2  8^3  7^5  4^3   2^5,  4)$,\\
IrMOA$(56^4,  20,            14^3  8^3  7^5  4^3   2^6,  4)$,\\
IrMOA$(56^4,  19,            14^3  8^3  7^5  4^4   2^4,  4)$,\\
IrMOA$(56^4,  21,       28^2       8^3  7^6  4^1   2^9,  4)$,\\
IrMOA$(56^4,  22,       28^1 14^1  8^3  7^6  4^12^{10},  4)$,\\
IrMOA$(56^4,  23,            14^2  8^3  7^6  4^12^{11},  4)$,\\
IrMOA$(56^4,  21,       28^1 14^1  8^3  7^6  4^2   2^8,  4)$,\\
IrMOA$(56^4,  22,            14^2  8^3  7^6  4^2   2^9,  4)$,\\
IrMOA$(56^4,  21,            14^2  8^3  7^6  4^3   2^7,  4)$,\\
IrMOA$(56^4,  23,       28^1       8^3  7^7  4^12^{11},  4)$,\\
IrMOA$(56^4,  24,            14^1  8^3  7^7  4^12^{12},  4)$,\\
IrMOA$(56^4,  23,            14^1  8^3  7^7  4^22^{10},  4)$,\\
IrMOA$(56^4,  25,                  8^3  7^8  4^12^{13},  4)$,\\
IrMOA$(56^4,  18,       28^3       8^3  7^5  4^2   2^5,  4)$,\\
IrMOA$(56^4,  19,       28^2 14^1  8^3  7^5  4^2   2^6,  4)$,\\
IrMOA$(56^4,  20,       28^1 14^2  8^3  7^5  4^2   2^7,  4)$,\\
IrMOA$(56^4,  21,            14^3  8^3  7^5  4^2   2^8,  4)$,\\
IrMOA$(56^4,  18,       28^2 14^1  8^3  7^5  4^3   2^4,  4)$,\\
IrMOA$(56^4,  19,       28^1 14^2  8^3  7^5  4^3   2^5,  4)$,\\
IrMOA$(56^4,  20,            14^3  8^3  7^5  4^3   2^6,  4)$,\\
IrMOA$(56^4,  18,       28^1 14^2  8^3  7^5  4^4   2^3,  4)$,\\
IrMOA$(56^4,  19,            14^3  8^3  7^5  4^4   2^4,  4)$,\\
IrMOA$(56^4,  18,            14^3  8^3  7^5  4^5   2^2,  4)$,\\
IrMOA$(56^4,  20,       28^2       8^3  7^6  4^2   2^7,  4)$,\\
IrMOA$(56^4,  21,       28^1 14^1  8^3  7^6  4^2   2^8,  4)$,\\
IrMOA$(56^4,  22,            14^2  8^3  7^6  4^2   2^9,  4)$,\\
IrMOA$(56^4,  20,       28^1 14^1  8^3  7^6  4^3   2^6,  4)$,\\
IrMOA$(56^4,  21,            14^2  8^3  7^6  4^3   2^7,  4)$,\\
IrMOA$(56^4,  20,            14^2  8^3  7^6  4^4   2^5,  4)$,\\
IrMOA$(56^4,  22,       28^1       8^3  7^7  4^2   2^9,  4)$,\\
IrMOA$(56^4,  23,            14^1  8^3  7^7  4^22^{10},  4)$,\\
IrMOA$(56^4,  22,            14^1  8^3  7^7  4^3   2^8,  4)$,\\
IrMOA$(56^4,  24,                  8^3  7^8  4^22^{11},  4)$,\\
IrMOA$(56^4,  19,       28^2       8^3  7^6  4^3   2^5,  4)$,\\
IrMOA$(56^4,  20,       28^1 14^1  8^3  7^6  4^3   2^6,  4)$,\\
IrMOA$(56^4,  21,            14^2  8^3  7^6  4^3   2^7,  4)$,\\
IrMOA$(56^4,  19,       28^1 14^1  8^3  7^6  4^4   2^4,  4)$,\\
IrMOA$(56^4,  20,            14^2  8^3  7^6  4^4   2^5,  4)$,\\
IrMOA$(56^4,  19,            14^2  8^3  7^6  4^5   2^3,  4)$,\\
IrMOA$(56^4,  21,       28^1       8^3  7^7  4^3   2^7,  4)$,\\
IrMOA$(56^4,  22,            14^1  8^3  7^7  4^3   2^8,  4)$,\\
IrMOA$(56^4,  21,            14^1  8^3  7^7  4^4   2^6,  4)$,\\
IrMOA$(56^4,  23,                  8^3  7^8  4^3   2^9,  4)$,\\
IrMOA$(56^4,  20,       28^1       8^3  7^7  4^4   2^5,  4)$,\\
IrMOA$(56^4,  21,            14^1  8^3  7^7  4^4   2^6,  4)$,\\
IrMOA$(56^4,  20,            14^1  8^3  7^7  4^5   2^4,  4)$,\\
IrMOA$(56^4,  22,                  8^3  7^8  4^4   2^7,  4)$,\\
IrMOA$(56^4,  21,                  8^3  7^8  4^5   2^5,  4)$,\\
IrMOA$(56^4,  16,       28^4       8^4  7^4        2^4,  4)$,\\
IrMOA$(56^4,  17,       28^3 14^1  8^4  7^4        2^5,  4)$,\\
IrMOA$(56^4,  18,       28^2 14^2  8^4  7^4        2^6,  4)$,\\
IrMOA$(56^4,  19,       28^1 14^3  8^4  7^4        2^7,  4)$,\\
IrMOA$(56^4,  20,            14^4  8^4  7^4        2^8,  4)$,\\
IrMOA$(56^4,  16,       28^3 14^1  8^4  7^4  4^1   2^3,  4)$,\\
IrMOA$(56^4,  17,       28^2 14^2  8^4  7^4  4^1   2^4,  4)$,\\
IrMOA$(56^4,  18,       28^1 14^3  8^4  7^4  4^1   2^5,  4)$,\\
IrMOA$(56^4,  19,            14^4  8^4  7^4  4^1   2^6,  4)$,\\
IrMOA$(56^4,  16,       28^2 14^2  8^4  7^4  4^2   2^2,  4)$,\\
IrMOA$(56^4,  17,       28^1 14^3  8^4  7^4  4^2   2^3,  4)$,\\
IrMOA$(56^4,  18,            14^4  8^4  7^4  4^2   2^4,  4)$,\\
IrMOA$(56^4,  16,       28^1 14^3  8^4  7^4  4^3   2^1,  4)$,\\
IrMOA$(56^4,  17,            14^4  8^4  7^4  4^3   2^2,  4)$,\\
IrMOA$(56^4,  16,            14^4  8^4  7^4  4^4      ,  4)$,\\
IrMOA$(56^4,  18,       28^3       8^4  7^5        2^6,  4)$,\\
IrMOA$(56^4,  19,       28^2 14^1  8^4  7^5        2^7,  4)$,\\
IrMOA$(56^4,  20,       28^1 14^2  8^4  7^5        2^8,  4)$,\\
IrMOA$(56^4,  21,            14^3  8^4  7^5        2^9,  4)$,\\
IrMOA$(56^4,  18,       28^2 14^1  8^4  7^5  4^1   2^5,  4)$,\\
IrMOA$(56^4,  19,       28^1 14^2  8^4  7^5  4^1   2^6,  4)$,\\
IrMOA$(56^4,  20,            14^3  8^4  7^5  4^1   2^7,  4)$,\\
IrMOA$(56^4,  18,       28^1 14^2  8^4  7^5  4^2   2^4,  4)$,\\
IrMOA$(56^4,  19,            14^3  8^4  7^5  4^2   2^5,  4)$,\\
IrMOA$(56^4,  18,            14^3  8^4  7^5  4^3   2^3,  4)$,\\
IrMOA$(56^4,  20,       28^2       8^4  7^6        2^8,  4)$,\\
IrMOA$(56^4,  21,       28^1 14^1  8^4  7^6        2^9,  4)$,\\
IrMOA$(56^4,  22,            14^2  8^4  7^6     2^{10},  4)$,\\
IrMOA$(56^4,  20,       28^1 14^1  8^4  7^6  4^1   2^7,  4)$,\\
IrMOA$(56^4,  21,            14^2  8^4  7^6  4^1   2^8,  4)$,\\
IrMOA$(56^4,  20,            14^2  8^4  7^6  4^2   2^6,  4)$,\\
IrMOA$(56^4,  22,       28^1       8^4  7^7     2^{10},  4)$,\\
IrMOA$(56^4,  23,            14^1  8^4  7^7     2^{11},  4)$,\\
IrMOA$(56^4,  22,            14^1  8^4  7^7  4^1   2^9,  4)$,\\
IrMOA$(56^4,  24,                  8^4  7^8     2^{12},  4)$,\\
IrMOA$(56^4,  17,       28^3       8^4  7^5  4^1   2^4,  4)$,\\
IrMOA$(56^4,  18,       28^2 14^1  8^4  7^5  4^1   2^5,  4)$,\\
IrMOA$(56^4,  19,       28^1 14^2  8^4  7^5  4^1   2^6,  4)$,\\
IrMOA$(56^4,  20,            14^3  8^4  7^5  4^1   2^7,  4)$,\\
IrMOA$(56^4,  17,       28^2 14^1  8^4  7^5  4^2   2^3,  4)$,\\
IrMOA$(56^4,  18,       28^1 14^2  8^4  7^5  4^2   2^4,  4)$,\\
IrMOA$(56^4,  19,            14^3  8^4  7^5  4^2   2^5,  4)$,\\
IrMOA$(56^4,  17,       28^1 14^2  8^4  7^5  4^3   2^2,  4)$,\\
IrMOA$(56^4,  18,            14^3  8^4  7^5  4^3   2^3,  4)$,\\
IrMOA$(56^4,  17,            14^3  8^4  7^5  4^4   2^1,  4)$,\\
IrMOA$(56^4,  19,       28^2       8^4  7^6  4^1   2^6,  4)$,\\
IrMOA$(56^4,  20,       28^1 14^1  8^4  7^6  4^1   2^7,  4)$,\\
IrMOA$(56^4,  21,            14^2  8^4  7^6  4^1   2^8,  4)$,\\
IrMOA$(56^4,  19,       28^1 14^1  8^4  7^6  4^2   2^5,  4)$,\\
IrMOA$(56^4,  20,            14^2  8^4  7^6  4^2   2^6,  4)$,\\
IrMOA$(56^4,  19,            14^2  8^4  7^6  4^3   2^4,  4)$,\\
IrMOA$(56^4,  21,       28^1       8^4  7^7  4^1   2^8,  4)$,\\
IrMOA$(56^4,  22,            14^1  8^4  7^7  4^1   2^9,  4)$,\\
IrMOA$(56^4,  21,            14^1  8^4  7^7  4^2   2^7,  4)$,\\
IrMOA$(56^4,  23,                  8^4  7^8  4^12^{10},  4)$,\\
IrMOA$(56^4,  18,       28^2       8^4  7^6  4^2   2^4,  4)$,\\
IrMOA$(56^4,  19,       28^1 14^1  8^4  7^6  4^2   2^5,  4)$,\\
IrMOA$(56^4,  20,            14^2  8^4  7^6  4^2   2^6,  4)$,\\
IrMOA$(56^4,  18,       28^1 14^1  8^4  7^6  4^3   2^3,  4)$,\\
IrMOA$(56^4,  19,            14^2  8^4  7^6  4^3   2^4,  4)$,\\
IrMOA$(56^4,  18,            14^2  8^4  7^6  4^4   2^2,  4)$,\\
IrMOA$(56^4,  20,       28^1       8^4  7^7  4^2   2^6,  4)$,\\
IrMOA$(56^4,  21,            14^1  8^4  7^7  4^2   2^7,  4)$,\\
IrMOA$(56^4,  20,            14^1  8^4  7^7  4^3   2^5,  4)$,\\
IrMOA$(56^4,  22,                  8^4  7^8  4^2   2^8,  4)$,\\
IrMOA$(56^4,  19,       28^1       8^4  7^7  4^3   2^4,  4)$,\\
IrMOA$(56^4,  20,            14^1  8^4  7^7  4^3   2^5,  4)$,\\
IrMOA$(56^4,  19,            14^1  8^4  7^7  4^4   2^3,  4)$,\\
IrMOA$(56^4,  21,                  8^4  7^8  4^3   2^6,  4)$,\\
IrMOA$(56^4,  20,                  8^4  7^8  4^4   2^4,  4)$,\\
IrMOA$(56^4,  16,       28^3       8^5  7^5        2^3,  4)$,\\
IrMOA$(56^4,  17,       28^2 14^1  8^5  7^5        2^4,  4)$,\\
IrMOA$(56^4,  18,       28^1 14^2  8^5  7^5        2^5,  4)$,\\
IrMOA$(56^4,  19,            14^3  8^5  7^5        2^6,  4)$,\\
IrMOA$(56^4,  16,       28^2 14^1  8^5  7^5  4^1   2^2,  4)$,\\
IrMOA$(56^4,  17,       28^1 14^2  8^5  7^5  4^1   2^3,  4)$,\\
IrMOA$(56^4,  18,            14^3  8^5  7^5  4^1   2^4,  4)$,\\
IrMOA$(56^4,  16,       28^1 14^2  8^5  7^5  4^2   2^1,  4)$,\\
IrMOA$(56^4,  17,            14^3  8^5  7^5  4^2   2^2,  4)$,\\
IrMOA$(56^4,  16,            14^3  8^5  7^5  4^3      ,  4)$,\\
IrMOA$(56^4,  18,       28^2       8^5  7^6        2^5,  4)$,\\
IrMOA$(56^4,  19,       28^1 14^1  8^5  7^6        2^6,  4)$,\\
IrMOA$(56^4,  20,            14^2  8^5  7^6        2^7,  4)$,\\
IrMOA$(56^4,  18,       28^1 14^1  8^5  7^6  4^1   2^4,  4)$,\\
IrMOA$(56^4,  19,            14^2  8^5  7^6  4^1   2^5,  4)$,\\
IrMOA$(56^4,  18,            14^2  8^5  7^6  4^2   2^3,  4)$,\\
IrMOA$(56^4,  20,       28^1       8^5  7^7        2^7,  4)$,\\
IrMOA$(56^4,  21,            14^1  8^5  7^7        2^8,  4)$,\\
IrMOA$(56^4,  20,            14^1  8^5  7^7  4^1   2^6,  4)$,\\
IrMOA$(56^4,  22,                  8^5  7^8        2^9,  4)$,\\
IrMOA$(56^4,  17,       28^2       8^5  7^6  4^1   2^3,  4)$,\\
IrMOA$(56^4,  18,       28^1 14^1  8^5  7^6  4^1   2^4,  4)$,\\
IrMOA$(56^4,  19,            14^2  8^5  7^6  4^1   2^5,  4)$,\\
IrMOA$(56^4,  17,       28^1 14^1  8^5  7^6  4^2   2^2,  4)$,\\
IrMOA$(56^4,  18,            14^2  8^5  7^6  4^2   2^3,  4)$,\\
IrMOA$(56^4,  17,            14^2  8^5  7^6  4^3   2^1,  4)$,\\
IrMOA$(56^4,  19,       28^1       8^5  7^7  4^1   2^5,  4)$,\\
IrMOA$(56^4,  20,            14^1  8^5  7^7  4^1   2^6,  4)$,\\
IrMOA$(56^4,  19,            14^1  8^5  7^7  4^2   2^4,  4)$,\\
IrMOA$(56^4,  21,                  8^5  7^8  4^1   2^7,  4)$,\\
IrMOA$(56^4,  18,       28^1       8^5  7^7  4^2   2^3,  4)$,\\
IrMOA$(56^4,  19,            14^1  8^5  7^7  4^2   2^4,  4)$,\\
IrMOA$(56^4,  18,            14^1  8^5  7^7  4^3   2^2,  4)$,\\
IrMOA$(56^4,  20,                  8^5  7^8  4^2   2^5,  4)$,\\
IrMOA$(56^4,  19,                  8^5  7^8  4^3   2^3,  4)$,\\
IrMOA$(56^4,  16,       28^2       8^6  7^6        2^2,  4)$,\\
IrMOA$(56^4,  17,       28^1 14^1  8^6  7^6        2^3,  4)$,\\
IrMOA$(56^4,  18,            14^2  8^6  7^6        2^4,  4)$,\\
IrMOA$(56^4,  16,       28^1 14^1  8^6  7^6  4^1   2^1,  4)$,\\
IrMOA$(56^4,  17,            14^2  8^6  7^6  4^1   2^2,  4)$,\\
IrMOA$(56^4,  16,            14^2  8^6  7^6  4^2      ,  4)$,\\
IrMOA$(56^4,  18,       28^1       8^6  7^7        2^4,  4)$,\\
IrMOA$(56^4,  19,            14^1  8^6  7^7        2^5,  4)$,\\
IrMOA$(56^4,  18,            14^1  8^6  7^7  4^1   2^3,  4)$,\\
IrMOA$(56^4,  20,                  8^6  7^8        2^6,  4)$,\\
IrMOA$(56^4,  17,       28^1       8^6  7^7  4^1   2^2,  4)$,\\
IrMOA$(56^4,  18,            14^1  8^6  7^7  4^1   2^3,  4)$,\\
IrMOA$(56^4,  17,            14^1  8^6  7^7  4^2   2^1,  4)$,\\
IrMOA$(56^4,  19,                  8^6  7^8  4^1   2^4,  4)$,\\
IrMOA$(56^4,  18,                  8^6  7^8  4^2   2^2,  4)$,\\
IrMOA$(56^4,  16,       28^1       8^7  7^7        2^1,  4)$,\\
IrMOA$(56^4,  17,            14^1  8^7  7^7        2^2,  4)$,\\
IrMOA$(56^4,  16,            14^1  8^7  7^7  4^1      ,  4)$,\\
IrMOA$(56^4,  18,                  8^7  7^8        2^3,  4)$,\\
IrMOA$(56^4,  17,                  8^7  7^8  4^1   2^1,  4)$,\\
IrMOA$(56^4,  16,                  8^8  7^8           ,  4)$.

\end{multicols}}